\documentclass[11pt]{article}
\usepackage[utf8]{inputenc}
\usepackage[a4paper, total={6in, 8in}]{geometry}
\usepackage{mathrsfs}
\usepackage{amssymb}
\usepackage{braket}
\usepackage{amsmath}
\usepackage{graphicx}
\usepackage{comment}

\usepackage[T1]{fontenc}
\usepackage{lmodern}
\usepackage{morefloats}
\usepackage[svgnames]{xcolor}
\usepackage{tikz}
\usetikzlibrary{calc,decorations.markings, shapes.misc, decorations.pathmorphing}
\tikzset{cross/.style={cross out, draw, 
		minimum size=2*(#1-\pgflinewidth), 
		inner sep=0pt, outer sep=0pt}} 
\usepackage{tqft}
\definecolor{lstbgcolor}{rgb}{0.9,0.9,0.9} 
\usepackage{listings}
\usepackage{fancyvrb}

\newcommand{\bea}{\begin{eqnarray}}
\newcommand{\eea}{\end{eqnarray}}
\newcommand{\bg}{\begin{gathered}}
\newcommand{\eg}{\end{gathered}}

\newcommand{\Perms}{ \hbox{Perms} }

\numberwithin{equation}{section}
\allowdisplaybreaks
\usepackage{subcaption}

\usepackage{amsthm}
\usepackage{rotating}
\usepackage{physics}
\makeatletter
\newcommand{\bigtimes}{%
	\DOTSB\mathop{\mathpalette\mattos@bigtimes\relax}\slimits@
}
\newcommand\mattos@bigtimes[2]{%
	\vcenter{\hbox{%
			\sbox\z@{$#1\sum$}%
			\resizebox{!}{0.9\dimexpr\ht\z@+\dp\z@}{\raisebox{\depth}{$\m@th#1\times$}}%
	}}%
	\vphantom{\sum}%
}
\makeatother

\DeclareMathOperator{\diag}{diag}

\newcommand{\Sym}{\mathrm{Sym}}

\definecolor{GREEN}{rgb}{0.0,0.70,0.24}
\definecolor{BLUE}{rgb}{0.0,0.24,0.70}

\def\ls[#1]{ {}_{#1}}
\def\us[#1]{ \mbox{\tiny{#1}}}

\usepackage{hyperref}

\begin{document}

\begin{flushright}
QMUL-PH-21-19
\end{flushright}

\bigskip

\begin{center}

{\Large \bf Permutation invariant Gaussian 2-matrix models  }

\bigskip


{George Barnes$^{a,}$\footnote{g.barnes@qmul.ac.uk}, Adrian Padellaro$^{a,}$\footnote{a.k.s.padellaro@qmul.ac.uk}, Sanjaye Ramgoolam$^{a , b,}$\footnote{s.ramgoolam@qmul.ac.uk}  }

\bigskip
$^{a}$ {\em  Centre for Research in String Theory}, {\em School of Physics and Astronomy}, \\
{\em Queen Mary University of London}, \\
{\em London E1 4NS, United Kingdom }\\

\medskip
$^{b}${\em  School of Physics and Mandelstam Institute for Theoretical Physics,} \\   
{\em University of Witwatersrand}, \\ 
{\em Wits, 2050, South Africa} \\
\medskip

\begin{abstract}

We construct the general  permutation invariant Gaussian 2-matrix model for matrices of arbitrary size $D$. The parameters of the model are given in terms of variables defined using the representation theory of the symmetric group $S_D$.  A correspondence is established between the 
permutation invariant polynomial functions of the matrix variables (the observables of the model) 
and directed colored graphs, which sheds light on stability properties in the large $D$ counting of these invariants. The refined counting of the graphs is given in terms of double cosets involving permutation groups defined by the 
local structure of the graphs. Linear and quadratic observables are transformed to  an $S_D$  representation theoretic basis and are used to define the convergent Gaussian measure.  The perturbative rules for the computation of expectation values of graph-basis observables of any degree are given in terms of the representation theoretic parameters. Explicit results for a number of observables of degree up to four are given along with a Sage programme that computes general expectation values.

\end{abstract}

\end{center}


\noindent 

\newpage 

\tableofcontents

\section{Introduction}

Gaussian matrix models have been studied as a universal structure which captures the statistics of a wide variety of complex systems. The initial applications by Wigner \cite{Wigner}  and Dyson \cite{Dyson}  focused on the statistics of  the energy levels of complex nuclei. Subsequent applications have included chaos, condensed matter physics, biological networks, feature-matrices in bio-statistics and financial correlations \cite{Beenakk,GMW,EW}. In quantum gravity,  matrix models played a key role in the early nineties  \cite{Klebanov1991,GinsMoore} and there has been a recent revival of interest \cite{SSS}. 

Traditional matrix theories have continuous symmetries and applications of these theories in data science tend to focus on eigenvalue distributions of the random matrices. The knowledge of these distributions  is typically equivalent to  that of 
the moments of  polynomial functions of matrices invariant under continuous symmetries. Novel constructions of matrix data from language have recently been found,  building on vector semantics \cite{Harris,Firth} and modelling grammatical composition with tensor contraction \cite{CSC,BBZ}.
In \cite{LMT} the study of the statistics of these matrices motivated the development of matrix theories where continuous symmetries are replaced by discrete permutation symmetries: for example, the constructions of vector semantics often use measures of similarity and dependence of word vectors which are permutation invariant but not invariant under continuous symmetries. Permutation invariant polynomial functions of matrices were proposed as the key observables for a linguistic matrix theory (LMT) programme. 
A 5-parameter permutation invariant Gaussian matrix model was solved and used to predict, with encouraging results,  cubic and quartic expectation values of matrix variables, using as input, linear and quadratic expectation values. In \cite{PIGMM} the general 13-parameter permutation invariant Gaussian matrix model  model was developed and solved to provide analytic formulae for low order correlators. This general 13-parameter model was found to be very good at predicting higher order expectation values \cite{GTMDS}. 

Permutation invariant observables are more general functions of matrices than functions invariant under continuous symmetries. Traditional arenas of application of random matrix theory (RMT) can be expected to be enriched and  extended by broadening the focus from eigenvalues to permutation invariant observables. The LMT programme initiated in \cite{LMT} also differs from traditional applications of RMT in focusing on Gaussianity tests using low order expectation values of observables, as opposed to eigenvalue distributions. This exploits  the fact that matrix integrals can be viewed as zero-dimensional quantum field theories (QFTs) and derives 
inspiration from   the use of correlators (generalizations of expectation values to higher dimensional QFTs), and their limits such as S-matrices,  in the applications of QFT to  particle physics and cosmology (see for example 
a survey of studies of Gaussianity in the cosmic microwave background radiation \cite{GaussCosmo}). 

The widespread applicability of Gaussian   matrix models with continuous symmetry 
in data sciences can be viewed as evidence for matrix Gaussianity. It is a natural question whether  this 
Gaussianity persists when the space of observables is enlarged to include those invariant under smaller symmetries such as permutation symmetries. The evidence from \cite{LMT,GTMDS} indicates that, in the context of type-driven compositional distributional semantics, this is the case. These applications of permutation invariant matrix  models  motivate the study of 2-matrix models. 
We develop results analogous to \cite{PIGMM} for this 2-matrix case. Here we are interested in $S_D$-invariant 
polynomial functions  of two $D \times D$ matrices $M$ and $N$: 
\begin{equation} \nonumber
f(M_{ij},N_{kl}) = f(M_{\sigma(i) \sigma(j)},N_{\sigma(k) \sigma(l)}), \quad \forall \sigma \in S_D
\end{equation}
 A weighted sum of all linear and quadratic observables will form the action of the most general $S_D$-invariant Gaussian model, $\mathcal{S}(M,N)$. Expectation values of the observables are given by
\begin{equation} \nonumber
\langle f(M,N) \rangle \equiv \frac{\int dM dN f(M,N) e^{-\mathcal{S}(M,N)}}{\int dM dN e^{-\mathcal{S}(M,N)}}.
\end{equation} 

The core results of this paper are  built around the description and solution of the most general permutation invariant Gaussian  2-matrix model. We derive a precise  correspondence between observables of the theory and colored 
directed graphs, which allows a graph-theoretic interpretation of the dependence on $D$ of the space of 2-matrix invariants. A group-theoretic description of the graphs allows for their enumeration and construction. The computation of expectation values of observables, derived using Wick contractions from the basic linear and quadratic expectation values of matrix observables,  produces sums involving products of projectors $F( i , j )$ for the hook representation  of dimension $( D-1) $ inside the natural $D$-dimensional  natural representation of $ S_D$. The different $F$-factors can have coincident indices, in a structure which is described in terms of an undirected graph, which we call an $F$-graph. We give a generating function for the analytic computation of these $F$-graph sums.  Building on the group-theoretic description of graphs and the F-graph generating function, we provide an algorithm, implemented in Sage, for the evaluation of  expectation values of general observables of the theory. 

The  paper is organised as follows: Section \ref{Section: Observables and multi-graphs} is concerned with the spectrum of observables; Section \ref{Section: The 2-matrix model} and \ref{Section: Feynman Graphs} with solving the model and in Section \ref{Section: Expectation values} we present some explicit expectation values as polynomials in $D$. Section \ref{Section: Observables and multi-graphs} extends the connection outlined in \cite{LMT,PIGMM}, between permutation invariant 1-matrix polynomials (observables) and directed  graphs to the case of   2-matrix observables. The extension involves directed {\it colored}  graphs. The correspondence was understood in the 1-matrix case by comparing the representation theoretic counting of invariants at large $D$, to the counting of directed uncolored graphs. In Section \ref{Section: Representation theory counting} we extend the representation theoretic counting to the 2-matrix case. The number of independent 2-matrix observables \eqref{Eqn: Counting invariants with k1, k2} stabilizes at large $D$, as previously observed for the 1-matrix case. These ideas motivate Section \ref{Multi-graph description}, where we develop a combinatoric formula \eqref{Countingformula1} for counting directed colored graphs. By exploiting connections between representation theory and combinatorics we prove that equation \eqref{Countingformula1} and \eqref{Eqn: Counting invariants with k1, k2} in fact count the same thing. This gives a geometrical interpretation of the large $D$ stability in equation \eqref{Eqn: Counting invariants with k1, k2} and explains the correspondence between observables and graphs at finite $D$. Having established the correspondence between graphs and observables, we turn to a systematic analysis of the counting and construction of directed colored graphs. In Section \ref{Double cosets and multi-graphs} we develop a group theoretical framework for constructing graphs, and by extension 2-matrix observables. The construction develops ideas from previous work \cite{TensorModelBranchedCovers, TensorModelsPermCentralAlgebras} on tensor model observables, Feynman graph counting \cite{deMelloKoch:2011uq}
and AdS/CFT \cite{deMelloKoch:2012ck}. Graphs are put into one-to-one correspondence with double cosets of permutation groups, which leads to refined counting formulas for observables and point toward efficient algorithms for their construction.

In Section \ref{Section: Desc of matrix model} we describe  the four linear and 37 quadratic $S_D$-invariant combinations of $M$ and $N$ which are used in defining the Gaussian model.  In Section \ref{Section: Solving 2 mat system} we use techniques from the representation theory of the symmetric group to define appropriate linear combinations of the matrix variables where the Gaussian model takes a simple form. We take 
advantage of the fact that the $D^2$ elements $M_{ij}$ (and similarly $N_{ ij}$)  transform as $V_D \otimes V_D$, where $V_D$ is the natural representation of the symmetric group. This is used to rewrite our action in block diagonal form in terms of variables corresponding to irreducible representations of $S_D$ leaving it amenable to methods of Gaussian integration. The 37 $S_D$-invariant quadratic terms are broken down into two copies of 11 terms of $M^2$ and $N^2$ type respectively, the diagonalization of these was performed in \cite{PIGMM}. The focus of this paper is the remaining 15 terms of mixed type, we find they can be parametrised by a $2 \times 2$ matrix $\Lambda_{V_0}^{\us[XY]}$, a $3 \times 3$ matrix $\Lambda_{V_H}^{\us[XY]}$ and two constants $\Lambda_{V_2}^{\us[XY]}$ and $\Lambda_{V_3}^{\us[XY]}$. Section \ref{Section: Linear and quadratic EVs} begins with the calculation of the first and second order expectation values of the representation theory variables. In terms of the representation theory variables the expectation values are simple.  Equation \eqref{Eqn: Two point function} gives the form of the quadratic expectation values of the original $M_{ij}$, $N_{ij}$ variables in terms of the expectation values of the representation theory variables.

Section \ref{Section: Feynman Graphs} extends the linear and quadratic results of the previous section to give methods of calculating expectation values of arbitrarily high degree observables. This is achieved 
using Wick's theorem to reduce these expectation values to sums of products of linear and quadratic expectation values. These basic  linear and quadratic expectation values themselves  have a graph-theoretic structure associated with the decomposition of the matrix variables into irreducible representations of $S_D$. Thus each 
term in the basic two point function 
 \eqref{Eqn: Two point function} is associated with a diagram involving solid and dotted lines in Section
  \ref{Subsec: Feynman decomp}. The dotted lines are associated with the trivial rep $V_0$ in $V_D$: they  
  can be removed and replaced with simple $D$-dependent coefficients. We are then left with an undirected graph with solid lines, each associated with a projector $F(i,j)$ for $V_H$ in $ V_D$. This is related to an important simplicity 
  in the Clebsch-Gordan decomposition of 
  \bea \nonumber
  V_D \otimes V_D = 2 V_0 \oplus 3 V_H \oplus V_2 \oplus V_3. 
  \eea
 The projectors for $V_2$ and $V_3$, which are irreducible representations of $S_D$ of dimension $ D ( D-3) /2 $
  and $ ( D  -1) ( D-2)/2$, can be constructed simply in terms of the projector $F(i,j)$.  
The outcome is that  every term in the computation of expectation values 
can be written as a weighted sum of the $S_D$-invariant tensor $F(i,j)$ and products thereof. Therefore, computing expectation values of observables ultimately reduces to computing general products of $F$ with some pattern of index coincidences and with all indices summed. We will introduce ``closed F-graphs'' to describe these sums of products and derive an algebraic expression \eqref{eq:kappagraphpolynomial} for their evaluation. Importantly, the computational complexity is independent of $D$. We give equation \eqref{eq:kappagraphpolynomial} a graph interpretation, in which it is manifest that the $D$-dependence has been traded for a dependence on the number of vertices in an undirected graph.

In Section \ref{Section: Expectation values} we use the results of the previous sections, most notably equations \eqref{Eqn: Two point function} and \eqref{Eqn: Two point function MM}, to calculate all quadratic expectation values comprised of one $M$ and one $N$. For the sake of brevity we show the full detail of the calculation in a single case $\sum_{i,j} \langle M_{ij} N_{ij} \rangle$, and content ourselves with listing the remaining 14 results. Using Wick's theorem we then calculate a selection of cubic and quartic expectation values. Again, in each case we list the details of only one calculation. In addition to the results listed in this section we provide user-friendly Sage code that can be used to calculate expectation values of any observable (\href{https://github.com/adrianpadellaro/PIG2MM}{\it Link to GitHub repository for this paper}). 

\section{Observables and graphs} \label{Section: Observables and multi-graphs}

In constructing the most general permutation invariant 2-matrix model the objects of central interest are the action and the observables of the theory. The observables are permutation invariant polynomials of the elements of matrices $M$ and $N$. The action is a linear combination of  observables of degree  one and two, with the coefficients in the linear combination being the parameters that define the model.  A basis for the space of observables is given by polynomials of homogeneous degree $m+n$ for which
\begin{align} \nonumber \label{Eqn: Observable definition}
f(M_{ij},N_{kl}) &= f(M_{\sigma(i)\sigma(j)},N_{\sigma(k)\sigma(l)}), \quad \forall \sigma \in S_D, \\
f(\lambda_1 M_{ij},\lambda_2 N_{kl}) &= \lambda_1^{m}\lambda_2^n f(M_{ij},N_{kl}).
\end{align}
In this section, we will see that this is a good basis for the problem of enumerating general observables. However, it will not be the most efficient basis for solving the partition function of the model. We delay the discuss of the action until Section \ref{Section: The 2-matrix model} and focus on observables for the remainder of this section.

In Section \ref{Section: Representation theory counting} we derive a formula \eqref{Eqn: Counting invariants with k1, k2} for the counting of observables as a function of the degrees $m,n$ of the polynomial and the matrix  size, $D$. Some low degree results of this formula are presented in Table \ref{tab: Table of invariant dimensions}. We observe that this counting stabilises at large enough  $D$: for fixed $m,n$ as $D$ is increased from low values and into the range $ D \ge 2m + 2n $, the number of invariants first increases and then stabilises, becoming independent of $D$.   This is similar to the simplification of  the counting of observables in $U(N)$ matrix models, in which the counting of degree $n$ observables  for $ N  \ge n$ is independent of $N$. For the case of  1-matrix $U(N)$ models, the counting gives the number of partitions of $n$ for $ n \le N$. The counting of observables and construction of physically useful bases of operators  \cite{CJR}   for $n >  N$ has implications for the stringy exclusion principle \cite{SEP} and CFT duals of giant gravitons \cite{susskind}  in AdS/CFT \cite{malda}. The review \cite{combfN} describes key results and  references for the $U(N)$  multi-matrix case.

This motivates Section \ref{Multi-graph description} where we find a precise connection  between graph counting and the counting of invariants which holds for  general $ D , m , n$. The counting of graphs with $ k$ vertices  here is done by labelling the vertices  with integers $ \{ 1, 2, \cdots , k \}$ and using ordered pairs of integers chosen from the labelling set to describe the directed edges. Considering the action of the symmetric group $S_k$ on the ordered pairs leads to the construction of formulae for the counting as a function of $ k $ and the numbers $m , n $ of edges of the two types. We find that $k$ can be identified with $D$ arising from the representation theory counting of Section \ref{Section: Representation theory counting}, thus establishing a direct connection between graphs and representation theoretic invariants in general: in the stable region of large $D$ as well as smaller $D$. This approach allows us to find the dimensions of invariant subspaces in general and also allows the explicit generation of combinatoric data for lists of graphs of length equal to the dimensions of invariant subspaces. The discussion is for the case of two matrices, but the method  generalises to higher numbers of matrices. This is a formalisation of an insight familiar from D-brane physics that matrices can be associated with strings between branes \cite{Polchinski,WittenBound} which has also been fruitfully connected to graph theory in  \cite{deMelloKoch:2012ck}. 

Having established the bijection between observables and graphs, we take advantage of the correspondence to count and construct observables, by counting and constructing graphs. To this end, Section \ref{Double cosets and multi-graphs} builds on the work in \cite{deMelloKoch:2011uq,deMelloKoch:2012ck} to provide a more refined double coset description of the graphs/observables. We extend the dictionary introduced for 1-matrix observables in \cite{LMT, PIGMM} to the 2-matrix case as follows: A directed blue edge going from a vertex $i$ to a vertex $j$, is associate with a factor of $M_{ij}$ and similarly for a green edge and $N_{ij}$. Each vertex is a sum over the vertex label. We illustrate with a few examples. There are 15 different directed graphs with one blue and one green edge on $k=1,2,3,4$ vertices. Equivalently, the space of degree $m=n=1$ observables has dimension 15 (assuming $D\geq 4$). The above dictionary gives the following basis
\begin{equation} \label{Eqn: Two matrix multi-graph diagrams} 
\begin{array}{ccccc}
\begin{tikzpicture}[baseline]
\begin{scope}[decoration={markings, mark=at position 0.45 with \arrow{latex}}]
\draw[draw=BLUE, postaction={decorate}] (0,0)node[circle, fill=black, inner sep=1pt, draw=black] {} to[in=180, out=180] (0,1) to[out=0, in=0] (0,0);
\draw[draw=GREEN, postaction={decorate}] (0,0)node[circle, fill=black, inner sep=1pt, draw=black] {} to[in=180, out=180] (0,.8) to[out=0, in=0] (0,0);
\end{scope}
\end{tikzpicture} &
\begin{tikzpicture}
\begin{scope}[decoration={markings, mark=at position 0.45 with \arrow{latex}}]
\draw[draw=BLUE, postaction={decorate}] (0,0)node[circle, fill=black, inner sep=1pt, draw=black] {} to[bend left] (1,0);
\draw[draw=GREEN, postaction={decorate}] (0,0) to[bend right] (1,0)node[circle, fill=black, inner sep=1pt, draw=black] {};
\end{scope}
\end{tikzpicture} &
\begin{tikzpicture}	
\begin{scope}[decoration={markings, mark=at position 0.45 with \arrow{latex}}]
\draw[draw=BLUE, postaction={decorate}] (1,0)node[circle, fill=black, inner sep=1pt, draw=black] {} to[bend right] (0,0);
\draw[draw=GREEN, postaction={decorate}] (0,0)node[circle, fill=black, inner sep=1pt, draw=black] {} to[bend right] (1,0);
\end{scope}
\end{tikzpicture} &
\begin{tikzpicture}	
\begin{scope}[decoration={markings, mark=at position 0.45 with \arrow{latex}}]
\draw[draw=BLUE, postaction={decorate}] (0,0)node[circle, fill=black, inner sep=1pt, draw=black] {} to[in=180, out=180] (0,1) to[out=0, in=0] (0,0);
\draw[draw=GREEN, postaction={decorate}] (.75,0)node[circle, fill=black, inner sep=1pt, draw=black] {} to[in=180, out=180] (.75,1) to[out=0, in=0] (.75,0);
\end{scope}
\end{tikzpicture}  &
\begin{tikzpicture}	
\begin{scope}[decoration={markings, mark=at position 0.45 with \arrow{latex}}]
\draw[draw=BLUE, postaction={decorate}] (0,0)node[circle, fill=black, inner sep=1pt, draw=black] {} to[bend left] (1,0)node[circle, fill=black, inner sep=1pt, draw=black] {};
\draw[draw=GREEN, postaction={decorate}] (0,0) to[in=180, out=180] (0,1) to[out=0, in=0] (0,0);
\end{scope}
\end{tikzpicture}  \\[0.1em]
\sum_i M_{ii} N_{ii}     &
\sum_{i,j} M_{ij} N_{ij} &
\sum_{i,j} M_{ji} N_{ij} &
\sum_{i,j} M_{ii} N_{jj} &
\sum_{i,j}  M_{ij} N_{ii}  \\[1em]
\begin{tikzpicture}	
\begin{scope}[decoration={markings, mark=at position 0.45 with \arrow{latex}}]
\draw[draw=GREEN, postaction={decorate}] (0,0)node[circle, fill=black, inner sep=1pt, draw=black] {} to[bend left] (1,0)node[circle, fill=black, inner sep=1pt, draw=black] {};
\draw[draw=BLUE, postaction={decorate}] (0,0) to[in=180, out=180] (0,1) to[out=0, in=0] (0,0);
\end{scope}
\end{tikzpicture} &
\begin{tikzpicture}	
\begin{scope}[decoration={markings, mark=at position 0.45 with \arrow{latex}}]
\draw[draw=GREEN, postaction={decorate}] (1,0)node[circle, fill=black, inner sep=1pt, draw=black] {} to[bend right] (0,0)node[circle, fill=black, inner sep=1pt, draw=black] {};
\draw[draw=BLUE, postaction={decorate}] (0,0) to[in=180, out=180] (0,1) to[out=0, in=0] (0,0);
\end{scope}
\end{tikzpicture} &
\begin{tikzpicture}	
\begin{scope}[decoration={markings, mark=at position 0.45 with \arrow{latex}}]
\draw[draw=GREEN, postaction={decorate}] (0,0)node[circle, fill=black, inner sep=1pt, draw=black] {} to[bend left] (1,0)node[circle, fill=black, inner sep=1pt, draw=black] {};
\draw[draw=BLUE, postaction={decorate}] (1,0) to[bend left] (2,0)node[circle, fill=black, inner sep=1pt, draw=black] {};
\end{scope}
\end{tikzpicture}  &
\begin{tikzpicture}	
\begin{scope}[decoration={markings, mark=at position 0.45 with \arrow{latex}}]
\draw[draw=BLUE, postaction={decorate}] (0,0)node[circle, fill=black, inner sep=1pt, draw=black] {} to[bend left] (1,0)node[circle, fill=black, inner sep=1pt, draw=black] {};
\draw[draw=GREEN, postaction={decorate}] (0,0) to[bend left] (2,0)node[circle, fill=black, inner sep=1pt, draw=black] {};
\end{scope}
\end{tikzpicture} &
\begin{tikzpicture}	
\begin{scope}[decoration={markings, mark=at position 0.45 with \arrow{latex}}]
\draw[draw=GREEN, postaction={decorate}] (0,0)node[circle, fill=black, inner sep=1pt, draw=black] {} to[bend left] (1,0)node[circle, fill=black, inner sep=1pt, draw=black] {};
\draw[draw=BLUE, postaction={decorate}] (2,0)node[circle, fill=black, inner sep=1pt, draw=black] {} to[bend right] (1,0)node[circle, fill=black, inner sep=1pt, draw=black] {};
\end{scope}
\end{tikzpicture} \\[0.1em]
\sum_{i,j} M_{ii} N_{ij} &
\sum_{i,j}  M_{ii}N_{ji} &
\sum_{i,j}  M_{jk} N_{ij} &
\sum_{i,j,k} M_{ij} N_{ik} &
\sum_{i,j,k} M_{kj} N_{ij}  \\[1em]
\begin{tikzpicture}	
\begin{scope}[decoration={markings, mark=at position 0.45 with \arrow{latex}}]
\draw[draw=BLUE, postaction={decorate}] (0,0)node[circle, fill=black, inner sep=1pt, draw=black] {} to[bend left] (1,0)node[circle, fill=black, inner sep=1pt, draw=black] {};
\draw[draw=GREEN, postaction={decorate}] (1,0) to[bend left] (2,0)node[circle, fill=black, inner sep=1pt, draw=black] {};
\end{scope}
\end{tikzpicture} &
\begin{tikzpicture}	
\begin{scope}[decoration={markings, mark=at position 0.45 with \arrow{latex}}]
\draw[draw=BLUE, postaction={decorate}] (1,0)node[circle, fill=black, inner sep=1pt, draw=black] {} to[bend right] (0,0)node[circle, fill=black, inner sep=1pt, draw=black] {};
\draw[draw=GREEN, postaction={decorate}] (0,0) to[in=180, out=180] (0,1) to[out=0, in=0] (0,0);
\end{scope}
\end{tikzpicture} &
\begin{tikzpicture}	
\begin{scope}[decoration={markings, mark=at position 0.45 with \arrow{latex}}]
\draw[draw=GREEN, postaction={decorate}] (0,0)node[circle, fill=black, inner sep=1pt, draw=black] {} to[bend left] (1,0)node[circle, fill=black, inner sep=1pt, draw=black] {};
\draw[draw=BLUE, postaction={decorate}] (1.5,0)node[circle, fill=black, inner sep=1pt, draw=black] {} to[in=180, out=180] (1.5,1) to[out=0, in=0] (1.5,0);
\end{scope}
\end{tikzpicture} &
\begin{tikzpicture}	
\begin{scope}[decoration={markings, mark=at position 0.45 with \arrow{latex}}]
\draw[draw=BLUE, postaction={decorate}] (0,0)node[circle, fill=black, inner sep=1pt, draw=black] {} to[bend left] (1,0)node[circle, fill=black, inner sep=1pt, draw=black] {};
\draw[draw=GREEN, postaction={decorate}] (1.5,0)node[circle, fill=black, inner sep=1pt, draw=black] {} to[in=180, out=180] (1.5,1) to[out=0, in=0] (1.5,0);
\end{scope}
\end{tikzpicture} &
\begin{tikzpicture}	
\begin{scope}[decoration={markings, mark=at position 0.45 with \arrow{latex}}]
\draw[draw=BLUE, postaction={decorate}] (0,0)node[circle, fill=black, inner sep=1pt, draw=black] {} to[bend left] (1,0)node[circle, fill=black, inner sep=1pt, draw=black] {};
\draw[draw=GREEN, postaction={decorate}] (1.5,0)node[circle, fill=black, inner sep=1pt, draw=black] {} to[bend left] (2.5,0)node[circle, fill=black, inner sep=1pt, draw=black] {};
\end{scope}
\end{tikzpicture} 
\\[0.1em]
\sum_{i,j,k} M_{ij} N_{jk} &
\sum_{i,j,k}  M_{ji}N_{ii} &
\sum_{i,j,k} M_{kk} N_{ij} &
\sum_{i,j,k}M_{ij} N_{kk} &
\sum_{i,j,k,l} M_{ij} N_{kl} 
\end{array}
\end{equation}
with all sums ranging from $1$ to $D$.

\subsection{Representation theory counting using characters of $V_D$ } \label{Section: Representation theory counting}\label{ObservsGraphs} 

There is a representation theoretic way of counting observables as a function of their degree $m,n$ and size $D$. Let $V_D$ be the natural representation of $S_D$. Each matrix forms a representation of $V_D \otimes V_D$ as each index transforms as $V_D$. Then a degree $m + n$ monomial with $m$ copies of $M$ and $n$ copies of $N$ transforms as
\bea \label{Eqn: Degree k monomial rep}
\text{Sym}^{m}(V_D \otimes V_D) \otimes \text{Sym}^{n}(V_D \otimes V_D)
\eea
due to the symmetry under permutations of the $M$'s or $N$'s. The counting of observables formed by $m$ copies of $M$ and $n$ copies of $N$ is equivalent to the multiplicity of the trivial (one-dimensional) representation $V_0$ in the irreducible decomposition of \eqref{Eqn: Degree k monomial rep}.

More generally the number of degree $m+n$ invariants is equivalent to the multiplicity of the trivial representation appearing in the irreducible decomposition of 
\bea \label{Eqn: Rep of two coloured monomial}
\bigoplus_{i=0}^{m+n} \text{Sym}^{m+n-i}(V_D \otimes V_D) \otimes \text{Sym}^{i}(V_D \otimes V_D).
\eea
In order to calculate this multiplicity we first average over all permutations, projecting the representation \eqref{Eqn: Rep of two coloured monomial} onto its trivial subspace and then take the trace to calculate the dimension of this subspace $\text{Dim}(D, m, n)$. For observables with $m$ copies of just one type of matrix $M$, this procedure was carried out in \cite{LMT}. It was found that the number of degree $m$ invariants was equal to
\begin{align} \label{Eqn: Non-coloured trivial counting} \nonumber 
\text{Dim}(D, m) &= \frac{1}{D!} \sum_{\sigma \in S_D} \text{tr}_{\text{Sym}^{m}(V_D \otimes V_D)} (\sigma) \\ \nonumber
&=  \frac{1}{D! m!} \sum_{\sigma \in S_D} \sum_{\tau \in S_m} \prod^{m}_{i=1} \big(\sum_{l|i} l C_l (\sigma) \big)^{2C_i (\tau)} \\
&= \frac{1}{D! m!} \sum_{p \vdash D} \sum_{q \vdash m}  \frac{D!}{\prod_{i=1}^D i^{p_i} p_i!} \frac{m!}{\prod_{i=1}^m i^{q_i} q_i!} \prod^{m}_{i=1} \big(\sum_{l|i} l p_l \big)^{2q_i}.
\end{align}
In the second line we have rewritten the trace in terms of the cycle structure of $\sigma$ and $\tau$, $C_l(\sigma)$ is the number of $l$-cycles in the permutation $\sigma$ and $l | i$ sums over the divisors of $i$. In the final line the sums over permutations have been reduced to sums over conjugacy classes labelled by partitions of $D$ and $m$, denoted by $p = \{p_1, p_2, \dots, p_D \}$, $q = \{ q_1, q_2, \dots, q_D \}$ obeying $\sum_i i p_i = D$, $\sum_i i q_i = m$ respectively. A more detailed derivation of this result is contained within the appendices of \cite{LMT}.

Following a similar procedure we generalise \eqref{Eqn: Non-coloured trivial counting} to find a counting formula for the dimension of the subspace of invariants with two different matrices
\begin{align} \label{Eqn: Counting invariants with k1, k2} \nonumber
\text{Dim}&(D, m, n) \\ \nonumber 
&= \frac{1}{D!} \sum_{\sigma \in S_D} \text{tr}_{\text{Sym}^{m}(V_D \otimes V_D) \otimes \text{Sym}^{n}(V_D \otimes V_D)} (\sigma) \\ \nonumber
&= \frac{1}{D!} \sum_{\sigma \in S_D} \text{tr}_{\text{Sym}^{m}(V_D \otimes V_D)} (\sigma) \cdot \text{tr}_{\text{Sym}^{n}(V_D \otimes V_D)} (\sigma) \\ \nonumber
&= \frac{1}{D! m! n!} \sum_{\sigma \in S_D} \sum_{\tau \in S_{m}} \sum_{\rho \in S_{n}} \prod^{m}_{i=1} \big(\sum_{l|i} l C_l (\sigma) \big)^{2C_i (\tau)} \prod^{n}_{j=1} \big(\sum_{l|j} l C_l (\sigma) \big)^{2C_j (\rho)} \\
&= \frac{1}{D! m! n!} \sum_{p \vdash D} \sum_{q \vdash m} \sum_{r \vdash n}  \frac{D!}{\prod_{i=1}^D i^{p_i} p_i!} \frac{m!}{\prod_{i=1}^m i^{q_i} q_i!} \frac{n!}{\prod_{i=1}^n i^{r_i} r_i!} \prod^{m}_{i=1} \big(\sum_{l|i} l p_l \big)^{2q_i} \prod^{n}_{j=1} \big(\sum_{l|j} l p_l \big)^{2r_j}.
\end{align}
With the use of Mathematica, we observe that the output of this formula for a given $m$ and $n$ remains constant for $D \geq 2m+2n$.

Plugging \eqref{Eqn: Counting invariants with k1, k2} into \eqref{Eqn: Rep of two coloured monomial} gives us the number of trivial representations in the decomposition of the representation of a general degree $m+n$ polynomial
\begin{equation}
\sum_{i=0}^{m+n} \text{Dim}(D,m+n-i,i).
\end{equation}
Running this through Mathematica we find that polynomials of degree $m+n =1, 2, 3, 4, 5, 6$ contain $4, 37, 338, 3598, 41200, 511444 $-dimensional invariant subspaces respectively in the stable $D$ limit. Table \ref{tab: Table of invariant dimensions} breaks down the number of invariants at each degree in powers of $M$ and $N$.
\begin{table}[h!]
	\begin{center}
		\caption{Number of invariants contained within a monomial of the form $M^m N^n$}
		\label{tab: Table of invariant dimensions}
		\begin{tabular}{c c c c}
			\textbf{Degree of monomial} & \textbf{\# of Ms} & \textbf{\# of N's} & \textbf{\# of invariants}\\
			$m+n$ & $m$ & $n$ & $\text{Dim}(D \geq 2m + 2n,m,n)$\\
			\hline
			1 & 1 & 0 & 2 \\
			2 & 2 & 0 & 11 \\
			2 & 1 & 1 & 15 \\
			3 & 3 & 0 & 52 \\
			3 & 2 & 1 & 117 \\
			4 & 4 & 0 & 296 \\
			4 & 3 & 1 & 877 \\
			4 & 2 & 2 & 1252 \\
			5 & 5 & 0 & 1724 \\
			5 & 4 & 1 & 6719 \\
			5 & 3 & 2 & 12157 \\
			6 & 6 & 0 & 11060 \\
			6 & 5 & 1 & 52505 \\
			6 & 4 & 2 & 117121 \\
			6 & 3 & 3 & 150072
		\end{tabular}
	\end{center}
\end{table}
We can read off that there are four invariants at degree one given by the two terms with $m=1$ and $n=1$, 
\begin{equation}
\sum_{i} M_{ii}, \quad \sum_{i} N_{ii}, \quad \sum_{i, j} M_{ij}, \quad \sum_{i, j} N_{ij}.
\end{equation}
The 37 invariants at degree two are given by the 15 mixed terms listed in \eqref{Eqn: Two matrix multi-graph diagrams} along with 11 invariants formed from just $M$ and another 11 formed from just $N$.

\subsection{Proof of equivalence between  observables and  directed colored graphs  } \label{Multi-graph description}

At $D\geq 2m+2n$, along the lines of \cite{LMT},  we expect that we can 
 enumerate observables by enumerating two-colored unlabeled directed graphs.
Here we will   count  graphs  having $k$ vertices by starting with graphs having labelled vertices, 
 with labels chosen from the set  $ \{ 1, 2, \cdots , k \}$. We will identify the unlabelled graphs as 
  orbits of $ S_k$ acting on lists of ordered pairs 
\begin{equation}
\begin{aligned}
&\left[(a_1^-, a_1^+), \dots, (a_m^-, a_m^+)\right], \left[(b_1^-, b_1^+), \dots, (b_n^-, b_n^+)\right],\label{Eqn: Graphs as list of tuples}
\end{aligned}
\end{equation}
for $m$ edges of color $M$ and $n$ edges of color $N$. A pair  $(a^-_i,a^+_i)$ corresponds to an $M$-colored edge from a vertex labeled $a^-_i$ to the vertex labeled $a_i^+$. Similarly $(b^-_i,b^+_i)$ corresponds to an $N$-colored edges. For a graph with $k$ vertices, the labels $a_i^\pm, b_i^\pm,$ take values from the set $\{1,\dots,k\}$.


Using the data in \eqref{Eqn: Graphs as list of tuples} we consider states in a Fock space
\bea\label{osc}  
&&\hspace{-0.75cm}| a_1^- ,  a_1^+ , \cdots , a_m^- , a_m^+ ; b_1^- , b_1^+ , \cdots , b_n^- , b_n^+ \rangle = 
A^{ \dagger}_{ a_1^- , a_1^+ } A^{ \dagger}_{a_2^- , a_2^+} \cdots A^{ \dagger}_{ a_m^- , a_m^+ } 
B^{ \dagger}_{ b_1^- , b_1^+} B^{ \dagger}_{ b_2^- , b_2^+ } \cdots B^{ \dagger}_{ b_n^- , b_n^+ } 
| 0 \rangle  \cr
&&
\eea
generated by commuting oscillators $ A^{ \dagger}_{ a^- , a^+ } , B^{ \dagger}_{ b^- , b^+} $, where $a^{ \pm} , b^{ \pm} $ take values in $ \{ 1, 2, \cdots ,  k \} $.   The commutativity of the oscillators reflects the fact that, as a way to describe the edges of a vertex-labelled graph,   the ordering of the pairs in \eqref{Eqn: Graphs as list of tuples} is immaterial. 
These oscillator states in a Fock space, in the sector with $ m$ oscillators of $A^{\dagger}$ type and $n$ oscillators of $B^{\dagger}$ type, form  a basis set for the vector space  
\begin{equation}
W_{m,n,k}=\Sym^m(V_k \otimes V_k) \otimes \Sym^n(V_k \otimes V_k).
\end{equation}
For the sake of brevity we will refer to this space at fixed $(m,n)$ as a Fock space. 
The basis states in \eqref{osc} form orbits under an action  of permutations $ \sigma \in S_k$ generated by 
\bea 
&& a_{ i }^{ \pm} \rightarrow \sigma ( a_i^{ \pm} ) \cr 
&& b_i^{ \pm} \rightarrow \sigma ( b_i^{ \pm} ) 
\eea 
using 
\bea 
&& | a_1^- ,  a_1^+ , \cdots , a_m^- , a_m^+ ; b_1^- , b_1^+ , \cdots , b_n^- , b_n^+ \rangle \cr 
&& \rightarrow  |\sigma (  a_1^- )  , \sigma (  a_1^+ )  , \cdots , \sigma ( a_m^- )  , \sigma ( a_m^+)  ;\sigma (  b_1^-)  , \sigma ( b_1^+ ) , \cdots , \sigma ( b_n^- )  , \sigma ( b_n^+ )  \rangle. 
\eea
The unlabelled graphs are in one-to-one correspondence with these orbits. The counting of these orbits can be done using 
Burnside's Lemma as 
\bea\label{CountBurns} 
&& \hbox{ Number of unlabelled graphs with $m$ $M$-colored edges and $n$ $N$-colored edges }  \cr 
&&  = \frac{1}{k!} \times \sum_{ \sigma \in S_k} 
\Bigg\{   \hbox{ Number of distinct Fock space basis-states of the form \eqref{osc} 
	such that } \cr 
&&  | \sigma ( a_1^- )  ,\sigma (  a_1^+ )  , \cdots , \sigma ( a_m^- )  ,\sigma (  a_m^+)  ; \sigma ( b_1^- )  , \sigma ( b_1^+ ) , \cdots , \sigma ( b_n^-)  ,\sigma (  b_n^+ ) \rangle \cr 
&& \hspace*{4cm}  =   |  a_1^- , a_1^+ , \cdots , a_m^- , a_m^+ ; b_1^- , b_1^+ , \cdots , b_n^- , b_n^+ \rangle   
\Bigg\} 
\eea
The action of permutations on the Fock space basis states extends by linearity to general vectors in the Fock space 
$ W_{ m , n , k }$. The Burnside Lemma calculation for group actions on sets which we are using above can also be recognised, in vector space language, as the computation of the trace of a projector
\bea 
P_{ V_0}^{ (S_k)}  = { 1 \over k! } \sum_{ \sigma \in S_k } \sigma 
\eea
in $W_{ m , n , k}$ for the trivial representation of $S_k$. We denote  the linear operator for $ \sigma $ in $ W_{ m , n, k}$ as  
\bea 
D^{ W_{ m , n , k } } ( \sigma )  
\eea
and recognise  the number of unlabelled graphs in \eqref{CountBurns} as 
\bea
\tr_{\phantom{}_{W_{ m , n , k }} } D^{ W_{ m , n , k } } ( P_{ V_0}^{ (S_k)}  ) .
\eea
This is precisely the formula for the counting of matrix invariants we arrived at in Section \ref{ObservsGraphs}, with the identification of $k$ the number of vertices, with $D$ the dimension of the natural representation of $S_D$. 
We conclude that 
\begin{align}\label{Countingformula1} 
\tr_{ W_{ m , n , k } } D^{ W_{ m , n , k } } ( &P_{ V_0}^{ (S_k)}  ) = \nonumber \\
\frac{1}{k! m! n!} \sum_{\substack{p \vdash k \\ q \vdash m \\ r \vdash m}} & \frac{k!}{\prod_{i=1}^k i^{p_i} p_i!} \frac{m!}{\prod_{i=1}^m i^{q_i} q_i!} \frac{n!}{\prod_{i=1}^n i^{r_i} r_i!} \prod^{m}_{i=1} \big(\sum_{l|i} l p_l \big)^{2q_i} \prod^{n}_{j=1} \big(\sum_{l|j} l p_l \big)^{2r_j}
\end{align}
which is just equation \eqref{Eqn: Counting invariants with k1, k2} with $D = k$. Very importantly, this equality between graph counting with $k$ vertices and invariants in the natural representation of $S_D$  which we have just derived, holds for any $D=k$, irrespective of the relative magnitudes of $D=k$ and $ m , n $. This is very  useful in giving a geometrical understanding of a stability property of \eqref{Countingformula1}. It is easy to verify by direct computation in examples that, as $k$ is increased for fixed $ m , n $, the counting increases at first and then stops increasing at $ k = 2m + 2n $. This is not immediately evident from directly looking at the expression.   The stability property  is explained by the fact that when $ k > 2m + 2n $, there are at least $ k - 2m - 2n $ nodes in the graph which have no incident (incoming or outgoing)  edges. Adding further unlabelled nodes  evidently does not change the counting. The connection between the counting of 1-matrix invariants and graphs was observed for large $D$  in \cite{LMT}. The fact that the connection continues to hold for small $D=k$ and that graph counting  informs the departures from the stable range in the counting of $S_k$ invariants in $W_{ m , n , k }$ is a new observation here, valid for the 2-matrix as well as the special case of  1-matrix invariants.

\subsection{Counting and construction using double cosets} \label{Double cosets and multi-graphs}
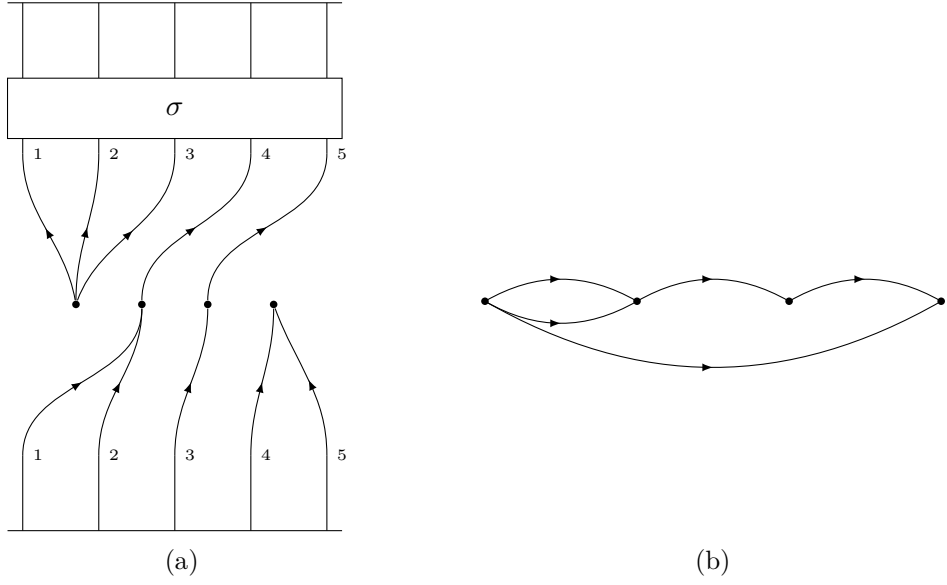
\begin{figure}[h!]
	\centering
	\subcaptionbox{\label{fig: One Colour Double Coset Graph _a}}[0.4\textwidth]
	{\begin{tikzpicture}[scale=2,baseline]
		\def \k {3}
		\def \m {4}
		\def \sep {0.5}
		\def \voffset {0.35}
		\pgfmathparse{(\sep*(\m)-2*\voffset)/\k};
		\pgfmathsetmacro{\vsep}{\pgfmathresult};
		\pgfmathint{\k-1};
		\pgfmathsetmacro{\kk}{\pgfmathresult};
		\foreach \v in {0,...,\k}
		{
			\pgfmathparse{\v*\vsep+\voffset}
			\node[circle, fill, inner sep=1pt](v\v) at (\pgfmathresult,0) {};
		}
		\foreach \eOutm in {0,...,\m}
		{
			\pgfmathint{\eOutm+1};
			\pgfmathsetmacro{\seOutm}{\pgfmathresult};
			\pgfmathparse{\eOutm*\sep};
			\coordinate (eom\seOutm) at (\pgfmathresult,2);
		}
		\foreach \eOutm in {0,...,\m}
		{
			\pgfmathint{\eOutm+1};
			\pgfmathsetmacro{\seOutm}{\pgfmathresult};
			\pgfmathparse{\eOutm*\sep};
			\coordinate (eomm\seOutm) at (\pgfmathresult,1);
			\node[right, node distance = 0pt and 0pt] at (eomm\seOutm) {\tiny \seOutm};
			\draw[] (eom\seOutm) -- (eomm\seOutm);
		}
		\foreach \eInm in {0,...,\m}
		{
			\pgfmathint{\eInm+1};
			\pgfmathsetmacro{\seInm}{\pgfmathresult};
			\pgfmathparse{\eInm*\sep};
			\coordinate (eim\seInm) at (\pgfmathresult,-1.5);
		}
		\foreach \eInm in {0,...,\m}
		{
			\pgfmathint{\eInm+1};
			\pgfmathsetmacro{\seInm}{\pgfmathresult};
			\pgfmathparse{\eInm*\sep};
			\coordinate (eimm\seInm) at (\pgfmathresult,-1);
			\node[right, node distance = 0pt and 0pt] at (eimm\seInm) {\tiny \seInm};
			\draw[] (eim\seInm) -- (eimm\seInm);
		}
		\begin{scope}[decoration={markings, mark=at position 0.5 with \arrow{latex}}]
		\draw[postaction={decorate}] (v0) to[out=100,in=-90] (eomm1);
		\draw[postaction={decorate}] (v0) to[out=90,in=-90] (eomm2);
		\draw[postaction={decorate}] (v0) to[out=70,in=-90] (eomm3);
		\draw[postaction={decorate}] (v1) to[out=90,in=-90] (eomm4);
		\draw[postaction={decorate}] (v2) to[out=90,in=-90] (eomm5);
		\draw[postaction={decorate}] (eimm1) to[out=90,in=-90] (v1);
		\draw[postaction={decorate}] (eimm2) to[out=90,in=-90] (v1);
		\draw[postaction={decorate}] (eimm3) to[out=90,in=-90] (v2);
		\draw[postaction={decorate}] (eimm4) to[out=90,in=-90] (v3);
		\draw[postaction={decorate}] (eimm5) to[out=90,in=-70] (v3);
		\end{scope}
		\draw[fill=white] ($(eomm1)+(-0.1,0.5)$) rectangle node{$\sigma$} ($(eomm5)+(0.1,0.1)$);
		\draw ($(eom1)-(0.1,0)$) -- ($(eom5)+(0.1,0)$);
		\draw ($(eim1)-(0.1,0)$) -- ($(eim5)+(0.1,0)$);
		\end{tikzpicture}}
	\subcaptionbox{\label{fig: One Colour Double Coset Graph _b}}[0.5\textwidth]
	{\begin{tikzpicture}[scale=2,baseline]
		\begin{scope}[decoration={markings, mark=at position 0.5 with \arrow{latex}}]
		\node[circle, fill, inner sep=1pt] at (0,0) {};
		\node[circle, fill, inner sep=1pt] at (1,0) {};	
		\node[circle, fill, inner sep=1pt] at (2,0) {};	
		\node[circle, fill, inner sep=1pt] at (3,0) {};
		\draw[postaction={decorate}] (0,0) to[bend left] (1,0);
		\draw[postaction={decorate}] (0,0) to[bend right] (1,0);
		\draw[postaction={decorate}] (0,0) to[bend right] (3,0);
		\draw[postaction={decorate}] (1,0) to[bend left] (2,0);
		\draw[postaction={decorate}] (2,0) to[bend left] (3,0);
		\end{scope}
		\node[circle, fill, inner sep=1pt,opacity=0] at (0,-1.5) {};			\end{tikzpicture}}
	\caption{Directed graphs of a fixed type (determined by a vector partition) correspond to a permutation $\sigma \in S_m$, where $m$ is the number of edges. (a) illustrates the correspondence with an example where the graph type is a vector partition $(5,5) = (3,0)+(1,2)+(1,1)+(0,2)$. (b) is the graph constructed from this vector partition with the permutation $\sigma = (3,4)$.}
	\label{fig: One Colour Double Coset Graph}
\end{figure}

By generalizing the double coset description of directed graphs introduced in \cite{deMelloKoch:2011uq, deMelloKoch:2012ck} we can enumerate invariants using appropriate equivalence classes of permutations, which define  double cosets. As a way to introduce the construction, we consider one-colored graphs. For a graph with $m$ edges and $k$ vertices, we have a vector partition
\begin{equation}
\label{eq: One Colour Graph Vector Partition}
(m,m) = (m^+_1, m^-_1)+ \dots +(m^+_k, m^-_k)\equiv (\vec{m}^+,\vec{m}^-), \quad 0 \leq m^\pm_i \leq m 
\end{equation}
where $m^+_i$($m^-_i$) describes the number of outgoing (incoming) edges at vertex $i$. For example, the graph in Figure \ref{fig: One Colour Double Coset Graph} is associated with the vector partition $(\vec{m}^+,\vec{m}^-)=(3,0)+(1,2)+(1,1)+(0,2)$. Edges are connected to vertices using the following rules.
\begin{itemize}
	\item Pick an order for the incoming edge labels. For example, we will use $\{1,2,\dots\}$ from left to right as seen in Figure \ref{fig: One Colour Double Coset Graph}.
	\item Assign every label to exactly one outgoing edge.
	\item Apply a permutation $ \sigma $ to the outgoing edges, which corresponds in Figure  
	\ref{fig: One Colour Double Coset Graph _a} to a re-ordering of the edges coming into the $ \sigma$-box from below before they emerge at the top. 
	\item Identify the end-points on the  top line which have incoming lines to the points on the bottom line directly below them, with outgoing lines. 
\end{itemize}
In Figure \ref{fig: One Colour Double Coset Graph _a} we take the incoming (and outgoing) edges as initially labeled $1,2,\dots, 5$ (from left to right). For $\sigma = (3,4)$ the third edge on the first vertex is swapped with the edge on the second vertex and we arrive at the graph in Figure \ref{fig: One Colour Double Coset Graph _b}.
By scanning over all $\sigma \in S_m$ we can construct all in-equivalent graphs of the type determined by the vector partition in equation \eqref{eq: One Colour Graph Vector Partition}.

The set of permutations $ \sigma \in S_m $ can be partitioned into equivalence classes: the permutations within an equivalence class lead to different labelings of the same graph. 
We now introduce some notation which will allow an efficient description of these equivalence classes. The labels for the outgoing  edges in  Figure \ref{fig: One Colour Double Coset Graph _a}
define lists 
\bea 
M_1^+  & = &  [ 1, 2, \cdots , m_1^+ ] \cr 
M_2^+  & = &  [ m_1^+ + 1, m_1^+ + 2, \cdots , m_1^+ + m_2^+ ] \cr  
& \vdots &  \cr 
M_i^+  & = &  [ m_1^+ + m_2^+ + \cdots + m_{i-1}^+ + 1 ,   m_1^+ + m_2^+ + \cdots + m_{i-1}^+ +2 , \cdots , m_1^+ + m_2^+ + \cdots + m_{ i-1}^+ +  m_{ i}^+  ] \cr 
& \vdots & \cr 
M_k^+  & = & [ m_1^+ + m_2^+ + \cdots + m_{k-1}^+ + 1 ,   m_1^+ + m_2^+ + \cdots + m_{k-1}^+ +2 , \cdots , m_1^+ + m_2^+ + \cdots + m_{ k-1}^+ +  m_{ k}^+  ] \cr 
&& 
\eea
The concatenation  of these lists is the set of numbers $ [ 1 ,  \cdots ,  m ] $.
\bea\label{concatM}  
[ M_1^+  , M_2^+ , \cdots , M_k^+ ] = [ 1, 2, \cdots , m ] 
\eea 
The permutation $ \sigma \in S_m $ can be viewed as re-arrangements of this list. 
The permutations within the sublists $ M_1^+ , M_2^+ , \cdots , M_{ k}^+ $ define
a subgroup isomorphic to 
\bea 
&& \hspace{-1.75cm} S_{ m_1^+ }  \times S_{ m_2^+ } \times \cdots \times S_{ m_k^+} \cr 
&& \hspace{-1cm} = \Perms (  [ 1, 2 , \cdots , m_1^+ ] ) \times \Perms ( [ 1, 2, \cdots , m_2^+] ) 
\times \cdots \times \Perms  ( [ 1 , 2 , \cdots , m_{ k}^+ ] )
\eea
There are  injective homomorphisms which we will denote $ \gamma_{  m_i^+ }  $ 
which map $ \nu^+_i \in S_{ m_i^+} $ to permutations in $ S_m$
\bea
\gamma^+_i ( \nu_i^+ )  : 
[ M_1^+ , M_2^+ , \cdots , M_i^+ , \cdots , M_k^+ ] \rightarrow [ M_1^+ , M_2^+ , \cdots , \nu_i^+ ( M_i^+ ) , \cdots , M_k^+ ] 
\eea
where
\bea 
\nu_i^+ ( M_i^+ ) = [ \sum_{ j =1}^{ i-1}  m_j^+ + \nu_i^+ ( 1 ) , 
\sum_{ j =1}^{ i-1}  m_j^+ + \nu_i^+ ( 2 ) , \cdots ,  \sum_{ j =1}^{ i-1}  m_j^+  + \nu_i^+ ( m_i^+ ) ]
\eea
There is a similar homomorphism from 
\bea 
\gamma^-  : S_{ m_1^-} \times S_{ m_2^-} \times \cdots \times S_{ m_k^- } \rightarrow S_m 
\eea
There is also a subgroup of $ S_{ k} = \Perms  ( [ 1, 2, \cdots , k ] ) $ which permutes 
vertices having the same number of incoming and outgoing vertices. These permutations $ \mu \in S_k$  are mapped to permutations in $ S_m $ as rearrangements of the concatenated lists 
\bea 
\rho^+(\mu) : [ M_1^+ , M_2^+ , \cdots , M_k^+ ] \rightarrow [ M_{ \mu ( 1) }^+ , M_{ \mu (2) }^+ , \cdots , 
M_{ \mu (k) }^+ ] 
\eea
and
\bea 
\rho^-(\mu) : [ M_1^- , M_2^- , \cdots , M_k^- ] \rightarrow [ M_{ \mu ( 1) }^- , M_{ \mu (2) }^- , \cdots , 
M_{ \mu (k) }^- ] 
\eea

\begin{figure}[h!]
	\centering	
	\caption{For any permutation $\sigma$ in $S_5$, the two diagrams correspond to the same graph for any $\nu_1^+ \in S_3, \nu_2^- \in S_2, \nu_4^- \in S_2$.}
	\raisebox{-0.5\height}{\begin{tikzpicture}[scale=2]
		\def \k {3}
		\def \m {4}
		\def \sep {0.5}
		\def \voffset {0.35}
		\pgfmathparse{(\sep*(\m)-2*\voffset)/\k};
		\pgfmathsetmacro{\vsep}{\pgfmathresult};
		\pgfmathint{\k-1};
		\pgfmathsetmacro{\kk}{\pgfmathresult};
		\foreach \v in {0,...,\k}
		{
			\pgfmathparse{\v*\vsep+\voffset}
			\node[circle, fill, inner sep=1pt](v\v) at (\pgfmathresult,0) {};
		}
		\foreach \eOutm in {0,...,\m}
		{
			\pgfmathint{\eOutm+1};
			\pgfmathsetmacro{\seOutm}{\pgfmathresult};
			\pgfmathparse{\eOutm*\sep};
			\coordinate (eom\seOutm) at (\pgfmathresult,2.2);
		}
		\foreach \eOutm in {0,...,\m}
		{
			\pgfmathint{\eOutm+1};
			\pgfmathsetmacro{\seOutm}{\pgfmathresult};
			\pgfmathparse{\eOutm*\sep};
			\coordinate (eomm\seOutm) at (\pgfmathresult,1);
			\node[right, node distance = 0pt and 0pt] at (eomm\seOutm) {\tiny \seOutm};
			\draw[] (eom\seOutm) -- (eomm\seOutm);
		}
		\foreach \eInm in {0,...,\m}
		{
			\pgfmathint{\eInm+1};
			\pgfmathsetmacro{\seInm}{\pgfmathresult};
			\pgfmathparse{\eInm*\sep};
			\coordinate (eim\seInm) at (\pgfmathresult,-2);
		}
		\foreach \eInm in {0,...,\m}
		{
			\pgfmathint{\eInm+1};
			\pgfmathsetmacro{\seInm}{\pgfmathresult};
			\pgfmathparse{\eInm*\sep};
			\coordinate (eimm\seInm) at (\pgfmathresult,-1);
			\node[right, node distance = 0pt and 0pt] at (eimm\seInm) {\tiny \seInm};
			\draw[] (eim\seInm) -- (eimm\seInm);
		}
		\begin{scope}[decoration={markings, mark=at position 0.5 with \arrow{latex}}]
		\draw[postaction={decorate}] (v0) to[out=100,in=-90] (eomm1);
		\draw[postaction={decorate}] (v0) to[out=90,in=-90] (eomm2);
		\draw[postaction={decorate}] (v0) to[out=70,in=-90] (eomm3);
		\draw[postaction={decorate}] (v1) to[out=90,in=-90] (eomm4);
		\draw[postaction={decorate}] (v2) to[out=90,in=-90] (eomm5);
		\draw[postaction={decorate}] (eimm1) to[out=90,in=-90] (v1);
		\draw[postaction={decorate}] (eimm2) to[out=90,in=-90] (v1);
		\draw[postaction={decorate}] (eimm3) to[out=90,in=-90] (v2);
		\draw[postaction={decorate}] (eimm4) to[out=90,in=-90] (v3);
		\draw[postaction={decorate}] (eimm5) to[out=90,in=-70] (v3);
		\end{scope}
		\draw[fill=white] ($(eomm1)+(-0.1,1)$) rectangle node{$\sigma$} ($(eomm5)+(0.1,0.6)$);
		\draw[fill=white] ($(eomm1)+(-0.1,0.5)$) rectangle node{$\nu_1^+$} ($(eomm3)+(0.1,0.1)$);
		\draw[fill=white] ($(eimm1)-(0.1,0.5)$) rectangle node{$(\nu_2^-)^{-1}$} ($(eimm2)-(-0.1,0.1)$);
		\draw[fill=white] ($(eimm4)-(0.1,0.5)$) rectangle node{$(\nu_4^-)^{-1}$} ($(eimm5)-(-0.1,0.1)$);
		\draw ($(eom1)-(0.1,0)$) -- ($(eom5)+(0.1,0)$);
		\draw ($(eim1)-(0.1,0)$) -- ($(eim5)+(0.1,0)$);
		\end{tikzpicture}} $\sim$
	\raisebox{-0.5\height}{\begin{tikzpicture}[scale=2]
		\def \k {3}
		\def \m {4}
		\def \sep {0.5}
		\def \voffset {0.35}
		\pgfmathparse{(\sep*(\m)-2*\voffset)/\k};
		\pgfmathsetmacro{\vsep}{\pgfmathresult};
		\pgfmathint{\k-1};
		\pgfmathsetmacro{\kk}{\pgfmathresult};
		\foreach \v in {0,...,\k}
		{
			\pgfmathparse{\v*\vsep+\voffset}
			\node[circle, fill, inner sep=1pt](v\v) at (\pgfmathresult,0) {};
		}
		\foreach \eOutm in {0,...,\m}
		{
			\pgfmathint{\eOutm+1};
			\pgfmathsetmacro{\seOutm}{\pgfmathresult};
			\pgfmathparse{\eOutm*\sep};
			\coordinate (eom\seOutm) at (\pgfmathresult,2.2);
		}
		\foreach \eOutm in {0,...,\m}
		{
			\pgfmathint{\eOutm+1};
			\pgfmathsetmacro{\seOutm}{\pgfmathresult};
			\pgfmathparse{\eOutm*\sep};
			\coordinate (eomm\seOutm) at (\pgfmathresult,1);
			\node[right, node distance = 0pt and 0pt] at (eomm\seOutm) {\tiny \seOutm};
			\draw[] (eom\seOutm) -- (eomm\seOutm);
		}
		\foreach \eInm in {0,...,\m}
		{
			\pgfmathint{\eInm+1};
			\pgfmathsetmacro{\seInm}{\pgfmathresult};
			\pgfmathparse{\eInm*\sep};
			\coordinate (eim\seInm) at (\pgfmathresult,-2);
		}
		\foreach \eInm in {0,...,\m}
		{
			\pgfmathint{\eInm+1};
			\pgfmathsetmacro{\seInm}{\pgfmathresult};
			\pgfmathparse{\eInm*\sep};
			\coordinate (eimm\seInm) at (\pgfmathresult,-1);
			\node[right, node distance = 0pt and 0pt] at (eimm\seInm) {\tiny \seInm};
			\draw[] (eim\seInm) -- (eimm\seInm);
		}
		\begin{scope}[decoration={markings, mark=at position 0.5 with \arrow{latex}}]
		\draw[postaction={decorate}] (v0) to[out=100,in=-90] (eomm1);
		\draw[postaction={decorate}] (v0) to[out=90,in=-90] (eomm2);
		\draw[postaction={decorate}] (v0) to[out=70,in=-90] (eomm3);
		\draw[postaction={decorate}] (v1) to[out=90,in=-90] (eomm4);
		\draw[postaction={decorate}] (v2) to[out=90,in=-90] (eomm5);
		\draw[postaction={decorate}] (eimm1) to[out=90,in=-90] (v1);
		\draw[postaction={decorate}] (eimm2) to[out=90,in=-90] (v1);
		\draw[postaction={decorate}] (eimm3) to[out=90,in=-90] (v2);
		\draw[postaction={decorate}] (eimm4) to[out=90,in=-90] (v3);
		\draw[postaction={decorate}] (eimm5) to[out=90,in=-70] (v3);
		\end{scope}
		\draw[fill=white] ($(eomm1)+(-0.1,1)$) rectangle node{$\sigma$} ($(eomm5)+(0.1,0.6)$);
		\draw ($(eom1)-(0.1,0)$) -- ($(eom5)+(0.1,0)$);
		\draw ($(eim1)-(0.1,0)$) -- ($(eim5)+(0.1,0)$);
		\end{tikzpicture}}
	\label{fig: One Color One Sigma Equivalence Relation Without Vertex Symmetry}
\end{figure}
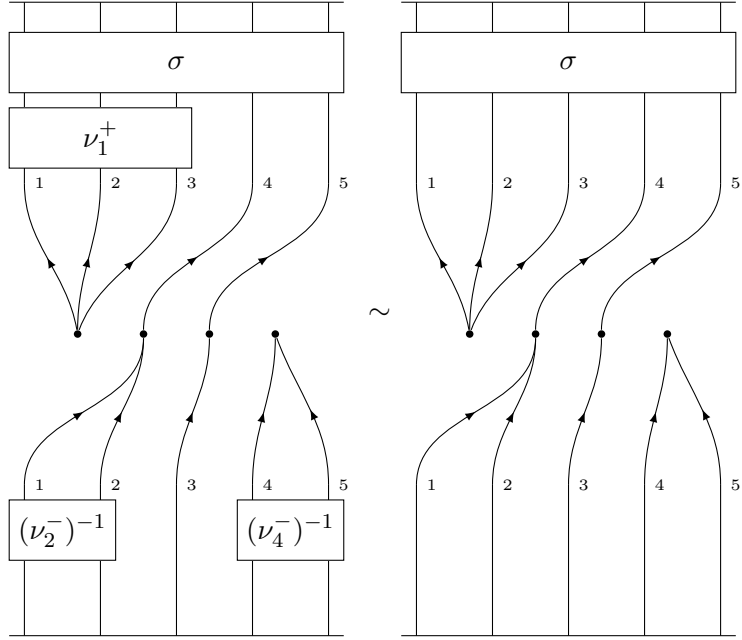

For example, consider the graph in Figure \ref{fig: One Color One Sigma Equivalence Relation Without Vertex Symmetry} where the first vertex has three outgoing edges labeled $1,2,3$. Here two permutations $\sigma \in S_m$, which are related by a permutation $\nu^+_1 $ in $S_3$ permuting the list $ [ 1,2,3]$, lead to equivalent graphs. From Figure \ref{fig: One Color One Sigma Equivalence Relation Without Vertex Symmetry} we see that this equivalence comes from left multiplication $\sigma \sim \gamma^+_1(\nu^+_1) \sigma $.
Similarly, for incoming edges we have equivalence under right multiplication $\sigma \sim \sigma \gamma^-_2((\nu^-_2)^{-1})\gamma^-_4((\nu^-_4)^{-1})$. In general, we have combined left and right equivalence
\begin{equation}
\sigma \sim \sigma' \qq{iff} \quad \exists \nu^+ \in S_{\vec{m}^+}, \nu^- \in S_{\vec{m}^-}, \quad \sigma = \gamma^+(\nu^+)  \sigma' \gamma^-((\nu^-)^{-1})
\end{equation}
where 
\begin{align} \nonumber
&S_{\vec{m}^+} \cong S_{m^+_1} \times \dots \times S_{m^+_k},\\
&S_{\vec{m}^-} \cong S_{m^-_1} \times \dots \times S_{m^-_k}.
\end{align}
are groups with elements
\begin{equation}
\nu^\pm = (\nu_1^\pm,\dots,\nu_k^\pm).
\end{equation}
The group $S_{\vec{m}^+}$ ($S_{\vec{m}^-}$) permutes outgoing (incoming) edges connected to the same vertices. The equivalence classes are in one-to-one correspondence with distinct graphs when the ordered pairs $(m^+_i, m^-_i)$ are all different.\footnote{Note that $m^\pm_i$ can be zero and one. We define $S_0$ to be the empty set and $S_1$ to be the trivial group, containing just the identity element.} 

\begin{figure}[h!]
	\centering
	\raisebox{-0.5\height}{\begin{tikzpicture}[scale=2]
		\def \k {2}
		\def \m {6}
		\def \sep {0.5}
		\def \voffset {0.35}
		\pgfmathparse{(\sep*(\m)-2*\voffset)/\k};
		\pgfmathsetmacro{\vsep}{\pgfmathresult};
		\pgfmathint{\k-1};
		\pgfmathsetmacro{\kk}{\pgfmathresult};
		\foreach \v in {0,...,\k}
		{
			\pgfmathparse{\v*\vsep+\voffset}
			\node[circle, fill, inner sep=1pt](v\v) at (\pgfmathresult,0) {};
		}
		\foreach \eOutm in {0,...,\m}
		{
			\pgfmathint{\eOutm+1};
			\pgfmathsetmacro{\seOutm}{\pgfmathresult};
			\pgfmathparse{\eOutm*\sep};
			\coordinate (eom\seOutm) at (\pgfmathresult,2);
		}
		\foreach \eOutm in {0,...,\m}
		{
			\pgfmathint{\eOutm+1};
			\pgfmathsetmacro{\seOutm}{\pgfmathresult};
			\pgfmathparse{\eOutm*\sep};
			\coordinate (eomm\seOutm) at (\pgfmathresult,1);
			\node[right, node distance = 0pt and 0pt] at (eomm\seOutm) {\tiny \seOutm};
			\draw[] (eom\seOutm) -- (eomm\seOutm);
		}
		\foreach \eInm in {0,...,\m}
		{
			\pgfmathint{\eInm+1};
			\pgfmathsetmacro{\seInm}{\pgfmathresult};
			\pgfmathparse{\eInm*\sep};
			\coordinate (eim\seInm) at (\pgfmathresult,-1.5);
		}
		\foreach \eInm in {0,...,\m}
		{
			\pgfmathint{\eInm+1};
			\pgfmathsetmacro{\seInm}{\pgfmathresult};
			\pgfmathparse{\eInm*\sep};
			\coordinate (eimm\seInm) at (\pgfmathresult,-1);
			\node[right, node distance = 0pt and 0pt] at (eimm\seInm) {\tiny \seInm};
			\draw[] (eim\seInm) -- (eimm\seInm);
		}
		\begin{scope}[decoration={markings, mark=at position 0.5 with \arrow{latex}}]
		\draw[postaction={decorate}] (v0) to[out=100,in=-90] (eomm1);
		\draw[postaction={decorate}] (v0) to[out=90,in=-90] (eomm2);
		\draw[postaction={decorate}] (v0) to[out=70,in=-90] (eomm3);
		\draw[postaction={decorate}] (v1) to[out=90,in=-90] (eomm4);
		\draw[postaction={decorate}] (v1) to[out=90,in=-90] (eomm5);
		\draw[postaction={decorate}] (v1) to[out=90,in=-90] (eomm6);
		\draw[postaction={decorate}] (v2) to[out=90,in=-90] (eomm7);
		\draw[postaction={decorate}] (eimm1) to[out=90,in=-90] (v0);
		\draw[postaction={decorate}] (eimm2) to[out=90,in=-90] (v0);
		\draw[postaction={decorate}] (eimm3) to[out=90,in=-90] (v1);
		\draw[postaction={decorate}] (eimm4) to[out=90,in=-90] (v1);
		\draw[postaction={decorate}] (eimm5) to[out=90,in=-70] (v2);
		\draw[postaction={decorate}] (eimm6) to[out=90,in=-70] (v2);
		\draw[postaction={decorate}] (eimm7) to[out=90,in=-70] (v2);
		\end{scope}
		\draw[fill=white] ($(eomm1)+(-0.1,0.5)$) rectangle node{$\sigma$} ($(eomm7)+(0.1,0.1)$);
		\draw ($(eom1)-(0.1,0)$) -- ($(eom7)+(0.1,0)$);
		\draw ($(eim1)-(0.1,0)$) -- ($(eim7)+(0.1,0)$);
		\end{tikzpicture}} $\sim$
	\raisebox{-0.5\height}{\begin{tikzpicture}[scale=2]
		\def \k {2}
		\def \m {6}
		\def \sep {0.5}
		\def \voffset {0.35}
		\pgfmathparse{(\sep*(\m)-2*\voffset)/\k};
		\pgfmathsetmacro{\vsep}{\pgfmathresult};
		\pgfmathint{\k-1};
		\pgfmathsetmacro{\kk}{\pgfmathresult};
		\foreach \v in {0,...,\k}
		{
			\pgfmathparse{\v*\vsep+\voffset}
			\node[circle, fill, inner sep=1pt](v\v) at (\pgfmathresult,0) {};
		}
		\foreach \eOutm in {0,...,\m}
		{
			\pgfmathint{\eOutm+1};
			\pgfmathsetmacro{\seOutm}{\pgfmathresult};
			\pgfmathparse{\eOutm*\sep};
			\coordinate (eom\seOutm) at (\pgfmathresult,3);
		}
		\foreach \eOutm in {0,...,\m}
		{
			\pgfmathint{\eOutm+1};
			\pgfmathsetmacro{\seOutm}{\pgfmathresult};
			\pgfmathparse{\eOutm*\sep};
			\coordinate (eomm\seOutm) at (\pgfmathresult,1);
			\node[right, node distance = 0pt and 0pt] at (eomm\seOutm) {\tiny \seOutm};
			\draw[] (eom\seOutm) -- (eomm\seOutm);
		}
		\foreach \eInm in {0,...,\m}
		{
			\pgfmathint{\eInm+1};
			\pgfmathsetmacro{\seInm}{\pgfmathresult};
			\pgfmathparse{\eInm*\sep};
			\coordinate (eim\seInm) at (\pgfmathresult,-2.5);
		}
		\foreach \eInm in {0,...,\m}
		{
			\pgfmathint{\eInm+1};
			\pgfmathsetmacro{\seInm}{\pgfmathresult};
			\pgfmathparse{\eInm*\sep};
			\coordinate (eimm\seInm) at (\pgfmathresult,-1);
			\node[right, node distance = 0pt and 0pt] at (eimm\seInm) {\tiny \seInm};
			\draw[] (eim\seInm) -- (eimm\seInm);
		}
		\begin{scope}[decoration={markings, mark=at position 0.5 with \arrow{latex}}]
		\draw[postaction={decorate}] (v0) to[out=100,in=-90] (eomm1);
		\draw[postaction={decorate}] (v0) to[out=90,in=-90] (eomm2);
		\draw[postaction={decorate}] (v0) to[out=70,in=-90] (eomm3);
		\draw[postaction={decorate}] (v1) to[out=90,in=-90] (eomm4);
		\draw[postaction={decorate}] (v1) to[out=90,in=-90] (eomm5);
		\draw[postaction={decorate}] (v1) to[out=90,in=-90] (eomm6);
		\draw[postaction={decorate}] (v2) to[out=90,in=-90] (eomm7);
		\draw[postaction={decorate}] (eimm1) to[out=90,in=-90] (v0);
		\draw[postaction={decorate}] (eimm2) to[out=90,in=-90] (v0);
		\draw[postaction={decorate}] (eimm3) to[out=90,in=-90] (v1);
		\draw[postaction={decorate}] (eimm4) to[out=90,in=-90] (v1);
		\draw[postaction={decorate}] (eimm5) to[out=90,in=-70] (v2);
		\draw[postaction={decorate}] (eimm6) to[out=90,in=-70] (v2);
		\draw[postaction={decorate}] (eimm7) to[out=90,in=-70] (v2);
		\end{scope}
		\draw[fill=white] ($(eomm1)+(-0.1,1.5)$) rectangle node{$\sigma$} ($(eomm7)+(0.1,1.1)$);
		\draw[fill=white] ($(eomm1)+(-0.1,1)$) rectangle node{$\nu_1^+$} ($(eomm3)+(0.1,.6)$);
		\draw[fill=white] ($(eomm4)+(-0.1,1)$) rectangle node{$\nu_2^+$} ($(eomm6)+(0.1,.6)$);
		\draw[fill=white] ($(eimm1)-(0.1,1)$) rectangle node{$(\nu_1^-)^{-1}$} ($(eimm2)-(-0.1,.6)$);
		\draw[fill=white] ($(eimm3)-(0.1,1)$) rectangle node{$(\nu_2^-)^{-1}$} ($(eimm4)-(-0.1,.6)$);
		\draw[fill=white] ($(eimm5)-(0.1,1)$) rectangle node{$(\nu_3^-)^{-1}$} ($(eimm7)-(-0.1,.6)$);
		\draw[fill=white] ($(eomm1)+(-0.1,0.5)$) rectangle node{$\mu$} ($(eomm6)+(0.1,0.1)$);
		\draw[fill=white] ($(eimm1)-(0.1,0.5)$) rectangle node{$\mu^{-1}$} ($(eimm4)-(-0.1,0.1)$);
		\draw ($(eom1)-(0.1,0)$) -- ($(eom7)+(0.1,0)$);
		\draw ($(eim1)-(0.1,0)$) -- ($(eim7)+(0.1,0)$);
		\end{tikzpicture}}	
	\caption{This graph has two identical vertices of type $(3,2)$. Therefore any $\mu \in S_k$ which swaps all the edges of the two vertices gives rise to the same graph.}
	\label{fig: One Color Graph Permutation Equivalence With Vertex Symmetry}
\end{figure}
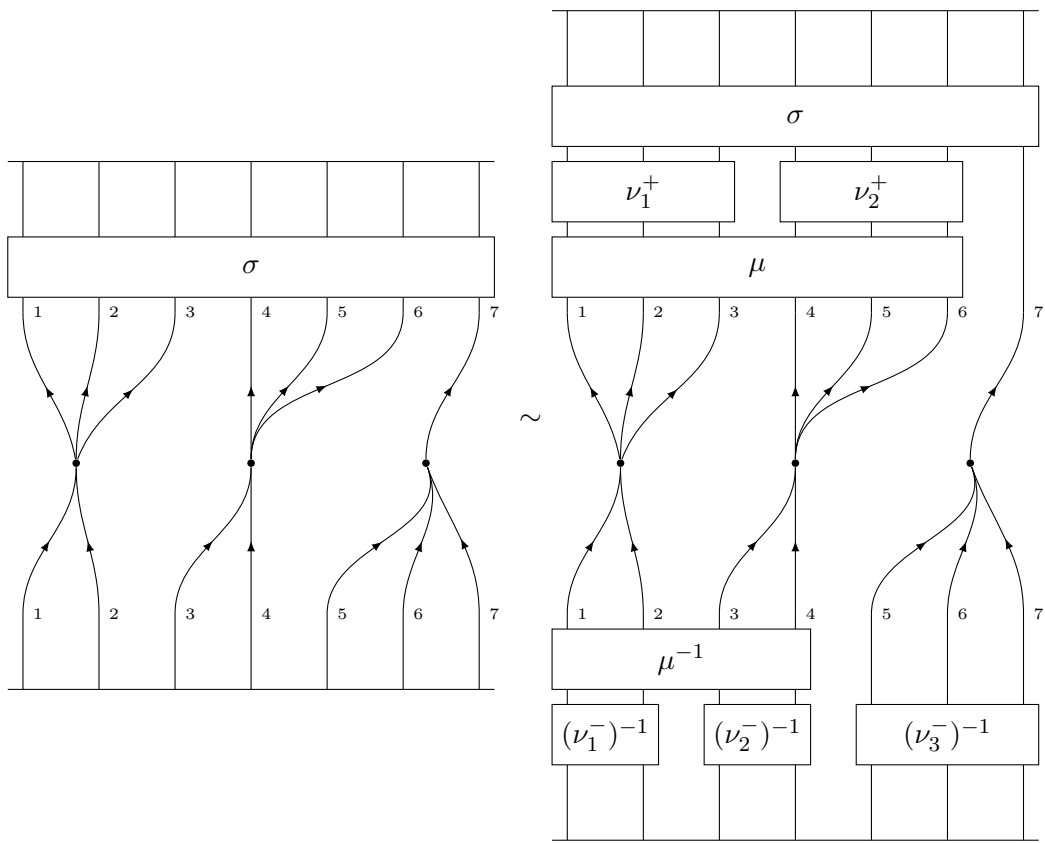

However, when $(m^+_i, m^-_i)=(m^+_j, m^-_j)$ for $i\neq j$, the symmetry is enhanced and permutations which are related by permuting indistinguishable vertices give equivalent graphs. For example, the graphs in Figure \ref{fig: One Color Graph Permutation Equivalence With Vertex Symmetry} have $(m^+_1, m^-_1) = (m^+_2, m^-_2) = (3,2)$.
A permutation $\mu \in S_2 \subset S_3$ which swaps the first two vertices
\begin{align} \nonumber
\rho^+((1,2)) :&[ M_1^+ , M_2^+ , M_3^+ ] \rightarrow [ M_{2 }^+ , M_{1 }^+ , M_{3}^+ ] \\
\rho^-((1,2)) :&[ M_1^- , M_2^- , M_3^- ] \rightarrow [ M_{2 }^- , M_{1 }^- , M_{3}^- ] 
\end{align}
gives back the same graph. More generally, if $(m^+_{i_{_1}}, m^-_{i_{_1}}) = \dots = (m^+_{i_{_l}}, m^-_{i_{_l}})$ for a set of vertex labels $\{i_1,\dots,i_l\} \subseteq \{1,\dots,k\}$, the subgroup $S_l \cong \Perms([i_1, \dots i_l]) \subseteq S_k$ will give equivalent graphs when acting on vertices. The full vertex symmetry group $G_V$ is a product of subgroups
\begin{equation}
G_V \cong S_{l_{_1}} \times S_{l_{_2}} \times \dots
\end{equation}
permuting equivalent vertices. Therefore, the full equivalence relation is
\begin{equation}
\sigma \sim \sigma' \qq{iff} \; \exists \nu^+ \in S_{\vec{m}^+}, \nu^- \in S_{\vec{m}^-}, \mu \in G_V, \;\; \sigma  = \rho^+(\mu)\gamma^+(\nu^+) \sigma' (\rho^-(\mu)\gamma^-(\nu^-))^{-1}. \label{eqn: One Color Graph Permutation Equivalence}
\end{equation}
which diagrammatically corresponds to the equivalence in Figure \ref{fig: One Color Graph Permutation Equivalence With Vertex Symmetry}.

The equivalence relation in \eqref{eqn: One Color Graph Permutation Equivalence} can be viewed as a gauge fixed version of a double quotient. A quotient
\begin{equation}
H_1 \left\backslash G \right/ H_2,
\end{equation}
is the set of equivalence classes of elements $g,g' \in G$ under the identification
\begin{equation}
g \sim g' \qq{iff} \quad \exists h_1 \in H_1, h_2 \in H_2, \quad g = h_1 g' h_2^{-1},
\end{equation}
where $H_1, H_2$ are subgroups of $G$. The equivalence classes are called double cosets.

\begin{figure}
	\centering
	\scalebox{0.75}[0.75]{\raisebox{-0.5\height}{\begin{tikzpicture}[scale=2]
			\def \k {2}
			\def \m {6}
			\def \sep {0.5}
			\def \voffset {0.35}
			\pgfmathparse{(\sep*(\m)-2*\voffset)/\k};
			\pgfmathsetmacro{\vsep}{\pgfmathresult};
			\pgfmathint{\k-1};
			\pgfmathsetmacro{\kk}{\pgfmathresult};
			\foreach \v in {0,...,\k}
			{
				\pgfmathparse{\v*\vsep+\voffset}
				\node[circle, fill, inner sep=1pt](v\v) at (\pgfmathresult,0) {};
			}
			\foreach \eOutm in {0,...,\m}
			{
				\pgfmathint{\eOutm+1};
				\pgfmathsetmacro{\seOutm}{\pgfmathresult};
				\pgfmathparse{\eOutm*\sep};
				\coordinate (eom\seOutm) at (\pgfmathresult,2);
			}
			\foreach \eOutm in {0,...,\m}
			{
				\pgfmathint{\eOutm+1};
				\pgfmathsetmacro{\seOutm}{\pgfmathresult};
				\pgfmathparse{\eOutm*\sep};
				\coordinate (eomm\seOutm) at (\pgfmathresult,1);
				\node[right, node distance = 0pt and 0pt] at (eomm\seOutm) {\tiny \seOutm};
				\draw[] (eom\seOutm) -- (eomm\seOutm);
			}
			\foreach \eInm in {0,...,\m}
			{
				\pgfmathint{\eInm+1};
				\pgfmathsetmacro{\seInm}{\pgfmathresult};
				\pgfmathparse{\eInm*\sep};
				\coordinate (eim\seInm) at (\pgfmathresult,-2);
			}
			\foreach \eInm in {0,...,\m}
			{
				\pgfmathint{\eInm+1};
				\pgfmathsetmacro{\seInm}{\pgfmathresult};
				\pgfmathparse{\eInm*\sep};
				\coordinate (eimm\seInm) at (\pgfmathresult,-1);
				\node[right, node distance = 0pt and 0pt] at (eimm\seInm) {\tiny \seInm};
				\draw[] (eim\seInm) -- (eimm\seInm);
			}
			\begin{scope}[decoration={markings, mark=at position 0.5 with \arrow{latex}}]
			\draw[postaction={decorate}] (v0) to[out=100,in=-90] (eomm1);
			\draw[postaction={decorate}] (v0) to[out=90,in=-90] (eomm2);
			\draw[postaction={decorate}] (v0) to[out=70,in=-90] (eomm3);
			\draw[postaction={decorate}] (v1) to[out=90,in=-90] (eomm4);
			\draw[postaction={decorate}] (v1) to[out=90,in=-90] (eomm5);
			\draw[postaction={decorate}] (v1) to[out=90,in=-90] (eomm6);
			\draw[postaction={decorate}] (v2) to[out=90,in=-90] (eomm7);
			\draw[postaction={decorate}] (eimm1) to[out=90,in=-90] (v0);
			\draw[postaction={decorate}] (eimm2) to[out=90,in=-90] (v0);
			\draw[postaction={decorate}] (eimm3) to[out=90,in=-90] (v1);
			\draw[postaction={decorate}] (eimm4) to[out=90,in=-90] (v1);
			\draw[postaction={decorate}] (eimm5) to[out=90,in=-70] (v2);
			\draw[postaction={decorate}] (eimm6) to[out=90,in=-70] (v2);
			\draw[postaction={decorate}] (eimm7) to[out=90,in=-70] (v2);
			\end{scope}
			\draw[fill=white] ($(eomm1)+(-0.1,0.5)$) rectangle node{$\sigma_1$} ($(eomm7)+(0.1,0.1)$);
			\draw[fill=white] ($(eimm1)-(0.1,0.5)$) rectangle node{$\sigma_2^{-1}$} ($(eimm7)-(-0.1,0.1)$);
			\draw ($(eom1)-(0.1,0)$) -- ($(eom7)+(0.1,0)$);
			\draw ($(eim1)-(0.1,0)$) -- ($(eim7)+(0.1,0)$);
			\end{tikzpicture}}
		$\sim$
		\raisebox{-0.5\height}{\begin{tikzpicture}[scale=2]
			\def \k {2}
			\def \m {6}
			\def \sep {0.5}
			\def \voffset {0.35}
			\pgfmathparse{(\sep*(\m)-2*\voffset)/\k};
			\pgfmathsetmacro{\vsep}{\pgfmathresult};
			\pgfmathint{\k-1};
			\pgfmathsetmacro{\kk}{\pgfmathresult};
			\foreach \v in {0,...,\k}
			{
				\pgfmathparse{\v*\vsep+\voffset}
				\node[circle, fill, inner sep=1pt](v\v) at (\pgfmathresult,0) {};
			}
			\foreach \eOutm in {0,...,\m}
			{
				\pgfmathint{\eOutm+1};
				\pgfmathsetmacro{\seOutm}{\pgfmathresult};
				\pgfmathparse{\eOutm*\sep};
				\coordinate (eom\seOutm) at (\pgfmathresult,3.5);
			}
			\foreach \eOutm in {0,...,\m}
			{
				\pgfmathint{\eOutm+1};
				\pgfmathsetmacro{\seOutm}{\pgfmathresult};
				\pgfmathparse{\eOutm*\sep};
				\coordinate (eomm\seOutm) at (\pgfmathresult,1);
				\node[right, node distance = 0pt and 0pt] at (eomm\seOutm) {\tiny \seOutm};
				\draw[] (eom\seOutm) -- (eomm\seOutm);
			}
			\foreach \eInm in {0,...,\m}
			{
				\pgfmathint{\eInm+1};
				\pgfmathsetmacro{\seInm}{\pgfmathresult};
				\pgfmathparse{\eInm*\sep};
				\coordinate (eim\seInm) at (\pgfmathresult,-3.5);
			}
			\foreach \eInm in {0,...,\m}
			{
				\pgfmathint{\eInm+1};
				\pgfmathsetmacro{\seInm}{\pgfmathresult};
				\pgfmathparse{\eInm*\sep};
				\coordinate (eimm\seInm) at (\pgfmathresult,-1);
				\node[right, node distance = 0pt and 0pt] at (eimm\seInm) {\tiny \seInm};
				\draw[] (eim\seInm) -- (eimm\seInm);
			}
			\begin{scope}[decoration={markings, mark=at position 0.5 with \arrow{latex}}]
			\draw[postaction={decorate}] (v0) to[out=100,in=-90] (eomm1);
			\draw[postaction={decorate}] (v0) to[out=90,in=-90] (eomm2);
			\draw[postaction={decorate}] (v0) to[out=70,in=-90] (eomm3);
			\draw[postaction={decorate}] (v1) to[out=90,in=-90] (eomm4);
			\draw[postaction={decorate}] (v1) to[out=90,in=-90] (eomm5);
			\draw[postaction={decorate}] (v1) to[out=90,in=-90] (eomm6);
			\draw[postaction={decorate}] (v2) to[out=90,in=-90] (eomm7);
			\draw[postaction={decorate}] (eimm1) to[out=90,in=-90] (v0);
			\draw[postaction={decorate}] (eimm2) to[out=90,in=-90] (v0);
			\draw[postaction={decorate}] (eimm3) to[out=90,in=-90] (v1);
			\draw[postaction={decorate}] (eimm4) to[out=90,in=-90] (v1);
			\draw[postaction={decorate}] (eimm5) to[out=90,in=-70] (v2);
			\draw[postaction={decorate}] (eimm6) to[out=90,in=-70] (v2);
			\draw[postaction={decorate}] (eimm7) to[out=90,in=-70] (v2);
			\end{scope}
			\draw[fill=white] ($(eomm1)+(-0.1,2)$) rectangle node{$\gamma$} ($(eomm7)+(0.1,1.6)$);
			\draw[fill=white] ($(eimm1)-(0.1,2)$) rectangle node{$\gamma^{-1}$} ($(eimm7)-(-0.1,1.6)$);
			\draw[fill=white] ($(eomm1)+(-0.1,1.5)$) rectangle node{$\sigma_1$} ($(eomm7)+(0.1,1.1)$);
			\draw[fill=white] ($(eimm1)-(0.1,1.5)$) rectangle node{$\sigma_2^{-1}$} ($(eimm7)-(-0.1,1.1)$);		
			\draw[fill=white] ($(eomm1)+(-0.1,1)$) rectangle node{$\nu^+_1$} ($(eomm3)+(0.1,0.6)$);
			\draw[fill=white] ($(eomm4)+(-0.1,1)$) rectangle node{$\nu^+_2$} ($(eomm6)+(0.1,0.6)$);
			\draw[fill=white] ($(eomm1)+(-0.1,0.5)$) rectangle node{$\mu$} ($(eomm6)+(0.1,0.1)$);
			\draw[fill=white] ($(eimm1)-(0.1,1)$) rectangle node{$(\nu^-_1)^{-1}$} ($(eimm2)-(-0.1,0.6)$);
			\draw[fill=white] ($(eimm3)-(0.1,1)$) rectangle node{$(\nu^-_2)^{-1}$} ($(eimm4)-(-0.1,0.6)$);
			\draw[fill=white] ($(eimm5)-(0.1,1)$) rectangle node{$(\nu^-_3)^{-1}$} ($(eimm7)-(-0.1,0.6)$);
			\draw[fill=white] ($(eimm1)-(0.1,0.5)$) rectangle node{$\mu^{-1}$} ($(eimm4)-(-0.1,0.1)$);		
			\draw ($(eom1)-(0.1,0)$) -- ($(eom7)+(0.1,0)$);
			\draw ($(eim1)-(0.1,0)$) -- ($(eim7)+(0.1,0)$);
			\end{tikzpicture}}}
	\caption{Diagrammatic description of the double coset equivalence in equation \eqref{eq: Double Coset Equivalence}.}
	\label{fig: Two Sigma One Color Double Coset Graph}
\end{figure}
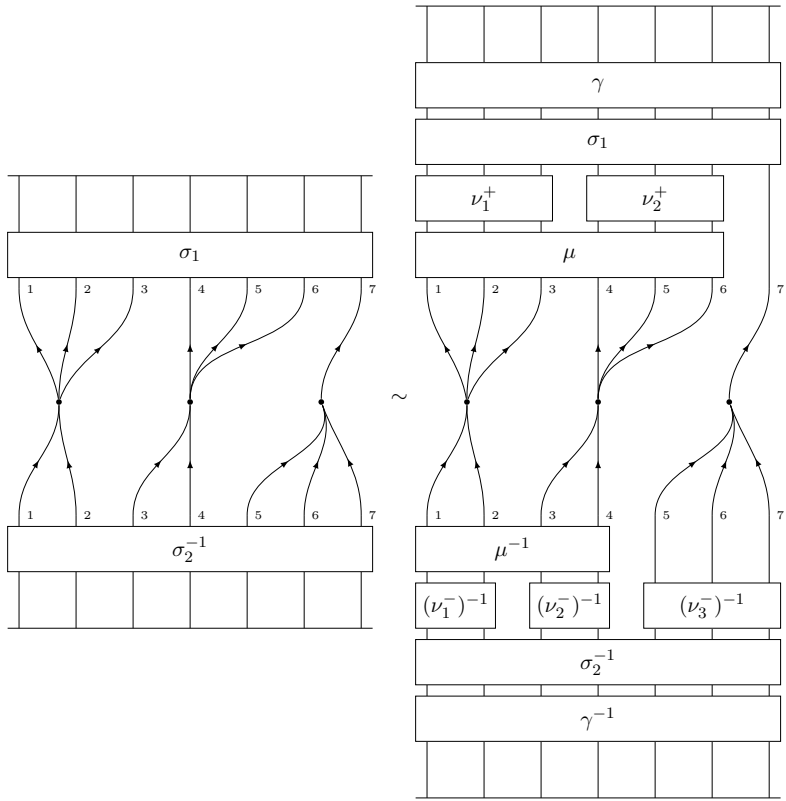

In our case, the quotient is
\begin{equation}
G(\vec{m}^+,\vec{m}^-) \left\backslash \qty( S_m^+ \times S_m^-) \right/ \diag(S_m). \label{eqn:1colordoublecoset}
\end{equation}
The diagrammatic equivalence to have in mind for the double coset is Figure \ref{fig: Two Sigma One Color Double Coset Graph}. Because the incoming edges at the top line are identified with the outgoing edges of the bottom line, it is effectively only the product $\sigma_1 \sigma_2^{-1}$ which acts on the edges in this picture. We have increased the redundancy in the picture by going from a single permutation to a pair $(\sigma_1, \sigma_2) \in S_m^+ \times S_m^-$. If $(\sigma_1, \sigma_2)$ is replaced by $(\sigma_1 \gamma, \sigma_2 \gamma)$ for $\gamma \in S_m$, then the combination $\sigma_1 \sigma_2^{-1} \mapsto \sigma_1 \gamma \gamma^{-1}\sigma_2^{-1} = \sigma_1\sigma_2^{-1}$ is unchanged. This is the origin of the quotient by $\diag(S_m)$, it describes the redundancy of using pairs of permutations.

The group $G(\vec{m}^+,\vec{m}^-)$ is a subgroup of $S_m^+ \times S_m^-$ generated by elements of the form
\begin{equation}
(\rho^+(\mu) , \rho^-(\mu)), \qq{and} (\gamma^+(\nu^+), \gamma^-(\nu^-))
\end{equation}
for $\mu \in G_V, \nu^+ \in S_{\vec{m}^+}, \nu^- \in S_{\vec{m}^-}$.
The double cosets are equivalence classes of the relation
\begin{align} \nonumber \label{eq: Double Coset Equivalence}
(\sigma_1, \sigma_2) \sim &(\sigma_1', \sigma_2') \qq{iff} \exists \nu^+ \in S_{\vec{m}^+}, \nu^- \in S_{\vec{m}^-}, \mu \in G_V, \gamma \in S_m, \\
&(\sigma_1, \sigma_2) = (\rho^+(\mu)\gamma^+(\nu^+)\sigma_1'\gamma^{-1}, \rho^-(\mu)\gamma^-(\nu^-)\sigma_2'\gamma^{-1}). 
\end{align}
To see how equation \eqref{eq: Double Coset Equivalence} relates to \eqref{eqn: One Color Graph Permutation Equivalence}, we count the number of equivalence classes. By Burnside's lemma, the number of double cosets is
\begin{align} \nonumber
N(\vec{m}^+,\vec{m}^-) &= \frac{1}{|G(\vec{m}^+,\vec{m}^-)||S_m|}\times \\ \nonumber
&\sum_{\substack{\mu \in G_V, \nu^+ \in S_{\vec{m}^+} \\ \nu^- \in S_{\vec{m}^-},\gamma \in S_m}} \sum_{\sigma_1, \sigma_2 \in S_m}
\begin{aligned}[t]
&\delta(\sigma_1^{-1}\rho^+(\mu)\gamma^+(\nu^+)\sigma_1\gamma^{-1}) \\ \nonumber
&\delta(\sigma_2^{-1}\rho^-(\mu)\gamma^-(\nu^-)\sigma_2\gamma^{-1})
\end{aligned} \label{eqn: Burnsides lemma double coset one color}\\ 
&=\frac{1}{|G(\vec{m}^+,\vec{m}^-)||S_m|}\times \nonumber\\
&\sum_{\substack{\mu \in G_V, \nu^+ \in S_{\vec{m}^+} \\ \nonumber \nu^- \in S_{\vec{m}^-}}} \sum_{\sigma_1, \sigma_2 \in S_m}
\begin{aligned}[t]
&\delta(\sigma_1^{-1}\rho^+(\mu)\gamma^+(\nu^+)\sigma_1\sigma_2^{-1}\gamma^-((\nu^-)^{-1})\rho^-(\mu^{-1})\sigma_2)
\end{aligned} \\ 
&=\frac{1}{|G(\vec{m}^+,\vec{m}^-)|}	\sum_{\substack{\mu \in G_V \\ \nu^+ \in S_{\vec{m}^+} \\ \nu^- \in S_{\vec{m}^-}}} \sum_{\tau \in S_m}
\begin{aligned}[t]
&\delta(\tau^{-1}\rho^+(\mu)\gamma^+(\nu^+)\tau\gamma^-((\nu^-)^{-1})\rho^-(\mu^{-1})).
\end{aligned}
\end{align}
The delta function of a group element $g \in G$ is defined as
\begin{equation}
\delta(g) = \begin{cases}
1, \qq{if} g=e \\
0, \qq{otherwise}.
\end{cases}
\end{equation}
In the second equality, we carried out the sum over $\gamma$ to impose the second delta function. In the third equality we renamed $\sigma_1\sigma_2^{-1} \equiv \tau$, this makes the summand independent of $\sigma_2$. Consequently the sum over $\sigma_2$ just gives a factor of $|S_m|$. From Burnside's lemma, we recognize the last line as the counting of equivalence classes of \eqref{eqn: One Color Graph Permutation Equivalence}. This shows the correspondence between the double coset and the counting of graphs under edge and vertex symmetry.

\begin{figure}[h!]
	\centering
	\raisebox{-0.5\height}{\scalebox{0.75}{\begin{tikzpicture}[scale=2]
			\definecolor{GREEN}{rgb}{0.0,0.70,0.24}
			\definecolor{BLUE}{rgb}{0.0,0.24,0.70}
			\def \k {3}
			\def \m {3}
			\def \n {3}
			\def \sep {0.5}
			\def \voffset {0.35}
			\pgfmathparse{(\sep*(\m+\n+2)-2*\voffset)/\k};
			\pgfmathsetmacro{\vsep}{\pgfmathresult};
			\pgfmathint{\k-1};
			\pgfmathsetmacro{\kk}{\pgfmathresult};
			\foreach \v in {0,...,\k}
			{
				\pgfmathparse{\v*\vsep+\voffset}
				\node[circle, fill, inner sep=1pt](v\v) at (\pgfmathresult,0) {};
			}
			\foreach \eOutm in {0,...,\m}
			{
				\pgfmathint{\eOutm+1};
				\pgfmathsetmacro{\seOutm}{\pgfmathresult};
				\pgfmathparse{\eOutm*\sep};
				\coordinate (eom\seOutm) at (\pgfmathresult,2);
				\coordinate (eomm\seOutm) at (\pgfmathresult,1);
				\node[right, node distance = 0pt and 0pt, text=BLUE] at (eomm\seOutm) {\tiny \seOutm};
				\draw[draw=BLUE] (eom\seOutm) -- (eomm\seOutm);
			}
			\foreach \eInm in {0,...,\m}
			{
				\pgfmathint{\eInm+1};
				\pgfmathsetmacro{\seInm}{\pgfmathresult};
				\pgfmathparse{\eInm*\sep};
				\coordinate (eim\seInm) at (\pgfmathresult,-2);
				\coordinate (eimm\seInm) at (\pgfmathresult,-1);
				\node[right, node distance = 0pt and 0pt, text=BLUE] at (eimm\seInm) {\tiny \seInm};
				\draw[draw=BLUE] (eim\seInm) -- (eimm\seInm);
			}
			\foreach \eOutn in {0,...,\n}
			{
				\pgfmathint{\m+\eOutn+2};
				\pgfmathsetmacro{\seOutn}{\pgfmathresult};
				\pgfmathparse{\seOutn*\sep};
				\coordinate (eon\seOutn) at (\pgfmathresult,2);
				\coordinate (eonn\seOutn) at (\pgfmathresult,1);
				\node[right, node distance = 0pt and 0pt, text=GREEN] at (eonn\seOutn) {\tiny \seOutn};
				\draw[draw=GREEN] (eon\seOutn) -- (eonn\seOutn);
			}
			\foreach \eInn in {0,...,\n}
			{
				\pgfmathint{\m+\eInn+2};
				\pgfmathsetmacro{\seInn}{\pgfmathresult};
				\pgfmathparse{\seInn*\sep};
				\coordinate (ein\seInn) at (\pgfmathresult,-2);
				\coordinate (einn\seInn) at (\pgfmathresult,-1);
				\node[right, node distance = 0pt and 0pt, text=GREEN] at (einn\seInn) {\tiny \seInn};
				\draw[draw=GREEN] (ein\seInn) -- (einn\seInn);
			}
			\begin{scope}[decoration={markings, mark=at position 0.4 with \arrow{latex}}]
			\draw[draw=BLUE, postaction={decorate}] (v0) to[out=90,in=-90] (eomm1);
			\draw[draw=BLUE, postaction={decorate}] (v0) to[out=70,in=-90] (eomm2);
			\draw[draw=BLUE, postaction={decorate}] (v1) to[out=90,in=-90] (eomm3);
			\draw[draw=BLUE, postaction={decorate}] (v2) to[out=140,in=-90] (eomm4);
			\draw[draw=GREEN, postaction={decorate}] (v1) to[out=40,in=-90] (eonn5);
			\draw[draw=GREEN, postaction={decorate}] (v2) to[out=90,in=-90] (eonn6);
			\draw[draw=GREEN, postaction={decorate}] (v3) to[out=90,in=-90] (eonn7);
			\draw[draw=GREEN, postaction={decorate}] (v3) to[out=70,in=-90] (eonn8);
			\draw[draw=BLUE, postaction={decorate}] (eimm1) to[out=90,in=-90] (v0);
			\draw[draw=BLUE, postaction={decorate}] (eimm2) to[out=90,in=-90] (v1);
			\draw[draw=BLUE, postaction={decorate}] (eimm3) to[out=90,in=-90] (v2);
			\draw[draw=BLUE, postaction={decorate}] (eimm4) to[out=90,in=-90] (v3);
			\draw[draw=GREEN, postaction={decorate}] (einn5) to[out=90,in=-90] (v1);
			\draw[draw=GREEN, postaction={decorate}] (einn6) to[out=90,in=-90] (v2);
			\draw[draw=GREEN, postaction={decorate}] (einn7) to[out=90,in=-90] (v3);
			\draw[draw=GREEN, postaction={decorate}] (einn8) to[out=90,in=-70] (v3);
			\end{scope}
			\draw[fill=white] ($(eomm1)+(-0.1,0.5)$) rectangle node{$\sigma_m$} ($(eomm4)+(0.1,0.1)$);
			\draw[fill=white] ($(eonn5)+(-0.1,0.5)$) rectangle node{$\sigma_n$} ($(eonn8)+(0.1,0.1)$);
			\draw ($(eom1)-(0.1,0)$) -- ($(eon8)+(0.1,0)$);
			\draw ($(eim1)-(0.1,0)$) -- ($(ein8)+(0.1,0)$);
			\end{tikzpicture}}}
	$\sim$
	\raisebox{-0.5\height}{\scalebox{0.75}{\begin{tikzpicture}[scale=2]
			\definecolor{GREEN}{rgb}{0.0,0.70,0.24}
			\definecolor{BLUE}{rgb}{0.0,0.24,0.70}
			\def \k {3}
			\def \m {3}
			\def \n {3}
			\def \sep {0.5}
			\def \voffset {0.35}
			\pgfmathparse{(\sep*(\m+\n+2)-2*\voffset)/\k};
			\pgfmathsetmacro{\vsep}{\pgfmathresult};
			\pgfmathint{\k-1};
			\pgfmathsetmacro{\kk}{\pgfmathresult};
			\foreach \v in {0,...,\k}
			{
				\pgfmathparse{\v*\vsep+\voffset}
				\node[circle, fill, inner sep=1pt](v\v) at (\pgfmathresult,0) {};
			}
			\foreach \eOutm in {0,...,\m}
			{
				\pgfmathint{\eOutm+1};
				\pgfmathsetmacro{\seOutm}{\pgfmathresult};
				\pgfmathparse{\eOutm*\sep};
				\coordinate (eom\seOutm) at (\pgfmathresult,3);
				\coordinate (eomm\seOutm) at (\pgfmathresult,1);
				\node[right, node distance = 0pt and 0pt, text=BLUE] at (eomm\seOutm) {\tiny \seOutm};
				\draw[draw=BLUE] (eom\seOutm) -- (eomm\seOutm);
			}
			\foreach \eInm in {0,...,\m}
			{
				\pgfmathint{\eInm+1};
				\pgfmathsetmacro{\seInm}{\pgfmathresult};
				\pgfmathparse{\eInm*\sep};
				\coordinate (eim\seInm) at (\pgfmathresult,-2.5);
				\coordinate (eimm\seInm) at (\pgfmathresult,-1);
				\node[right, node distance = 0pt and 0pt, text=BLUE] at (eimm\seInm) {\tiny \seInm};
				\draw[draw=BLUE] (eim\seInm) -- (eimm\seInm);
			}
			\foreach \eOutn in {0,...,\n}
			{
				\pgfmathint{\m+\eOutn+2};
				\pgfmathsetmacro{\seOutn}{\pgfmathresult};
				\pgfmathparse{\seOutn*\sep};
				\coordinate (eon\seOutn) at (\pgfmathresult,3);
				\coordinate (eonn\seOutn) at (\pgfmathresult,1);
				\node[right, node distance = 0pt and 0pt, text=GREEN] at (eonn\seOutn) {\tiny \seOutn};
				\draw[draw=GREEN] (eon\seOutn) -- (eonn\seOutn);
			}
			\foreach \eInn in {0,...,\n}
			{
				\pgfmathint{\m+\eInn+2};
				\pgfmathsetmacro{\seInn}{\pgfmathresult};
				\pgfmathparse{\seInn*\sep};
				\coordinate (ein\seInn) at (\pgfmathresult,-2.5);
				\coordinate (einn\seInn) at (\pgfmathresult,-1);
				\node[right, node distance = 0pt and 0pt, text=GREEN] at (einn\seInn) {\tiny \seInn};
				\draw[draw=GREEN] (ein\seInn) -- (einn\seInn);
			}
			\begin{scope}[decoration={markings, mark=at position 0.4 with \arrow{latex}}]
			\draw[draw=BLUE, postaction={decorate}] (v0) to[out=90,in=-90] (eomm1);
			\draw[draw=BLUE, postaction={decorate}] (v0) to[out=70,in=-90] (eomm2);
			\draw[draw=BLUE, postaction={decorate}] (v1) to[out=90,in=-90] (eomm3);
			\draw[draw=BLUE, postaction={decorate}] (v2) to[out=140,in=-90] (eomm4);
			\draw[draw=GREEN, postaction={decorate}] (v1) to[out=40,in=-90] (eonn5);
			\draw[draw=GREEN, postaction={decorate}] (v2) to[out=90,in=-90] (eonn6);
			\draw[draw=GREEN, postaction={decorate}] (v3) to[out=90,in=-90] (eonn7);
			\draw[draw=GREEN, postaction={decorate}] (v3) to[out=70,in=-90] (eonn8);
			\draw[draw=BLUE, postaction={decorate}] (eimm1) to[out=90,in=-90] (v0);
			\draw[draw=BLUE, postaction={decorate}] (eimm2) to[out=90,in=-90] (v1);
			\draw[draw=BLUE, postaction={decorate}] (eimm3) to[out=90,in=-90] (v2);
			\draw[draw=BLUE, postaction={decorate}] (eimm4) to[out=90,in=-90] (v3);
			\draw[draw=GREEN, postaction={decorate}] (einn5) to[out=90,in=-90] (v1);
			\draw[draw=GREEN, postaction={decorate}] (einn6) to[out=90,in=-90] (v2);
			\draw[draw=GREEN, postaction={decorate}] (einn7) to[out=90,in=-90] (v3);
			\draw[draw=GREEN, postaction={decorate}] (einn8) to[out=90,in=-70] (v3);
			\end{scope}
			\draw[fill=white] ($(eomm1)+(-0.1,1.5)$) rectangle node{$\sigma_m$} ($(eomm4)+(0.1,1.1)$);
			\draw[fill=white] ($(eonn5)+(-0.1,1.5)$) rectangle node{$\sigma_n$} ($(eonn8)+(0.1,1.1)$);
			\draw[fill=white] ($(eomm1)+(-0.1,1)$) rectangle node{$\nu_{m,1}^+$} ($(eomm2)+(0.1,.6)$);
			\draw[fill=white] ($(eonn7)+(-0.1,1)$) rectangle node{$\nu_{n,4}^+$} ($(eonn8)+(0.1,.6)$);
			\draw[fill=white] ($(einn7)-(0.1,1)$) rectangle node{$(\nu_{n,4}^-)^{-1}$} ($(einn8)-(-0.1,.6)$);
			\draw[fill=white] ($(eomm3)+(-0.1,0.5)$) rectangle node{$\mu$} ($(eomm4)+(0.1,0.1)$);
			\draw[fill=white] ($(eonn5)+(-0.1,0.5)$) rectangle node{$\mu$} ($(eonn6)+(0.1,0.1)$);
			\draw[fill=white] ($(eimm2)-(0.1,0.5)$) rectangle node{$\mu^{-1}$} ($(eimm3)-(-0.1,0.1)$);
			\draw[fill=white] ($(einn5)-(0.1,0.5)$) rectangle node{$\mu^{-1}$} ($(einn6)-(-0.1,0.1)$);
			\draw ($(eom1)-(0.1,0)$) -- ($(eon8)+(0.1,0)$);
			\draw ($(eim1)-(0.1,0)$) -- ($(ein8)+(0.1,0)$);
			\end{tikzpicture}}}
	\caption{Diagrammatic description of the permutation equivalence in equation \eqref{eqn: Two Color Graph Permutation Equivalence}.}
	\label{fig: Two Colour Double Coset Graph}
\end{figure}
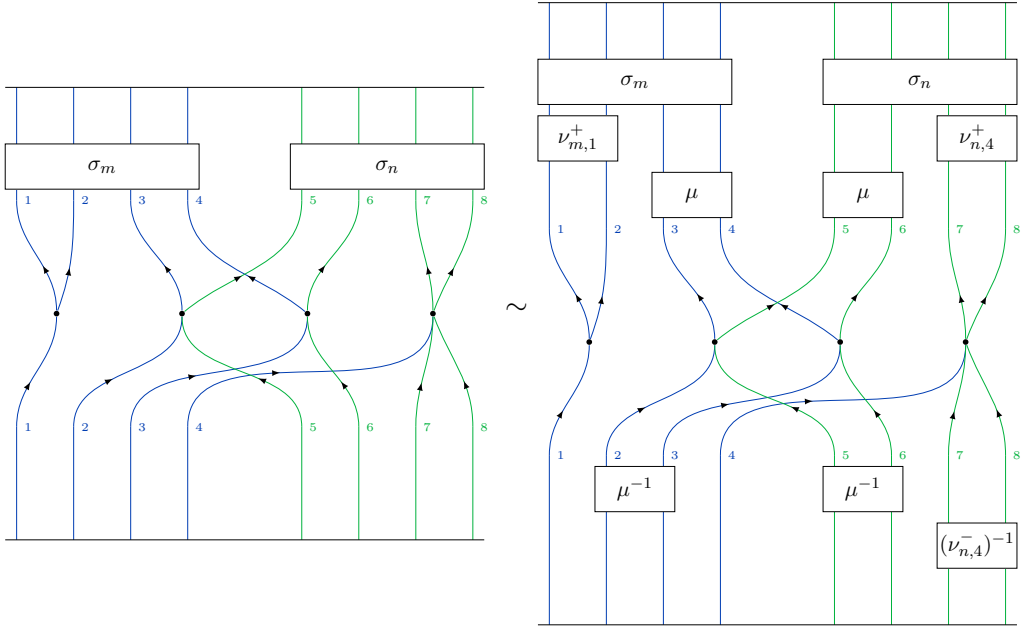

The same story generalizes to two-colored graphs. For a graph with $m$ blue edges, $n$ green edges and $k$ vertices we have a vector partition
\begin{align} \nonumber
(m, m; n, n) = (m_1^+,m_1^-; n_1^+, n_1^-) + \dots +  &(m_k^+,  m_k^-; n_k^+ , n_k^-) \equiv (\vec{m}^+,\vec{m}^-;\vec{n}^+,\vec{n}^-) \\
& 0 \leq m^\pm_i \leq m, \; 0 \leq n^\pm_i \leq n. 
\end{align}
As before, the labels in Figure \ref{fig: Two Colour Double Coset Graph} define a set of lists
\begin{align} \nonumber
M_1^\pm &= [1,2, \dots m_1^\pm] \\ \nonumber
M_2^\pm &= [m_1^\pm +1, \dots, m_1^\pm + m_2^\pm]\\ \nonumber
&\phantom{=}\vdots \\ \nonumber
M_i^\pm&= [m_1^\pm + m_2^\pm + \dots + m_{i-1}^\pm + 1, \dots, m_1^\pm + \dots + m_{i-1}^\pm+m_i^\pm] \\ \nonumber
&\phantom{=}\vdots\\
M_k^\pm &=[m_1^\pm + m_2^\pm + \dots + m_{k-1}^\pm + 1, \dots, \underbrace{m_1^\pm + \dots + m_k^\pm}_{=m}]
\end{align}
and
\begin{align} \nonumber
N_1^\pm &= [m+1,m+2, \dots,m+ n_1^\pm] \\ \nonumber
N_2^\pm &= [m+n_1^\pm +1, \dots, m+n_1^\pm + n_2^\pm]\\ \nonumber
&\phantom{=}\vdots \\ \nonumber
N_i^\pm&= [m+n_1^\pm + n_2^\pm + \dots + n_{i-1}^\pm + 1, \dots,m+ n_1^\pm + \dots + n_{i-1}^\pm+n_i^\pm] \\ \nonumber
&\phantom{=}\vdots\\
N_k^\pm &=[m+n_1^\pm + n_2^\pm + \dots + n_{k-1}^\pm + 1, \dots, \underbrace{m+n_1^\pm + \dots + n_k^\pm}_{=m+n}].
\end{align}
A permutation $\sigma_m \in S_m$ is a re-arrangement of the list $[M_1^\pm, \dots, M_k^\pm] = [1,\dots,m]$ and a permutation $\sigma_n \in S_n$ is a re-arrangement of $[N_1^\pm, \dots, N_k^\pm]=[m+1,\dots,m+n]$. The permutations within sublists define subgroups isomorphic to
\begin{align} \nonumber
S_{\vec{m}^\pm} &= \Perms([1,2\dots,m_1^\pm]) \times \Perms([1,2\dots,m_2^\pm]) \times \dots \times \Perms([1,2\dots,m_k^\pm]), \\ \nonumber
S_{\vec{n}^\pm} &= S_{n_1^\pm} \times S_{n_2^\pm} \times \dots \times S_{n_k^\pm} \\
&=\Perms([1,2\dots,n_1^\pm]) \times \Perms([1,2\dots,n_2^\pm]) \times \dots \times \Perms([1,2\dots,n_k^\pm]).
\end{align}
The homomorphisms into $S_m$ and $S_n$ are
\begin{align} \nonumber
\gamma_m^\pm(\nu_m^\pm): &[M_1^\pm, M_2^\pm, \dots, M_k^\pm] \rightarrow [\nu_{m,1}^\pm(M_1^\pm), \nu_{m,2}^\pm(M_2^\pm), \dots, \nu_{m,k}^\pm(M_k^\pm)] \\ \nonumber
&\qq{for} \nu_m^\pm = (\nu_{m,1}^\pm,\dots,\nu_{m,k}^\pm) \in S_{\vec{m}^\pm} \\ \nonumber
\gamma_n^\pm(\nu_n^\pm): &[N_1^\pm, N_2^\pm, \dots, N_k^\pm] \rightarrow [\nu_{n,1}^\pm(N_1^\pm), \nu_{n,2}^\pm(N_2^\pm), \dots, \nu_{n,k}^\pm(N_k^\pm)] \\
&\qq{for} \nu_n^\pm = (\nu_{n,1}^\pm,\dots,\nu_{n,k}^\pm) \in S_{\vec{n}^\pm}
\end{align}
where
\begin{equation}
\nu_{n,i}^\pm(N_i^\pm) = [m+\sum_{j=1}^{i-1}n_j^\pm+\nu_{n,i}^\pm(1),m+\sum_{j=1}^{i-1}n_j^\pm +\nu_{n,i}^\pm(2),\dots,m+\sum_{j=1}^{i-1}n_j^\pm + \nu_{n,i}^\pm(n_i^\pm)].
\end{equation}
The homomorphism for permutations of vertices is completely analogous to the one-colored case. The equivalence in Figure \ref{fig: Two Colour Double Coset Graph} is more precisely written
\begin{align} \nonumber \label{eqn: Two Color Graph Permutation Equivalence}
&(\sigma_m, \sigma_n) \sim (\sigma_m', \sigma_n') \qq{iff} 
\begin{aligned}[t]
\exists &\nu_m^+ \in S_{\vec{m}^+}, \nu_m^- \in S_{\vec{m}^-}, \\ &\nu_n^+ \in S_{\vec{n}^+}, \nu_n^- \in S_{\vec{n}^-}, \mu \in G_V 
\end{aligned} \\ 
&(\sigma_m, \sigma_n) = (\rho^+_m(\mu)\gamma^+_m(\nu_m^+)\sigma_m'\gamma^-_m((\nu_m^-)^{-1})\rho^-_m(\mu^{-1}),\nonumber\\
&\phantom{(\sigma_m, \sigma_n) =}\rho^+_n(\mu)\gamma^+_n(\nu_n^+)\sigma_n'\gamma^-_n((\nu_n^-)^{-1})\rho^-_n(\mu^{-1})).
\end{align}

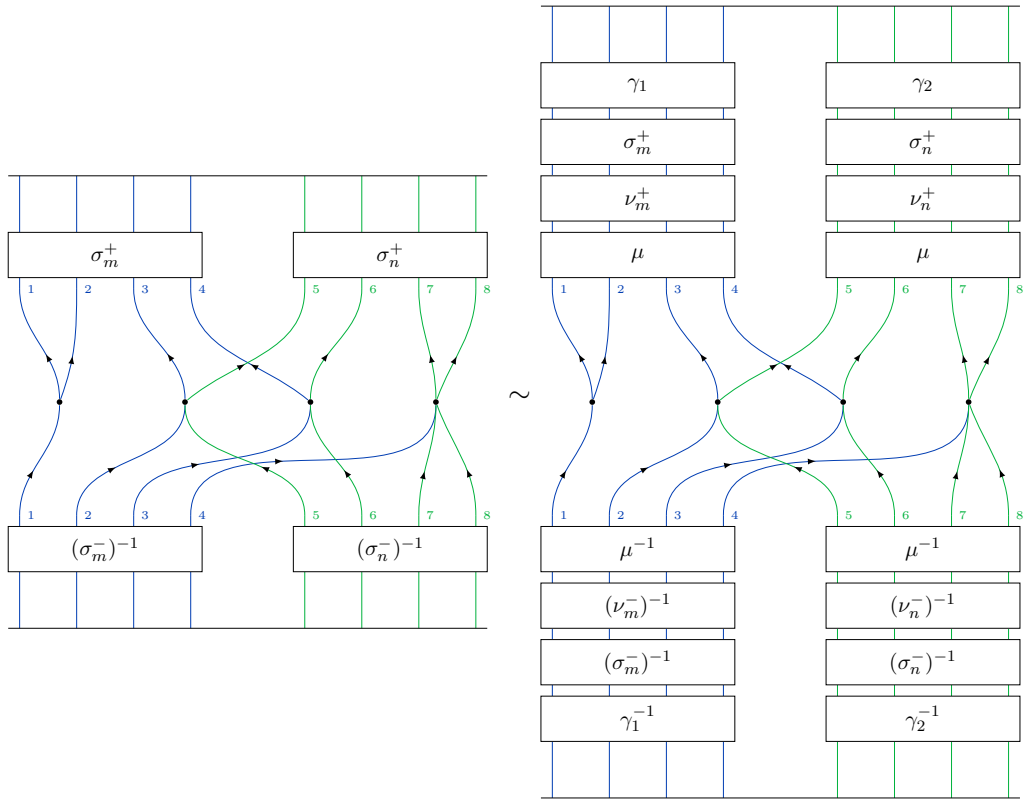
\begin{figure}
	\centering
	\raisebox{-0.5\height}{\scalebox{0.75}{\begin{tikzpicture}[scale=2]
			\definecolor{GREEN}{rgb}{0.0,0.70,0.24}
			\definecolor{BLUE}{rgb}{0.0,0.24,0.70}
			\def \k {3}
			\def \m {3}
			\def \n {3}
			\def \sep {0.5}
			\def \voffset {0.35}
			\pgfmathparse{(\sep*(\m+\n+2)-2*\voffset)/\k};
			\pgfmathsetmacro{\vsep}{\pgfmathresult};
			\pgfmathint{\k-1};
			\pgfmathsetmacro{\kk}{\pgfmathresult};
			\foreach \v in {0,...,\k}
			{
				\pgfmathparse{\v*\vsep+\voffset}
				\node[circle, fill, inner sep=1pt](v\v) at (\pgfmathresult,0) {};
			}
			\foreach \eOutm in {0,...,\m}
			{
				\pgfmathint{\eOutm+1};
				\pgfmathsetmacro{\seOutm}{\pgfmathresult};
				\pgfmathparse{\eOutm*\sep};
				\coordinate (eom\seOutm) at (\pgfmathresult,2);
				\coordinate (eomm\seOutm) at (\pgfmathresult,1);
				\node[right, node distance = 0pt and 0pt, text=BLUE] at (eomm\seOutm) {\tiny \seOutm};
				\draw[draw=BLUE] (eom\seOutm) -- (eomm\seOutm);
			}
			\foreach \eInm in {0,...,\m}
			{
				\pgfmathint{\eInm+1};
				\pgfmathsetmacro{\seInm}{\pgfmathresult};
				\pgfmathparse{\eInm*\sep};
				\coordinate (eim\seInm) at (\pgfmathresult,-2);
				\coordinate (eimm\seInm) at (\pgfmathresult,-1);
				\node[right, node distance = 0pt and 0pt, text=BLUE] at (eimm\seInm) {\tiny \seInm};
				\draw[draw=BLUE] (eim\seInm) -- (eimm\seInm);
			}
			\foreach \eOutn in {0,...,\n}
			{
				\pgfmathint{\m+\eOutn+2};
				\pgfmathsetmacro{\seOutn}{\pgfmathresult};
				\pgfmathparse{\seOutn*\sep};
				\coordinate (eon\seOutn) at (\pgfmathresult,2);
				\coordinate (eonn\seOutn) at (\pgfmathresult,1);
				\node[right, node distance = 0pt and 0pt, text=GREEN] at (eonn\seOutn) {\tiny \seOutn};
				\draw[draw=GREEN] (eon\seOutn) -- (eonn\seOutn);
			}
			\foreach \eInn in {0,...,\n}
			{
				\pgfmathint{\m+\eInn+2};
				\pgfmathsetmacro{\seInn}{\pgfmathresult};
				\pgfmathparse{\seInn*\sep};
				\coordinate (ein\seInn) at (\pgfmathresult,-2);
				\coordinate (einn\seInn) at (\pgfmathresult,-1);
				\node[right, node distance = 0pt and 0pt, text=GREEN] at (einn\seInn) {\tiny \seInn};
				\draw[draw=GREEN] (ein\seInn) -- (einn\seInn);
			}
			\begin{scope}[decoration={markings, mark=at position 0.4 with \arrow{latex}}]
			\draw[draw=BLUE, postaction={decorate}] (v0) to[out=90,in=-90] (eomm1);
			\draw[draw=BLUE, postaction={decorate}] (v0) to[out=70,in=-90] (eomm2);
			\draw[draw=BLUE, postaction={decorate}] (v1) to[out=90,in=-90] (eomm3);
			\draw[draw=BLUE, postaction={decorate}] (v2) to[out=140,in=-90] (eomm4);
			\draw[draw=GREEN, postaction={decorate}] (v1) to[out=40,in=-90] (eonn5);
			\draw[draw=GREEN, postaction={decorate}] (v2) to[out=90,in=-90] (eonn6);
			\draw[draw=GREEN, postaction={decorate}] (v3) to[out=90,in=-90] (eonn7);
			\draw[draw=GREEN, postaction={decorate}] (v3) to[out=70,in=-90] (eonn8);
			\draw[draw=BLUE, postaction={decorate}] (eimm1) to[out=90,in=-90] (v0);
			\draw[draw=BLUE, postaction={decorate}] (eimm2) to[out=90,in=-90] (v1);
			\draw[draw=BLUE, postaction={decorate}] (eimm3) to[out=90,in=-90] (v2);
			\draw[draw=BLUE, postaction={decorate}] (eimm4) to[out=90,in=-90] (v3);
			\draw[draw=GREEN, postaction={decorate}] (einn5) to[out=90,in=-90] (v1);
			\draw[draw=GREEN, postaction={decorate}] (einn6) to[out=90,in=-90] (v2);
			\draw[draw=GREEN, postaction={decorate}] (einn7) to[out=90,in=-90] (v3);
			\draw[draw=GREEN, postaction={decorate}] (einn8) to[out=90,in=-70] (v3);
			\end{scope}
			\draw[fill=white] ($(eomm1)+(-0.1,0.5)$) rectangle node{$\sigma_m^+$} ($(eomm4)+(0.1,0.1)$);
			\draw[fill=white] ($(eonn5)+(-0.1,0.5)$) rectangle node{$\sigma_n^+$} ($(eonn8)+(0.1,0.1)$);
			\draw[fill=white] ($(eimm1)-(0.1,0.5)$) rectangle node{$(\sigma_m^-)^{-1}$} ($(eimm4)-(-0.1,0.1)$);
			\draw[fill=white] ($(einn5)-(0.1,0.5)$) rectangle node{$(\sigma_n^-)^{-1}$} ($(einn8)-(-0.1,0.1)$);
			\draw ($(eom1)-(0.1,0)$) -- ($(eon8)+(0.1,0)$);
			\draw ($(eim1)-(0.1,0)$) -- ($(ein8)+(0.1,0)$);
			\end{tikzpicture}}}
	$\sim$
	\raisebox{-0.5\height}{\scalebox{0.75}{\begin{tikzpicture}[scale=2]
			\definecolor{GREEN}{rgb}{0.0,0.70,0.24}
			\definecolor{BLUE}{rgb}{0.0,0.24,0.70}
			\def \k {3}
			\def \m {3}
			\def \n {3}
			\def \sep {0.5}
			\def \voffset {0.35}
			\pgfmathparse{(\sep*(\m+\n+2)-2*\voffset)/\k};
			\pgfmathsetmacro{\vsep}{\pgfmathresult};
			\pgfmathint{\k-1};
			\pgfmathsetmacro{\kk}{\pgfmathresult};
			\foreach \v in {0,...,\k}
			{
				\pgfmathparse{\v*\vsep+\voffset}
				\node[circle, fill, inner sep=1pt](v\v) at (\pgfmathresult,0) {};
			}
			\foreach \eOutm in {0,...,\m}
			{
				\pgfmathint{\eOutm+1};
				\pgfmathsetmacro{\seOutm}{\pgfmathresult};
				\pgfmathparse{\eOutm*\sep};
				\coordinate (eom\seOutm) at (\pgfmathresult,3.5);
				\coordinate (eomm\seOutm) at (\pgfmathresult,1);
				\node[right, node distance = 0pt and 0pt, text=BLUE] at (eomm\seOutm) {\tiny \seOutm};
				\draw[draw=BLUE] (eom\seOutm) -- (eomm\seOutm);
			}
			\foreach \eInm in {0,...,\m}
			{
				\pgfmathint{\eInm+1};
				\pgfmathsetmacro{\seInm}{\pgfmathresult};
				\pgfmathparse{\eInm*\sep};
				\coordinate (eim\seInm) at (\pgfmathresult,-3.5);
				\coordinate (eimm\seInm) at (\pgfmathresult,-1);
				\node[right, node distance = 0pt and 0pt, text=BLUE] at (eimm\seInm) {\tiny \seInm};
				\draw[draw=BLUE] (eim\seInm) -- (eimm\seInm);
			}
			\foreach \eOutn in {0,...,\n}
			{
				\pgfmathint{\m+\eOutn+2};
				\pgfmathsetmacro{\seOutn}{\pgfmathresult};
				\pgfmathparse{\seOutn*\sep};
				\coordinate (eon\seOutn) at (\pgfmathresult,3.5);
				\coordinate (eonn\seOutn) at (\pgfmathresult,1);
				\node[right, node distance = 0pt and 0pt, text=GREEN] at (eonn\seOutn) {\tiny \seOutn};
				\draw[draw=GREEN] (eon\seOutn) -- (eonn\seOutn);
			}
			\foreach \eInn in {0,...,\n}
			{
				\pgfmathint{\m+\eInn+2};
				\pgfmathsetmacro{\seInn}{\pgfmathresult};
				\pgfmathparse{\seInn*\sep};
				\coordinate (ein\seInn) at (\pgfmathresult,-3.5);
				\coordinate (einn\seInn) at (\pgfmathresult,-1);
				\node[right, node distance = 0pt and 0pt, text=GREEN] at (einn\seInn) {\tiny \seInn};
				\draw[draw=GREEN] (ein\seInn) -- (einn\seInn);
			}
			\begin{scope}[decoration={markings, mark=at position 0.4 with \arrow{latex}}]
			\draw[draw=BLUE, postaction={decorate}] (v0) to[out=90,in=-90] (eomm1);
			\draw[draw=BLUE, postaction={decorate}] (v0) to[out=70,in=-90] (eomm2);
			\draw[draw=BLUE, postaction={decorate}] (v1) to[out=90,in=-90] (eomm3);
			\draw[draw=BLUE, postaction={decorate}] (v2) to[out=140,in=-90] (eomm4);
			\draw[draw=GREEN, postaction={decorate}] (v1) to[out=40,in=-90] (eonn5);
			\draw[draw=GREEN, postaction={decorate}] (v2) to[out=90,in=-90] (eonn6);
			\draw[draw=GREEN, postaction={decorate}] (v3) to[out=90,in=-90] (eonn7);
			\draw[draw=GREEN, postaction={decorate}] (v3) to[out=70,in=-90] (eonn8);
			\draw[draw=BLUE, postaction={decorate}] (eimm1) to[out=90,in=-90] (v0);
			\draw[draw=BLUE, postaction={decorate}] (eimm2) to[out=90,in=-90] (v1);
			\draw[draw=BLUE, postaction={decorate}] (eimm3) to[out=90,in=-90] (v2);
			\draw[draw=BLUE, postaction={decorate}] (eimm4) to[out=90,in=-90] (v3);
			\draw[draw=GREEN, postaction={decorate}] (einn5) to[out=90,in=-90] (v1);
			\draw[draw=GREEN, postaction={decorate}] (einn6) to[out=90,in=-90] (v2);
			\draw[draw=GREEN, postaction={decorate}] (einn7) to[out=90,in=-90] (v3);
			\draw[draw=GREEN, postaction={decorate}] (einn8) to[out=90,in=-70] (v3);
			\end{scope}
			\draw[fill=white] ($(eomm1)+(-0.1,2)$) rectangle node{$\gamma_1$} ($(eomm4)+(0.1,1.6)$);
			\draw[fill=white] ($(eimm1)-(0.1,2)$) rectangle node{$\gamma_1^{-1}$} ($(eimm4)-(-0.1,1.6)$);
			\draw[fill=white] ($(eomm1)+(-0.1,1.5)$) rectangle node{$\sigma_m^+$} ($(eomm4)+(0.1,1.1)$);
			\draw[fill=white] ($(eimm1)-(0.1,1.5)$) rectangle node{$(\sigma_m^-)^{-1}$} ($(eimm4)-(-0.1,1.1)$);
			\draw[fill=white] ($(eomm1)+(-0.1,1)$) rectangle node{$\nu_m^+$} ($(eomm4)+(0.1,.6)$);
			\draw[fill=white] ($(eimm1)-(0.1,1)$) rectangle node{$(\nu_m^-)^{-1}$} ($(eimm4)-(-0.1,0.6)$);
			\draw[fill=white] ($(eomm1)+(-0.1,0.5)$) rectangle node{$\mu$} ($(eomm4)+(0.1,0.1)$);
			\draw[fill=white] ($(eimm1)-(0.1,0.5)$) rectangle node{$\mu^{-1}$} ($(eimm4)-(-0.1,0.1)$);
			
			\draw[fill=white] ($(eonn5)+(-0.1,2)$) rectangle node{$\gamma_2$} ($(eonn8)+(0.1,1.6)$);
			\draw[fill=white] ($(einn5)-(0.1,2)$) rectangle node{$\gamma_2^{-1}$} ($(einn8)-(-0.1,1.6)$);
			\draw[fill=white] ($(eonn5)+(-0.1,1.5)$) rectangle node{$\sigma_n^+$} ($(eonn8)+(0.1,1.1)$);
			\draw[fill=white] ($(einn5)-(0.1,1.5)$) rectangle node{$(\sigma_n^-)^{-1}$} ($(einn8)-(-0.1,1.1)$);
			\draw[fill=white] ($(eonn5)+(-0.1,1)$) rectangle node{$\nu_n^+$} ($(eonn8)+(0.1,0.6)$);
			\draw[fill=white] ($(einn5)-(0.1,1)$) rectangle node{$(\nu_n^-)^{-1}$} ($(einn8)-(-0.1,0.6)$);
			\draw[fill=white] ($(eonn5)+(-0.1,.5)$) rectangle node{$\mu$} ($(eonn8)+(0.1,0.1)$);
			\draw[fill=white] ($(einn5)-(0.1,.5)$) rectangle node{$\mu^{-1}$} ($(einn8)-(-0.1,0.1)$);
			\draw ($(eom1)-(0.1,0)$) -- ($(eon8)+(0.1,0)$);
			\draw ($(eim1)-(0.1,0)$) -- ($(ein8)+(0.1,0)$);
			\end{tikzpicture}}}
	\caption{Diagrammatic description of the permutation equivalence defining the double cosets in \eqref{eq: Two Color Double Coset Quotient}.}
	\label{fig: Two Colour Four Sigma Double Coset Graph}
\end{figure}

As before, the counting of equivalence classes \eqref{eqn: Two Color Graph Permutation Equivalence} is equivalent to the counting of double cosets. In this case the relevant double coset is
\begin{equation}
G(\vec{m}^+,\vec{m}^-;\vec{n}^+,\vec{n}^-) \left\backslash \qty(S_m^+ \times S_m^- \times S_n^+ \times S_n^- )\right/ (\diag(S_m) \times \diag(S_n)), \label{eq: Two Color Double Coset Quotient}
\end{equation}
where $G(\vec{m}^+,\vec{m}^-;\vec{n}^+,\vec{n}^-)$ is the subgroup generated by elements of the form
\begin{align}
(\rho^+_m(\mu),\rho^-_m(\mu), \rho^+_n(\mu), \rho^-_n(\mu)) \qq{and} (\gamma^+_m(\nu^+_m),\gamma^-_m(\nu^-_m),\gamma^+_n(\nu^+_n),\gamma^-_n(\nu^-_n)).
\end{align}
The associated picture (Figure \ref{fig: Two Colour Four Sigma Double Coset Graph}) of the equivalence has four permutations $\sigma_m^+, \sigma_m^-, \sigma_n^+, \sigma_n^-$. The number of double cosets is counted by
\begin{align}
N(\vec{m}^+,\vec{m}^-;\vec{n}^+,\vec{n}^-) &=\frac{1}{|G(\vec{m}^+,\vec{m}^-;\vec{n}^+,\vec{n}^-)||S_n||S_m|} \times \nonumber\\
&\sum_{\substack{\mu \in G_V\\{(\nu^+_m, \nu^-_m, \nu^+_n,\nu^-_n) \in S_{\vec{m}^+}\times S_{\vec{m}^-}\times S_{\vec{n}^+}\times S_{\vec{n}^-}}\\ \gamma_1 \in S_m, \gamma_2 \in S_n\\(\sigma_m^+, \sigma_m^-, \sigma_n^+, \sigma_n^-) \in S_m^+ \times S_m^- \times S_n^+ \times S_n -}} \begin{aligned}[t]
&\delta((\sigma_m^+)^{-1}\rho^+_m(\mu)\gamma^+_m(\nu_m^+)\sigma_m^+\gamma_1^{-1}) \\
&\delta((\sigma_m^-)^{-1}\rho^-_m(\mu)\gamma^-_m(\nu_m^-)\sigma_m^-\gamma_1^{-1}) \\
&\delta((\sigma_n^+)^{-1}\rho^+_n(\mu)\gamma^+_n(\nu_n^+)\sigma_n^+\gamma_2^{-1}) \\
&\delta((\sigma_n^-)^{-1}\rho^-_n(\mu)\gamma^-_n(\nu_n^-)\sigma_n^-\gamma_2^{-1})
\end{aligned} \label{eqn: Burnsides lemma double coset twocolor}
\end{align}
and the connection to the equivalence in \eqref{eqn: Two Color Graph Permutation Equivalence} can be derived in a fashion completely analogous to the one-color case.

In general, the number of double cosets in $H_1 \left\backslash G \right/ H_2$ can be written \cite{TensorModelBranchedCovers}
\begin{equation}
|H_1 \left\backslash G \right/ H_2| = \frac{1}{|H_1||H_2|}\sum_{C} Z_C^{H_1\rightarrow G}Z_C^{H_2\rightarrow G}\Sym(C),
\end{equation}
where the sum is over conjugacy classes of $G$. The symbols $Z_C^{G\rightarrow H_1}, Z_C^{G\rightarrow H_2}$ denote the number of elements of $H_1$ and $H_2$ in the conjugacy class $C$ of $G$, respectively. $\Sym(C)$ is the number of elements in $G$ which commute with an element in $C$. For the double coset relevant to the one-colored graphs we have
\begin{align} \nonumber \label{eq: N(m+,m-) Cycle Index Formula}
N(\vec{m}^+,\vec{m}^-) &= \frac{1}{|G(\vec{m}^+,\vec{m}^-)||S_m|} \sum_{p \vdash m} Z_{p,p}^{G(\vec{m}^+,\vec{m}^-)}Z_{p,p}^{\diag(S_m)} \Sym(p)^2 \\
&=\frac{1}{|G(\vec{m}^+,\vec{m}^-)|} \sum_{p \vdash m} Z_{p,p}^{G(\vec{m}^+,\vec{m}^-)} \Sym(p).
\end{align} 
with
\begin{equation}
\Sym(p) = \prod_{i=1}^m p_i!i^{p_i}, \quad \sum_i ip_i = m.
\end{equation}
For a permutation subgroup $H \subset G_1 \times G_2$, $Z_{p,q}^{H}$ is the number of permutations $(h_1, h_2) \in H$ with cycle structure $p$ in the first slot and $q$ in the second slot. The conjugacy classes of permutation groups have fixed cycle structures and are labeled by integer partitions $p$ of $m$. The last equality follows from
\begin{equation}
Z_{p,p}^{\diag(S_m)}=Z_{p}^{S_m} = \frac{|S_m|}{\Sym(p)}.
\end{equation}
That this counting formula is equivalent to the counting due to Burnside's lemma \eqref{eqn: Burnsides lemma double coset one color} is understood as follows. Organize the sum $\sum_{\gamma \in \diag(S_m)}$ into a sum over conjugacy classes of $S_m^+ \times S_m^-$, and a sum over elements in the conjugacy class,
\begin{equation}
\sum_{\gamma \in \diag(S_m)} = \sum_{p \vdash m} \hspace{4pt} \sum_{\gamma \in C_p \times C_p}.
\end{equation}
The Kronecker delta $\delta(\sigma_1^{-1}\rho^+(\mu)\gamma^+(\nu^+)\sigma_1 \gamma^{-1})$ vanishes unless $\rho^+(\mu)\gamma^+(\nu^+)$ is in the same conjugacy class as $\gamma^{-1}$. Similarly for the the second Kronecker delta. The number of elements $(\rho^+(\mu)\gamma^+(\nu^+),\rho^-(\mu)\gamma^-(\nu^-))$ in the conjugacy class $C_p \times C_p$ of $S_m^+ \times S_m^-$ is the definition of the coefficients $Z_{p,p}^{G(\vec{m}^+,\vec{m}^-)}$. The number of elements $(\gamma,\gamma)$ in the conjugacy class $C_p \times C_p$ is $Z_{p,p}^{\diag(S_m)}$. Given an element in $G(\vec{m}^+,\vec{m}^-)$ and an element in $\diag(S_m)$ in the same conjugacy class, there exists at least one element $(\sigma_1,\sigma_2)$ which relates the two by conjugation. Therefore, the Kronecker delta is non-zero at least $ Z_{p,p}^{G(\vec{m}^+,\vec{m}^-)}Z_{p,p}^{\diag(S_m)}$ times for each conjugacy class $C_p \times C_p$. In equations we have
\begin{align} \nonumber
&\sum_{\substack{\mu \in G_V, \nu^+ \in S_{\vec{m}^+} \\ \nu^- \in S_{\vec{m}^-},\gamma \in S_m}} \sum_{\sigma_1, \sigma_2 \in S_m}
\begin{aligned}[t]
&\delta(\sigma_1^{-1}\rho^+(\mu)\gamma^+(\nu^+)\sigma_1\gamma^{-1})
\delta(\sigma_2^{-1}\rho^-(\mu)\gamma^-(\nu^-)\sigma_2\gamma^{-1})
\end{aligned}\\ \nonumber
&= \sum_{p \vdash m} \sum_{\sigma_1, \sigma_2 \in S_m}
\delta(\sigma_1^{-1}G_p^+\sigma_1\gamma^{-1}_p)
\delta(\sigma_2^{-1}G_p^-\sigma_2\gamma^{-1}_p)
Z_{p,p}^{G(\vec{m}^+,\vec{m}^-)}Z_{p,p}^{\diag(S_m)} \\
&= \sum_{p \vdash m} Z_{p,p}^{G(\vec{m}^+,\vec{m}^-)}Z_{p,p}^{\diag(S_m)} \Sym(p)^2.
\end{align}
where $(G_p^+,G_p^-)$ is an arbitrary element of $G(\vec{m}^+,\vec{m}^-)$ in the conjugacy class $C_p \times C_p$ and similarly for $\gamma^{-1}_p$ in $\diag(S_m)$. To understand the last equality, consider the case where $(\sigma_1^{-1}G_p^+\sigma_1, \sigma_2^{-1}G_p^-\sigma_2) = (\gamma_p, \gamma_p)$. If $\sigma_1',\sigma_2'$ commute with $G_p^+$ and $G_p^-$ respectively, then
\begin{equation}
((\sigma_1'\sigma_1)^{-1}G_p^+\sigma_1'\sigma_1,(\sigma_2'\sigma_2)^{-1}G_p^-\sigma_2'\sigma_2 ) = (\sigma_1^{-1}G_p^+\sigma_1, \sigma_2^{-1}G_p^-\sigma_2) = (\gamma_p, \gamma_p).
\end{equation}
The function $\Sym(p)$ is the number of elements in $S_m$ which commute with $G_p$. This only depends on the conjugacy class $C_p$, or equivalently, the partition $p$.

Similarly, for the two-colored graphs, the number of double cosets is
\begin{align} \nonumber
N(&\vec{m}^+,\vec{m}^-;\vec{n}^+,\vec{n}^-) \\ \nonumber
&= \frac{1}{|G(\vec{m}^+,\vec{m}^-;\vec{n}^+,\vec{n}^-)||S_m||S_n|} \sum_{\substack{p \vdash m \\ q \vdash n}} \begin{aligned}[t]
Z_{p,p,q,q}^{G(\vec{m}^+,\vec{m}^-;\vec{n}^+,\vec{n}^-)}&Z_{p,p,q,q}^{\diag(S_m)\times \diag(S_n)} \times \\
&\Sym(p)^2\Sym(q)^2,
\end{aligned} \\
&=\frac{1}{|G(\vec{m}^+,\vec{m}^-;\vec{n}^+,\vec{n}^-)|}\sum_{\substack{p \vdash m \\ q \vdash n}} Z_{p,p,q,q}^{G(\vec{m}^+,\vec{m}^-;\vec{n}^+,\vec{n}^-)}\Sym(q) \Sym(p).
\end{align}
and the connection to the counting in \eqref{eqn: Burnsides lemma double coset twocolor} is completely analogous. Here $Z_{p_1,p_2,q_3,q_4}^{H}$ is the number of permutations $(h_1,h_2,h_3,h_4) \in H \subset G_1 \times G_2 \times G_3 \times G_4$ with cycle structure given by $p_1,p_2,p_3,p_4$, respectively. The last equality follows from
\begin{equation}
Z_{p,p,q,q}^{\diag(S_m)\times \diag(S_n)} = Z_{p,q}^{S_m \times S_n} = \frac{|S_m||S_n|}{\Sym(p)\Sym(q)}.
\end{equation}

In both cases, the number of double cosets is determined by functions $Z^H_{p}$, which count the number of elements in the conjugacy class labeled by $p$. They are commonly packaged into generating functions, called cycle indices,
\begin{equation}
Z^H(x_1,x_2,\dots) = \frac{1}{|H|} \sum_p Z_p^H \mathbf{x}^p, 
\end{equation}
such that
\begin{equation}
\frac{1}{|H|}Z^H_p = \text{Coefficient}(Z^H(x_1,x_2,\dots), \mathbf{x}^p).
\end{equation}
For a partition $p \vdash l$, $\mathbf{x}^p$ is shorthand for the degree $l$ monomial $x_1^{p_1}x_2^{p_2}\dots$, where $\sum_j jp_j = l$. We are interested in the cycle indices $Z^{G(\vec{m}^+,\vec{m}^-)}$ and $Z^{G(\vec{m}^+,\vec{m}^-;\vec{n}^+,\vec{n}^-)}$.

To efficiently describe the cycle indices we define the following compact notation. A vector partition
\begin{equation}
(\vec{m}^+,\vec{m}^-;\vec{n}^+,\vec{n}^-) = (m_1^+, m_1^-;n_1^+, n_1^-) + \dots +  (m_k^+, m_k^-; n_k^+ , n_k^-).
\end{equation}
can equivalently be described using a generalization of exponential notation for partitions,
\begin{equation}
(\vec{m}^+,\vec{m}^-; \vec{n}^+,\vec{n}^-) = p_{0001}(0,0;0,1) + p_{0010}(0,0;1,0) + \dots = \sum_{v^{(4)}} p_{v^{(4)}} v^{(4)},
\end{equation}
where the sum is over ordered lists of four integers $v^{(4)}$ with at least one non-zero entry and $p_{v^{(4)}}$ is the number of times it appears in the vector partition. Using this notation, we can write the symmetry group as
\begin{equation}
G(\vec{m}^+,\vec{m}^-;\vec{n}^+,\vec{n}^-) = \bigtimes_{v^{(4)}} S_{p_{v^{(4)}}}[S_{v^{(4)}}], \quad v^{(4)} \in \mathbb{N}^{\times 4}\left\backslash \{0,0,0,0\}\right.
\end{equation}
where
\begin{equation}
S_{v^{(4)}} = S_{v^{(4)}_1} \times S_{v^{(4)}_3} \times S_{v^{(4)}_2} \times S_{v^{(4)}_4}.
\end{equation}
and $S_{p_{v^{(4)}}}[S_{v^{(4)}}]$ is the wreath product. For direct product groups the cycle index factorizes,
\begin{equation}
Z^{G_1 \times G_2}(\mathbf{x}) = Z^{{G_1}}(\mathbf{x})Z^{{G_2}}(\mathbf{x}). \label{eqn: Cycle Index Product Group}
\end{equation}
It is convenient to formally think of $\mathbf{x} = (x_1,x_2,\dots)$ as a countably infinite number of variables. In practice it truncates at $x_c$, where $c$ is the size of the largest cycle in ${G_1 \times G_2}$.

A general wreath product $S_l[S_v]$ is a semi-direct product 
\begin{equation}
S_l \ltimes \underbrace{(S_v \times \dots \times S_v)}_{\text{l factors}},
\end{equation}
which is naturally viewed as a subgroup of $S_{lv}$.
For example, elements of $S_4[S_2]$ correspond to diagrams
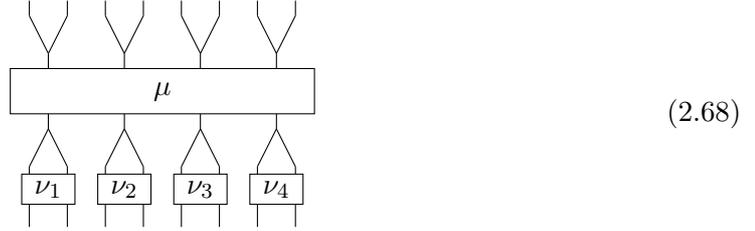
\begin{equation}
\vcenter{\hbox{\begin{tikzpicture}[scale=1]
		\def \k {3}
		\def \m {7}
		\def \sep {0.5}
		\def \voffset {0.25}
		\pgfmathparse{2*\sep};
		\pgfmathsetmacro{\vsep}{\pgfmathresult};
		\pgfmathint{\k-1};
		\pgfmathsetmacro{\kk}{\pgfmathresult};
		\foreach \v in {0,...,\k}
		{
			\pgfmathparse{\v*\vsep+\voffset}
			\coordinate (v\v) at (\pgfmathresult,.5) {};
		}
		\foreach \v in {0,...,\k}
		{
			\pgfmathparse{\v*\vsep+\voffset}
			\coordinate (w\v) at (\pgfmathresult,-.5) {};
			\draw[] (w\v) -- (v\v);
		}
		\foreach \eOutm in {0,...,\m}
		{
			\pgfmathint{\eOutm+1};
			\pgfmathsetmacro{\seOutm}{\pgfmathresult};
			\pgfmathparse{\eOutm*\sep};
			\coordinate (eom\seOutm) at (\pgfmathresult,1.2);
		}
		\foreach \eOutm in {0,...,\m}
		{
			\pgfmathint{\eOutm+1};
			\pgfmathsetmacro{\seOutm}{\pgfmathresult};
			\pgfmathparse{\eOutm*\sep};
			\coordinate (eomm\seOutm) at (\pgfmathresult,1);
			\draw[] (eom\seOutm) -- (eomm\seOutm);
		}
		\foreach \eInm in {0,...,\m}
		{
			\pgfmathint{\eInm+1};
			\pgfmathsetmacro{\seInm}{\pgfmathresult};
			\pgfmathparse{\eInm*\sep};
			\coordinate (eim\seInm) at (\pgfmathresult,-1.8);
		}
		\foreach \eInm in {0,...,\m}
		{
			\pgfmathint{\eInm+1};
			\pgfmathsetmacro{\seInm}{\pgfmathresult};
			\pgfmathparse{\eInm*\sep};
			\coordinate (eimm\seInm) at (\pgfmathresult,-1);
			\draw[] (eim\seInm) -- (eimm\seInm);
		}
		\begin{scope}[decoration={markings}]
		\draw[postaction={decorate}] (v0) -- (eomm1);
		\draw[postaction={decorate}] (v0) -- (eomm2);
		\draw[postaction={decorate}] (v1) -- (eomm3);
		\draw[postaction={decorate}] (v1) -- (eomm4);
		\draw[postaction={decorate}] (v2) -- (eomm5);
		\draw[postaction={decorate}] (v2) -- (eomm6);
		\draw[postaction={decorate}] (v3) -- (eomm7);
		\draw[postaction={decorate}] (v3) -- (eomm8);
		\draw[postaction={decorate}] (eimm1)  -- (w0);
		\draw[postaction={decorate}] (eimm2)  -- (w0);
		\draw[postaction={decorate}] (eimm3)  -- (w1);
		\draw[postaction={decorate}] (eimm4)  -- (w1);
		\draw[postaction={decorate}] (eimm5)  -- (w2);
		\draw[postaction={decorate}] (eimm6)  -- (w2);
		\draw[postaction={decorate}] (eimm7)  -- (w3);
		\draw[postaction={decorate}] (eimm8)  -- (w3);
		\end{scope}
		\draw[fill=white] ($(v0)+(-0.5,-0.2)$) rectangle node{$\mu$} ($(w3)+(0.5,0.2)$);
		\draw[fill=white] ($(eimm1)-(0.1,.5)$) rectangle node{$\nu_1$} ($(eimm2)-(-0.1,0.1)$);
		\draw[fill=white] ($(eimm3)-(0.1,.5)$) rectangle node{$\nu_2$} ($(eimm4)-(-0.1,0.1)$);
		\draw[fill=white] ($(eimm5)-(0.1,.5)$) rectangle node{$\nu_3$} ($(eimm6)-(-0.1,0.1)$);
		\draw[fill=white] ($(eimm7)-(0.1,.5)$) rectangle node{$\nu_4$} ($(eimm8)-(-0.1,0.1)$);
		\end{tikzpicture}}}\label{eqn: wreath_diagram}
\end{equation}
with $\nu_i \in S_2, \mu \in S_4$. The vertices are concatenations of edges and $\mu$ permutes the resulting collections
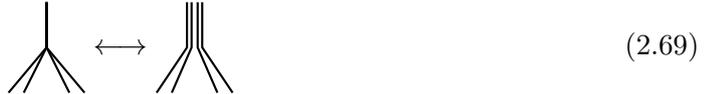
\begin{equation}
\vcenter{\hbox{\scalebox{2}[2]{\begin{tikzpicture}[scale=1]
			\draw (-0.25,-0.3) -- (0,0) -- (-0.0,0.3);
			\draw (-0.15,-0.3) -- (0,0) -- (-0.0,0.3);
			\draw (0.15,-0.3) -- (0,0) -- (0.0,0.3);
			\draw (0.25,-0.3) -- (0,0) -- (0.0,0.3);
			\end{tikzpicture}}}}
\longleftrightarrow
\vcenter{\hbox{\scalebox{2}[2]{\begin{tikzpicture}[scale=1]
			\draw (-0.25,-0.3) -- (-0.05,0) -- (-0.05,0.3);
			\draw (-0.15,-0.3) -- (-0.02,0) -- (-0.02,0.3);
			\draw (0.15,-0.3) -- (0.02,0) -- (0.02,0.3);
			\draw (0.25,-0.3) -- (0.05,0) -- (0.05,0.3);
			\end{tikzpicture}}}}
\end{equation}
The cycle index of a wreath product $S_l[S_v]$ is
\begin{equation}
Z^{S_l[S_v]}(x_1,\dots,x_{lv}) = Z^{S_l}(Z_1^{S_v}(\mathbf{x}), \dots, Z_{l}^{S_v}(\mathbf{x})),
\end{equation}
where
\begin{equation}
Z_i^{S_v}(\mathbf{x}) = Z^{S_v}(x_{1\cdot i}, x_{2 \cdot i}, \dots,x_{v\cdot i}),
\end{equation}
is given by multiplying the labels on $x_1,x_2, \dots$ by $i$ in the cycle index. This result originally proved by Pólya in \cite{Polya} says: for a permutation \eqref{eqn: wreath_diagram} with $\mu$ fixed to have cycle structure $p \vdash l$, the contribution to the cycle index as we sum over $S_v^{\times l}$ is \cite{Constantine}
\begin{equation}
\frac{1}{\abs{S_l}}Z_1^{S_v}(\mathbf{x})^{p_1} \dots Z_l^{S_v}(\mathbf{x})^{p_l}.
\end{equation}

In the one-color case we are interested in counting cycles of wreath products of the form $S_l[S_{v^+} \times S_{v^-}]$. This wreath product is most naturally thought of as a subgroup of $S_{lv^+ + lv^-}$. However, elements in $S_l[S_{v^+} \times S_{v^-}]$ are determined by $\mu \in S_l, \nu^+_i \in S_{v^+}, \nu^-_i \in S_{v^-}$ according to the diagram (in the case of $S_4[S_2 \times S_2]$)
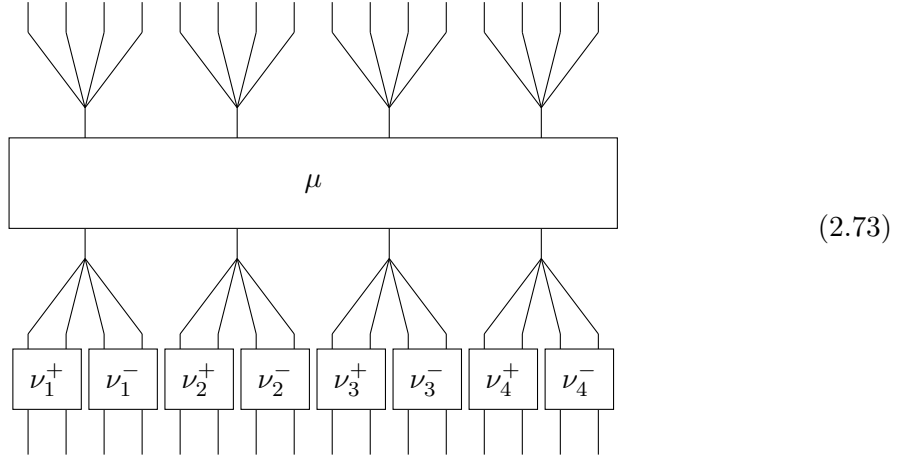
\begin{equation}
\vcenter{\hbox{\begin{tikzpicture}[scale=2]
		\def \k {3}
		\def \m {15}
		\def \sep {0.25}
		\def \voffset {0.375}
		\pgfmathparse{(\sep*(\m)-2*\voffset)/\k};
		\pgfmathsetmacro{\vsep}{\pgfmathresult};
		\pgfmathint{\k-1};
		\pgfmathsetmacro{\kk}{\pgfmathresult};
		\foreach \v in {0,...,\k}
		{
			\pgfmathparse{\v*\vsep+\voffset}
			\coordinate (v\v) at (\pgfmathresult,.5) {};
		}
		\foreach \v in {0,...,\k}
		{
			\pgfmathparse{\v*\vsep+\voffset}
			\coordinate (w\v) at (\pgfmathresult,-.5) {};
			\draw[] (w\v) -- (v\v);
		}
		\foreach \eOutm in {0,...,\m}
		{
			\pgfmathint{\eOutm+1};
			\pgfmathsetmacro{\seOutm}{\pgfmathresult};
			\pgfmathparse{\eOutm*\sep};
			\coordinate (eom\seOutm) at (\pgfmathresult,1.2);
		}
		\foreach \eOutm in {0,...,\m}
		{
			\pgfmathint{\eOutm+1};
			\pgfmathsetmacro{\seOutm}{\pgfmathresult};
			\pgfmathparse{\eOutm*\sep};
			\coordinate (eomm\seOutm) at (\pgfmathresult,1);
			\draw[] (eom\seOutm) -- (eomm\seOutm);
		}
		\foreach \eInm in {0,...,\m}
		{
			\pgfmathint{\eInm+1};
			\pgfmathsetmacro{\seInm}{\pgfmathresult};
			\pgfmathparse{\eInm*\sep};
			\coordinate (eim\seInm) at (\pgfmathresult,-1.8);
		}
		\foreach \eInm in {0,...,\m}
		{
			\pgfmathint{\eInm+1};
			\pgfmathsetmacro{\seInm}{\pgfmathresult};
			\pgfmathparse{\eInm*\sep};
			\coordinate (eimm\seInm) at (\pgfmathresult,-1);
			\draw[] (eim\seInm) -- (eimm\seInm);
		}
		\begin{scope}[decoration={markings}]
		\draw[postaction={decorate}] (v0) -- (eomm1);
		\draw[postaction={decorate}] (v0) -- (eomm2);
		\draw[postaction={decorate}] (v0) -- (eomm3);
		\draw[postaction={decorate}] (v0) -- (eomm4);
		\draw[postaction={decorate}] (v1) -- (eomm5);
		\draw[postaction={decorate}] (v1) -- (eomm6);
		\draw[postaction={decorate}] (v1) -- (eomm7);
		\draw[postaction={decorate}] (v1) -- (eomm8);
		\draw[postaction={decorate}] (v2) -- (eomm9);
		\draw[postaction={decorate}] (v2) -- (eomm10);
		\draw[postaction={decorate}] (v2) -- (eomm11);
		\draw[postaction={decorate}] (v2) -- (eomm12);
		\draw[postaction={decorate}] (v3) -- (eomm13);
		\draw[postaction={decorate}] (v3) -- (eomm14);
		\draw[postaction={decorate}] (v3) -- (eomm15);
		\draw[postaction={decorate}] (v3) -- (eomm16);
		\draw[postaction={decorate}] (eimm1)  -- (w0);
		\draw[postaction={decorate}] (eimm2)  -- (w0);
		\draw[postaction={decorate}] (eimm3)  -- (w0);
		\draw[postaction={decorate}] (eimm4)  -- (w0);
		\draw[postaction={decorate}] (eimm5)  -- (w1);
		\draw[postaction={decorate}] (eimm6)  -- (w1);
		\draw[postaction={decorate}] (eimm7)  -- (w1);
		\draw[postaction={decorate}] (eimm8)  -- (w1);
		\draw[postaction={decorate}] (eimm9)  -- (w2);
		\draw[postaction={decorate}] (eimm10)  -- (w2);
		\draw[postaction={decorate}] (eimm11)  -- (w2);
		\draw[postaction={decorate}] (eimm12)  -- (w2);
		\draw[postaction={decorate}] (eimm13)  -- (w3);
		\draw[postaction={decorate}] (eimm14)  -- (w3);
		\draw[postaction={decorate}] (eimm15)  -- (w3);
		\draw[postaction={decorate}] (eimm16)  -- (w3);
		\end{scope}
		\draw[fill=white] ($(v0)+(-0.5,-0.2)$) rectangle node{$\mu$} ($(w3)+(0.5,0.2)$);
		\draw[fill=white] ($(eimm1)-(0.1,.5)$) rectangle node{$\nu_1^+$} ($(eimm2)-(-0.1,0.1)$);
		\draw[fill=white] ($(eimm3)-(0.1,.5)$) rectangle node{$\nu_1^-$} ($(eimm4)-(-0.1,0.1)$);
		\draw[fill=white] ($(eimm5)-(0.1,.5)$) rectangle node{$\nu_2^+$} ($(eimm6)-(-0.1,0.1)$);
		\draw[fill=white] ($(eimm7)-(0.1,.5)$) rectangle node{$\nu_2^-$} ($(eimm8)-(-0.1,0.1)$);
		\draw[fill=white] ($(eimm9)-(0.1,.5)$) rectangle node{$\nu_3^+$} ($(eimm10)-(-0.1,0.1)$);
		\draw[fill=white] ($(eimm11)-(0.1,.5)$) rectangle node{$\nu_3^-$} ($(eimm12)-(-0.1,0.1)$);
		\draw[fill=white] ($(eimm13)-(0.1,.5)$) rectangle node{$\nu_4^+$} ($(eimm14)-(-0.1,0.1)$);
		\draw[fill=white] ($(eimm15)-(0.1,.5)$) rectangle node{$\nu_4^-$} ($(eimm16)-(-0.1,0.1)$);
		\end{tikzpicture}}}\label{eqn: wreath_diagram2}
\end{equation}
which can be factorized as
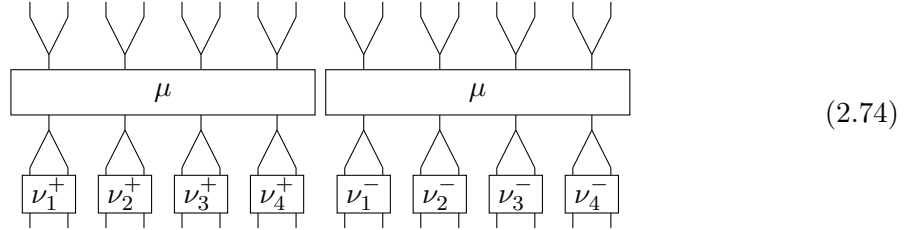
\begin{equation}
\vcenter{\hbox{{\begin{tikzpicture}[scale=1]
			\def \k {3}
			\def \m {7}
			\def \sep {0.5}
			\def \voffset {0.25}
			\pgfmathparse{(\sep*(\m)-2*\voffset)/\k};
			\pgfmathsetmacro{\vsep}{\pgfmathresult};
			\pgfmathint{\k-1};
			\pgfmathsetmacro{\kk}{\pgfmathresult};
			\foreach \v in {0,...,\k}
			{
				\pgfmathparse{\v*\vsep+\voffset}
				\coordinate (v\v) at (\pgfmathresult,.5) {};
			}
			\foreach \v in {0,...,\k}
			{
				\pgfmathparse{\v*\vsep+\voffset}
				\coordinate (w\v) at (\pgfmathresult,-.5) {};
				\draw[] (w\v) -- (v\v);
			}
			\foreach \eOutm in {0,...,\m}
			{
				\pgfmathint{\eOutm+1};
				\pgfmathsetmacro{\seOutm}{\pgfmathresult};
				\pgfmathparse{\eOutm*\sep};
				\coordinate (eom\seOutm) at (\pgfmathresult,1.2);
			}
			\foreach \eOutm in {0,...,\m}
			{
				\pgfmathint{\eOutm+1};
				\pgfmathsetmacro{\seOutm}{\pgfmathresult};
				\pgfmathparse{\eOutm*\sep};
				\coordinate (eomm\seOutm) at (\pgfmathresult,1);
				\draw[] (eom\seOutm) -- (eomm\seOutm);
			}
			\foreach \eInm in {0,...,\m}
			{
				\pgfmathint{\eInm+1};
				\pgfmathsetmacro{\seInm}{\pgfmathresult};
				\pgfmathparse{\eInm*\sep};
				\coordinate (eim\seInm) at (\pgfmathresult,-1.8);
			}
			\foreach \eInm in {0,...,\m}
			{
				\pgfmathint{\eInm+1};
				\pgfmathsetmacro{\seInm}{\pgfmathresult};
				\pgfmathparse{\eInm*\sep};
				\coordinate (eimm\seInm) at (\pgfmathresult,-1);
				\draw[] (eim\seInm) -- (eimm\seInm);
			}
			\begin{scope}[decoration={markings}]
			\draw[postaction={decorate}] (v0) -- (eomm1);
			\draw[postaction={decorate}] (v0) -- (eomm2);
			\draw[postaction={decorate}] (v1) -- (eomm3);
			\draw[postaction={decorate}] (v1) -- (eomm4);
			\draw[postaction={decorate}] (v2) -- (eomm5);
			\draw[postaction={decorate}] (v2) -- (eomm6);
			\draw[postaction={decorate}] (v3) -- (eomm7);
			\draw[postaction={decorate}] (v3) -- (eomm8);
			\draw[postaction={decorate}] (eimm1)  -- (w0);
			\draw[postaction={decorate}] (eimm2)  -- (w0);
			\draw[postaction={decorate}] (eimm3)  -- (w1);
			\draw[postaction={decorate}] (eimm4)  -- (w1);
			\draw[postaction={decorate}] (eimm5)  -- (w2);
			\draw[postaction={decorate}] (eimm6)  -- (w2);
			\draw[postaction={decorate}] (eimm7)  -- (w3);
			\draw[postaction={decorate}] (eimm8)  -- (w3);
			\end{scope}
			\draw[fill=white] ($(v0)+(-0.5,-0.2)$) rectangle node{$\mu$} ($(w3)+(0.5,0.2)$);
			\draw[fill=white] ($(eimm1)-(0.1,.6)$) rectangle node{$\nu_1^+$} ($(eimm2)-(-0.1,0.1)$);
			\draw[fill=white] ($(eimm3)-(0.1,.6)$) rectangle node{$\nu_2^+$} ($(eimm4)-(-0.1,0.1)$);
			\draw[fill=white] ($(eimm5)-(0.1,.6)$) rectangle node{$\nu_3^+$} ($(eimm6)-(-0.1,0.1)$);
			\draw[fill=white] ($(eimm7)-(0.1,.6)$) rectangle node{$\nu_4^+$} ($(eimm8)-(-0.1,0.1)$);
			\end{tikzpicture}}~{\begin{tikzpicture}[scale=1]
			\def \k {3}
			\def \m {7}
			\def \sep {0.5}
			\def \voffset {0.25}
			\pgfmathparse{(\sep*(\m)-2*\voffset)/\k};
			\pgfmathsetmacro{\vsep}{\pgfmathresult};
			\pgfmathint{\k-1};
			\pgfmathsetmacro{\kk}{\pgfmathresult};
			\foreach \v in {0,...,\k}
			{
				\pgfmathparse{\v*\vsep+\voffset}
				\coordinate (v\v) at (\pgfmathresult,.5) {};
			}
			\foreach \v in {0,...,\k}
			{
				\pgfmathparse{\v*\vsep+\voffset}
				\coordinate (w\v) at (\pgfmathresult,-.5) {};
				\draw[] (w\v) -- (v\v);
			}
			\foreach \eOutm in {0,...,\m}
			{
				\pgfmathint{\eOutm+1};
				\pgfmathsetmacro{\seOutm}{\pgfmathresult};
				\pgfmathparse{\eOutm*\sep};
				\coordinate (eom\seOutm) at (\pgfmathresult,1.2);
			}
			\foreach \eOutm in {0,...,\m}
			{
				\pgfmathint{\eOutm+1};
				\pgfmathsetmacro{\seOutm}{\pgfmathresult};
				\pgfmathparse{\eOutm*\sep};
				\coordinate (eomm\seOutm) at (\pgfmathresult,1);
				\draw[] (eom\seOutm) -- (eomm\seOutm);
			}
			\foreach \eInm in {0,...,\m}
			{
				\pgfmathint{\eInm+1};
				\pgfmathsetmacro{\seInm}{\pgfmathresult};
				\pgfmathparse{\eInm*\sep};
				\coordinate (eim\seInm) at (\pgfmathresult,-1.8);
			}
			\foreach \eInm in {0,...,\m}
			{
				\pgfmathint{\eInm+1};
				\pgfmathsetmacro{\seInm}{\pgfmathresult};
				\pgfmathparse{\eInm*\sep};
				\coordinate (eimm\seInm) at (\pgfmathresult,-1);
				\draw[] (eim\seInm) -- (eimm\seInm);
			}
			\begin{scope}[decoration={markings}]
			\draw[postaction={decorate}] (v0) -- (eomm1);
			\draw[postaction={decorate}] (v0) -- (eomm2);
			\draw[postaction={decorate}] (v1) -- (eomm3);
			\draw[postaction={decorate}] (v1) -- (eomm4);
			\draw[postaction={decorate}] (v2) -- (eomm5);
			\draw[postaction={decorate}] (v2) -- (eomm6);
			\draw[postaction={decorate}] (v3) -- (eomm7);
			\draw[postaction={decorate}] (v3) -- (eomm8);
			\draw[postaction={decorate}] (eimm1)  -- (w0);
			\draw[postaction={decorate}] (eimm2)  -- (w0);
			\draw[postaction={decorate}] (eimm3)  -- (w1);
			\draw[postaction={decorate}] (eimm4)  -- (w1);
			\draw[postaction={decorate}] (eimm5)  -- (w2);
			\draw[postaction={decorate}] (eimm6)  -- (w2);
			\draw[postaction={decorate}] (eimm7)  -- (w3);
			\draw[postaction={decorate}] (eimm8)  -- (w3);
			\end{scope}
			\draw[fill=white] ($(v0)+(-0.5,-0.2)$) rectangle node{$\mu$} ($(w3)+(0.5,0.2)$);
			\draw[fill=white] ($(eimm1)-(0.1,.6)$) rectangle node{$\nu_1^-$} ($(eimm2)-(-0.1,0.1)$);
			\draw[fill=white] ($(eimm3)-(0.1,.6)$) rectangle node{$\nu_2^-$} ($(eimm4)-(-0.1,0.1)$);
			\draw[fill=white] ($(eimm5)-(0.1,.6)$) rectangle node{$\nu_3^-$} ($(eimm6)-(-0.1,0.1)$);
			\draw[fill=white] ($(eimm7)-(0.1,.6)$) rectangle node{$\nu_4^-$} ($(eimm8)-(-0.1,0.1)$);
			\end{tikzpicture}}}}\label{eqn: wreath_diagram3}
\end{equation}
This amounts to embedding $S_l[S_{v^+} \times S_{v^-}]$ as a subgroup of $S_{lv^+} \times S_{lv^-}$.
From the double coset \eqref{eqn:1colordoublecoset} we can see that this is the type of embedding we are interested in. By re-using the result for the cycle index of a wreath product, we can separately keep track of the cycle structure of the left and right diagram in \eqref{eqn: wreath_diagram3}. For $\mu$ with fixed cycle structure $p \vdash l$, the contribution of the cycle index for $S_l[S_{v^+} \times S_{v^-}]$ as embedded into $S_{lv^+} \times S_{lv^-}$ is simply the product of the contribution from each. That is
\begin{equation}
\frac{1}{\abs{S_l}}\qty[Z_1^{S_{v^+}}(\mathbf{x})^{p_1}\dots Z_l^{S_{v^+}}(\mathbf{x})^{p_l}]\qty[Z_1^{S_{v^-}}(\mathbf{y})^{p_1}\dots Z_l^{S_{v^-}}(\mathbf{y})^{p_l}].
\end{equation}
If we sum over all $\mu \in S_k$ we get the generating function
\begin{equation}
Z^{S_l[S_{v^+} \times S_{v^-}]}(x_1,\dots,x_{lv^+ + lv^-}) = Z^{S_l}(Z_1^{S_{v^+}}(\mathbf{x})Z_1^{S_{v^-}}(\mathbf{y}), \dots, Z_{l}^{S_{v^+}}(\mathbf{x})Z_l^{S_{v^-}}(\mathbf{y})).
\end{equation}

Returning to the case at hand, $S_{p_{v^{(4)}}}\qty[S_{v^{(4)}}]$ is considered as a subgroup of $S_m^+\times S_m^- \times S_n^+ \times S_n^-$. Our goal is to count the number of elements $(\sigma_m^+, \sigma_m^-,\sigma_n^+,\sigma_n^-) \in S_{p_{v^{(4)}}}\qty[S_{v^{(4)}}]$ with cycle structure $p_1,p_2,q_1, q_2$, respectively. To that end, we construct the refined version as
\begin{align}
Z^{S_{p_{v^{(4)}}}\qty[S_{v^{(4)}}]}(\mathbf{x},\mathbf{y},\mathbf{z},\mathbf{w}) =Z^{S_{p_{v^{(4)}}}}\Big(&Z^{S_{v^{(4)}_1}}_1(\mathbf{x})Z^{S_{v^{(4)}_2}}_1(\mathbf{y})Z^{S_{v^{(4)}_3}}_1(\mathbf{z}) Z^{S_{v^{(4)}_4}}_1(\mathbf{w}), \dots, \nonumber\\
&Z^{S_{v^{(4)}_1}}_{p_{v^{(4)}}}(\mathbf{x})Z^{S_{v^{(4)}_2}}_{p_{v^{(4)}}}(\mathbf{y})Z^{S_{v^{(4)}_3}}_{p_{v^{(4)}}}(\mathbf{z}) Z^{S_{v^{(4)}_4}}_{p_{v^{(4)}}}(\mathbf{w})\Big)
\end{align}
Then the number of elements in $S_{p_{v^{(4)}}}\qty[S_{v^{(4)}}]$ with cycle structure $p_1,p_2,q_1,q_2$ is
\begin{equation}
\frac{1}{|S_{p_{v^{(4)}}}[S_{v^{(4)}}]|} Z^{S_{p_{v^{(4)}}}\qty[S_{v^{(4)}}]}_{p_1,p_2,q_1,q_2} = \text{Coefficient}(Z^{S_{p_{v^{(4)}}}\qty[S_{v^{(4)}}]}(\mathbf{x},\mathbf{y},\mathbf{z},\mathbf{w}), \mathbf{x}^{p_1} \mathbf{y}^{p_2} \mathbf{z}^{q_1} \mathbf{w}^{q_2}).
\end{equation}
For products of wreath products we can use the factorization property \eqref{eqn: Cycle Index Product Group}. Consequently, the full cycle index of $G(\vec{m}^+,\vec{m}^-;\vec{n}^+,\vec{n}^-)$ is given by a product
\begin{equation}
Z^{G(\vec{m}^+,\vec{m}^-;\vec{n}^+,\vec{n}^-)}(\mathbf{x},\mathbf{y},\mathbf{z}, \mathbf{w}) = \prod_{v^{(4)}} Z^{S_{p_{v^{(4)}}}\qty[S_{v^{(4)}}]}(\mathbf{x},\mathbf{y},\mathbf{z}, \mathbf{w}).
\end{equation}

It is instructive to calculate $N(\vec{m}^+,\vec{m}^-)$ for Figure \ref{fig: Two Sigma One Color Double Coset Graph}, where $(\vec{m}^+,\vec{m}^-) = (3,2) + (3,2) + (1,3)= 2(3,2)+(1,3)$. The first step is to write down $G(\vec{m}^+,\vec{m}^-)$ as a product of wreath products,
\begin{equation}
G(\vec{m}^+,\vec{m}^-) = S_2[S_3 \times S_2] \times S_1 \times S_3.
\end{equation}
Using the factorization property \eqref{eqn: Cycle Index Product Group} for cycle indices we have
\begin{equation}
Z^{G(\vec{m}^+,\vec{m}^-)}(\mathbf{x},\mathbf{y}) = Z^{S_2[S_3 \times S_2]}(\mathbf{x},\mathbf{y})Z^{S_1}(\mathbf{x})Z^{S_3}(\mathbf{y}),
\end{equation}
where
\begin{equation}
Z^{S_2[S_3 \times S_2]}(\mathbf{x},\mathbf{y}) = Z^{S_2}(Z_1^{S_3}(\mathbf{x})Z_1^{S_2}(\mathbf{y}), Z_2^{S_3}(\mathbf{x})Z_2^{S_2}(\mathbf{y})).
\end{equation}
The four relevant cycle indices are
\begin{equation}
Z^{S_0}(\mathbf{x})=1,\quad Z^{S_1}(\mathbf{x}) = x_1, \quad Z^{S_2}(\mathbf{x}) = \frac{1}{2}(x_1^2 + x_2), \quad Z^{S_3}(\mathbf{x}) = \frac{1}{6}(x_1^3 + 3x_2x_1+2x_3).
\end{equation}
Explicitly, the cycle index for the wreath product is
\begin{equation}
Z^{S_2[S_3 \times S_2]}(\mathbf{x},\mathbf{y}) = \frac{1}{2}\qty(\frac{(x_1^3 + 3x_2x_1+2x_3)}{6})^2\qty(\frac{(y_1^2 + y_2)}{2})^2 + \frac{1}{2}\frac{(x_2^3 + 3x_4x_2+2x_6)}{6}\frac{(y_2^2 + y_4)}{2}.
\end{equation}
To perform the sum in equation \eqref{eq: N(m+,m-) Cycle Index Formula}, we need to pick out the coefficients
\begin{equation}
\text{Coefficient}(Z^{G(\vec{m}^+,\vec{m}^-)}, \mathbf{x}^p\mathbf{y}^p), \quad p \vdash 7.
\end{equation}
There are seven non-zero coefficients of this form,
\begin{align} \nonumber
\text{Coefficient}(Z^{G(\vec{m}^+,\vec{m}^-)}, x_1^7y_1^7) &= \frac{1}{1728}, \\ \nonumber
\text{Coefficient}(Z^{G(\vec{m}^+,\vec{m}^-)}, x_1^5x_2y_1^5 y_2) &=  \frac{5}{288}, \\ \nonumber
\text{Coefficient}(Z^{G(\vec{m}^+,\vec{m}^-)}, x_1^4x_3y_1^4 y_3) &=  \frac{1}{216}, \\ \nonumber
\text{Coefficient}(Z^{G(\vec{m}^+,\vec{m}^-)}, x_1^3x_2^2y_1^3 y_2^2) &=  \frac{7}{192}, \\ \nonumber
\text{Coefficient}(Z^{G(\vec{m}^+,\vec{m}^-)}, x_1^2x_2x_3y_1^2 y_2y_3) &=  \frac{1}{36}, \\ \nonumber
\text{Coefficient}(Z^{G(\vec{m}^+,\vec{m}^-)}, x_1^1x_2^3y_1^1 y_2^3) &=  \frac{1}{48}, \\
\text{Coefficient}(Z^{G(\vec{m}^+,\vec{m}^-)}, x_1^1x_2^1x_4^1y_1^1y_2^1y_4) &=  \frac{1}{16}.
\end{align}
We find
\begin{equation}
N(\vec{m}^+,\vec{m}^-) = \frac{1}{1728}7!+\frac{5}{288}5!2+\frac{1}{216}4!3+\frac{7}{192}3!2!2^2+\frac{1}{36}2!2\cdot 3+\frac{1}{48}3!2^3+\frac{1}{16}2\cdot 4 = 11.
\end{equation}

A similar analysis for Figure \ref{fig: Two Colour Four Sigma Double Coset Graph} gives
\begin{equation}
G(\vec{m}^+,\vec{m}^-;\vec{n}^+,\vec{n}^-) = (S_2 \times S_0 \times S_1 \times S_0 ) \times S_2[S_1 \times S_1 \times S_1 \times S_1] \times (S_0 \times S_2 \times S_1 \times S_2).
\end{equation}
The cycle polynomial is
\begin{align}
Z^{G(\vec{m}^+,\vec{m}^-;\vec{n}^+,\vec{n}^-)} &= Z^{S_2}(\mathbf{x}) Z^{S_1}(\mathbf{z}) Z^{S_2[S_1\times S_1 \times S_1 \times S_1]}(\mathbf{x},\mathbf{y},\mathbf{z}, \mathbf{w}) Z^{S_1}(\mathbf{z})Z^{S_2}(\mathbf{w}) \nonumber\\
&=\frac{(x_1^2+x_2)}{2}z_1\frac{(x_1y_1z_1w_1)^2+x_2y_2z_2w_2}{2}\frac{(y_1^2+y_2)}{2}z_1\frac{(w_1^2+w_2)}{2}.
\end{align}
The relevant non-zero coefficients are
\begin{align} \nonumber
\text{Coefficient}(Z^{G(\vec{m}^+,\vec{m}^-;\vec{n}^+,\vec{n}^-)}, x_1^4z_1^4 y_1^4 w_1^4) &= \frac{1}{16}, \\ \nonumber
\text{Coefficient}(Z^{G(\vec{m}^+,\vec{m}^-;\vec{n}^+,\vec{n}^-)}, x_1^4z_1^4 y_1^2y_2^1 w_1^2w_2^1) &= \frac{1}{16}, \\ \nonumber
\text{Coefficient}(Z^{G(\vec{m}^+,\vec{m}^-;\vec{n}^+,\vec{n}^-)}, x_1^2x_2^1 z_1^2 z_2^1 y_1^2 y_2^1 w_1^2 w_2^1) &= \frac{1}{16}, \\
\text{Coefficient}(Z^{G(\vec{m}^+,\vec{m}^-;\vec{n}^+,\vec{n}^-)}, x_1^2x_2^1 z_1^2 z_2^1 y_2^2 w_2^2) &= \frac{1}{16}.
\end{align}
The number of double cosets is
\begin{equation}
N(\vec{m}^+,\vec{m}^-;\vec{n}^+,\vec{n}^-) = \frac{1}{16}\qty(4!^2+4!2!2+2!^2 2^2+2!^22^3) = 45.
\end{equation}

In this section we have discussed three ways of counting observables, with increasing level of refinement. Because the double coset counting is the most granular of the three, we expect appropriate sums over $N(\vec{m}^+,\vec{m}^-;\vec{n}^+,\vec{n}^-)$ to reproduce previous counting formulae. Let $\text{Count}(m,n,k)$ be the number of graphs with $m+n$ edges and exactly $k$ vertices. It is given by a sum over those vector partitions which have exactly $k$ parts,
\begin{align} \nonumber
\text{Count}(m,n,k) &= \sum_{\substack{(\vec{m}^+,\vec{m}^-;\vec{n}^+,\vec{n}^-)\vdash (m,m;n,n) \\ \text{with}~ k ~\text{parts.}}} N(\vec{m}^+,\vec{m}^-;\vec{n}^+,\vec{n}^-)\\ \nonumber
&= \sum_{\substack{(\vec{m}^+,\vec{m}^-;\vec{n}^+,\vec{n}^-)\vdash (m,m;n,n) \\ \text{with} ~k~\text{ parts.}}} \sum_{\substack{p\vdash m \\ q\vdash n}} \frac{Z^{G(\vec{m}^+,\vec{m}^-;\vec{n}^+,\vec{n}^-)}_{p,p,q,q} \Sym(p)\Sym(q) }{|G(\vec{m}^+,\vec{m}^-;\vec{n}^+,\vec{n}^-)|} \\
&= \tr_{\phantom{}_{W_{ m , n , k }} } D^{ W_{ m , n , k } } ( P_{ V_0}^{ (S_k)}  ) - \tr_{\phantom{}_{W_{ m , n , k-1 }} } D^{ W_{ m , n , k-1 } } ( P_{ V_0}^{ (S_{k-1})}  ).
\end{align}
The last equality follows from the realization in Section \ref{Multi-graph description} that $\tr_{\phantom{}_{W_{ m , n , k }} } D^{ W_{ m , n , k } } ( P_{ V_0}^{ (S_k)}  )$ counts the number of graphs with $m+n$ edges with up to $k$ vertices. This is a refinement of the counting in Section \ref{Section: Representation theory counting} (Table \ref{tab: Table of invariant dimensions}), as can be seen from Table \ref{tab: Table of invariant dimensions refined}.
\begin{table}[t!]
	\begin{center}
		\caption{Number of invariants contained within a monomial of the form $M^{m} N^n$ with refinement on the number of vertices $k$.}
		\label{tab: Table of invariant dimensions refined}
		\begin{tabular}{c c c c}
			\textbf{m} & \textbf{n} & \textbf{\# graphs} &\textbf{$k=1,2,\dots,2(m+n)$}\\
			\hline
			1 & 0 & 2 & 1,1 \\
			2 & 0 & 11 & 1,5,4,1\\
			2 & 1 & 15 & 1, 7, 6, 1\\
			3 & 0 & 52 & 1, 9, 21, 16, 4, 1\\
			3 & 1 & 117& 1, 19, 50, 37, 9, 1 \\
			4 & 0 & 296 & 1, 18, 71, 108, 71, 22, 4, 1\\
			4 & 1 & 877 & 1, 39, 210, 340, 217, 60, 9, 1\\
			4 & 2 & 1252&1, 51, 298, 493, 310, 86, 12, 1 \\
			5 & 0 & 1724&1, 27, 194, 491, 557, 326, 101, 22, 4, 1 \\
			5 & 1 & 6719&1, 69, 680, 1952, 2287, 1283, 371, 66, 9, 1 \\
			5 & 2 & 12157&1, 99, 1150, 3552, 4234, 2341, 658, 109, 12, 1 \\
			6 & 0 & 11060&1, 43, 476, 1903, 3353, 3062, 1587, 497, 111, 22, 4, 1 \\
			6 & 1 & 52505&1, 111, 1826, 8660, 16438, 15174, 7589, 2205, 425, 66, 9, 1 \\
			6 & 2 & 117121&1, 177, 3572, 18716, 37178, 34647, 17046, 4796, 860, 115, 12, 1 \\
			6 & 3 & 150072&1, 199, 4353, 23687, 47882, 44763, 21902, 6078, 1062, 132, 12, 1
		\end{tabular}
	\end{center}
\end{table}
Similarly, we can count the total number of graphs with $m+n$ edges by summing $\text{Count}(m,n,k)$ from $k=1$ to $k=2m+2n$,
\begin{align} \nonumber
\text{Count}(m,n) &= \sum_{k=1}^{2m+2n} \text{Count}(m,n,k) \\ \nonumber
&=\sum_{k=1}^{2m+2n} \sum_{\substack{(\vec{m}^+,\vec{m}^-;\vec{n}^+,\vec{n}^-)\vdash (m,m;n,n) \\ \text{with} ~k~ \text{parts.}}} \sum_{\substack{p\vdash m \\ q\vdash n}} \frac{Z^{G(\vec{m}^+,\vec{m}^-;\vec{n}^+,\vec{n}^-)}_{p,p,q,q} \Sym(p)\Sym(q) }{|G(\vec{m}^+,\vec{m}^-;\vec{n}^+,\vec{n}^-)|} \\ \nonumber
&=\sum_{\substack{(\vec{m}^+,\vec{m}^-;\vec{n}^+,\vec{n}^-)\vdash (m,m;n,n)}} \sum_{\substack{p\vdash m \\ q\vdash n}} \frac{Z^{G(\vec{m}^+,\vec{m}^-;\vec{n}^+,\vec{n}^-)}_{p,p,q,q} \Sym(p)\Sym(q) }{|G(\vec{m}^+,\vec{m}^-;\vec{n}^+,\vec{n}^-)|} \\
&=\tr_{\phantom{}_{W_{ m , n , 2m+2n }} } D^{ W_{ m , n , 2m+2n } } ( P_{ V_0}^{ (S_{2m+2n})}  ).
\end{align}

\section{ Permutation invariant Gaussian 2-matrix models }  \label{Section: The 2-matrix model}

Equipped with detailed descriptions of $S_D$-invariant observables we turn our attention to the construction and solution of the 2-matrix model. We begin this section in \ref{Section: Desc of matrix model} by writing down the most general linear and quadratic permutation invariant Gaussian model. We observe that representation theory of the symmetric group provides the technology needed to transform the partition function such that it factorises to a form amenable to the techniques of Gaussian integration. In \ref{Section: Linear and quadratic EVs} these techniques are applied in order to extract the linear and quadratic expectation values of the model's observables.

\subsection{Gaussian 2-matrix  models in the graph basis } \label{Section: Desc of matrix model}
The most general Gaussian action consistent with $S_D$ symmetry is composed of a sum of all linear and quadratic $S_D$ invariants, each with an arbitrary coefficient
\begin{equation} \label{Eqn: naive action}
S= c_1 \sum_i M_{ii} + c_2 \sum_i N_{ii} + \dots  + c_5 \sum_{i}M_{ii}N_{ii} +  \dots +  c_{41} \sum_{i, j, k, l} M_{ij}N_{kl}.
\end{equation}
This action has four linear and 37 quadratic terms as per the counting in the previous section.

To make contact with usual Gaussian integration (see Appendix \ref{Section: Gaussian integration of matrices}) vectors of the $D^2$ independent variables for both $M_{ij}$ and $N_{ij}$ are formed
\begin{equation}
\textbf{x} = 
\begin{bmatrix}
M_{11} \\
M_{12} \\
\vdots \\
M_{21} \\
\vdots \\
M_{DD}
\end{bmatrix},
\quad
\textbf{y} = 
\begin{bmatrix}
N_{11} \\
N_{12} \\
\vdots \\
N_{21} \\
\vdots \\
N_{DD}
\end{bmatrix}.
\end{equation}
Four $D^2 \times D^2$ coupling matrices $\Lambda^x$, $\Lambda^y$, $\widetilde{\Lambda}^{xy}$ and $\widetilde{\Lambda}^{yx}$ are constructed containing the weights of the quadratic terms of types $M^2$, $N^2$, $MN$ and $NM$ respectively. This recasts the quadratic piece of the action as
\begin{equation}
S_{\text{quad}} = \frac{1}{2} \big( \textbf{x}^T \Lambda^x \textbf{x} + \textbf{y}^T \Lambda^y \textbf{y} + \textbf{x}^T \widetilde{\Lambda}^{xy} \textbf{y} + \textbf{y}^T \widetilde{\Lambda}^{yx} \textbf{x} \big).
\end{equation}
The $\Lambda$ matrices reconstruct each term in Equation \eqref{Eqn: naive action} and so have 41 independent parameters. It is clear that $\Lambda^x$ and $\Lambda^y$ are symmetric:
\begin{align} \nonumber
\textbf{x}^T  \Lambda^x \textbf{x} &= \frac{1}{2} \big( \textbf{x}^T  \Lambda^x \textbf{x} + \textbf{x}^T ( \Lambda^x)^T \textbf{x} \big) = \frac{1}{2} \textbf{x}^T  \big( \Lambda^x + ( \Lambda^x)^T \big) \textbf{x} \\
&\Rightarrow (\Lambda^x)^T = \Lambda^x
\end{align}
and similarly for $\Lambda^y$. Further, we see that it is possible to redefine the cross-term couplings $ \widetilde{\Lambda}^{xy}$ and $ \widetilde{\Lambda}^{yx}$ such that they are related by transposition
\begin{align} \nonumber \label{Eqn: Cross coupling relation}
\textbf{x}^T \widetilde{\Lambda}^{xy} \textbf{y} + \textbf{y}^T \widetilde{\Lambda}^{yx} \textbf{x} &= \frac{1}{2} \big( \textbf{x}^T \widetilde{\Lambda}^{xy} \textbf{y} + \textbf{y}^T (\widetilde{\Lambda}^{xy})^T \textbf{x} \big) + \frac{1}{2} \big( \textbf{y}^T \widetilde{\Lambda}^{yx} \textbf{x} + \textbf{x}^T (\widetilde{\Lambda}^{yx})^T \textbf{y} \big) \\ \nonumber
&= \frac{1}{2} \textbf{x}^T \big( \widetilde{\Lambda}^{xy} + (\widetilde{\Lambda}^{yx})^T \big) \textbf{y} + \frac{1}{2} \textbf{y}^T \big( (\widetilde{\Lambda}^{xy})^T +  \widetilde{\Lambda}^{yx} \big) \textbf{x} \\
&= \textbf{x}^T  \Lambda^{xy}  \textbf{y} + \textbf{y}^T (\Lambda^{xy} )^T \textbf{x}
\end{align}
where in the last line we have defined $ \Lambda^{xy} \equiv \frac{1}{2} \big( \widetilde{\Lambda}^{xy} + (\widetilde{\Lambda}^{yx})^T \big)$.

We now construct a vector
\begin{equation}
\textbf{z} = 
\begin{bmatrix}
\textbf{x} \\
\textbf{y}
\end{bmatrix}
=
\begin{bmatrix}
M_{11} \\
M_{12} \\
\vdots \\
M_{DD} \\
N_{11} \\
N_{12} \\
\vdots \\
N_{DD}
\end{bmatrix}
\end{equation}
and use the above properties of the coupling matrices to rewrite the quadratic action in terms of a symmetric matrix $\Lambda$ defined by the following 
\begin{equation}
S_{\text{quad}} =
\frac{1}{2} \textbf{z}^T \Lambda \textbf{z} =
\frac{1}{2}
\begin{bmatrix}
\textbf{x}^T & \textbf{y}^T
\end{bmatrix}
\begin{bmatrix}
\Lambda^x & \Lambda^{xy} \\
(\Lambda^{xy})^T & \Lambda^{y}
\end{bmatrix}
\begin{bmatrix}
\textbf{x} \\
\textbf{y}
\end{bmatrix}.
\end{equation}
Similarly, we can rewrite the linear terms
\begin{equation}
S_{\text{linear}} = - \big( \textbf{J}_1^T \textbf{x} + \textbf{J}_2^T \textbf{y} \big) = 
-
\begin{bmatrix}
\textbf{J}_1^T & \textbf{J}_2^T 
\end{bmatrix}
\begin{bmatrix}
 \textbf{x} \\
 \textbf{y} 
\end{bmatrix}
= - \textbf{J}^T \textbf{z}
\end{equation}
where $\textbf{J}$ contains all linear couplings, such that the action reads
\begin{equation} \label{Eqn: Naive action}
S = \frac{1}{2} \textbf{z}^T \Lambda \textbf{z} - \textbf{J}^T \textbf{z}.
\end{equation}

To find expectation values of this model we need to solve the partition function
\begin{equation} \label{Eqn: Naive partition function}
\mathcal{Z} = \int dMdN e^{ - \frac{1}{2} \textbf{z}^T \Lambda \textbf{z} + \textbf{J}^T \textbf{z}}.
\end{equation}
The measure $ dMdN$ over the matrix variables $M_{ ij}, N_{ ij} $ is taken to be the Euclidean 
measure on $ {\mathbb{R}}^{2D^2}$ parametrised by $2D^2$ variables  
\begin{equation}
dMdN \equiv \prod_i dM_{ii} dN_{ii} \prod_{i \neq j} dM_{ij}dN_{ij}.
\end{equation}
The remaining problem is that the $2D^2 \times 2D^2$ matrix $\Lambda$ in Equation \eqref{Eqn: Naive partition function} mixes the $M_{ij}$ and $N_{ij}$ variables in some non-trivial way. The solution of this partition function will require the inversion of this coupling matrix. Directly inverting  this matrix would be computationally intractable for large $D$. By using a representation theoretic change of variables for $M_{ij}$ and $N_{ ij}$, we are able to express the action in terms of a partially diagonalised quadratic form:  the only non-trivial mixing remaining  then involves a $ 2 \times 2$ and a $ 3 \times 3$ matrix of parameters coupling representation theoretic multiplicities.

\subsection{Solving the 2-matrix models} \label{Section: Solving 2 mat system}
We saw in the previous section that the key to solving the 2-matrix system is the diagonalisation of the partition function. This can be achieved with the help of $S_D$ representation theory. We sketch the appropriate steps needed to factorise the integral in Equation \eqref{Eqn: Naive partition function} following the discussion in \cite{PIGMM}.   

\subsubsection{Transforming the action}
We begin by irreducibly decomposing the natural representation $V_D$ of $S_D$. The natural representation is isomorphic to the direct sum of two irreducible representations: the trivial representation $V_0$ and the Hook or standard representation $V_H$
\begin{equation}
V_D = V_0 \oplus V_H.
\end{equation}
The two sets of $D^2$ elements $M_{ij}$ and $N_{ij}$ both transform as $V_D \otimes V_D$. This can be decomposed into irreducible representations of the diagonal $S_D$ as
\begin{equation} \label{Eqn:Vd Vd irrep decomp}
V_D \otimes V_D = 2V_0 \oplus 3V_H \oplus V_2 \oplus V_3.
\end{equation}

The terms in the action quadratic in $M$ or $N$ alone transform as $\text{Sym}^2(V_D \otimes V_D)$ due to the symmetry under exchange of two $M$'s or two $N$'s respectively. An important fact about representations of the symmetric group is that the product of two irreps contains a copy of the trivial representation if and only if the irreps are the same, and in this case they contain exactly one copy of the trivial representation. From this, and the multiplicities in Equation \eqref{Eqn:Vd Vd irrep decomp} it follows that the corresponding 22 quadratic parameters can be decomposed as two copies of
\begin{equation}
11 = 3 + 6 + 1 + 1, 
\end{equation} 
one copy for $M$ and another for $N$. A  symmetric $2 \times 2$ matrix is parametrized by $3$  numbers
and a  symmetric $3 \times 3$ matrix is parametrised by $6$ numbers. 

The quadratic terms involving both $M$ and $N$ variables transform under $(V_D \otimes V_D)^{\otimes 2}$, as a result the corresponding 15 quadratic parameters can be decomposed as
\begin{equation}
15 = 4 + 9 + 1 + 1.
\end{equation}
A general $2 \times 2$ matrix is parametrised by $4$ numbers 
 and a general $3 \times 3$ matrix is parameterised by $9$ numbers. We will restrict these parameters to be real, and there are convergence conditions on the parameters which we discuss in Section \ref{sssec:convergence}. 


Rewriting Equation \eqref{Eqn:Vd Vd irrep decomp} with multiplicity labels $\alpha$ for each irrep that appears more than once we have
\begin{equation} \label{Eqn: Vd Vd irrep with mult}
V_D \otimes V_D = \bigoplus_{\alpha = 1}^2 V_0^{(\alpha)} \bigoplus_{\alpha = 1}^3 V_H^{(\alpha)} \oplus V_2 \oplus V_3.
\end{equation}
We define variables $X^{V_i}_{\alpha , a}$ transforming according to this decomposition into irreducible representations of the diagonal subgroup $\text{Diag}(S_D) \subset S_D \times S_D$. As in \eqref{Eqn: Vd Vd irrep with mult} the $\alpha$ index of these variables tracks the multiplicity of the irrep $V_i$ and the $a$ index runs over a basis of the irrep. This exercise is performed for both $M$ and $N$ independently, leaving us with two copies of these variables $X^{V_i}_{\alpha , a}$ for $M$ and $Y^{V_i}_{\dot{\alpha} , a}$ for $N$, with $\dot{\alpha}$ running over the $V_i$ multiplicity index of $Y$. 

To summarise, the matrix variables $M_{ij}$ can be linearly transformed to the following variables, organised according to representations of the diagonal $S_D$
\begin{align} \nonumber \label{Eqn: Sd diag variables}
\text{Trivial Rep:}&\quad X^{V_0}_{1}, X^{V_0}_{2} \\ \nonumber
\text{Hook Rep:}&\quad X^{V_H}_{1,a}, X^{V_H}_{ 2,a}, X^{V_H}_{3, a} \\ \nonumber
\text{The rep $V_2$:}&\quad X^{V_2}_a \\
\text{The rep $V_3$:}&\quad X^{V_3}_a .
\end{align}
Similarly the variables $N_{ij}$ give a second copy of \eqref{Eqn: Sd diag variables} with $X \rightarrow Y$.

\subsubsection{Representation theoretic description of quadratic invariants} \label{Section: Representation description of quadratic invariants}

From here we can introduce a set of representation theoretic parameters for the 41-parameter Gaussian matrix model. In terms of these parameters the linear and quadratic expectation values of $X^{V_i}_{\alpha , a}$ and $Y^{V_i}_{\dot{\alpha} , a}$ are simple.

We start with the description of quadratic invariants of just the $M_{ij}$ variables. Quadratic invariant functions of $M_{ij}$ form the invariant subspace of $\text{Sym}^2(V_D \otimes V_D)$. There are two copies of $V_0$ in the decomposition of $V_D \otimes V_D$,
\begin{equation} \label{Eqn: Vd decomp}
V_D \otimes V_D = V_{0,0}^0 \oplus V_{0,H}^H \oplus V_{H,0}^H \oplus V_{H,H}^0 \oplus V_{H,H}^H \oplus V_{H,H}^{V_2} \oplus V_{H,H}^{V_3}
\end{equation}
where the subscripts track the origin of each irrep within the decomposition of $V_D \otimes V_D$. The symmetric product of the trivial representations, $\text{Sym}^2(V_{0,0}^0 \oplus V_{H,H}^0)$, contains three invariants
\begin{align} \nonumber
&(X^{V_0}_{1})^2 \\ \nonumber
&X^{V_0}_{1} X^{V_0}_{2} = X^{V_0}_{2} X^{V_0}_{1} \\
& (X^{V_0}_{2})^2.
\end{align}
The general invariant quadratic function of the $X^{V_0}_{\alpha}$ variables is
\begin{equation}
\sum_{\alpha, \beta = 1}^2 (\Lambda_{V_0}^{\us[X]})_{\alpha \beta} X^{V_0}_{\alpha} X^{V_0}_{\beta},
\end{equation}
with $\Lambda_{V_0}^{\us[X]}$ a $2 \times 2$ symmetric matrix controlling the coupling of these variables.
Repeating this exercise for $V_H$ we find six invariants and write the most general invariant quadratic function of these variables
\begin{equation}
\sum_{\alpha, \beta = 1}^3 (\Lambda_{V_H}^{\us[X]})_{\alpha \beta} \sum_a^{\text{Dim}(V_H)} X^{V_H}_{\alpha, a} X^{V_H}_{\beta,a},
\end{equation}
with $\Lambda_{V_H}^{\us[X]}$ a $3 \times 3$ symmetric matrix.
 The quadratic invariants constructed from $V_2$ and $V_3$ are
\begin{align} \nonumber
&\Lambda_{V_2}^{\us[X]} \sum_a^{\text{Dim}V_2} X^{V_2}_a X^{V_2}_a \\
&\Lambda_{V_3}^{\us[X]} \sum_a^{\text{Dim}V_3} X^{V_3}_a X^{V_3}_a
\end{align}
with $\Lambda_{V_2}^{\us[X]}$ and $\Lambda_{V_3}^{\us[X]}$ single parameters. This analysis goes through identically for $N$ with $X \rightarrow Y$ leading to the definition of analogous coupling matrices $\Lambda_{V_0}^{\us[Y]} , \Lambda_{V_H}^{\us[Y]} , \Lambda_{V_2}^{\us[Y]}$ and $\Lambda_{V_3}^{\us[Y]}$.

Lastly we consider invariant quadratic functions of $M$ and $N$. Quadratic invariant functions of $M_{ij}$ and $N_{ij}$ form the invariant subspace of $(V_D \otimes V_D) \otimes (V_D \otimes V_D)$, as there is no longer a symmetry under $M \leftrightarrow N$. Again, we begin with the invariants formed from the diagonal variables transforming trivially. We see $(V_{0,0}^0 \oplus V_{H,H}^0) \otimes (V_{0,0}^0 \oplus V_{H,H}^0)$ contains four invariants
\begin{align}
X^{V_0}_{1} Y^{V_0}_{1}, \quad X^{V_0}_{1} Y^{V_0}_{2}, \quad X^{V_0}_{2} Y^{V_0}_{1}, \quad X^{V_0}_{2} Y^{V_0}_{2},
\end{align}
and the general invariant quadratic function of the $X^{V_0}_{\alpha}$, $Y^{V_0}_{\dot{\beta}}$ variables is
\begin{equation}
\sum_{\alpha, \dot{\beta} = 1}^2 (\Lambda_{V_0}^{\us[XY]})_{\alpha \dot{\beta}} X^{V_0}_{\alpha} Y^{V_0}_{\dot{\beta}},
\end{equation}
with $\Lambda_{V_0}^{\us[XY]}$ a $2 \times 2$ general matrix. There are three copies of $V_H$ in \eqref{Eqn: Vd decomp} from which we get 9 invariants
\begin{align} \nonumber
&\sum_a X^{V_H}_{1,a} Y^{V_H}_{1,a}, \quad \sum_a X^{V_H}_{2,a} Y^{V_H}_{2,a}, \quad \sum_a X^{V_H}_{3,a} Y^{V_H}_{3,a}, \quad \sum_a X^{V_H}_{1,a} Y^{V_H}_{2,a}, \quad \sum_a X^{V_H}_{1,a} Y^{V_H}_{3,a}, \\ 
&\sum_a X^{V_H}_{2,a} Y^{V_H}_{3,a},  \quad \sum_a X^{V_H}_{2,a} Y^{V_H}_{1,a}, \quad \sum_a X^{V_H}_{3,a} Y^{V_H}_{1,a}, \quad \sum_a X^{V_H}_{3,a} Y^{V_H}_{2,a}
\end{align}
The general invariant quadratic function of these variables is
\begin{equation}
\sum_{\alpha, \dot{\beta} = 1}^3 (\Lambda_{V_H}^{\us[XY]})_{\alpha \dot{\beta}} \sum_a^{\text{Dim}(V_H)} X^{V_H}_{\alpha,a} Y^{V_H}_{\dot{\beta},a}
\end{equation}
with $\Lambda_{V_H}^{\us[XY]}$ a $3 \times 3$ general matrix. Similarly to before, we find just one quadratic invariant constructed from each of the $V_2$ and $V_3$ variables 
\begin{align} \nonumber
&\Lambda_{V_2}^{\us[XY]} \sum_a^{\text{Dim}V_2} X^{V_2}_a Y^{V_2}_a \\
&\Lambda_{V_3}^{\us[XY]} \sum_a^{\text{Dim}V_3} X^{V_3}_a Y^{V_3}_a.
\end{align}
If we repeat the above, swapping the order of $X$ and $Y$ we would find the following invariant quadratic functions 
\begin{align} \nonumber
&\sum_{\dot{\alpha}, \beta = 1}^2 (\Lambda_{V_0}^{\us[YX]})_{\dot{\alpha} \beta} Y^{V_0}_{\dot{\alpha}} X^{V_0}_{\beta} \\ \nonumber
&\sum_{\dot{\alpha}, \beta = 1}^3 (\Lambda_{V_H}^{\us[YX]})_{\dot{\alpha} \beta} \sum_a^{\text{Dim}(V_H)} Y^{V_H}_{\dot{\alpha},a} X^{V_H}_{\beta,a} \\ \nonumber
&\Lambda_{V_2}^{\us[YX]} \sum_a^{\text{Dim}V_2} Y^{V_2}_a X^{V_2}_a \\
&\Lambda_{V_3}^{\us[YX]} \sum_a^{\text{Dim}V_3} Y^{V_3}_a X^{V_3}_a.
\end{align}
with analogous couplings $\Lambda_{V_0}^{\us[YX]}, \Lambda_{V_H}^{\us[YX]} , \Lambda_{V_2}^{\us[YX]}$ and $\Lambda_{V_3}^{\us[YX]}$. We showed in \eqref{Eqn: Cross coupling relation} that it is possible to rewrite these couplings such that $\Lambda_{V_A}^{\us[XY]} = (\Lambda_{V_A}^{\us[YX]})^T$. With the aim of applying standard Gaussian integration techniques, performing the same exercise here allows us to construct larger coupling matrices $\Lambda_{V_A}$, one for each irrep $V_A \in \{V_0 , V_H , V_2 , V_3 \}$,
\begin{equation} \label{Eqn: Large lambda matrix}
\Lambda_{V_A} = 
\begingroup
\renewcommand*{\arraystretch}{1.5}
\begin{bmatrix}
\Lambda_{V_A}^{\us[X]} & \Lambda_{V_A}^{\us[XY]} \\
(\Lambda_{V_A}^{\us[XY]})^T & \Lambda_{V_A}^{\us[Y]}
\end{bmatrix}.
\endgroup
\end{equation}
Importantly, these larger matrices are symmetric. 

We also package the $X^{V_A}_{\alpha , a}$ and $Y^{V_A}_{\dot{\alpha} , a}$ into a single vector
\begin{align} \label{Eqn: Z diag definition}
Z_{\tilde{\alpha} , a}^{V_A} &= 
\begin{bmatrix}
X^{V_A}_{\alpha , a} & Y^{V_A }_{\dot{\alpha} , a} 
\end{bmatrix}, \qquad \tilde{\alpha} = (\alpha, \dot{\alpha})
\end{align}
where $\tilde{\alpha}$ runs over twice the multiplicity of the irrep $V_A$, once for each of $X^{V_A}$ and $Y^{V_A}$. The quadratic terms in the action can then be rewritten in terms of \eqref{Eqn: Large lambda matrix} and \eqref{Eqn: Z diag definition}
\begin{align} \nonumber  \label{Eqn: Quadratic action diagonal Sd} 
\mathcal{S}_{\text{quad}} &= \frac{1}{2} \Big[ \sum_{V_A} \sum_a \sum_{\alpha , \beta} (\Lambda_{V_A}^{\us[X]})_{\alpha \beta} X^{V_A}_{\alpha,a} X^{V_A}_{\beta,a} + \sum_{V_A} \sum_a \sum_{\alpha , \beta} (\Lambda_{V_A}^{\us[Y]})_{\alpha \beta} Y^{V_A}_{\alpha,a} Y^{V_A}_{\beta,a} \\ \nonumber
&\quad + \sum_{V_A} \sum_a \sum_{\alpha , \dot{\beta}} (\Lambda_{V_A}^{\us[XY]})_{\alpha \dot{\beta}} X^{V_A}_{\alpha,a} Y^{V_A}_{\dot{\beta},a} + \sum_{V_A} \sum_a \sum_{\alpha , \dot{\beta}} (\Lambda_{V_A}^{\us[XY]})_{\dot{\beta} \alpha} X^{V_A}_{\alpha,a} Y^{V_A}_{\dot{\beta},a} \Big] \\
& = \frac{1}{2} \sum_{V_A} \sum_a \sum_{\tilde{\alpha} , \tilde{\beta}} Z^{V_A}_{\tilde{\alpha},a} (\Lambda_{V_A})_{\tilde{\alpha} \tilde{\beta}}  Z^{V_A }_{\tilde{\beta},a}. 
\end{align}
The linear terms in our action can be rewritten as
\begin{align} \nonumber \label{Eqn: Linear action diagonal Sd} 
\mathcal{S}_{\text{Linear}} &=  - \sum_{V_A} \sum_a \sum_{\alpha , \dot{\alpha}} 
\begin{bmatrix}
X^{V_A}_{\alpha , a} & Y^{V_A }_{\dot{\alpha} , a} 
\end{bmatrix}
\begin{bmatrix}
\rho_{\alpha , a}^{V_A, \us[X]}  \\
\rho_{\dot{\alpha} , a}^{V_A, \us[Y]}
\end{bmatrix} \\
&= - \sum_{V_A} \sum_a \sum_{\tilde{\alpha}} Z^{V_A}_{\tilde{\alpha} ,a} \rho^{V_A}_{\tilde{\alpha} , a}
\end{align}
where $\rho^{V_A}_{\tilde{\alpha} , a}$ is the linear coupling of the $a$ state in the $V_A$ irrep with multiplicity label $\tilde{\alpha}$, for example $\rho^{V_0}_{1}$ is the coupling of $X^{V_0}_{1}$ and $\rho^{V_0}_{3}$ is the coupling of $Y^{V_0}_{1}$. The expression \eqref{Eqn: Linear action diagonal Sd} includes the possibility of linear contributions from all variables contained within $Z^{V_A}_{\tilde{\alpha},a}$ - even those that do not transform trivially under permutations. The couplings $\rho^{V_A}_{\tilde{\alpha} , a} $ serve as source terms in the computation of expectation values: we take derivatives with respect to these and then evaluate,  setting to zero the couplings for the cases where  $ V_A $ is not the trivial representation $V_0$ of $S_D$. 
 Putting \eqref{Eqn: Linear action diagonal Sd} and \eqref{Eqn: Quadratic action diagonal Sd} together gives our total action
\begin{align} \nonumber \label{Eqn: Block diag action}
\mathcal{S} &= - \sum_{V_A} \sum_a \sum_{\tilde{\alpha}} Z^{V_A}_{\tilde{\alpha} ,a} \rho^{V_A}_{\tilde{\alpha} , a} + \frac{1}{2} \sum_{V_A} \sum_a^{\text{Dim}V_A} \sum_{\tilde{\alpha} , \tilde{\beta}} Z^{V_A }_{\tilde{\alpha},a} (\Lambda_{V_A})_{\tilde{\alpha} \tilde{\beta}}  Z^{V_A}_{\tilde{\beta},a} \\ \nonumber
&= - \sum_{V_A} \sum_a \sum_{\tilde{\alpha}} Z^{V_A}_{\tilde{\alpha} ,a} \rho^{V_A}_{\tilde{\alpha} , a} + \frac{1}{2} \sum_{\tilde{\alpha} , \tilde{\beta} = 1}^4 Z^{V_0}_{\tilde{\alpha}} (\Lambda_{V_0})_{\tilde{\alpha} \tilde{\beta}}  Z^{V_0}_{\tilde{\beta}} + \frac{1}{2} \sum_{a=1}^{D-1} \sum_{\tilde{\alpha} , \tilde{\beta} = 1}^6 Z^{V_H}_{\tilde{\alpha},a} (\Lambda_{V_H})_{\tilde{\alpha} \tilde{\beta}}  Z^{V_H}_{\tilde{\beta},a} \\ 
&\quad + \frac{1}{2} \sum_{a=1}^{(D-1)(D-2)/2} \sum_{\tilde{\alpha} , \tilde{\beta} = 1}^2 Z^{V_2}_{\tilde{\alpha},a} (\Lambda_{V_2})_{\tilde{\alpha} \tilde{\beta}}  Z^{V_2}_{\tilde{\beta},a} + \frac{1}{2} \sum_{a=1}^{D(D-3)/2} \sum_{\tilde{\alpha} , \tilde{\beta} = 1}^2 Z^{V_3}_{\tilde{\alpha},a} (\Lambda_{V_3})_{\tilde{\alpha} \tilde{\beta}}  Z^{V_3}_{\tilde{\beta},a}.
\end{align}

\subsubsection{Convergence conditions and the measure}\label{sssec:convergence}
Since the $X^{V_A}_{\alpha , a}, Y^{V_A}_{\dot{\alpha} , a}$ variables are given by an orthogonal change of basis \cite{PIGMM}, the measure is transformed to
\begin{align} \nonumber
dMdN & = dX^{V_0}_1 dX^{V_0}_2 dY^{V_0}_1 dY^{V_0}_2 \prod_{a=1}^{\text{Dim}V_H} dX^{V_H}_{1,a} dX^{V_H}_{2,a} dX^{V_H}_{3,a} dY^{V_H}_{1,a} dY^{V_H}_{2,a} dY^{V_H}_{3,a} \\  \nonumber
&\prod_{a=1}^{\text{Dim}V_2} dX^{V_2}_{a} dY^{V_2}_{a} \prod_{a=1}^{\text{Dim}V_3} dX^{V_3}_{a} dY^{V_3}_{a} \\
&= dXdY.
\end{align}

In order that the partition function be well defined we demand that the determinants of the $\Lambda $ matrices should be greater than or equal to zero
\begin{align} \nonumber
& 2 \times 2 ~ symmetric ~ matrix  : \det   (\Lambda_{V_0}^{\us[X]})_{\alpha \beta}  \ge 0 \cr \nonumber
& 2 \times 2 ~ symmetric ~ matrix  : \det   (\Lambda_{V_0}^{\us[Y]})_{\alpha \beta}  \ge 0 \cr \nonumber
& 2 \times 2 ~ matrix  : \det   (\Lambda_{V_0}^{\us[XY]})_{\alpha \dot{\beta}}  \ge 0 \cr \nonumber
& 3 \times 3  ~  symmetric ~ matrix  : \det   (\Lambda_{V_H}^{\us[X]})_{\alpha \beta}  \ge 0 \cr \nonumber
& 3 \times 3  ~  symmetric ~ matrix  : \det   (\Lambda_{V_H}^{\us[Y]})_{\alpha \beta}  \ge 0 \cr  \nonumber
& 3 \times 3  ~ matrix  : \det   (\Lambda_{V_H}^{\us[XY]})_{\alpha \dot{\beta}}  \ge 0 \cr \nonumber
& \Lambda_{  V_2 }^{\us[X]} \ge 0  ~~~~~~~~~~~\Lambda_{  V_3 }^{\us[X]} \ge 0  \cr \nonumber
& \Lambda_{ V_2}^{\us[Y]} \ge 0    ~~~~~~~~~~~\Lambda_{  V_3 }^{\us[Y]} \ge 0  \\
& \Lambda_{ V_2 }^{\us[XY]} \ge 0 ~~~~~~~~~~~\Lambda_{  V_3 }^{\us[XY]} \ge 0.
\end{align}

\subsection{Linear and quadratic expectation values} \label{Section: Linear and quadratic EVs}
We we will find it useful to define $(\Lambda_{V_A}^{-1})^{\us[X]}$, $(\Lambda_{V_A}^{-1})^{\us[Y]}$, $(\Lambda_{V_A}^{-1})^{\us[XY]}$ and $(\Lambda_{V_A}^{-1})^{\us[YX]}$ as the matrices appearing in $\Lambda_{V_A}^{-1}$ e.g.
\begin{equation} \label{Eqn: Large lambda matrix inverse}
\begingroup
\renewcommand*{\arraystretch}{1.5}
\begin{bmatrix}
(\Lambda_{V_A}^{-1})^{\us[X]} & (\Lambda_{V_A}^{-1})^{\us[XY]} \\
(\Lambda_{V_A}^{-1})^{\us[YX]} & (\Lambda_{V_A}^{-1})^{\us[Y]}
\end{bmatrix}
\endgroup
\equiv \Lambda_{V_A}^{-1}. 
\end{equation}
We also define $\Lambda$ as the block diagonal matrix containing all quadratic couplings
\begin{equation} \label{Eqn: Very large lambda matrix}
\Lambda \equiv 
\begin{bmatrix}
\Lambda_{V_0} & \\
& \Lambda_{V_H} \\
&& \Lambda_{V_2} \\
&&& \Lambda_{V_3} 
\end{bmatrix}
\end{equation}
and $\rho$ as the vector containing all linear couplings
\begin{equation} 
\rho_{\tilde{\alpha} , a} \equiv 
\begin{bmatrix}
\rho^{V_0}_{\tilde{\alpha}} & \rho^{V_H}_{\tilde{\alpha} , a} & \rho^{V_2}_{\tilde{\alpha}, a} & \rho^{V_3}_{\tilde{\alpha}, a}
\end{bmatrix}.
\end{equation}

Equipped with the transformed action in block diagonal form \eqref{Eqn: Block diag action} we can write the solution to the partition function
\begin{align} \nonumber \label{Eqn: Partition function result}
\mathcal{Z} &= \int dXdY e^{ - \frac{1}{2} \sum_{V_A} \sum_{\tilde{\alpha}, \tilde{\beta}} \sum_a^{\text{Dim}V_A} Z^{V_A}_{\tilde{\alpha} ,a} (\Lambda_{V_A})_{\tilde{\alpha} \tilde{\beta}}  Z^{V_A}_{\tilde{\beta},a} + \sum_{V_A} \sum_{\tilde{\alpha}} \sum_a \rho^{V_A}_{\tilde{\alpha} , a} Z^{V_A}_{\tilde{\alpha} ,a} } \\ 
&= \frac{(2 \pi)^{D^2}}{ (\text{det}\Lambda)^{1/2}} e^{ \frac{1}{2} \sum_{V_A} \sum_{\tilde{\alpha} \tilde{\beta}} \sum_a \rho_{\tilde{\alpha} , a} (\Lambda^{-1}_{V_A})_{\tilde{\alpha} \tilde{\beta}} \rho_{\tilde{\beta} , a} }
\end{align}
where we have used the standard Gaussian integration result Equation \eqref{Eqn: multi-dim gaussian result}. As usual, taking derivatives of this partition function allows us to generate expectation values, for example
\begin{equation}
\langle X^{V_A}_{\alpha,a} Y^{V_B}_{\dot{\beta},b} \rangle = \int dX dY e^{-S} X^{V_A}_{\alpha,a} Y^{V_B}_{\dot{\beta},b}.
\end{equation}
In order to evaluate expectation values of this form it is enough to take appropriate derivatives of the RHS of \eqref{Eqn: Partition function result}. This exercise is performed in the following sections, firstly for expectation values of linear order observables, and then for quadratic order observables. As our theory is Gaussian, higher-order expectation values can be reduced to sums of products of these linear and quadratic expectation values with the application of Wick's theorem.

\subsubsection{Linear expectation values}
We first consider first order expectation values of $X$ and $Y$
\begin{align} \nonumber \label{Eqn: Full X one point function}
\langle X^{V_A}_{\alpha,a} \rangle &= \frac{1}{\mathcal{Z}} \int dX dY X^{V_A}_{\alpha,a} e^{- \mathcal{S}} = \frac{1}{\mathcal{Z}} \frac{\partial \mathcal{Z}}{\partial \rho_{\alpha , a}^{V_A, \us[X]}} \bigg\rvert_{\rho^{V_H} = \rho^{V_2} = \rho^{V_3} = 0} \\ \nonumber
&= \frac{1}{\mathcal{Z}} (\Lambda^{-1}_{V_A})_{\ls[\alpha \tilde{\beta}]} \rho_{\ls[\tilde{\beta} , a]}^{V_A} \frac{(2 \pi)^{D^2}}{(\text{det} \Lambda )^{1/2}} e^{\frac{1}{2}  \sum_{V_B} \sum_{\tilde{\gamma} \tilde{\mu}} \sum_b \rho_{\tilde{\gamma} , b} (\Lambda^{-1}_{V_B})_{\tilde{\gamma} \tilde{\mu}} \rho_{\tilde{\mu} , b} } \bigg\rvert_{\rho^{V_H} = \rho^{V_2} = \rho^{V_3} = 0} \\
&= (\Lambda^{-1}_{V_A})_{\ls[\alpha \tilde{\beta}]} \rho_{\ls[\tilde{\beta} , a]}^{V_A} \delta(V_A , V_0).
\end{align}
For $V_A = \{V_H, V_2, V_3 \} $ this expression is zero due to the vanishing linear coupling of the representations that transform non-trivially under $S_D$ and $\Lambda$ being block diagonal. For $V_0$ we have
\begin{align} \label{Eqn: X one point function} \nonumber
\langle X^{V_0}_{\alpha} \rangle &= (\Lambda_{V_0}^{-1})_{\ls[\alpha \tilde{\beta}]} \rho_{\ls[\tilde{\beta}]} = (\Lambda_{V_0}^{-1})_{\ls[\alpha \beta]}^{\us[X]} \rho_{\ls[\beta]}^{V_0 , \us[X]} +(\Lambda_{V_0}^{-1})_{\ls[\alpha \dot{\beta}]}^{\us[XY]} \rho_{\ls[\dot{\beta}]}^{V_0 , \us[Y]} \\ 
&= (\Lambda_{V_0}^{-1})_{\ls[\alpha 1]} \rho_{\ls[1]}^{V_0} + (\Lambda_{V_0}^{-1})_{\ls[\alpha 2]} \rho_{\ls[2]}^{V_0}  + (\Lambda_{V_0}^{-1})_{\ls[\alpha 3]} \rho_{\ls[3]}^{V_0} + (\Lambda_{V_0}^{-1})_{\ls[\alpha 4]} \rho_{\ls[4]}^{V_0}
\end{align}
where $\alpha = \{1,2\}$.
Similarly
\begin{align} \nonumber
\langle Y^{V_A}_{\dot{\alpha},a} \rangle &= \frac{1}{\mathcal{Z}} \int dX dY Y^{V_A}_{\dot{\alpha},a} e^{- \mathcal{S}} = \frac{1}{\mathcal{Z}} \frac{\partial \mathcal{Z}}{\partial \rho_{\dot{\alpha} , a}^{V_A, \us[Y]}} \bigg\rvert_{\rho^{V_H} = \rho^{V_2} = \rho^{V_3} = 0} \\ \nonumber
&= \frac{1}{\mathcal{Z}} (\Lambda^{-1}_{V_A})_{\ls[\dot{\alpha} \tilde{\beta}]} \rho_{\ls[\tilde{\beta} , a]}^{V_A } \frac{(2 \pi)^{D^2}}{(\text{det} \Lambda )^{1/2}} e^{\frac{1}{2}  \sum_{V_B} \sum_{\tilde{\gamma} \tilde{\mu}} \sum_b \rho_{\tilde{\gamma} , b} (\Lambda^{-1}_{V_B})_{\tilde{\gamma} \tilde{\mu}} \rho_{\tilde{\mu} , b} } \bigg\rvert_{\rho^{V_H} = \rho^{V_2} = \rho^{V_3} = 0} \\  
&= (\Lambda^{-1}_{V_A})_{\ls[\dot{\alpha} \tilde{\beta}]} \rho_{\ls[\tilde{\beta} , a]}^{V_A } \delta(V_A, V_0)
\end{align}
the $\dot{\alpha}$ index runs over $V_A, Y$ inside $\Lambda$. Again for $V_A = \{ V_H, V_2, V_3 \}$ this expression is zero. For $V_0$ we have
\begin{align} \label{Eqn: Y one point function} \nonumber
\langle Y^{V_0}_{\dot{\alpha}} \rangle &= (\Lambda_{V_0}^{-1})_{\ls[\dot{\alpha} \tilde{\beta}]} \rho_{\ls[\tilde{\beta}]} = (\Lambda_{V_0}^{-1})_{\ls[\dot{\alpha} \beta]}^{\us[YX]} \rho_{\ls[\beta]}^{V_0 , \us[X]} +(\Lambda_{V_0}^{-1})_{\ls[\dot{\alpha} \dot{\beta}]}^{\us[Y]} \rho_{\ls[\dot{\beta}]}^{V_0 , \us[Y]} \\
&= (\Lambda_{V_0}^{-1})_{\ls[\dot{\alpha} 1]} \rho_{\ls[1]}^{V_0} + (\Lambda_{V_0}^{-1})_{\ls[\dot{\alpha} 2]} \rho_{\ls[2]}^{V_0}  + (\Lambda_{V_0}^{-1})_{\ls[\dot{\alpha} 3]} \rho_{\ls[3]}^{V_0} + (\Lambda_{V_0}^{-1})_{\ls[\dot{\alpha} 4]} \rho_{\ls[4]}^{V_0}
\end{align}
where $\dot{\alpha} = \{3,4\}$ on the RHS of the first equality. For later convenience we define
\begin{equation}
\tilde{\rho}_{\tilde{\alpha}}^{V_0} = (\Lambda^{-1}_{V_0})_{\ls[\tilde{\alpha} \tilde{\beta}]} \rho_{\ls[\tilde{\beta}]}^{V_0}.
\end{equation}

Of course what we are really interested in calculating are expectation values of the original $M_{ij}, N_{ij}$ variables. At linear order Equation \eqref{Eqn: M diagonal variable expansion} allows us to write these in terms of the expectation values of the representation theory variables that we have just calculated,
\begin{align} \nonumber \label{Eqn: M one point function}
\langle M_{ij} \rangle &= \frac{1}{D} \langle X^{V_0}_{1} \rangle + \frac{1}{\sqrt{D-1}} \sum_{a=1}^{D-1} C_{a,i} C_{a,j} \langle X^{V_0}_{2} \rangle \\ \nonumber 
&= \frac{(\Lambda_{V_0}^{-1})_{\ls[1 \tilde{\beta}]} \rho_{\ls[\tilde{\beta}]}^{V_0}}{D} + \frac{(\Lambda_{V_0}^{-1})_{\ls[2 \tilde{\beta}]} \rho_{\ls[\tilde{\beta}]}^{V_0}}{\sqrt{D-1}} F(i,j) \\
&= \frac{\tilde{\rho}_{\ls[1]}^{V_0}}{D} + \frac{\tilde{\rho}_{\ls[2]}^{V_0}}{\sqrt{D-1}} F(i,j)
\end{align}
where we have substituted our previous result \eqref{Eqn: Full X one point function} for the expectation values of $X^{V_A}$.  We have also made use of \eqref{Eqn: f(i,j)} to rewrite the sum over the overlap coefficients $C_{a,i}$ in terms of $F(i,j)$, the projector for $V_H$ in $V_D$. A more detailed discussion of this rewriting is the subject of Appendix \ref{Appendix: Sup formulae}. Similarly, using \eqref{Eqn: Y one point function} the linear expectation value of $N_{ij}$ is given by
\begin{align} \nonumber \label{Eqn: N one point function}
\langle N_{ij} \rangle &= \frac{1}{D} \langle Y^{V_0}_{1} \rangle + \frac{1}{\sqrt{D-1}} \sum_{a=1}^{D-1} C_{a,i} C_{a,j} \langle Y^{V_0}_{2} \rangle \\ \nonumber
&= \frac{(\Lambda_{V_0}^{-1})_{\ls[3 \tilde{\beta}]} \rho_{\ls[\tilde{\beta}]}^{V_0}}{D} + \frac{(\Lambda_{V_0}^{-1})_{\ls[4 \tilde{\beta}]} \rho_{\ls[\tilde{\beta}]}^{V_0}}{\sqrt{D-1}} F(i,j) \\
&= \frac{\tilde{\rho}_{\ls[3]}^{V_0}}{D} + \frac{\tilde{\rho}_{\ls[4]}^{V_0}}{\sqrt{D-1}} F(i,j).
\end{align}

\subsubsection{Quadratic expectation values}
To generate quadratic expectation values we take two derivatives of $\mathcal{Z}$ as follows
\begin{align} \nonumber
\langle X^{V_A }_{\alpha,a} X^{V_B}_{\beta,b} \rangle &= \frac{1}{\mathcal{Z}} \int dX dY X^{V_A}_{\alpha,a} X^{V_B}_{\beta,b} e^{- \mathcal{S}} = \frac{1}{\mathcal{Z}} \frac{\partial}{\partial \rho_{\beta , b}^{V_B, \us[X]}} \frac{\partial \mathcal{Z}}{\partial \rho_{\alpha , a}^{V_A, \us[X]}}\bigg\rvert_{\rho^{V_H} = \rho^{V_2} = \rho^{V_3} = 0} \\ \nonumber
&= \frac{1}{\mathcal{Z}} \frac{(2 \pi)^{D^2}}{(\text{det} \Lambda )^{1/2}} \frac{\partial}{\partial \rho_{\ls[\beta], b}^{V_B, \us[X]}} \Big( (\Lambda^{-1}_{V_A})_{\alpha \tilde{\gamma}} \rho_{\tilde{\gamma} ,a}^{V_A } e^{\frac{1}{2}  \sum_{V_C} \sum_{\tilde{\gamma} \tilde{\mu}} \sum_c \rho_{\tilde{\gamma} , c} (\Lambda^{-1}_{V_C})_{\tilde{\gamma} \tilde{\mu}} \rho_{\tilde{\mu} , c} } \Big)\bigg\rvert_{\rho^{V_H} = \rho^{V_2} = \rho^{V_3} = 0} \\ 
&= \delta(V_A , V_B) (\Lambda_{V_A}^{-1})_{\ls[\alpha \beta]}^{\us[X]} \delta_{a b} + \langle X^{V_A}_{\alpha,a} \rangle \langle X^{V_B}_{\beta,b} \rangle.
\end{align}
Again, this expression is evaluated with all linear couplings of representation variables transforming non-trivially set to zero. The other quadratic terms are calculated in a similar manner
\begin{align}
\langle Y^{V_A}_{\dot{\alpha},a} Y^{V_B}_{\dot{\beta},b} \rangle &= \delta(V_A , V_B) (\Lambda_{V_A}^{-1})_{\ls[\dot{\alpha} \dot{\beta}]}^{\us[Y]} \delta_{a b} + \langle Y^{V_A}_{\dot{\alpha},a} \rangle \langle Y^{V_B}_{\dot{\beta},b} \rangle, \\
\langle X^{V_A}_{\alpha,a} Y^{V_B}_{\dot{\beta},b} \rangle &= \delta(V_A , V_B) (\Lambda_{V_A}^{-1})_{\ls[\alpha \dot{\beta}]}^{\us[XY]} \delta_{a b} + \langle X^{V_A}_{\alpha,a} \rangle \langle Y^{V_B}_{\dot{\beta},b} \rangle, \\ 
\langle Y^{V_A}_{\dot{\alpha},a} X^{V_B}_{\beta,b} \rangle &= \delta(V_A , V_B) (\Lambda_{V_A}^{-1})_{\ls[\dot{\alpha} \beta]}^{\us[YX]} \delta_{a b} + \langle Y^{V_A}_{\dot{\alpha},a} \rangle \langle X^{V_B}_{\beta,b} \rangle.
\end{align}
Defining the connected part of the expectation value 
\begin{equation}
\langle X^{V_A}_{\alpha,a} Y^{V_B}_{\dot{\beta},b} \rangle_{\text{conn}} \equiv \langle X^{V_A}_{\alpha,a} Y^{V_B}_{\dot{\beta},b} \rangle - \langle X^{V_A}_{\alpha,a} \rangle \langle Y^{V_B}_{\dot{\beta},b} \rangle
\end{equation}
we see
\begin{align} \nonumber \label{eqn: Rep basis two point}
\langle X^{V_A}_{\alpha,a} Y^{V_B}_{\dot{\beta},b} \rangle_{\text{conn}} &= \delta(V_A , V_B) (\Lambda_{V_A}^{-1})_{\ls[\alpha \dot{\beta}]}^{\us[XY]} \delta_{a b} = \delta(V_A , V_B) (\Lambda_{V_A}^{-1})_{\ls[\dot{\beta} \alpha]}^{\us[YX]} \delta_{a b} \\
&= \langle Y^{V_A}_{\alpha,a} X^{V_B}_{\dot{\beta},b} \rangle_{\text{conn}}
\end{align}
so that
\begin{equation}
\langle X^{V_A}_{\alpha,a} Y^{V_B}_{\dot{\beta},b} \rangle = \langle Y^{V_A}_{\alpha,a} X^{V_B}_{\dot{\beta},b} \rangle
\end{equation}
for $V_A, V_B = \{V_H,V_2,V_3 \}$.

We now consider quadratic expectation values of the original $M_{ij}$, $N_{ij}$ variables. Writing $\langle M_{ij} N_{kl} \rangle^{V_A}$ to denote the contribution to $\langle M_{ij} N_{kl} \rangle$ dependent on $\Lambda^{-1}_{V_A}$ we have
\begin{equation} \label{Eqn: Channel decomposition}
\langle M_{ij} N_{kl} \rangle_{\text{conn}} = \langle M_{ij} N_{kl} \rangle_{\text{conn}}^{V_0} + \langle M_{ij} N_{kl} \rangle_{\text{conn}}^{V_H} + \langle M_{ij} N_{kl} \rangle_{\text{conn}}^{V_2} + \langle M_{ij} N_{kl} \rangle_{\text{conn}}^{V_3}
\end{equation}
as well as similar relations for $\langle M_{ij} M_{kl} \rangle_{\text{conn}}$ and $\langle N_{ij} N_{kl} \rangle_{\text{conn}}$. Using \eqref{Eqn: M diagonal variable expansion} we expand the quadratic expectation values in $X^{V_A}$ and $Y^{V_A}$, in a similar fashion to the expansion performed in \cite{PIGMM}, as follows
\begin{align} \nonumber
\langle M_{ij} N_{kl} \rangle_{\text{conn}} &= \frac{1}{D^2} \langle X^{V_0}_{1} Y^{V_0}_{1} \rangle_{\text{conn}} + \frac{1}{D-1} \sum_{a_1, a_2 = 1}^{D-1} C_{a_1 , i} C_{a_1 , j} C_{a_2 , k} C_{a_2 , l} \langle X^{V_0}_{2} Y^{V_0}_{2} \rangle_{\text{conn}} \\ \nonumber
&+ \frac{1}{D \sqrt{D-1}} \sum_{a = 1}^{D-1} C_{a , k} C_{a , l} \langle X^{V_0}_{1} Y^{V_0}_{2} \rangle_{\text{conn}} + \frac{1}{D \sqrt{D-1}} \sum_{a = 1}^{D-1} C_{a , i} C_{a , j} \langle X^{V_0}_{2} Y^{V_0}_{1} \rangle_{\text{conn}}  \\ \nonumber
&+ \frac{1}{D} \sum_{a_1, a_2 = 1}^{D-1} C_{a_1 , j} C_{a_2 , l} \langle X^{V_H}_{1, a_1} Y^{V_H}_{1, a_2} \rangle_{\text{conn}} + \frac{1}{D} \sum_{a_1, a_2 = 1}^{D-1} C_{a_1 , i} C_{a_2 , k} \langle X^{V_H}_{2, a_1} Y^{V_H}_{2, a_2} \rangle_{\text{conn}} \\ \nonumber
&+ \sum_{a_1, b_1, c_1, a_2, b_2, c_2 = 1}^{D-1} C_{a_1 , i} C_{b_1 , j} C_{a_2 , k} C_{b_2 , l} C_{a_1, b_1; \quad c_1}^{V_H V_H \rightarrow V_H} C_{a_2, b_2; \quad c_2}^{V_H V_H \rightarrow V_H}  \langle X^{V_H}_{3, c_1} Y^{V_H}_{3, c_2} \rangle_{\text{conn}} \\ \nonumber
&+ \frac{1}{D} \sum_{a_1, a_2 = 1}^{D-1} C_{a_1 , j} C_{a_2 , k} \langle X^{V_H}_{1, a_1} Y^{V_H}_{2, a_2} \rangle_{\text{conn}} + \frac{1}{D} \sum_{a_1, a_2 = 1}^{D-1} C_{a_1 , i} C_{a_2 , l} \langle X^{V_H}_{2, a_1} Y^{V_H}_{1, a_2} \rangle_{\text{conn}} \\ \nonumber
&+ \frac{1}{\sqrt{D}} \sum_{a_1 = 1}^{D-1} \sum_{a_2, b_2, c_2 = 1}^{D-1} C_{a_1 , j} C_{a_2 , k} C_{b_2 , l}  C_{a_2, b_2; \quad c_2}^{V_H V_H \rightarrow V_H} \langle X^{V_H}_{1, a_1} Y^{V_H}_{3, c_2} \rangle_{\text{conn}} \\ \nonumber
&+ \frac{1}{\sqrt{D}} \sum_{a_1,b_1,c_1 = 1}^{D-1} \sum_{a_2 = 1}^{D-1} C_{a_1 , i} C_{b_1 , j} C_{a_1, b_1; \quad c_1}^{V_H V_H \rightarrow V_H} C_{a_2 , l} \langle X^{V_H}_{3, c_1} Y^{V_H}_{1, a_2} \rangle_{\text{conn}} \\ \nonumber
&+ \frac{1}{\sqrt{D}} \sum_{a_1 = 1}^{D-1} \sum_{a_2, b_2, c_2 = 1}^{D-1} C_{a_1 , i} C_{a_2 , k} C_{b_2 , l}  C_{a_2, b_2; \quad c_2}^{V_H V_H \rightarrow V_H} \langle X^{V_H}_{2, a_1} Y^{V_H}_{3, c_2} \rangle_{\text{conn}} \\ \nonumber
&+ \frac{1}{\sqrt{D}} \sum_{a_1,b_1,c_1 = 1}^{D-1} \sum_{a_2 = 1}^{D-1} C_{a_1 , i} C_{b_1 , j} C_{a_1, b_1; \quad c_1}^{V_H V_H \rightarrow V_H} C_{a_2 , k} \langle X^{V_H}_{3, c_1} Y^{V_H}_{2, a_2} \rangle_{\text{conn}} \\ \nonumber
&+ \sum_{a_1,b_1 = 1}^{D-1} \sum_{a_2, b_2 = 1}^{D-1} \sum_{c_1, c_2 = 1}^{\text{Dim}V_2} C_{a_1 , i} C_{b_1 , j} C_{a_1, b_1; \quad c_1}^{V_H V_H \rightarrow V_2} C_{a_2 , k} C_{b_2 , l} C_{a_2, b_2; \quad c_2}^{V_H V_H \rightarrow V_2} \langle X_{c_1}^{V_2} Y_{c_2}^{V_2} \rangle_{\text{conn}} \\
&+ \sum_{a_1,b_1 = 1}^{D-1} \sum_{a_2, b_2 = 1}^{D-1} \sum_{c_1, c_2 = 1}^{\text{Dim}V_3} C_{a_1 , i} C_{b_1 , j} C_{a_1, b_1; \quad c_1}^{V_H V_H \rightarrow V_3} C_{a_2 , k} C_{b_2 , l} C_{a_2, b_2; \quad c_2}^{V_H V_H \rightarrow V_3} \langle X_{c_1}^{V_2} Y_{c_2}^{V_3} \rangle_{\text{conn}}.
\end{align}
Again, utilising the results presented in Appendix \ref{Appendix: Sup formulae} we can write this in terms of the $V_H$ projector in $V_D$ as
\begin{align} \nonumber \label{Eqn: Two point function}
&\langle M_{ij} N_{kl} \rangle_{\text{conn}} = \frac{1}{D^2} (\Lambda_{V_0}^{-1})_{\ls[11]}^{\us[XY]} + \frac{(\Lambda_{V_0}^{-1})_{\ls[22]}^{\us[XY]}}{D-1} F(i,j) F(k,l) + \frac{(\Lambda_{V_0}^{-1})_{\ls[12]}^{\us[XY]}}{D \sqrt{D-1}} F(k,l) +  \frac{(\Lambda_{V_0}^{-1})_{\ls[21]}^{\us[XY]}}{D \sqrt{D-1}} F(i,j) \\ \nonumber
&+ \frac{(\Lambda_{V_H}^{-1})_{\ls[11]}^{\us[XY]}}{D} F(j,l) + \frac{(\Lambda_{V_H}^{-1})_{\ls[22]}^{\us[XY]}}{D} F(i,k) + \frac{D(\Lambda_{V_H}^{-1})_{\ls[33]}^{\us[XY]}}{(D-2)} \sum_{p,q = 1}^{D} F(i,p) F(j,p) F(k,q) F(l,q) F(p,q) \\ \nonumber
&+ \frac{(\Lambda_{V_H}^{-1})_{\ls[12]}^{\us[XY]}}{D} F(j,k) + \frac{(\Lambda_{V_H}^{-1})_{\ls[21]}^{\us[XY]}}{D} F(i,l) + \frac{(\Lambda_{V_H}^{-1})_{\ls[13]}^{\us[XY]}}{\sqrt{D-2}} \sum_{p = 1}^{D} F(j,p) F(k,p) F(l,p) \\ \nonumber
&+ \frac{(\Lambda_{V_H}^{-1})_{\ls[31]}^{\us[XY]}}{\sqrt{D-2}} \sum_{p = 1}^{D} F(i,p) F(j,p) F(l,p) + \frac{(\Lambda_{V_H}^{-1})_{\ls[23]}^{\us[XY]}}{\sqrt{D-2}} \sum_{p = 1}^{D} F(i,p) F(k,p) F(l,p) \\ \nonumber
&+ \frac{(\Lambda_{V_H}^{-1})_{\ls[32]}^{\us[XY]}}{\sqrt{D-2}} \sum_{p = 1}^{D} F(i,p) F(j,p) F(k,p) + (\Lambda_{V_2}^{-1})^{\us[XY]} \Big( \frac{1}{2} F(i,k) F(j,l) + \frac{1}{2} F(i,l) F(j,k) \\ \nonumber
&- \frac{D}{D-2} \sum_{p,q = 1}^D F(i,p) F(j,p) F(k,q) F(l,q) F(p,q) - \frac{1}{(D-1)} F(i,j) F(k,l) \Big) \\ 
&+ \frac{(\Lambda_{V_3}^{-1})^{\us[XY]}}{2} \big( F(i,k) F(j,l) - F(i,l) F(j,k) \big).
\end{align}
We will use this result extensively throughout the rest of the paper to write expectation values of the original $M_{ij}, N_{ij}$ variables in terms of expectation values of the simpler representation theory variables. As originally presented in \cite{PIGMM} the analogue of \eqref{Eqn: Two point function} for the $\langle M_{ij} M_{kl} \rangle_{\text{conn}}$ case is
\begin{align} \nonumber \label{Eqn: Two point function MM}
&\langle M_{ij} M_{kl} \rangle_{\text{conn}} = \frac{1}{D^2} (\Lambda_{V_0}^{-1})_{\ls[11]}^{\us[X]} + \frac{(\Lambda_{V_0}^{-1})_{\ls[22]}^{\us[X]}}{D-1} F(i,j) F(k,l) + \frac{(\Lambda_{V_0}^{-1})_{\ls[12]}^{\us[X]}}{D \sqrt{D-1}} \big( F(k,l) +  F(i,j) \big) \\ \nonumber
&+ \frac{(\Lambda_{V_H}^{-1})_{\ls[11]}^{\us[X]}}{D} F(j,l) + \frac{(\Lambda_{V_H}^{-1})_{\ls[22]}^{\us[X]}}{D} F(i,k) + \frac{D(\Lambda_{V_H}^{-1})_{\ls[33]}^{\us[X]}}{(D-2)} \sum_{p,q = 1}^{D} F(i,p) F(j,p) F(k,q) F(l,q) F(p,q) \\ \nonumber
&+ \frac{(\Lambda_{V_H}^{-1})_{\ls[12]}^{\us[X]}}{D} \big( F(j,k) + F(i,l) \big) + \frac{(\Lambda_{V_H}^{-1})_{\ls[13]}^{\us[X]}}{\sqrt{D-2}} \sum_{p = 1}^{D} \big( F(j,p) F(k,p) F(l,p) + F(i,p) F(j,p) F(l,p) \big)  \\ \nonumber
&+ \frac{(\Lambda_{V_H}^{-1})_{\ls[23]}^{\us[X]}}{\sqrt{D-2}} \sum_{p = 1}^{D} \big( F(i,p) F(k,p) F(l,p) +F(i,p) F(j,p) F(k,p) \big) \\ \nonumber
&+ (\Lambda_{V_2}^{-1})^{\us[X]} \Big( \frac{1}{2} F(i,k) F(j,l) + \frac{1}{2} F(i,l) F(j,k) \\ \nonumber
&- \frac{D}{D-2} \sum_{p,q = 1}^D F(i,p) F(j,p) F(k,q) F(l,q) F(p,q) - \frac{1}{(D-1)} F(i,j) F(k,l) \Big) \\ 
&+ \frac{(\Lambda_{V_3}^{-1})^{\us[X]}}{2} \big( F(i,k) F(j,l) - F(i,l) F(j,k) \big).
\end{align}
The result for quadratic expectation values of two $N$'s, $\langle N_{ij} N_{kl} \rangle_{\text{conn}}$ is given by the same expression with the $M$ couplings replaced by those for $N$ i.e. $(\Lambda_{V_A}^{-1})^{\us[X]} \rightarrow (\Lambda_{V_A}^{-1})^{\us[Y]}$.

The next section outlines a Feynman graph like interpretation of \eqref{Eqn: Two point function}, \eqref{Eqn: Two point function MM} and its $N$ counterpart that can be used to keep track of the representation theoretic origin of each of the terms on the right hand side. We associate rules to each of these graphs, from which it is possible to write down expressions for each quadratic observable directly - without reference to \eqref{Eqn: Two point function} and \eqref{Eqn: Two point function MM}.

\section{Evaluating expectation values: Wick's theorem and $F$-graphs}
\label{Section: Feynman Graphs}
In this section we use Wick's theorem to write degree $k \geq 3$ expectation values in terms of the basic two and one-point functions derived in the previous section. Based on the Wick contraction combinatorics and the $S_D$ representation theoretic structure of the model we develop diagrammatic Feynman rules for the evaluation of expectation values of any degree. In the first step of the Feynman rules,  we have solid lines, dotted lines and simple functions of $D$ arising from the basic two-point function  \eqref{Eqn: Two point function}. The dotted lines are related to $V_0$ while the solid lines are related to $V_H$.  The non-trivial subsequent evaluation involves  sums of products of  factors $ F ( i , j )$, which are  projectors to $V_H$ in $V_D$. By associating $F(i,j)$ with an undirected edge in a graph, and distinct indices with distinct vertices, we get a description of contributions to the expectation values in terms of graphs, which we call ``$F$-graphs''. Expectation values of monomials (with some indices not summed) are described as ``open $F$-graphs'' while expectation values of observables are associated with ``closed $F$-graphs'', which  correspond to products of $F$ with all indices summed
\begin{equation}
\sum F\dots F.
\end{equation}
When evaluated, the sums produce Laurent polynomials in $D$. Section \ref{subsection: F-graph Evaluation} describes a graph algorithm for computing these sums. In Section \ref{subsection: F-graph Derivation} we derive the formula corresponding to the algorithm. We end this section with a discussion of the Feynman rules as an operation 
relating the double cosets describing observable-graphs in Section \ref{Section: Observables and multi-graphs} and
double cosets corresponding to  the undirected uncolored ``F-graphs''. 

\subsection{Wick's theorem and Matrix Feynman rules }
\label{subsection: wicks theorem}
Wick's theorem for distributions with non-zero mean states 
\begin{equation} \label{Eqn: Wicks theorem}
\langle M_{i_1 j_1 }\dots M_{i_m j_m} \rangle = \sum_{p \in P_m^{1,2}} \prod_{c \in p} \prod _{(a,b) \in p}  \langle M_{i_a j_a} M_{i_b j_b} \rangle_{\text{conn}} \langle M_{i_c j_c} \rangle,
\end{equation}
where the sum is over the set partitions $ P_m^{1,2} $ of $\{1,\dots,m\}$ with blocks of size one or two, corresponding to integers $c$ or unordered integer pairs $(a,b)$ in the formula above,  and the products are over the blocks. The rule continues to hold when any number of $M$'s are replaced with $N$'s. The combinatoric structure is encoded in graphs by means of Feynman rules. We assign the following diagrams to the expectation values in Wick's theorem,
\begin{align}
\langle M_{ij} \rangle& = \vcenter{\hbox{\begin{tikzpicture}[scale=2]
		\begin{scope}[decoration={markings, mark=at position 0.6 with \arrow{latex}}]
		\draw[draw=black, line width=2pt, dashed] (0.25,-0.25) -- (0.5,-0.25) node[circle, fill=white, inner sep=2pt, draw=black, solid ]{};;
		\draw[draw=BLUE, postaction={decorate}] (0,0) node[label={left:$i$}] {} -- (0.25,-0.25);
		\draw[draw=BLUE, postaction={decorate}] (0.25,-0.25)--(0,-0.5) node[label={left:$j$}] {};
		\end{scope}
		\end{tikzpicture}}}\\
\langle N_{ij} \rangle &= \vcenter{\hbox{\begin{tikzpicture}[scale=2]
		\begin{scope}[decoration={markings, mark=at position 0.6 with \arrow{latex}}]
		\draw[draw=black, line width=2pt, dashed] (0.25,-0.25) -- (0.5,-0.25)node[circle, fill=white, inner sep=2pt, draw=black, solid ]{};;
		\draw[draw=GREEN, postaction={decorate}] (0,0) node[label={left:$i$}] {} -- (0.25,-0.25);
		\draw[draw=GREEN, postaction={decorate}] (0.25,-0.25)--(0,-0.5) node[label={left:$j$}] {};
		\end{scope}
		\end{tikzpicture}}}\\
\langle M_{ij} M_{kl} \rangle_{\text{conn}}& = \vcenter{\hbox{\begin{tikzpicture}[scale=2]
		\begin{scope}[decoration={markings, mark=at position 0.65 with \arrow{latex}}]
		\draw[draw=BLUE, postaction={decorate}] (0,0) node[label={left:$i$}] {} -- (0.25,-0.25);
		\draw[draw=BLUE, postaction={decorate}] (1,0) node[label={right:$k$}] {} -- (0.75,-0.25);
		\draw[draw=BLUE, postaction={decorate}]  (0.75,-0.25) -- (1,-.5) node[label={right:$l$}] {};
		\draw[draw=BLUE, postaction={decorate}] (0.25,-0.25) -- (0,-.5) node[label={left:$j$}] {};
		\draw[draw=black, line width=2.5pt] (0.25,-0.25) -- (0.5,-0.25);
		\draw[draw=black, line width=2.5pt] (0.5,-0.25) -- (0.75,-0.25);
		\end{scope}
		\end{tikzpicture}}} \\
\langle M_{ij} N_{kl} \rangle_{\text{conn}}& =\vcenter{\hbox{\begin{tikzpicture}[scale=2]
		\begin{scope}[decoration={markings, mark=at position 0.65 with \arrow{latex}}]
		\draw[draw=BLUE, postaction={decorate}] (0,0) node[label={left:$i$}] {} -- (0.25,-0.25);
		\draw[draw=GREEN, postaction={decorate}] (1,0) node[label={right:$k$}] {} -- (0.75,-0.25);
		\draw[draw=GREEN, postaction={decorate}]  (0.75,-0.25) -- (1,-.5) node[label={right:$l$}] {};
		\draw[draw=BLUE, postaction={decorate}] (0.25,-0.25) -- (0,-.5) node[label={left:$j$}] {};
		\draw[draw=black, line width=2.5pt] (0.25,-0.25) -- (0.5,-0.25);
		\draw[draw=black, line width=2.5pt] (0.5,-0.25) -- (0.75,-0.25);
		\end{scope}
		\end{tikzpicture}}} \label{Eqn: 2pt MN diagram EV}\\ \label{Eqn: 2pt diagram EV}
\langle N_{ij} N_{kl} \rangle_{\text{conn}}& = \vcenter{\hbox{\begin{tikzpicture}[scale=2]
		\begin{scope}[decoration={markings, mark=at position 0.65 with \arrow{latex}}]
		\draw[draw=GREEN, postaction={decorate}] (0,0) node[label={left:$i$}] {} -- (0.25,-0.25);
		\draw[draw=GREEN, postaction={decorate}] (1,0) node[label={right:$k$}] {} -- (0.75,-0.25);
		\draw[draw=GREEN, postaction={decorate}]  (0.75,-0.25) -- (1,-.5) node[label={right:$l$}] {};
		\draw[draw=GREEN, postaction={decorate}] (0.25,-0.25) -- (0,-.5) node[label={left:$j$}] {};
		\draw[draw=black, line width=2.5pt] (0.25,-0.25) -- (0.5,-0.25);
		\draw[draw=black, line width=2.5pt] (0.5,-0.25) -- (0.75,-0.25);
		\end{scope}
		\end{tikzpicture}}}
\end{align}
Then Wick's theorem for the $k=3$ expectation values takes the form
\begin{align} \nonumber
\langle &M_{i_1 j_1 }M_{i_2 j_2 } N_{i_3 j_3 } \rangle = \\ \nonumber 
&\langle M_{i_1 j_1 }M_{i_2 j_2 } \rangle_{\text{conn}} \langle N_{i_3 j_3 }\rangle + \langle M_{i_1 j_1 } N_{i_3 j_3 } \rangle_{\text{conn}} \langle M_{i_2 j_2 } \rangle + \langle M_{i_2 j_2 } N_{i_3 j_3 } \rangle_{\text{conn}} \langle M_{i_1 j_1 } \rangle \\ \nonumber
& \hspace{234pt}+ \langle M_{i_1 j_1 } \rangle \langle M_{i_2 j_2 } \rangle \langle N_{i_3 j_3 }\rangle = \\& 
\begin{aligned}
&\vcenter{\hbox{\begin{tikzpicture}[scale=2]
		\begin{scope}[decoration={markings, mark=at position 0.65 with \arrow{latex}}]
		\draw[draw=BLUE, postaction={decorate}] (0,0) node[label={left:$i_1$}] {} -- (0.25,-0.25);
		\draw[draw=BLUE, postaction={decorate}] (1,0) node[label={right:$i_2$}] {} -- (0.75,-0.25);
		\draw[draw=BLUE, postaction={decorate}]  (0.75,-0.25) -- (1,-.5) node[label={right:$j_2$}] {};
		\draw[draw=BLUE, postaction={decorate}] (0.25,-0.25) -- (0,-.5) node[label={left:$j_1$}] {};
		\draw[draw=black, line width=2.5pt] (0.25,-0.25) -- (0.5,-0.25);
		\draw[draw=black, line width=2.5pt] (0.5,-0.25) -- (0.75,-0.25);
		\end{scope}
		\end{tikzpicture}}} \\
&\vcenter{\hbox{\begin{tikzpicture}[scale=2]
		\begin{scope}[decoration={markings, mark=at position 0.6 with \arrow{latex}}]
		\draw[draw=black, line width=2pt, dashed] (0.25,-0.25) -- (0.5,-0.25)node[circle, fill=white, inner sep=2pt, draw=black, solid ]{};
		\draw[draw=GREEN, postaction={decorate}] (0,0) node[label={left:$i_3$}] {} -- (0.25,-0.25);
		\draw[draw=GREEN, postaction={decorate}] (0.25,-0.25)--(0,-0.5) node[label={left:$j_3$}] {};
		\end{scope}
		\end{tikzpicture}}}
\end{aligned}
+
\begin{aligned}
&\vcenter{\hbox{\begin{tikzpicture}[scale=2]
		\begin{scope}[decoration={markings, mark=at position 0.65 with \arrow{latex}}]
		\draw[draw=BLUE, postaction={decorate}] (0,0) node[label={left:$i_1$}] {} -- (0.25,-0.25);
		\draw[draw=GREEN, postaction={decorate}] (1,0) node[label={right:$i_3$}] {} -- (0.75,-0.25);
		\draw[draw=GREEN, postaction={decorate}]  (0.75,-0.25) -- (1,-.5) node[label={right:$j_3$}] {};
		\draw[draw=BLUE, postaction={decorate}] (0.25,-0.25) -- (0,-.5) node[label={left:$j_1$}] {};
		\draw[draw=black, line width=2.5pt] (0.25,-0.25) -- (0.5,-0.25);
		\draw[draw=black, line width=2.5pt] (0.5,-0.25) -- (0.75,-0.25);
		\end{scope}
		\end{tikzpicture}}} \\
&\vcenter{\hbox{\begin{tikzpicture}[scale=2]
		\begin{scope}[decoration={markings, mark=at position 0.6 with \arrow{latex}}]
		\draw[draw=black, line width=2pt, dashed] (0.25,-0.25) -- (0.5,-0.25)node[circle, fill=white, inner sep=2pt, draw=black, solid ]{};
		\draw[draw=BLUE, postaction={decorate}] (0,0) node[label={left:$i_2$}] {} -- (0.25,-0.25);
		\draw[draw=BLUE, postaction={decorate}] (0.25,-0.25)--(0,-0.5) node[label={left:$j_2$}] {};
		\end{scope}
		\end{tikzpicture}}}
\end{aligned}
+
\begin{aligned}
&\vcenter{\hbox{\begin{tikzpicture}[scale=2]
		\begin{scope}[decoration={markings, mark=at position 0.65 with \arrow{latex}}]
		\draw[draw=BLUE, postaction={decorate}] (0,0) node[label={left:$i_2$}] {} -- (0.25,-0.25);
		\draw[draw=GREEN, postaction={decorate}] (1,0) node[label={right:$i_3$}] {} -- (0.75,-0.25);
		\draw[draw=GREEN, postaction={decorate}]  (0.75,-0.25) -- (1,-.5) node[label={right:$j_3$}] {};
		\draw[draw=BLUE, postaction={decorate}] (0.25,-0.25) -- (0,-.5) node[label={left:$j_2$}] {};
		\draw[draw=black, line width=2.5pt] (0.25,-0.25) -- (0.5,-0.25);
		\draw[draw=black, line width=2.5pt] (0.5,-0.25) -- (0.75,-0.25);
		\end{scope}
		\end{tikzpicture}}} \\
&\vcenter{\hbox{\begin{tikzpicture}[scale=2]
		\begin{scope}[decoration={markings, mark=at position 0.6 with \arrow{latex}}]
		\draw[draw=BLUE, postaction={decorate}] (0,0) node[label={left:$i_1$}] {} -- (0.25,-0.25);
		\draw[draw=BLUE, postaction={decorate}] (0.25,-0.25)--(0,-0.5) node[label={left:$j_1$}] {};
		\draw[draw=black, line width=2pt, dashed] (0.25,-0.25) -- (0.5,-0.25)node[circle, fill=white, inner sep=2pt, draw=black, solid ]{};
		\end{scope}
		\end{tikzpicture}}}
\end{aligned}
+
\begin{aligned}
&\vcenter{\hbox{\begin{tikzpicture}[scale=2]
		\begin{scope}[decoration={markings, mark=at position 0.6 with \arrow{latex}}]
		\draw[draw=black, line width=2pt, dashed] (0.25,-0.25) -- (0.5,-0.25)node[circle, fill=white, inner sep=2pt, draw=black, solid ]{};
		\draw[draw=BLUE, postaction={decorate}] (0,0) node[label={left:$i_1$}] {} -- (0.25,-0.25);
		\draw[draw=BLUE, postaction={decorate}] (0.25,-0.25)--(0,-0.5) node[label={left:$j_1$}] {};
		\end{scope}
		\end{tikzpicture}}} \\
&\vcenter{\hbox{\begin{tikzpicture}[scale=2]
		\begin{scope}[decoration={markings, mark=at position 0.6 with \arrow{latex}}]
		\draw[draw=black, line width=2pt, dashed] (0.25,-0.25) -- (0.5,-0.25)node[circle, fill=white, inner sep=2pt, draw=black, solid ]{};
		\draw[draw=BLUE, postaction={decorate}] (0,0) node[label={left:$i_2$}] {} -- (0.25,-0.25);
		\draw[draw=BLUE, postaction={decorate}] (0.25,-0.25)--(0,-0.5) node[label={left:$j_2$}] {};
		\end{scope}
		\end{tikzpicture}}} \\
&\vcenter{\hbox{\begin{tikzpicture}[scale=2]
		\begin{scope}[decoration={markings, mark=at position 0.6 with \arrow{latex}}]
		\draw[draw=black, line width=2pt, dashed] (0.25,-0.25) -- (0.5,-0.25)node[circle, fill=white, inner sep=2pt, draw=black, solid ]{};
		\draw[draw=GREEN, postaction={decorate}] (0,0) node[label={left:$i_3$}] {} -- (0.25,-0.25);
		\draw[draw=GREEN, postaction={decorate}] (0.25,-0.25)--(0,-0.5) node[label={left:$j_3$}] {};
		\end{scope}
		\end{tikzpicture}}}
\end{aligned}
\end{align}

%

\subsection{Representation theoretic decomposition of Feynman rules} \label{Subsec: Feynman decomp}
Equations \eqref{Eqn: M one point function}, \eqref{Eqn: N one point function} and \eqref{Eqn: Two point function} describe the expectation values of matrix elements $M_{ij}$, $N_{ij}$ in terms of the couplings of variables that transform irreducibly under $S_D$. Each term can be given a Feynman rule, which captures the associated representation theory.

The degree one Feynman rules can be decomposed into two diagrams corresponding to each term in \eqref{Eqn: M one point function} or \eqref{Eqn: N one point function}.
\begin{align}
\langle M_{ij} \rangle = &\vcenter{\hbox{\begin{tikzpicture}[scale=2]
		\begin{scope}[decoration={markings, mark=at position 0.6 with \arrow{latex}}]
		\draw[draw=BLUE, postaction={decorate}] (0,0) node[label={left:$i$}] {} -- (0.25,-0.25);
		\draw[draw=BLUE, postaction={decorate}] (0.25,-0.25)--(0,-0.5) node[label={left:$j$}] {};
		\draw[draw=black, line width=2pt, dashed] (0.25,-0.25) -- (0.5,-0.25)node[circle, fill=white, inner sep=2pt, draw=black, solid ]{};
		\end{scope}
		\end{tikzpicture}}} \,= \vcenter{\hbox{\begin{tikzpicture}[scale=2]\begin{scope}[decoration={markings, mark=at position 0.65 with \arrow{latex}}]
		\node[label={left:$i$}] at (0,0) {};
		\node[label={left:$j$}] at (0,-.5) {};
		\draw[draw=BLUE, dashed] (0,0) --(0.25,-.25) -- (0,-.5);	
		\draw[draw=black, dashed] (.25,-0.25) -- (.5,-0.25) node[circle, fill=white, inner sep=2pt, draw=black, solid] {};			
		\node[circle,fill=BLUE, inner sep=1pt, draw=BLUE] at (0.25,-.25) {};
		\end{scope}
		\end{tikzpicture}}} \; + \vcenter{\hbox{\begin{tikzpicture}[scale=2]\begin{scope}[decoration={markings, mark=at position 0.65 with \arrow{latex}}]
		\node[label={left:$i$}] at (0,0) {};
		\node[label={left:$j$}] at (0,-.5) {};
		\draw[draw=BLUE ] (0,0) -- (0.25,-.25) -- (0,-.5);	
		\draw[draw=black, dashed] (.25,-0.25) -- (.5,-0.25)node[circle, fill=white, inner sep=2pt, draw=black, solid ]{};;
		\node[circle, fill=BLUE, inner sep=1pt, draw=BLUE] at (0.25,-.25) {};
		\end{scope}
		\end{tikzpicture}}},\\
&\vcenter{\hbox{\begin{tikzpicture}[scale=2]\begin{scope}[decoration={markings, mark=at position 0.65 with \arrow{latex}}]
		\node[label={left:$i$}] at (0,0) {};
		\node[label={left:$j$}] at (0,-.5) {};
		\draw[draw=BLUE, dashed] (0,0) --(0.25,-.25) -- (0,-.5);	
		\draw[draw=black, dashed] (.25,-0.25) -- (.5,-0.25) node[circle, fill=white, inner sep=2pt, draw=black, solid] {};			
		\node[circle,fill=BLUE, inner sep=1pt, draw=BLUE] at (0.25,-.25) {};
		\end{scope}
		\end{tikzpicture}}} = \frac{\tilde{\rho}_{\ls[1]}^{V_0}}{D}, \label{Eqn: V_0V_0 M One Point Projection}\\
&\vcenter{\hbox{\begin{tikzpicture}[scale=2]\begin{scope}[decoration={markings, mark=at position 0.65 with \arrow{latex}}]
		\node[label={left:$i$}] at (0,0) {};
		\node[label={left:$j$}] at (0,-.5) {};
		\draw[draw=BLUE ] (0,0) -- (0.25,-.25) -- (0,-.5);	
		\draw[draw=black, dashed] (.25,-0.25) -- (.5,-0.25)node[circle, fill=white, inner sep=2pt, draw=black, solid ]{};;
		\node[circle, fill=BLUE, inner sep=1pt, draw=BLUE] at (0.25,-.25) {};
		\end{scope}
		\end{tikzpicture}}} = \frac{\tilde{\rho}_{\ls[2]}^{V_0}}{\sqrt{D-1}} F(i,j). \label{Eqn: V_HV_H M One Point Projection}
\end{align}

\begin{align}
\langle N_{ij} \rangle = &\vcenter{\hbox{\begin{tikzpicture}[scale=2]
		\begin{scope}[decoration={markings, mark=at position 0.6 with \arrow{latex}}]
		\draw[draw=GREEN, postaction={decorate}] (0,0) node[label={left:$i$}] {} -- (0.25,-0.25);
		\draw[draw=GREEN, postaction={decorate}] (0.25,-0.25)--(0,-0.5) node[label={left:$j$}] {};
		\draw[draw=black, line width=2pt, dashed] (0.25,-0.25) -- (0.5,-0.25)node[circle, fill=white, inner sep=2pt, draw=black, solid ]{};
		\end{scope}
		\end{tikzpicture}}} \,= \vcenter{\hbox{\begin{tikzpicture}[scale=2]\begin{scope}[decoration={markings, mark=at position 0.65 with \arrow{latex}}]
		\node[label={left:$i$}] at (0,0) {};
		\node[label={left:$j$}] at (0,-.5) {};
		\draw[draw=GREEN, dashed] (0,0) --(0.25,-.25) -- (0,-.5);	
		\draw[draw=black, dashed] (.25,-0.25) -- (.5,-0.25) node[circle, fill=white, inner sep=2pt, draw=black, solid] {};			
		\node[circle,fill=GREEN, inner sep=1pt, draw=GREEN] at (0.25,-.25) {};
		\end{scope}
		\end{tikzpicture}}} \; + \vcenter{\hbox{\begin{tikzpicture}[scale=2]\begin{scope}[decoration={markings, mark=at position 0.65 with \arrow{latex}}]
		\node[label={left:$i$}] at (0,0) {};
		\node[label={left:$j$}] at (0,-.5) {};
		\draw[draw=GREEN ] (0,0) -- (0.25,-.25) -- (0,-.5);	
		\draw[draw=black, dashed] (.25,-0.25) -- (.5,-0.25)node[circle, fill=white, inner sep=2pt, draw=black, solid ]{};;
		\node[circle, fill=GREEN, inner sep=1pt, draw=GREEN] at (0.25,-.25) {};
		\end{scope}
		\end{tikzpicture}}},\\
&\vcenter{\hbox{\begin{tikzpicture}[scale=2]\begin{scope}[decoration={markings, mark=at position 0.65 with \arrow{latex}}]
		\node[label={left:$i$}] at (0,0) {};
		\node[label={left:$j$}] at (0,-.5) {};
		\draw[draw=GREEN, dashed] (0,0) --(0.25,-.25) -- (0,-.5);	
		\draw[draw=black, dashed] (.25,-0.25) -- (.5,-0.25) node[circle, fill=white, inner sep=2pt, draw=black, solid] {};			
		\node[circle,fill=GREEN, inner sep=1pt, draw=GREEN] at (0.25,-.25) {};
		\end{scope}
		\end{tikzpicture}}} = \frac{\tilde{\rho}_{\ls[3]}^{V_0}}{D},\\
&\vcenter{\hbox{\begin{tikzpicture}[scale=2]\begin{scope}[decoration={markings, mark=at position 0.65 with \arrow{latex}}]
		\node[label={left:$i$}] at (0,0) {};
		\node[label={left:$j$}] at (0,-.5) {};
		\draw[draw=GREEN ] (0,0) -- (0.25,-.25) -- (0,-.5);	
		\draw[draw=black, dashed] (.25,-0.25) -- (.5,-0.25)node[circle, fill=white, inner sep=2pt, draw=black, solid ]{};;
		\node[circle, fill=GREEN, inner sep=1pt, draw=GREEN] at (0.25,-.25) {};
		\end{scope}
		\end{tikzpicture}}} = \frac{\tilde{\rho}_{\ls[4]}^{V_0}}{\sqrt{D-1}} F(i,j).
\end{align}
The dashed external lines in equation \eqref{Eqn: V_0V_0 M One Point Projection} each give a representation theoretical factor $C_{0i}$, which picks out $V_0$ in the decomposition $V_D \cong V_0 \oplus V_H$. Similarly, the solid lines in equation \eqref{Eqn: V_HV_H M One Point Projection} each come with a factor $C_{ai}$, picking out $V_H$. The solid lines are contracted with a Clebsch-Gordan coefficient which picks out the trivial representation in $V_H \otimes V_H$ (see Appendix C in \cite{PIGMM}),
\begin{equation}
C_{ab,0}^{V_H, V_H \rightarrow V_0} = \frac{\delta_{ab}}{\sqrt{D-1}}.
\end{equation}

Similarly one-color and two-color degree two moments can be decomposed into 11 and 15 independent contributions, respectively. For the two-color case $\langle M N \rangle_{\text{conn}}$, given by \eqref{Eqn: Two point function}, we have the term-wise diagrammatic descriptions
\begin{flalign}
&\vcenter{\hbox{\begin{tikzpicture}[scale=2]\begin{scope}[decoration={markings, mark=at position 0.65 with \arrow{latex}}]
		\node[label={left:$i$}] at (0,0) {};
		\node[label={right:$k$}] at (1,0) {};
		\node[label={left:$j$}] at (0,-.5) {};
		\node[label={right:$l$}] at (1,-.5) {};
		\draw[draw=BLUE, dashed] (0,-.5) -- (0.25,-.25);
		\draw[draw=BLUE, dashed] (.25,-0.25) -- (0,0);
		\draw[draw=GREEN, dashed] (.75,-0.25) -- (1,0);	
		\draw[draw=GREEN, dashed] (.75,-0.25) -- (1,-.5);	
		\draw[draw=black, dashed] (.25,-0.25) -- (.5,-0.25) node[label=above:{\tiny $V_0$}]{};
		\draw[draw=black, dashed] (.5,-0.25) -- (.75,-0.25);
		\node[circle, fill=BLUE, inner sep=1pt, draw=BLUE] at (0.25,-.25) {};
		\node[circle, fill=GREEN, inner sep=1pt, draw=GREEN] at (0.75,-.25) {};
		\end{scope}
		\end{tikzpicture}}} = \frac{(\Lambda_{V_0}^{-1})_{\ls[11]}^{\us[XY]}}{D^2},
&&
\vcenter{\hbox{\begin{tikzpicture}[scale=2]\begin{scope}[decoration={markings, mark=at position 0.65 with \arrow{latex}}]
		\node[label={left:$i$}] at (0,0) {};
		\node[label={right:$k$}] at (1,0) {};
		\node[label={left:$j$}] at (0,-.5) {};
		\node[label={right:$l$}] at (1,-.5) {};
		\draw[draw=BLUE ] (0,0) -- (0.25,-.25) -- (0,-.5);
		\draw[draw=GREEN ] (1,0) -- (0.75, -.25)--(1,-.5);
		\draw[draw=black, dashed] (.25,-0.25) -- (.5,-0.25) node[label=above:{\tiny $V_0$}]{};
		\draw[draw=black, dashed] (.5,-0.25) -- (.75,-0.25);
		\node[circle, fill=BLUE, inner sep=1pt, draw=BLUE] at (0.25,-.25) {};
		\node[circle, fill=GREEN, inner sep=1pt, draw=GREEN] at (0.75,-.25) {};
		\end{scope}
		\end{tikzpicture}}} =\frac{(\Lambda_{V_0}^{-1})_{\ls[22]}^{\us[XY]}}{D-1} F(i,j) F(k,l),
\\[2em]
&\vcenter{\hbox{\begin{tikzpicture}[scale=2]\begin{scope}[decoration={markings, mark=at position 0.65 with \arrow{latex}}]
		\node[label={left:$i$}] at (0,0) {};
		\node[label={right:$k$}] at (1,0) {};
		\node[label={left:$j$}] at (0,-.5) {};
		\node[label={right:$l$}] at (1,-.5) {};
		\draw[draw=BLUE,dashed] (0,0) -- (0.25,-.25) -- (0,-.5);
		\draw[draw=GREEN ] (1,0) -- (0.75, -.25)--(1,-.5);
		\draw[draw=black, dashed] (.25,-0.25) -- (.5,-0.25) node[label=above:{\tiny $V_0$}]{};
		\draw[draw=black, dashed] (.5,-0.25) -- (.75,-0.25);
		\node[circle, fill=BLUE, inner sep=1pt, draw=BLUE] at (0.25,-.25) {};
		\node[circle, fill=GREEN, inner sep=1pt, draw=GREEN] at (0.75,-.25) {};
		\end{scope}
		\end{tikzpicture}}} =\frac{(\Lambda_{V_0}^{-1})_{\ls[12]}^{\us[XY]}}{D \sqrt{D-1}} F(k,l),
&&
\vcenter{\hbox{\begin{tikzpicture}[scale=2]\begin{scope}[decoration={markings, mark=at position 0.65 with \arrow{latex}}]
		\node[label={left:$i$}] at (0,0) {};
		\node[label={right:$k$}] at (1,0) {};
		\node[label={left:$j$}] at (0,-.5) {};
		\node[label={right:$l$}] at (1,-.5) {};
		\draw[draw=BLUE ] (0,0) -- (0.25,-.25) -- (0,-.5);
		\draw[draw=GREEN, dashed] (1,0) -- (0.75, -.25)--(1,-.5);
		\draw[draw=black, dashed] (.25,-0.25) -- (.5,-0.25) node[label=above:{\tiny $V_0$}]{};
		\draw[draw=black, dashed] (.5,-0.25) -- (.75,-0.25);
		\node[circle, fill=BLUE, inner sep=1pt, draw=BLUE] at (0.25,-.25) {};
		\node[circle, fill=GREEN, inner sep=1pt, draw=GREEN] at (0.75,-.25) {};
		\end{scope}
		\end{tikzpicture}}} = \frac{(\Lambda_{V_0}^{-1})_{\ls[21]}^{\us[XY]}}{D \sqrt{D-1}} F(i,j),
\\[2em]
&\vcenter{\hbox{\begin{tikzpicture}[scale=2]\begin{scope}[decoration={markings, mark=at position 0.65 with \arrow{latex}}]
		\node[label={left:$i$}] at (0,0) {};
		\node[label={right:$k$}] at (1,0) {};
		\node[label={left:$j$}] at (0,-.5) {};
		\node[label={right:$l$}] at (1,-.5) {};
		\draw[draw=BLUE ] (0,-.5) -- (0.25,-.25);
		\draw[draw=BLUE,dashed ] (.25,-0.25) -- (0,0);
		\draw[draw=GREEN,dashed ] (.75,-0.25) -- (1,0);	
		\draw[draw=GREEN ] (.75,-0.25) -- (1,-.5);	
		\draw[draw=black] (.25,-0.25) -- (.5,-0.25) node[label=above:{\tiny $V_H$}]{};
		\draw[draw=black] (.5,-0.25) -- (.75,-0.25);
		\node[circle, fill=BLUE, inner sep=1pt, draw=BLUE] at (.25,-.25) {};
		\node[circle, fill=GREEN, inner sep=1pt, draw=GREEN] at (.75,-.25) {};
		\end{scope}
		\end{tikzpicture}}} =\frac{(\Lambda_{V_H}^{-1})_{\ls[11]}^{\us[XY]}}{D} F(j,l),
&&
\vcenter{\hbox{\begin{tikzpicture}[scale=2]\begin{scope}[decoration={markings, mark=at position 0.65 with \arrow{latex}}]
		\node[label={left:$i$}] at (0,0) {};
		\node[label={right:$k$}] at (1,0) {};
		\node[label={left:$j$}] at (0,-.5) {};
		\node[label={right:$l$}] at (1,-.5) {};
		\draw[draw=BLUE,dashed] (0,-.5) -- (0.25,-.25);
		\draw[draw=BLUE] (.25,-0.25) -- (0,0);
		\draw[draw=GREEN ] (.75,-0.25) -- (1,0);	
		\draw[draw=GREEN, dashed] (.75,-0.25) -- (1,-.5);	
		\draw[draw=black] (.25,-0.25) -- (.5,-0.25) node[label=above:{\tiny $V_H$}]{};
		\draw[draw=black] (.5,-0.25) -- (.75,-0.25);
		\node[circle, fill=BLUE, inner sep=1pt, draw=BLUE] at (.25,-.25) {};
		\node[circle, fill=GREEN, inner sep=1pt, draw=GREEN] at (.75,-.25) {};
		\end{scope}
		\end{tikzpicture}}} =\frac{(\Lambda_{V_H}^{-1})_{\ls[22]}^{\us[XY]}}{D} F(i,k),
\\[2em]
&\vcenter{\hbox{\begin{tikzpicture}[scale=2]\begin{scope}[decoration={markings, mark=at position 0.65 with \arrow{latex}}]
		\node[label={left:$i$}] at (0,0) {};
		\node[label={right:$k$}] at (1,0) {};
		\node[label={left:$j$}] at (0,-.5) {};
		\node[label={right:$l$}] at (1,-.5) {};
		\draw[draw=BLUE, dashed] (0,-.5) -- (0.25,-.25);
		\draw[draw=BLUE ] (.25,-0.25) -- (0,0);
		\draw[draw=GREEN, dashed] (.75,-0.25) -- (1,0);	
		\draw[draw=GREEN ] (.75,-0.25) -- (1,-.5);	
		\draw[draw=black] (.25,-0.25) -- (.5,-0.25) node[label=above:{\tiny $V_H$}]{};
		\draw[draw=black] (.5,-0.25) -- (.75,-0.25);
		\node[circle, fill=BLUE, inner sep=1pt, draw=BLUE] at (.25,-.25) {};
		\node[circle, fill=GREEN, inner sep=1pt, draw=GREEN] at (.75,-.25) {};
		\end{scope}
		\end{tikzpicture}}} =\frac{(\Lambda_{V_H}^{-1})_{\ls[21]}^{\us[XY]}}{D} F(i,l),
&&
\vcenter{\hbox{\begin{tikzpicture}[scale=2]\begin{scope}[decoration={markings, mark=at position 0.65 with \arrow{latex}}]
		\node[label={left:$i$}] at (0,0) {};
		\node[label={right:$k$}] at (1,0) {};
		\node[label={left:$j$}] at (0,-.5) {};
		\node[label={right:$l$}] at (1,-.5) {};
		\draw[draw=BLUE] (0,-.5) -- (0.25,-.25);
		\draw[draw=BLUE, dashed] (.25,-0.25) -- (0,0);
		\draw[draw=GREEN ] (.75,-0.25) -- (1,0);	
		\draw[draw=GREEN, dashed] (.75,-0.25) -- (1,-.5);	
		\draw[draw=black] (.25,-0.25) -- (.5,-0.25) node[label=above:{\tiny $V_H$}]{};
		\draw[draw=black] (.5,-0.25) -- (.75,-0.25);
		\node[circle, fill=BLUE, inner sep=1pt, draw=BLUE] at (.25,-.25) {};
		\node[circle, fill=GREEN, inner sep=1pt, draw=GREEN] at (.75,-.25) {};
		\end{scope}
		\end{tikzpicture}}} =\frac{(\Lambda_{V_H}^{-1})_{\ls[12]}^{\us[XY]}}{D} F(j,k).
\end{flalign}
For diagrams with an internal $V_H$ propagator there is one new ingredient. It follows from equation \eqref{eqn: Rep basis two point} with $V_A=V_B=V_H$ that the internal $V_H$ propagator is $\delta_{ab}$. This explains the factor
\begin{equation}
	\sum_{a,b} C_{0j}C_{0l}C_{ai}C_{bk}\delta_{ab}=\frac{1}{D}F(i,k).
\end{equation}
For the diagrams where three solid edges meet at a vertex, such as \eqref{Eqn: V_0V_HV_HV_H Two Point Projection}, there is a Clebsch-Gordan coefficient (see Appendix C in \cite{PIGMM})
\begin{equation}
C_{ab,c}^{V_H, V_H \rightarrow V_H} = \sqrt{\frac{D}{D-2}} \sum_p C_{ap}C_{bp}C_{cp},
\end{equation}
which picks out $V_H$ in $V_H \otimes V_H$.
In total they give (for example equation \eqref{Eqn: V_0V_HV_HV_H Two Point Projection}),
\begin{equation}
	\sum_{a,b,c,d} C_{0i}C_{aj} \delta_{ab} C_{cd,b}^{V_H, V_H \rightarrow V_H} C_{ck}C_{dl} = \frac{1}{\sqrt{D-2}}\sum_{p=1}^DF(j,p)F(k,p)F(l,p).
\end{equation}

\begin{align} 
\vcenter{\hbox{\begin{tikzpicture}[scale=2]\begin{scope}[decoration={markings, mark=at position 0.65 with \arrow{latex}}]
		\node[label={left:$i$}] at (0,0) {};
		\node[label={right:$k$}] at (1,0) {};
		\node[label={left:$j$}] at (0,-.5) {};
		\node[label={right:$l$}] at (1,-.5) {};
		\draw[draw=BLUE ] (0,-.5) -- (0.25,-.25);
		\draw[draw=BLUE, dashed] (.25,-0.25) -- (0,0);
		\draw[draw=GREEN ] (.75,-0.25) -- (1,0);	
		\draw[draw=GREEN ] (.75,-0.25) -- (1,-.5);	
		\draw[draw=black] (.25,-0.25) -- (.5,-0.25) node[label=above:{\tiny $V_H$}]{};
		\draw[draw=black] (.5,-0.25) -- (.75,-0.25);
		\node[circle, fill=BLUE, inner sep=1pt, draw=BLUE] at (.25,-.25) {};
		\node[circle, fill=GREEN, inner sep=1pt, draw=GREEN] at (.75,-.25) {};
		\end{scope}
		\end{tikzpicture}}} &= \frac{(\Lambda_{V_H}^{-1})_{\ls[13]}^{\us[XY]}}{\sqrt{D-2}} \sum_{p = 1}^{D} F(j,p) F(k,p) F(l,p) \label{Eqn: V_0V_HV_HV_H Two Point Projection}\\
\vcenter{\hbox{\begin{tikzpicture}[scale=2]\begin{scope}[decoration={markings, mark=at position 0.65 with \arrow{latex}}]
		\node[label={left:$i$}] at (0,0) {};
		\node[label={right:$k$}] at (1,0) {};
		\node[label={left:$j$}] at (0,-.5) {};
		\node[label={right:$l$}] at (1,-.5) {};
		\draw[draw=BLUE ] (0,-.5) -- (0.25,-.25);
		\draw[draw=BLUE ] (.25,-0.25) -- (0,0);
		\draw[draw=GREEN, dashed] (.75,-0.25) -- (1,0);	
		\draw[draw=GREEN ] (.75,-0.25) -- (1,-.5);	
		\draw[draw=black] (.25,-0.25) -- (.5,-0.25) node[label=above:{\tiny $V_H$}]{};
		\draw[draw=black] (.5,-0.25) -- (.75,-0.25);
		\node[circle, fill=BLUE, inner sep=1pt, draw=BLUE] at (.25,-.25) {};
		\node[circle, fill=GREEN, inner sep=1pt, draw=GREEN] at (.75,-.25) {};
		\end{scope}
		\end{tikzpicture}}} &=\frac{(\Lambda_{V_H}^{-1})_{\ls[31]}^{\us[XY]}}{\sqrt{D-2}} \sum_{p = 1}^{D} F(i,p) F(j,p) F(l,p) \\
\vcenter{\hbox{\begin{tikzpicture}[scale=2]\begin{scope}[decoration={markings, mark=at position 0.65 with \arrow{latex}}]
		\node[label={left:$i$}] at (0,0) {};
		\node[label={right:$k$}] at (1,0) {};
		\node[label={left:$j$}] at (0,-.5) {};
		\node[label={right:$l$}] at (1,-.5) {};
		\draw[draw=BLUE,dashed] (0,-.5) -- (0.25,-.25);
		\draw[draw=BLUE ] (.25,-0.25) -- (0,0);
		\draw[draw=GREEN ] (.75,-0.25) -- (1,0);	
		\draw[draw=GREEN ] (.75,-0.25) -- (1,-.5);	
		\draw[draw=black] (.25,-0.25) -- (.5,-0.25) node[label=above:{\tiny $V_H$}]{};
		\draw[draw=black] (.5,-0.25) -- (.75,-0.25);
		\node[circle, fill=BLUE, inner sep=1pt, draw=BLUE] at (.25,-.25) {};
		\node[circle, fill=GREEN, inner sep=1pt, draw=GREEN] at (.75,-.25) {};
		\end{scope}
		\end{tikzpicture}}} &=\frac{(\Lambda_{V_H}^{-1})_{\ls[23]}^{\us[XY]}}{\sqrt{D-2}} \sum_{p = 1}^{D} F(i,p) F(k,p) F(l,p) \\
\vcenter{\hbox{\begin{tikzpicture}[scale=2]\begin{scope}[decoration={markings, mark=at position 0.65 with \arrow{latex}}]
		\node[label={left:$i$}] at (0,0) {};
		\node[label={right:$k$}] at (1,0) {};
		\node[label={left:$j$}] at (0,-.5) {};
		\node[label={right:$l$}] at (1,-.5) {};
		\draw[draw=BLUE ] (0,-.5) -- (0.25,-.25);
		\draw[draw=BLUE ] (.25,-0.25) -- (0,0);
		\draw[draw=GREEN ] (.75,-0.25) -- (1,0);	
		\draw[draw=GREEN,dashed] (.75,-0.25) -- (1,-.5);	
		\draw[draw=black] (.25,-0.25) -- (.5,-0.25) node[label=above:{\tiny $V_H$}]{};
		\draw[draw=black] (.5,-0.25) -- (.75,-0.25);
		\node[circle, fill=BLUE, inner sep=1pt, draw=BLUE] at (.25,-.25) {};
		\node[circle, fill=GREEN, inner sep=1pt, draw=GREEN] at (.75,-.25) {};
		\end{scope}
		\end{tikzpicture}}} &=\frac{(\Lambda_{V_H}^{-1})_{\ls[32]}^{\us[XY]}}{\sqrt{D-2}} \sum_{p = 1}^{D} F(i,p) F(j,p) F(k,p) \\
\vcenter{\hbox{\begin{tikzpicture}[scale=2]\begin{scope}[decoration={markings, mark=at position 0.65 with \arrow{latex}}]
		\node[label={left:$i$}] at (0,0) {};
		\node[label={right:$k$}] at (1,0) {};
		\node[label={left:$j$}] at (0,-.5) {};
		\node[label={right:$l$}] at (1,-.5) {};
		\draw[draw=BLUE ] (0,-.5) -- (0.25,-.25);
		\draw[draw=BLUE ] (.25,-0.25) -- (0,0);
		\draw[draw=GREEN ] (.75,-0.25) -- (1,0);	
		\draw[draw=GREEN ] (.75,-0.25) -- (1,-.5);	
		\draw[draw=black] (.25,-0.25) -- (.5,-0.25) node[label=above:{\tiny $V_H$}]{};
		\draw[draw=black] (.5,-0.25) -- (.75,-0.25);
		\node[circle, fill=BLUE, inner sep=1pt, draw=BLUE] at (.25,-.25) {};
		\node[circle, fill=GREEN, inner sep=1pt, draw=GREEN] at (.75,-.25) {};
		\end{scope}
		\end{tikzpicture}}} &=\frac{D(\Lambda_{V_H}^{-1})_{\ls[33]}^{\us[XY]}}{(D-2)} \sum_{p,q = 1}^{D} F(i,p) F(j,p) F(k,q) F(l,q) F(p,q) \label{Eqn: V_H V_H Two Point Projection}\\
\vcenter{\hbox{\begin{tikzpicture}[scale=2]\begin{scope}[decoration={zigzag,segment length=3,amplitude=2,post length=0pt}]
		\node[label={left:$i$}] at (0,0) {};
		\node[label={right:$k$}] at (1,0) {};
		\node[label={left:$j$}] at (0,-.5) {};
		\node[ label={right:$l$}] at (1,-.5) {};
		\draw[draw=BLUE ] (0,-.5) -- (0.25,-.25);
		\draw[draw=BLUE ] (.25,-0.25) -- (0,0);
		\draw[draw=GREEN ] (.75,-0.25) -- (1,0);	
		\draw[draw=GREEN ] (.75,-0.25) -- (1,-.5);	
		\draw[draw=black] (.25,-0.25) -- (.5,-0.25) node[label=above:{\tiny $V_3$}]{};
		\draw[draw=black] (.5,-0.25) -- (.75,-0.25);
		\node[circle, fill=BLUE, inner sep=1pt, draw=BLUE] at (.25,-.25) {};
		\node[circle, fill=GREEN, inner sep=1pt, draw=GREEN] at (.75,-.25) {};
		\end{scope}
		\end{tikzpicture}}}&=\frac{(\Lambda_{V_3}^{-1})^{\us[XY]}}{2} \Big( F(i,k) F(j,l) - F(i,l) F(j,k) \Big)\label{Eqn: V_3 V_3 Two Point Projection}\\
\vcenter{\hbox{\begin{tikzpicture}[scale=2]\begin{scope}[decoration={snake,segment length=3,amplitude=2, post length=0pt}]
		\node[label={left:$i$}] at (0,0) {};
		\node[label={right:$k$}] at (1,0) {};
		\node[label={left:$j$}] at (0,-.5) {};
		\node[label={right:$l$}] at (1,-.5) {};
		\draw[draw=BLUE ] (0,-.5) -- (0.25,-.25);
		\draw[draw=BLUE ] (.25,-0.25) -- (0,0);
		\draw[draw=GREEN ] (.75,-0.25) -- (1,0);	
		\draw[draw=GREEN ] (.75,-0.25) -- (1,-.5);	
		\draw[draw=black] (.25,-0.25) -- (.5,-0.25) node[label=above:{\tiny $V_2$}]{};
		\draw[draw=black] (.5,-0.25) -- (.75,-0.25);
		\node[circle, fill=BLUE, inner sep=1pt, draw=BLUE] at (.25,-.25) {};
		\node[circle, fill=GREEN, inner sep=1pt, draw=GREEN] at (.75,-.25) {};
		\end{scope}
		\end{tikzpicture}}} &=\begin{aligned}[t]
&(\Lambda_{V_2}^{-1})^{\us[XY]}\Big(\frac{F(i,k) F(j,l)}{2}  + \frac{F(i,l) F(j,k)}{2} - \frac{F(i,j) F(k,l)}{(D-1)}\\
& -\frac{D}{D-2} \sum_{p,q = 1}^D F(i,p) F(j,p) F(k,q) F(l,q) F(p,q) \Big) 
\end{aligned} \label{Eqn: V_2 V_2 Two Point Projection}
\end{align}
The right-hand side of the diagrams \eqref{Eqn: V_3 V_3 Two Point Projection} and \eqref{Eqn: V_2 V_2 Two Point Projection} rely on the fact that while the decomposition $C^{V_D,V_D \rightarrow V_{2,3}}$ requires a choice of basis in $V_2$ or $V_3$, the projectors $P^{V_D,V_D \rightarrow V_{2,3}}$ can be written in terms of $V_H$ projectors without reference to a $V_2$ or $V_3$ basis. This can be understood from the following two observations. The representation $V_3$ is isomorphic to the anti-symmetric  subspace of the tensor product of two copies of $V_H$
\begin{align} \label{Eqn: V3 decomp}
V_3 =  \Lambda^2(V_H),
\end{align}
and $V_2$ is the orthogonal complement to $V_0 \oplus V_H$ in the symmetric product of $V_H$,
\begin{align} \label{Eqn: V2 decomp}
\text{Sym}^2 (V_H) = V_0 \oplus V_H \oplus V_2.
\end{align}
Therefore, $P_{V_3}$ can be written using an anti-symmetrizer $\tfrac{1}{2}(1-\tau)$, where $\tau \in S_2$ acts by permuting the factors in a tensor product, 
\begin{align} \nonumber
\langle e_i& \otimes e_j | (P_{V_H} \otimes P_{V_H}) P_{V_3}^2(P_{V_H} \otimes P_{V_H}) | e_k \otimes e_l \rangle \\ \nonumber
&= \langle e_i \otimes e_j | (P_{V_H} \otimes P_{V_H}) \Big(\frac{1-\tau}{2}\Big) \Big(\frac{1-\tau}{2}\Big) (P_{V_H} \otimes P_{V_H}) | e_k \otimes e_l \rangle \\ \nonumber
&= \langle E_i \otimes E_j | \Big(\frac{1-\tau}{2}\Big) | E_k \otimes E_l \rangle \\  \nonumber
&= \frac{1}{2} \Big( \langle E_i \otimes E_j | E_k \otimes E_l \rangle - \langle E_j \otimes E_i | E_k \otimes E_l \rangle \Big) \\
&=\frac{1}{2}\Big(F(i,k)F(j,l)-F(j,k)F(i,l)\Big),
\end{align}
with $e_i$ forming an orthonormal basis of $V_D$ and $E_i$ forming an orthonormal basis over $V_H$, the key properties of which can be found in Appendix \ref{Appendix: Sup formulae} and a more detailed discussion is contained within \cite{PIGMM}. Similarly, using the above $V_2$ decomposition \eqref{Eqn: V2 decomp} we have 
\begin{align} \nonumber
\langle e_i& \otimes e_j | (P_{V_H} \otimes P_{V_H}) P_{V_2}^2(P_{V_H} \otimes P_{V_H}) | e_k \otimes e_l \rangle \\ \nonumber
&= \langle e_i \otimes e_j | (P_{V_H} \otimes P_{V_H}) \Big(\frac{1+\tau}{2} - P_{V_0} - P_{V_H} \Big) \Big(\frac{1+\tau}{2} - P_{V_0} - P_{V_H} \Big) (P_{V_H} \otimes P_{V_H}) | e_k \otimes e_l \rangle \\ \nonumber
&= \langle E_i \otimes E_j | \Big(\frac{1+\tau}{2} - P_{V_0} - P_{V_H} \Big) \Big(\frac{1+\tau}{2} - P_{V_0} - P_{V_H} \Big) | E_k \otimes E_l \rangle \\ \nonumber
&= \langle E_i \otimes E_j | \Big(\frac{1+\tau}{2} - (1 + \tau)(P_{V_0} + P_{V_H}) + P_{V_0}^2 + P_{V_H}^2 \Big) | E_k \otimes E_l \rangle \\ \nonumber
&= \langle E_i \otimes E_j | \Big(\frac{1+\tau}{2} - P_{V_0} - P_{V_H} \Big) | E_k \otimes E_l \rangle \\ \nonumber
&= \frac{1}{2} \Big( \langle E_i \otimes E_j | E_k \otimes E_l \rangle + \langle E_j \otimes E_i | E_k \otimes E_l \rangle \Big) - \langle E_i \otimes E_j | P_{V_0} | E_k \otimes E_l \rangle 
\\ &\quad- \langle E_i \otimes E_j | P_{V_H} | E_k \otimes E_l \rangle \, , 
\end{align}
where we have used the fact that $P^2 = P$ and both $P_{V_0}$ and $P_{V_H}$ commute with the action of $\tau$. Evaluating the matrix elements gives the Feynman rule for diagram \eqref{Eqn: V_2 V_2 Two Point Projection}. The expectation values $\langle MM \rangle_{\text{conn}}$ and $\langle NN \rangle_{\text{conn}}$ have similar term-wise diagrammatic descriptions using one-colored graphs, with the $XY$ label replaced by $X$ and $Y$, respectively.

The $F(i,j)$ portions of the Feynman rules can be given their own graph interpretation. From now on, we refer to these graphs as $F$-graphs, to distinguish from the directed colored graphs related to observables. A factor of $F(i,j)$ corresponds to an undirected uncolored edge with end points labeled $i$ and $j$. More generally, for expressions involving sums over the indices of $F$ we have the following rules
\begin{itemize}
	\item Every factor $F(i_a,i_b)$ is associated with an edge whose endpoints are labeled $i_a$, $i_b$,
	\begin{equation}
	F(i_a,i_b) = \vcenter{\hbox{\begin{tikzpicture}
			\draw (0,0) node[label={left:\small $i_a$}] {} -- (1,0) node[label={right:\small $i_b$}] {};
			\end{tikzpicture}}}
	\end{equation}
	\item To each sum $\sum_{i_a}$, associate a vertex whose incident edges are those with endpoints $i_a$,
\end{itemize}
For example, we have the following correspondences
\begin{align}
&\vcenter{\hbox{\begin{tikzpicture}[scale=2]\begin{scope}[decoration={markings, mark=at position 0.65 with \arrow{latex}}]
		\node[label={left:$i$}] at (0,0) {};
		\node[label={right:$k$}] at (1,0) {};
		\node[label={left:$j$}] at (0,-.5) {};
		\node[label={right:$l$}] at (1,-.5) {};
		\draw[draw=BLUE ] (0,-.5) -- (0.25,-.25);
		\draw[draw=BLUE ] (.25,-0.25) -- (0,0);
		\draw[draw=GREEN ] (.75,-0.25) -- (1,0);	
		\draw[draw=GREEN ] (.75,-0.25) -- (1,-.5);	
		\draw[draw=black] (.25,-0.25) -- (.5,-0.25) node[label=above:{\tiny $V_H$}]{};
		\draw[draw=black] (.5,-0.25) -- (.75,-0.25);
		\node[circle, fill=BLUE, inner sep=1pt, draw=BLUE] at (.25,-.25) {};
		\node[circle, fill=GREEN, inner sep=1pt, draw=GREEN] at (.75,-.25) {};
		\end{scope}
		\end{tikzpicture}}} = (\Lambda_{V_H}^{-1})_{\ls[33]}^{\us[XY]}\frac{D}{D-2} \vcenter{\hbox{\begin{tikzpicture}[scale=2]\begin{scope}[decoration={markings, mark=at position 0.65 with \arrow{latex}}]
		\node[label={left:$i$}] at (0,0) {};
		\node[label={right:$k$}] at (1,0) {};
		\node[label={left:$j$}] at (0,-.5) {};
		\node[label={right:$l$}] at (1,-.5) {};
		\draw[draw=black ] (0,-.5) -- (0.25,-.25);
		\draw[draw=black ] (.25,-0.25) -- (0,0);
		\draw[draw=black ] (.75,-0.25) -- (1,0);	
		\draw[draw=black ] (.75,-0.25) -- (1,-.5);	
		\draw[draw=black] (.25,-0.25) -- (.5,-0.25);
		\draw[draw=black] (.5,-0.25) -- (.75,-0.25);
		\node[circle, fill=black, inner sep=1pt, draw=black] at (.25,-.25) {};
		\node[circle, fill=black, inner sep=1pt, draw=black] at (.75,-.25) {};
		\end{scope}
		\end{tikzpicture}}} \\
&\vcenter{\hbox{\begin{tikzpicture}[scale=2]\begin{scope}[decoration={zigzag,segment length=3,amplitude=2,post length=0pt}]
		\node[label={left:$i$}] at (0,0) {};
		\node[label={right:$k$}] at (1,0) {};
		\node[label={left:$j$}] at (0,-.5) {};
		\node[ label={right:$l$}] at (1,-.5) {};
		\draw[draw=BLUE ] (0,-.5) -- (0.25,-.25);
		\draw[draw=BLUE ] (.25,-0.25) -- (0,0);
		\draw[draw=GREEN ] (.75,-0.25) -- (1,0);	
		\draw[draw=GREEN ] (.75,-0.25) -- (1,-.5);	
		\draw[draw=black] (.25,-0.25) -- (.5,-0.25) node[label=above:{\tiny $V_3$}]{};
		\draw[draw=black] (.5,-0.25) -- (.75,-0.25);
		\node[circle, fill=BLUE, inner sep=1pt, draw=BLUE] at (.25,-.25) {};
		\node[circle, fill=GREEN, inner sep=1pt, draw=GREEN] at (.75,-.25) {};
		\end{scope}
		\end{tikzpicture}}} = \frac{(\Lambda_{V_3}^{-1})^{\us[XY]}}{2}\qty(\vcenter{\hbox{\begin{tikzpicture}[scale=2]
		\node[label={left:$i$}] at (0,0) {};
		\node[label={right:$k$}] at (1,0) {};
		\node[label={left:$j$}] at (0,-.5) {};
		\node[label={right:$l$}] at (1,-.5) {};
		\draw (0,0) -- (1,0);
		\draw (0,-.5) -- (1,-.5);
		\end{tikzpicture}}} - \vcenter{\hbox{\begin{tikzpicture}[scale=2]
		\node[label={left:$i$}] at (0,0) {};
		\node[label={right:$k$}] at (1,0) {};
		\node[label={left:$j$}] at (0,-.5) {};
		\node[label={right:$l$}] at (1,-.5) {};
		\draw (0,0) -- (1,-.5);
		\draw (0,-.5) -- (0.4,-0.3);
		\draw (0.6,-.2) -- (1,0);
		\end{tikzpicture}}}) \label{Eqn: V_3 Diagram Decomposition}\\
&\vcenter{\hbox{\begin{tikzpicture}[scale=2]\begin{scope}[decoration={snake,segment length=3,amplitude=2, post length=0pt}]
		\node[label={left:$i$}] at (0,0) {};
		\node[label={right:$k$}] at (1,0) {};
		\node[label={left:$j$}] at (0,-.5) {};
		\node[label={right:$l$}] at (1,-.5) {};
		\draw[draw=BLUE ] (0,-.5) -- (0.25,-.25);
		\draw[draw=BLUE ] (.25,-0.25) -- (0,0);
		\draw[draw=GREEN ] (.75,-0.25) -- (1,0);	
		\draw[draw=GREEN ] (.75,-0.25) -- (1,-.5);	
		\draw[draw=black] (.25,-0.25) -- (.5,-0.25) node[label=above:{\tiny $V_2$}]{};
		\draw[draw=black] (.5,-0.25) -- (.75,-0.25);
		\node[circle, fill=BLUE, inner sep=1pt, draw=BLUE] at (.25,-.25) {};
		\node[circle, fill=GREEN, inner sep=1pt, draw=GREEN] at (.75,-.25) {};
		\end{scope}
		\end{tikzpicture}}} = \begin{aligned}[t] (\Lambda_{V_2}^{-1})^{\us[XY]}&\left(\frac{1}{2}\vcenter{\hbox{\begin{tikzpicture}[scale=2]
		\node[label={left:$i$}] at (0,0) {};
		\node[label={right:$k$}] at (1,0) {};
		\node[label={left:$j$}] at (0,-.5) {};
		\node[label={right:$l$}] at (1,-.5) {};
		\draw (0,0) -- (1,0);
		\draw (0,-.5) -- (1,-.5);
		\end{tikzpicture}}} +\frac{1}{2} \vcenter{\hbox{\begin{tikzpicture}[scale=2]
		\node[label={left:$i$}] at (0,0) {};
		\node[label={right:$k$}] at (1,0) {};
		\node[label={left:$j$}] at (0,-.5) {};
		\node[label={right:$l$}] at (1,-.5) {};
		\draw (0,0) -- (1,-.5);
		\draw (0,-.5) -- (0.4,-0.3);
		\draw (0.6,-.2) -- (1,0);
		\end{tikzpicture}}} \right.\\
&\left.-\frac{1}{D-1}\vcenter{\hbox{\begin{tikzpicture}[scale=2]\begin{scope}[decoration={markings, mark=at position 0.65 with \arrow{latex}}]
		\node[label={left:$i$}] at (0,0) {};
		\node[label={right:$k$}] at (1,0) {};
		\node[label={left:$j$}] at (0,-.5) {};
		\node[label={right:$l$}] at (1,-.5) {};
		\draw[draw=black] (0,0) -- (0,-.5);
		\draw[draw=black] (1,0) -- (1,-.5);
		\end{scope}
		\end{tikzpicture}}}-\frac{D}{D-2}\vcenter{\hbox{\begin{tikzpicture}[scale=2]\begin{scope}[decoration={markings, mark=at position 0.65 with \arrow{latex}}]
		\node[label={left:$i$}] at (0,0) {};
		\node[label={right:$k$}] at (1,0) {};
		\node[label={left:$j$}] at (0,-.5) {};
		\node[label={right:$l$}] at (1,-.5) {};
		\draw[draw=black ] (0,-.5) -- (0.25,-.25);
		\draw[draw=black ] (.25,-0.25) -- (0,0);
		\draw[draw=black ] (.75,-0.25) -- (1,0);	
		\draw[draw=black ] (.75,-0.25) -- (1,-.5);	
		\draw[draw=black] (.25,-0.25) -- (.5,-0.25);
		\draw[draw=black] (.5,-0.25) -- (.75,-0.25);
		\node[circle, fill=black, inner sep=1pt, draw=black] at (.25,-.25) {};
		\node[circle, fill=black, inner sep=1pt, draw=black] at (.75,-.25) {};
		\end{scope}
		\end{tikzpicture}}}\right) \label{Eqn: V_2 Diagram Decomposition}
\end{aligned}
\end{align}
From the definition of $F(i,j)$ as the matrix elements of a projector $P_{V_H}: V_D \rightarrow V_H$,
\begin{equation}
F(i,j) = \qty(\delta_{ij}-\frac{1}{D}),
\end{equation}
we find the properties
\begin{align} \nonumber
&\sum_{p} F(i,p)F(p,j) = F(i,j), \\ \nonumber
&\sum_i F(i,i) = (D-1),\\
&\sum_i F(i,j) = \sum_j F(i,j) = 0,
\end{align}
which correspond to the operator statements $P_{V_H}^2 = P_{V_H}$, $\Tr P_{V_H} = \dim(V_H)$ and $P_{V_0} P_{V_H} = 0$ respectively.
It also satisfies the identity
\begin{align} \nonumber \label{eq: H-Projector Identity}
\sum_{i,j}F(i_1,i) \dots F(i_{k_1},i)&F(i,j)^nF(j_1,j) \dots F(j_{k_2},j) \\
&=\begin{aligned}[t]&\frac{(1-D)^n-1}{(-D)^n}\sum_{i}F(i_1,i) \dots F(i_{k_1},i)F(j_1,i) \dots F(j_{k_2},i) \\
&+ \frac{1}{(-D)^n}\sum_{i,j}F(i_1,i) \dots F(i_{k_1},i) F(j_1,j) \dots F(j_{k_2},j).\end{aligned} 
\end{align}
which in terms of $F$-graphs is
\begin{align} \nonumber
\vcenter{\hbox{\begin{tikzpicture}
		\coordinate (i1) at (150:2);
		\coordinate (ik) at (210:2);
		\coordinate (o1) at (30:2);
		\coordinate (ok) at (-30:2);
		\draw (i1) node[label={\small $i_1$}] {} -- (-1,0) node[circle, fill,inner sep=1pt,label={\small $i$}] {} to[bend left] node[midway, label={\small $1$}] {} (1,0) node[circle, fill,inner sep=1pt, label={\small $j$}] {} -- (o1) node[label={\small $j_1$}] {};
		\draw (-1,0) to[bend right] node[midway, label={below:\small$n$}] {} (1,0);
		\draw (ik) node[label={below:\small $i_{k_1}$}] {} -- (-1,0);
		\draw (1,0) -- (ok) node[label={below:\small $j_{k_2}$}] {};
		\draw[loosely dotted] (i1) to[bend right] (ik);
		\draw[loosely dotted] (o1) to[bend left] (ok);
		\end{tikzpicture}}} = \frac{(1-D)^n-1}{(-D)^n}\vcenter{\hbox{\begin{tikzpicture}
		\coordinate (i1) at (150:2);
		\coordinate (ik) at (210:2);
		\coordinate (o1) at (30:2);
		\coordinate (ok) at (-30:2);
		\draw (i1) node[ label={\small $i_1$}] {} -- (0,0) node[circle, fill,inner sep=1pt,label={\small $i$}] {} -- (o1) node[label={\small $j_1$}] {};
		\draw (ik) node[label={below:\small $i_{k_1}$}] {} -- (0,0);
		\draw (0,0) -- (ok) node[label={below:\small $j_{k_2}$}] {};
		\draw[loosely dotted] (i1) to[bend right] (ik);
		\draw[loosely dotted] (o1) to[bend left] (ok);
		\end{tikzpicture}}}& \\
+ \frac{1}{(-D)^n}\vcenter{\hbox{\begin{tikzpicture}
		\coordinate (i1) at (150:2);
		\coordinate (ik) at (210:2);
		\coordinate (o1) at (30:2);
		\coordinate (ok) at (-30:2);
		\draw (i1) node[label={\small $i_1$}] {}  -- (-.5,0) node[circle, fill,inner sep=1pt,label={\small $i$}] {};
		\draw (.5,0) node[circle, fill,inner sep=1pt, label={\small $j$}] {} -- (o1) node[label={\small $j_1$}] {};
		\draw (ik) node[label={below:\small $i_{k_1}$}] {} -- (-.5,0);
		\draw (.5,0) -- (ok) node[label={below:\small $j_{k_2}$}] {};
		\draw[loosely dotted] (i1) to[bend right] (ik);
		\draw[loosely dotted] (o1) to[bend left] (ok);
		\end{tikzpicture}}}&
\end{align}

When we consider expectation values of observables, we sum over all the external indices. This turns every product of $F$'s into products with all indices summed. In terms of $F$-graphs, we will only get graphs which have all edges connected to vertices. We call such graphs closed $F$-graphs, while $F$-graphs where at least one edge is not connected to a vertex is called an open $F$-graph. Evaluation of expectation values is ultimately reduced to evaluating generic products of fully summed $F$'s. We call this evaluating the closed $F$-graph. In general, the evaluation gives a Laurent polynomial in $D$. We now move on to develop techniques for evaluating general closed $F$-graphs.

\subsection{Evaluating closed  $F$-graphs: a graph algorithm with parameter $D$. }
\label{subsection: F-graph Evaluation}
The main result of this section will be a graph algorithm for evaluating closed $F$-graphs. We start by outlining the algorithm and in the next section we provide a proof of the formula corresponding to the algorithm.

For the purposes of outlining the algorithm, start with a general closed $F$-graph $G$ with $p$ vertices. When evaluated, it will give a Laurent polynomial $\mathcal{F}(G)$. If $G$ has $L$ loops, we remove the loops to construct the graph $\widetilde{G}$. $L$ loops in $G$ give a multiplicative constant
\begin{equation}
	F(i,i)^L = \qty(1 - \frac{1}{D})^L,
\end{equation}
that is
\begin{equation}
	\mathcal{F}(G) = \qty(1 - \frac{1}{D})^L \mathcal{F}(\widetilde{G}).
\end{equation}
For example,
\begin{equation}
	\mathcal{F}\qty(\begin{tikzpicture}[scale=2,baseline]
	\draw (-0.5,0) node[circle, fill,inner sep=1pt] {}  to[bend left] (0,0) node[circle, fill,inner sep=1pt] {};
	\draw (-0.5,0)  to[bend right] (0,0);
	\draw (0,0) to[bend left] (0.5,0) node[circle, fill,inner sep=1pt] {};
	\draw (0,0) to[bend right] (0.5,0);
	\draw (0,0) to[bend left] (0,-0.5) to[bend left] (0,0);
	\draw (-0.5,0) to[bend left] (-0.5,-0.5) to[bend left] (-0.5,0);
	\end{tikzpicture}) = \qty(1 - \frac{1}{D})^2 \mathcal{F}\qty(\begin{tikzpicture}[scale=2,baseline]
	\draw (-0.5,0) node[circle, fill,inner sep=1pt] {}  to[bend left] (0,0) node[circle, fill,inner sep=1pt] {};
	\draw (-0.5,0)  to[bend right] (0,0);
	\draw (0,0) to[bend left] (0.5,0) node[circle, fill,inner sep=1pt] {};
	\draw (0,0) to[bend right] (0.5,0);
	\end{tikzpicture})
\end{equation}
We now assume that $\widetilde{G}$ is a graph where all vertices have edges and there are no loops.\footnote{If we have $v$ vertices with no edges, they can be removed to give a multiplicative contribution of $D^v$ from the fact that $(\sum_{i=1}^D 1)^v = D^v$.}

The second step is to compute $\mathcal{F}(\widetilde{G})$. We start by labeling the vertices using $[p] = \{1,\dots,p\}$. The goal is to partition the vertices using set partitions of $[p]$. The definition of a set partition is easily understood after having seen an example. Consider the set $[3]=\{1,2,3\}$, it can be partitioned into five different set partitions,
	\begin{align} \nonumber
	&\{\{1,2,3\}\}, \\ \nonumber
	&\{ \{1,2\}, \{3 \}\}, \\ \nonumber
	&\{ \{1,3\}, \{2 \}\}, \\ \nonumber
	&\{ \{2,3\}, \{1 \}\}, \\
	&\{\{1 \},\{2 \}, \{3 \}\}.
	\end{align}
The definition of a set partition is the following. A set partition $P$ of $[p]$, is a set of subsets $P_i \subseteq [p]$ called blocks. The blocks have the property that the union of all blocks is $[p]$ and the intersection of two distinct blocks is empty. The number of blocks in a set partition is denoted $\abs{P}$. In set notation we have
\begin{equation}
	P = \Big\{ P_i \subseteq [p] \, \Big\vert \, \bigcup_i^{\abs{P}} P_i = [p] \qq{and} P_i \cap P_j = 0 \qq{for } i\neq j.\Big\}.
\end{equation}
A set with these properties is written $P \, \vdash \, [p]$. The shape of a set partition $P \, \vdash \, [p]$ is an integer partition describing the block structure of $P$. For example, in the previous example we have the shapes $3 = 3$, $3 = 2 + 1$, $3 = 2 + 1$, $3 = 2 + 1$ and $3 = 1+ 1 +1$. Instead of the set of subsets notation, we will use vertical lines to separate blocks. In this compact notation the five set partitions of $[3]$ are written
\begin{equation}
	123, \quad 12|3, \quad 13|2, \quad 23|1, \quad 1|2|3.
\end{equation}

Coming back to the graph with $p$ vertices labeled by $[p]$, each set partition $P \, \vdash \, [p]$ defines a reduced graph $\widetilde{G}_{P}$. The reduced graph is constructed from $\widetilde{G}$ by keeping only the edges connecting vertices within the same block of $P$. For example, consider the following graph with three labeled vertices
\begin{equation}
	\widetilde{G} = \begin{tikzpicture}[scale=2,baseline]
	\draw (-0.5,0) node[circle, fill,inner sep=1pt, label={\small $1$}] {}  to[bend left] (0,0) node[circle, fill,inner sep=1pt,label={\small $2$}] {};
	\draw (-0.5,0)  to[bend right] (0,0);
	\draw (0,0) to[bend left] (0.5,0) node[circle, fill,inner sep=1pt,label={\small $3$}] {};
	\draw (0,0) to[bend right] (0.5,0);
	\end{tikzpicture}. \label{eq: Example F-Graph on 3 vertices}
\end{equation}
The corresponding five reduced graphs are
\begin{align} \nonumber \label{eq: Vertex Partitions of the First Kind}
&\widetilde{G}_{123} \leftrightarrow \begin{tikzpicture}[scale=2,baseline]
\draw (-0.5,0) node[circle, fill,inner sep=1pt, label={\small $1$}] {}  to[bend left] (0,0) node[circle, fill,inner sep=1pt,label={\small $2$}] {};
\draw (-0.5,0)  to[bend right] (0,0);
\draw (0,0) to[bend left] (0.5,0) node[circle, fill,inner sep=1pt,label={\small $3$}] {};
\draw (0,0) to[bend right] (0.5,0);
\end{tikzpicture} \\ \nonumber
&\widetilde{G}_{12|3} \leftrightarrow \begin{tikzpicture}[scale=2,baseline]
\draw (-0.5,0) node[circle, fill,inner sep=1pt, label={\small $1$}] {}  to[bend left] (0,0) node[circle, fill,inner sep=1pt,label={\small $2$}] {};
\draw (-0.5,0)  to[bend right] (0,0);
\draw (0.5,0) node[circle, fill,inner sep=1pt,label={\small $3$}] {};
\end{tikzpicture} \\ \nonumber
&\widetilde{G}_{13|2} \leftrightarrow \begin{tikzpicture}[scale=2,baseline]
\draw (-0.5,0) node[circle, fill,inner sep=1pt, label={\small $1$}] {};
\draw (0,0) node[circle, fill,inner sep=1pt,label={\small $2$}] {};
\draw (0.5,0) node[circle, fill,inner sep=1pt,label={\small $3$}] {};
\end{tikzpicture} \\ \nonumber
&\widetilde{G}_{23|1} \leftrightarrow \begin{tikzpicture}[scale=2,baseline]
\draw (0,0) node[circle, fill,inner sep=1pt, label={\small $2$}] {}  to[bend left] (.5,0) node[circle, fill,inner sep=1pt,label={\small $3$}] {};
\draw (0,0)  to[bend right] (.5,0);
\draw (-0.5,0) node[circle, fill,inner sep=1pt,label={\small $1$}] {};
\end{tikzpicture} \\
&\widetilde{G}_{1|2|3}  \leftrightarrow \begin{tikzpicture}[scale=2,baseline]
\draw (-0.5,0) node[circle, fill,inner sep=1pt, label={\small $1$}] {};
\draw (0,0) node[circle, fill,inner sep=1pt,label={\small $2$}] {};
\draw (0.5,0) node[circle, fill,inner sep=1pt,label={\small $3$}] {};
\end{tikzpicture}
\end{align}

Each reduced graph gives a contribution to $\mathcal{F}(\widetilde{G})$ by the following rule
\begin{equation}
	F(\widetilde{G}_{P}) = \frac{1}{(-D)^{E(\widetilde{G})}}\frac{D!}{(D-|P|)!}\qty(1-D)^{E(\widetilde{G}_{P})},
\end{equation}
where $E(G)$ is the number of edges in the graph $G$. Summing over partitions we get
\begin{equation}
	\mathcal{F}(\widetilde{G}) = \sum_{P \, \vdash \, [p]} F(\widetilde{G}_{P}) =  \frac{1}{(-D)^{E(\widetilde{G})}}\sum_{P \, \vdash \, [p]}\frac{D!}{(D-|P|)!}\qty(1-D)^{E(\widetilde{G}_{P})}. \label{eq: Evaluating F-graph}
\end{equation}
For the $F$-graph in \eqref{eq: Example F-Graph on 3 vertices} we find the polynomial
\begin{equation}
\begin{aligned}
	\mathcal{F}\qty(\begin{tikzpicture}[scale=2,baseline]
	\draw (-0.5,0) node[circle, fill,inner sep=1pt, label={\small $1$}] {}  to[bend left] (0,0) node[circle, fill,inner sep=1pt,label={\small $2$}] {};
	\draw (-0.5,0)  to[bend right] (0,0);
	\draw (0,0) to[bend left] (0.5,0) node[circle, fill,inner sep=1pt,label={\small $3$}] {};
	\draw (0,0) to[bend right] (0.5,0);
	\end{tikzpicture}) &=
	\frac{1}{(-D)^4}\Big[\frac{D!}{(D-1)!}(1-D)^4+\frac{D!}{(D-2)!}(1-D)^2+\\&\phantom{=}\frac{D!}{(D-2)!}(1-D)^0+\frac{D!}{(D-2)!}(1-D)^2+\frac{D!}{(D-3)!}(1-D)^0\Big]\\
	&=\frac{(D-1)^2}{D}.
\end{aligned}
\end{equation}
Algebraically, this corresponds to the identity
\begin{equation}
	\sum_{i,j,k} F(i,j)^2F(j,k)^2 = \frac{(D-1)^2}{D}.
\end{equation}
Having described the graph algorithm, we now turn to the derivation of equation \eqref{eq: Evaluating F-graph}.

\subsection{Derivation of algorithm.  } \label{subsection: F-graph Derivation}
To derive equation \eqref{eq: Evaluating F-graph} we consider the sequence of all possible polynomials (all graphs $\widetilde{G}$ on $p$ vertices)
\begin{equation}
\mathcal{F}(A) = \sum_{i_1, \dots, i_p} \prod_{1\leq a < b \leq p} F(i_a, i_b)^{A_{ab}}, \label{eq: Sequence of F Polynomials}
\end{equation}
where $A_{ab}$ is a matrix with non-negative integer entries. The corresponding exponential generating function (EGF) is
\begin{equation}
\mathcal{F}(x_{12},\dots,x_{p-1 p}) = \sum_{i_1, \dots, i_p} \prod_{1 \leq a<b\leq p} \qty(\sum_{A_{ab}=0}^\infty \frac{x_{ab}^{A_{ab}}}{A_{ab}!} F(i_a,i_b)^{A_{ab}}).
\end{equation}
The condition $a < b$ takes into account the redundancy due to $F(i,j) = F(j,i)$.

To evaluate the EGF it will be useful to introduce some notation. We denote the set of pairs $\sigma=(\sigma_1, \sigma_2)$, with $\sigma_1 < \sigma_2$ and $\sigma_1, \sigma_2 \in S \subset [p]$, by $\mathcal{P}(S)$. Then our generating function is parametrised by elements in $\mathcal{P}([p])$ and the EGF can be written as
\begin{equation}
\mathcal{F} = \sum_{i_1, \dots, i_p} \prod_{\sigma \in \mathcal{P}([p])} \sum_{A_\sigma}\frac{x_\sigma^{A_\sigma}}{A_\sigma !} F_{i^\sigma}^{A_\sigma} = \sum_{i_1, \dots, i_p} \exp(\sum_{ \sigma \in \mathcal{P}([p])} x_\sigma F_{i^\sigma}).
\end{equation}
Here we have introduced the notation
\begin{equation}
x_\sigma = x_{\sigma_1 \sigma_2}, \quad F_{i^\sigma} =  F(i_{\sigma_1}, i_{\sigma_2}) \equiv \delta_{i^\sigma} - \frac{1}{D}
\end{equation}
and $\delta_{i^\sigma} = \delta_{i_{\sigma_1} i_{\sigma_2}}$. This gives
\begin{align} \nonumber
\mathcal{F} &=\sum_{i_1, \dots, i_p} \exp(\sum_{ \sigma \in \mathcal{P}([p]) } x_\sigma \delta_{i^\sigma} - x_{\sigma}/D) \\
&= \exp(-\sum_{ \sigma \in \mathcal{P}([p]) } x_{\sigma}/D) \sum_{i_1, \dots, i_p} \exp(\sum_{ \sigma \in \mathcal{P}([p]) } x_\sigma \delta_{i^\sigma}).
\end{align}

To evaluate $\mathcal{F}$, observe that we can separate the sum $\sum_{i_1, \dots, i_p}$ into restricted sums determined by set partitions of $\{i_1, \dots,i_p\}$. For example, with $p=3$, we have the following set partitions
\begin{align} \nonumber
i_1 i_2 i_3 &\leftrightarrow i_1 = i_2 = i_3, \\ \nonumber
i_1 i_2|i_3&\leftrightarrow i_1 = i_2 \neq i_3, \\ \nonumber
i_1 i_3|i_2&\leftrightarrow i_1 = i_3 \neq i_2, \\ \nonumber
i_2 i_3|i_1&\leftrightarrow i_2 = i_3 \neq i_1, \\
i_1|i_2|i_3&\leftrightarrow i_1 \neq i_2 \neq i_3.
\end{align}
which translates into the following separation of the sum
\begin{equation}
\sum_{i_1,i_2,i_3} = \sum_{i_1=i_2=i_3} + \sum_{i_1=i_2\neq i_3} + \sum_{i_1 = i_3 \neq i_2} + \sum_{i_2 = i_3 \neq i_1} +\sum_{i_1 \neq i_2 \neq i_3}.
\end{equation}
The summand
\begin{equation}
\exp(\sum_{ \sigma \in \mathcal{P}([p]) } x_\sigma \delta_{i^\sigma}),
\end{equation}
is constant on each restricted sum. The number of terms in a restricted sum is determined by the number of blocks in the associated set partition. For example, there are $D(D-1)=\tfrac{D!}{(D-2)!}$ values of $i_1, i_2, i_3$ satisfying $i_1 = i_2 \neq i_3 \leftrightarrow i_1 i_2|i_3$, which has two blocks. The set partitions are more succinctly written by omitting the $i$. That is,
\begin{equation}
i_1 i_2|i_3 \equiv 12|3.
\end{equation}
The Kronecker delta $\delta_{i^\sigma}$ is $1$ if $\sigma$ is a pair of integers from the same block in the partition $P$. Equivalently,
\begin{equation}
\delta_{i^\sigma} = \begin{cases}
1 \quad \sigma \in \bigcup_{i=1}^{|P|}\mathcal{P}(P_i) \\
0 \quad \text{otherwise}.
\end{cases}
\end{equation}
For instance, $\delta_{i^{(1,2)}} = 1$ for $12|3$ since $i_1 = i_2$.

In this language the generating function is
\begin{equation}
\mathcal{F} = \exp(-\sum_{ \sigma \in \mathcal{P}([p])} x_{\sigma}/D) \sum_{P \vdash [p]} \frac{D!}{(D-|P|)!} \exp(\sum_{ \sigma \in \cup \mathcal{P}(P_i)} x_{\sigma}),
\end{equation}
where $\cup \mathcal{P}(P_i)$ is shorthand for
\begin{equation}
\bigcup_{i=1}^{|P|}\mathcal{P}(P_i).
\end{equation}
To prepare the EGF for the last step we will further rewrite it by splitting $\mathcal{P}([p])$ into $\bigcup \mathcal{P}(P_i)$ and its complement $(\bigcup \mathcal{P}(P_i))_c$,
\begin{equation}
\mathcal{F} = \sum_{P \vdash [p]} \frac{D!}{(D-|P|)!} \exp(\sum_{ \sigma \in \cup \mathcal{P}(P_i) } \frac{D-1}{D}x_{\sigma} - \frac{1}{D}\sum_{ \sigma \in (\cup \mathcal{P}(P_i))_c} x_{\sigma} )
\end{equation}

To extract a particular polynomial, we apply derivatives. Introduce
\begin{equation}
\partial_\tau = \frac{\partial}{\partial x_\tau}, \quad \tau \in \mathcal{P}([p]).
\end{equation}
Note that a derivative acting on $\mathcal{F}$ will either hit $x_\sigma$ for $\sigma \in \bigcup P_i^2$ and give a factor $\tfrac{D-1}{D}$ or $x_\sigma$ for $\sigma \in \bigcup (P^2_i)_c$ and give a factor $\tfrac{-1}{D}$. 
Then
\begin{align} \nonumber \label{eq:kappagraphpolynomial}
\prod_{\tau \in \mathcal{P}([p])} \partial_\tau^{A_\tau} \mathcal{F}\vert_{0} ={}&\sum_{P \vdash [p]} \frac{D!}{(D-|P|)!} \prod_{\tau \in \mathcal{P}([p])} \partial_\tau^{A_\tau} \exp(\sum_{ \sigma \in \cup \mathcal{P}(P_i) } \frac{D-1}{D}x_{\sigma} - \frac{1}{D}\sum_{ \sigma \in (\cup \mathcal{P}(P_i))_c} x_{\sigma} )\vert_{0} \nonumber \\
={}&\sum_{P \vdash [p]} \frac{D!}{(D-|P|)!} \prod_{\tau \in \cup \mathcal{P}(P_i) } \left(\frac{D-1}{D}\right)^{A_\tau} \prod_{\tau \in (\cup \mathcal{P}(P_i))_c} \frac{1}{(-D)^{A_\tau}} \nonumber \\
={}&\frac{1}{(-D)^{\sum_{\tau \in \mathcal{P}([p])} A_\tau}}\sum_{P \vdash [p]} \frac{D!}{(D-|P|)!} \left(1-D\right)^{\sum_{\tau \in \cup \mathcal{P}(P_i)} A_\tau}. 
\end{align}
The second equality follows by splitting the product over the set $\mathcal{P}([p])$ into two products, one for each complementary set. In summary, we have found
\begin{equation}
	\sum_{i_1,\dots,i_p}\prod_{\tau \in \mathcal{P}([p])} F_{i^\tau}^{A_\tau} = \frac{1}{(-D)^{\sum_{\tau \in \mathcal{P}([p])} A_\tau}}\sum_{P \vdash [p]} \frac{D!}{(D-|P|)!} \left(1-D\right)^{\sum_{\tau \in \cup \mathcal{P}(P_i)} A_\tau}.
\end{equation}
We get to the graph algorithm \eqref{eq: Evaluating F-graph} by interpreting the elements $A_\tau$ as the matrix elements of a symmetric adjacency matrix defining an $F$-graph.

\subsection{Feynman rules as maps between double cosets}
\label{subsection: maps between graphs}
\begin{figure}[h!]
	\centering
	\begin{tikzpicture}[scale=5]
	\draw (-0.5,0) node[circle, fill,inner sep=1pt] {}  to[bend left] node[midway, cross=2pt] {} (0,0) node[circle, fill,inner sep=1pt] {};
	\draw (-0.5,0)  to[bend right] node[midway, cross=2pt] {} (0,0);
	\draw (0,0) to[bend left] node[midway, cross=2pt] {} (0.5,0) node[circle, fill,inner sep=1pt] {};
	\draw (0,0) to[bend right] node[midway, cross=2pt] {} (0.5,0);
	\end{tikzpicture}
	\caption{An example of a graph with auxiliary vertices at the midpoint of every edge.}
	\label{fig: Undirected Graph Edge Cutting}
\end{figure}
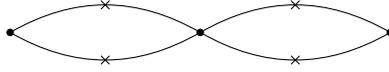

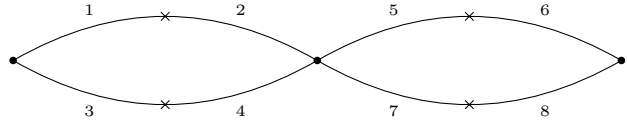
\begin{figure}[h!]
	\centering
	\begin{tikzpicture}[scale=8]
	\draw (-0.5,0) node[circle, fill,inner sep=1pt] {}  to[bend left] node[midway, cross=2pt] {} node[near start,inner sep=0pt, label={above:\tiny$1$}] {}  node[near end,inner sep=0pt, label={above:\tiny$2$}] {}(0,0) node[circle, fill,inner sep=1pt] {};
	\draw (-0.5,0)  to[bend right] node[midway, cross=2pt] {} node[near start, inner sep=0pt,label={below:\tiny$3$}] {} node[near end, inner sep=0pt,label={below:\tiny$4$}] {} (0,0);
	\draw (0,0) to[bend left] node[midway, cross=2pt] {} node[near start,inner sep=0pt, label={above:\tiny$5$}] {} node[near end, inner sep=0pt,label={above:\tiny$6$}] {} (0.5,0) node[circle, fill,inner sep=1pt] {};
	\draw (0,0) to[bend right] node[midway, cross=2pt] {} node[near start,inner sep=0pt, label={below:\tiny$7$}] {} node[near end, inner sep=0pt,label={below:\tiny$8$}] {} (0.5,0);
	\end{tikzpicture}
	\caption{Example of labeling the half edges of a graph to give an explicit combinatorial description of the graph.}
	\label{fig: Undirected Graph Edge Cutting Labeled}
\end{figure}

\begin{figure}
	\centering
	\begin{tikzpicture}[scale=4]
	\def \k {2}
	\def \m {3}
	\def \sep {.5}
	\def \voffset {0}
	\pgfmathparse{(\sep*(\m)-2*\voffset)/\k};
	\pgfmathsetmacro{\vsep}{\pgfmathresult};
	\pgfmathint{\k-1};
	\pgfmathsetmacro{\kk}{\pgfmathresult};
	\foreach \v in {0,...,\k}
	{
		\pgfmathparse{\v*\vsep+\voffset}
		\node[circle, fill, inner sep=1pt](v\v) at (\pgfmathresult,0) {};
	}
	\foreach \eOutm in {0,...,\m}
	{
		\pgfmathint{\eOutm+1};
		\pgfmathsetmacro{\seOutm}{\pgfmathresult};
		\pgfmathparse{\eOutm*\sep};
		\coordinate (eom\seOutm) at (\pgfmathresult,1.5);
		\node[cross=2pt] at (eom\seOutm) {};
	}
	\draw (eom1) to[out=-110, in=110] node[very near start,inner sep=0pt, label={left:\tiny$1$}] {} node[very near end,inner sep=0pt, label={left:\tiny$1$}] {}(v0);
	\draw (eom1) to[out=-70, in=70] node[very near start,inner sep=0pt, label={right:\tiny$2$}] {}node[very near end,inner sep=0pt, label={right:\tiny$2$}] {}(v0);
	\draw (eom2) to[out=-110, in=150] node[very near start,inner sep=0pt, label={left:\tiny$3$}]{} node[very near end, inner sep=0pt,label={left:\tiny$3$}]{} (v1);
	\draw (eom2) to[out=-70, in=125]node[very near start, inner sep=0pt,label={right:\tiny$4$}] {}node[very near end,inner sep=0pt, label={right:\tiny$4$}] {} (v1);
	\draw (eom3) to[out=-110, in=55] node[very near start, inner sep=0pt,label={left:\tiny$5$}]{} node[very near end,inner sep=0pt, label={left:\tiny$5$}]{} (v1);
	\draw (eom3) to[out=-70, in=35]node[very near start,inner sep=0pt, label={right:\tiny$6$}] {}node[very near end, inner sep=0pt,label={right:\tiny$6$}] {} (v1);
	\draw (eom4) to[out=-110, in=110] node[very near start,inner sep=0pt, label={left:\tiny$7$}] {}node[very near end,inner sep=0pt, label={left:\tiny$7$}] {}(v2);
	\draw (eom4) to[out=-70, in=70]node[very near start,inner sep=0pt, label={right:\tiny$8$}] {}node[very near end, inner sep=0pt,label={right:\tiny$8$}] {} (v2);
	\draw[fill=white] ($(eom1)-(.3,0.3)$) rectangle node{$\sigma_1$} ($(eom4)-(-.3,.5)$);
	\draw[fill=white] ($(v0)+(-.3,0.3)$) rectangle node{$\sigma_2^{-1}$} ($(v2)+(.3,.5)$);
	\end{tikzpicture}
	\caption{In the figure, permutations act on the half edges to give new graphs.}
	\label{fig: Undirected Double Coset Picture}
\end{figure}
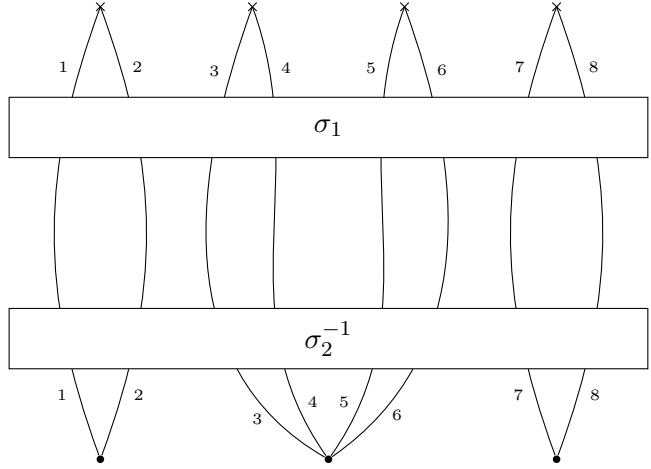

Another description of undirected graphs is in terms of two pairs of sets $\Sigma_{0}, \Sigma_1$, where $\Sigma_{0}$ holds data about edges and $\Sigma_1$ contains data about vertices (see \cite{deMelloKoch:2011uq} for the 
description of Feynman graphs using such pairs and the connection to double cosets). To go from a graph to the pair $(\Sigma_{0}, \Sigma_1)$ we introduce auxiliary vertices at the midpoint of every edge as in Figure \ref{fig: Undirected Graph Edge Cutting}.
By introducing labels for the half edges (see Figure \ref{fig: Undirected Graph Edge Cutting Labeled}) we get the following description of the graph in terms of pairs
\begin{align} \nonumber
\Sigma_0 &= \langle 1,2 \rangle \langle 3,4 \rangle \langle 5,6 \rangle \langle 7,8 \rangle, \\
\Sigma_1&= \langle 1,3 \rangle \langle 2,4,5,7 \rangle \langle 6,8 \rangle. 
\end{align}
We use angled brackets to convey that these are sets, where order does not matter.
This construction is the analog of the double coset described in Section \ref{Section: Observables and multi-graphs}, adapted to undirected graphs. The similarity is apparent from the picture in Figure \ref{fig: Undirected Double Coset Picture}. Any graph can be generated from a pair of permutations $(\sigma_1, \sigma_2)$ and a pair $(\Sigma_0,\Sigma_1)$ through the action $(\sigma_1 (\Sigma_0), \sigma_2 (\Sigma_1))$, where $\sigma_1,\sigma_2 \in S_{2m}$ for a graph with $m$ edges. The graph in Figure \ref{fig: Undirected Graph Edge Cutting Labeled} corresponds to $(\sigma_1, \sigma_2)=(e, (23)(67))$, where $e$ is the identity. A simultaneous relabeling $(\sigma (\Sigma_0), \sigma (\Sigma_1))$, with $\sigma \in S_{2m}$, will give the same graph. Furthermore, the vertices in $\Sigma_0$ are symmetric under permutations in the wreath product $S_m[S_2]$. The vertices in $\Sigma_1$ are symmetric under
\begin{equation}
S_{p} = S_{p_1}[S_1] \times \dots \times S_{p_{2m}}[S_{2m}], \quad \sum_{l=1}^{2m} l p_l = 2m,
\end{equation}
where $p$ describes the shape of the set partition $\Sigma_1$, in analogy with the double coset discussed in Section \ref{Section: Observables and multi-graphs}. Therefore, distinct undirected graphs are in correspondence with double cosets in
\begin{equation}
\diag(S_m)\left\backslash \qty(S_m \times S_m) \right/ (S_{p} \times S_m[S_2]).
\end{equation}

If we import the graph interpretation of equation \eqref{eq:kappagraphpolynomial} into this setting we get a description in terms of set partitions. This description is related to the implementation (algorithm) in Appendix \ref{apx: The Algorithm}. Note that we can use the simultaneous relabeling invariance $(\Sigma_0, \Sigma_1) \rightarrow (\sigma (\Sigma_0), \sigma (\Sigma_1))$ to restrict ourselves to pairs where 
\begin{equation}
\Sigma_0 = \langle 1,2\rangle \dots \langle 2m-1, 2m\rangle.
\end{equation}
The vertex partitions corresponding to equations \eqref{eq: Vertex Partitions of the First Kind} are
\begin{align}
P=123 &&\rightarrow&& P(\Sigma_1) &&=&& \langle 1,3 \rangle \langle 2,4,5,7 \rangle \langle 6,8 \rangle
&&\leftrightarrow &&\begin{tikzpicture}[scale=2,baseline]
\draw (-0.5,0) node[circle, fill,inner sep=1pt,label={\tiny $1$}] {}  to[bend left] (0,0) node[circle, fill,inner sep=1pt,label={\tiny $2$}] {};
\draw (-0.5,0)  to[bend right] (0,0);
\draw (0,0) to[bend left] (0.5,0) node[circle, fill,inner sep=1pt,label={\tiny $3$}] {};
\draw (0,0) to[bend right] (0.5,0);
\end{tikzpicture} \nonumber\\
P=12|3 &&\rightarrow&& P(\Sigma_1)&&=&& \langle 1,3 \rangle \langle 2,4,5,7 \rangle \vert \langle 6,8 \rangle &&\leftrightarrow &&\begin{tikzpicture}[scale=2,baseline]
\draw (-0.5,0) node[circle, fill,inner sep=1pt,label={\tiny $1$}] {}  to[bend left] (0,0) node[circle, fill,inner sep=1pt,label={\tiny $2$}] {};
\draw (-0.5,0)  to[bend right] (0,0);
\draw (0.5,0) node[circle, fill,inner sep=1pt,label={\tiny $3$}] {};
\end{tikzpicture} \nonumber\\ 
P=13|2 &&\rightarrow&&P(\Sigma_1)&&=&&\langle 1,3 \rangle  \langle 6,8 \rangle \vert \langle 2,4,5,7 \rangle &&\leftrightarrow &&\begin{tikzpicture}[scale=2,baseline]
\draw (-0.5,0) node[circle, fill,inner sep=1pt,label={\tiny $1$}] {};
\draw (0,0) node[circle, fill,inner sep=1pt,label={\tiny $2$}] {};
\draw (0.5,0) node[circle, fill,inner sep=1pt,label={\tiny $3$}] {};
\end{tikzpicture} \nonumber\\
P=23|1 &&\rightarrow&& P(\Sigma_1)&&=&& \langle 2,4,5,7 \rangle \langle 6,8 \rangle \vert \langle 1,3 \rangle  &&\leftrightarrow &&\begin{tikzpicture}[scale=2,baseline]
\draw (0,0) node[circle, fill,inner sep=1pt,label={\tiny $1$}] {}  to[bend left] (.5,0) node[circle, fill,inner sep=1pt,label={\tiny $2$}] {};
\draw (0,0)  to[bend right] (.5,0);
\draw (-0.5,0) node[circle, fill,inner sep=1pt,label={\tiny $3$}] {};
\end{tikzpicture} \nonumber\\
P=1|2|3 &&\rightarrow&& P(\Sigma_1)&&=&&\langle 1,3 \rangle \vert \langle 2,4,5,7 \rangle \vert \langle 6,8 \rangle  &&\leftrightarrow &&\begin{tikzpicture}[scale=2,baseline]
\draw (-0.5,0) node[circle, fill,inner sep=1pt,label={\tiny $1$}] {};
\draw (0,0) node[circle, fill,inner sep=1pt,label={\tiny $2$}] {};
\draw (0.5,0) node[circle, fill,inner sep=1pt,label={\tiny $3$}] {};
\end{tikzpicture}
\end{align}
As before, each set partition corresponds to one term in equation \eqref{eq:kappagraphpolynomial}. The map is
\begin{equation}
P(\Sigma_1) \mapsto \frac{1}{(-D)^{\#\text{Pairs in $\Sigma_0$}}}(D)_{\#\text{Parts in P}}\prod_{p \in P(\Sigma_1)}(1-D)^{\#\text{Connected edges in p}},
\end{equation}
where $(D)_k$ is the falling factorial $D\dots (D-k+1)$ and the product is over the parts in $P(\Sigma_1)$. In general, the graph function is a sum over all vertex set partitions, and each term is determined by the map above. 

In Section \ref{Section: Observables and multi-graphs} we found that observables correspond to two-colored directed graphs, which in turn correspond to double cosets in
\begin{equation}
G(\vec{m}^+,\vec{m}^-;\vec{n}^+,\vec{n}^-) \left\backslash\qty( S_m^+ \times S_m^- \times S_n^+ \times S_n^-) \right/ (\diag(S_m) \times \diag(S_n)).
\end{equation}
The expectation value of any observable is an evaluation of uncolored graphs, as just described. Since uncolored graphs also correspond to double cosets, it is natural to ask if the evaluation of observables can be described using maps between the two types of double cosets.

Let $P(\sigma)$ be the observable corresponding to an element $\sigma$ in a coset. Define the vector space 
\begin{equation}
F = \oplus_{m=0}^\infty F_m,
\end{equation}
where $F_0 = \mathbb{C}$ and $F_m$ is a vector space with basis labeled by representatives of 
\begin{equation}
\diag(S_m)\left\backslash (S_m \times S_m) \right/ (S_{\vec{p}} \times S_m[S_2]). 
\end{equation}
Then, the expectation value
\begin{equation}
\langle P(\sigma) \rangle \in \mathbb{C}[D,D^{-1}],
\end{equation}
can be decomposed into two maps
\begin{align}
&\phi: G(\vec{m}^+,\vec{m}^-;\vec{n}^+,\vec{n}^-) \left\backslash \qty(S_m^+ \times S_m^- \times S_n^+ \times S_n^-) \right/ (\diag(S_m) \times \diag(S_n) ) \rightarrow F, \\
&\mathcal{F}: F \rightarrow \mathbb{C}[D,D^{-1}].
\end{align}
The map $\mathcal{F}$ is the graph evaluation discussed in the previous section. $\phi$ can be understood through cutting and gluing of observable graphs with Feynman graphs. For example, consider the observable graph
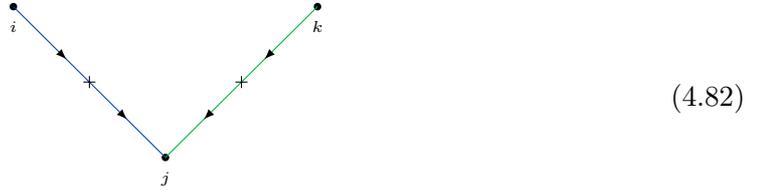
\begin{equation}
\vcenter{\hbox{\begin{tikzpicture}[scale=4]
\definecolor{GREEN}{rgb}{0.0,0.70,0.24}
\definecolor{BLUE}{rgb}{0.0,0.24,0.70}
\coordinate (v1) at (0,0);
\coordinate (v2) at (1,0);
\coordinate (v3) at (0.5,-0.5);
\node[circle, fill, inner sep=1pt, label={below: \tiny $i$}] at (v1) {};
\node[circle, fill, inner sep=1pt, label={below: \tiny $k$}] at (v2) {};
\node[circle, fill, inner sep=1pt, label={below: \tiny $j$}] at (v3) {};
\draw[draw=BLUE, postaction={decorate,decoration={markings, mark=at position 0.35 with {\arrow{latex}}}}, postaction={decorate,decoration={markings, mark=at position 0.75 with {\arrow{latex}}}}] (v1)  -- node[pos=0.5, cross=2pt, sloped] {} (v3);
\draw[draw=GREEN, postaction={decorate,decoration={markings, mark=at position 0.35 with {\arrow{latex}}}}, postaction={decorate,decoration={markings, mark=at position 0.75 with {\arrow{latex}}}}] (v2)  -- node[pos=0.5, cross=2pt, sloped] {} (v3);
\end{tikzpicture}}}
\end{equation}
with auxiliary vertices added mid-way through every edge (see Appendix \ref{apx: The Algorithm} for more details). $\phi$ connects the two-valent crosses with a line that corresponds to the diagrammatic rule in equation \eqref{Eqn: 2pt MN diagram EV}
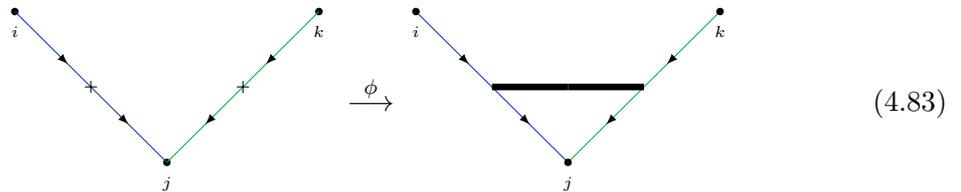
\begin{equation}
\vcenter{\hbox{\begin{tikzpicture}[scale=4]
		\definecolor{GREEN}{rgb}{0.0,0.70,0.24}
		\definecolor{BLUE}{rgb}{0.0,0.24,0.70}
		\coordinate (v1) at (0,0);
		\coordinate (v2) at (1,0);
		\coordinate (v3) at (0.5,-0.5);
		\node[circle, fill, inner sep=1pt, label={below: \tiny $i$}] at (v1) {};
		\node[circle, fill, inner sep=1pt, label={below: \tiny $k$}] at (v2) {};
		\node[circle, fill, inner sep=1pt, label={below: \tiny $j$}] at (v3) {};
		\draw[draw=BLUE, postaction={decorate,decoration={markings, mark=at position 0.35 with {\arrow{latex}}}}, postaction={decorate,decoration={markings, mark=at position 0.75 with {\arrow{latex}}}}] (v1)  -- node[pos=0.5, cross=2pt, sloped] {} (v3);
		\draw[draw=GREEN, postaction={decorate,decoration={markings, mark=at position 0.35 with {\arrow{latex}}}}, postaction={decorate,decoration={markings, mark=at position 0.75 with {\arrow{latex}}}}] (v2)  -- node[pos=0.5, cross=2pt, sloped] {} (v3);
		\end{tikzpicture}}} \, \overset{\phi}{\longrightarrow}
\vcenter{\hbox{\begin{tikzpicture}[scale=4]
		\draw[draw=BLUE, postaction={decorate, decoration={markings, mark=at position 0.35 with \arrow{latex}}},postaction={decorate, decoration={markings, mark=at position 0.75 with \arrow{latex}}}] (0,0) node[circle, fill, inner sep=1pt,label={below: \tiny $i$}] {} -- (0.25,-0.25) -- (0.5,-0.5);
		\draw[draw=GREEN, postaction={decorate, decoration={markings, mark=at position 0.35 with \arrow{latex}}},postaction={decorate, decoration={markings, mark=at position 0.75 with \arrow{latex}}}] (1,0) node[circle, fill, inner sep=1pt,label={below: \tiny $k$}] {} -- (0.75,-0.25) -- (0.5,-0.5);
		\draw[draw=black, line width=2.5pt] (0.25,-0.25) -- (0.5,-0.25);
		\draw[draw=black, line width=2.5pt] (0.5,-0.25) -- (0.75,-0.25);
		\node[circle, fill=black, inner sep=1pt,label={below: \tiny $j$}] at (0.5,-0.5) {};
		\end{tikzpicture}}}
\end{equation}
The propagator can be expanded using the term-wise Feynman rules in Section \ref{Subsec: Feynman decomp} to give a map between colored directed graphs and uncolored undirected graphs. The exploration of the existence of these maps and their explicit realizations is an interesting future direction.

\section{Expectation values of permutation invariant observables} \label{Section: Expectation values}
Armed with the technology developed in the previous sections we turn our attention to the evaluation of permutation invariant expectation values in the 2-matrix model. We begin by considering quadratic observables, of which there are 37, with corresponding graphs possessing a maximum of four vertices. We evaluate all 15 quadratic observables of mixed type. The following two subsections consider a selection of the 338 cubic and 3598 quartic observables respectively. In each subsection we complete a detailed example calculation of the appropriate order before presenting a collection of results. 

For the quadratic calculations we first use \eqref{Eqn: Two point function} to write the observables in terms of a sum over $F$s with the relevant indices. The sums over products of $F$s are then performed using the technology developed in Section \ref{Section: Feynman Graphs}. The higher order expectation values are calculated by first applying Wick's theorem, \eqref{Eqn: Wicks theorem}, and then evaluating the resultant products of linear and quadratic expectation values.

\subsection{Quadratic expectation values} \label{Subsection: Example quad calc}

We consider the calculation of the following quadratic expectation value as an example
\begin{align} \nonumber
\sum_{i,j =1}^D \langle M_{ij} N_{ij} \rangle &= \sum_{i,j =1}^D \vcenter{\hbox{\begin{tikzpicture}[scale=2]
		\begin{scope}[decoration={markings, mark=at position 0.65 with \arrow{latex}}]
		\draw[draw=BLUE, postaction={decorate}] (0,0) node[label={left:$i$}] {} -- (0.25,-0.25);
		\draw[draw=GREEN, postaction={decorate}] (1,0) node[label={right:$i$}] {} -- (0.75,-0.25);
		\draw[draw=GREEN, postaction={decorate}]  (0.75,-0.25) -- (1,-.5) node[label={right:$j$}] {};
		\draw[draw=BLUE, postaction={decorate}] (0.25,-0.25) -- (0,-.5) node[label={left:$j$}] {};
		\draw[draw=black, line width=2.5pt] (0.25,-0.25) -- (0.5,-0.25);
		\draw[draw=black, line width=2.5pt] (0.5,-0.25) -- (0.75,-0.25);
		\end{scope}
		\end{tikzpicture}}}
	+ \sum_{i,j =1}^D \Bigg( \vcenter{\hbox{\begin{tikzpicture}[scale=2]
		\begin{scope}[decoration={markings, mark=at position 0.6 with \arrow{latex}}]
		\draw[draw=black, line width=2pt, dashed] (0.25,-0.25) -- (0.5,-0.25) node[circle, fill=white, inner sep=2pt, draw=black, solid ]{};;
		\draw[draw=BLUE, postaction={decorate}] (0,0) node[label={left:$i$}] {} -- (0.25,-0.25);
		\draw[draw=BLUE, postaction={decorate}] (0.25,-0.25)--(0,-0.5) node[label={left:$j$}] {};
		\end{scope}
		\end{tikzpicture}}} \Bigg)
	\Bigg(\vcenter{\hbox{\begin{tikzpicture}[scale=2]
		\begin{scope}[decoration={markings, mark=at position 0.6 with \arrow{latex}}]
		\draw[draw=black, line width=2pt, dashed] (0.25,-0.25) -- (0.5,-0.25)node[circle, fill=white, inner sep=2pt, draw=black, solid ]{};;
		\draw[draw=GREEN, postaction={decorate}] (0,0) node[label={left:$i$}] {} -- (0.25,-0.25);
		\draw[draw=GREEN, postaction={decorate}] (0.25,-0.25)--(0,-0.5) node[label={left:$j$}] {};
		\end{scope}
		\end{tikzpicture}}} \Bigg)
	\\
&= \sum_{i,j =1}^D \langle M_{ij} N_{ij} \rangle_{\text{conn}} + \sum_{i,j = 1}^D \langle M_{ij} \rangle \langle N_{ij} \rangle.
\end{align}
After expanding as per \eqref{Eqn: Wicks theorem} it is possible to calculate the connected piece of this expression either by direct application of the general two-point function \eqref{Eqn: Two point function}, or equivalently by evaluating the Feynman rules associated with the diagrammatic form of \eqref{Eqn: Two point function}.  In both cases we must specialise to the appropriate index structure and then perform the relevant sums over products of $F$s:
\begin{align} \nonumber
\sum_{i,j =1}^D \langle M_{ij} N_{ij} \rangle_{\text{conn}} &= \sum_{i,j =1}^D\vcenter{\hbox{\begin{tikzpicture}[scale=2]
		\begin{scope}[decoration={markings, mark=at position 0.65 with \arrow{latex}}]
		\draw[draw=BLUE, postaction={decorate}] (0,0) node[label={left:$i$}] {} -- (0.25,-0.25);
		\draw[draw=GREEN, postaction={decorate}] (1,0) node[label={right:$i$}] {} -- (0.75,-0.25);
		\draw[draw=GREEN, postaction={decorate}]  (0.75,-0.25) -- (1,-.5) node[label={right:$j$}] {};
		\draw[draw=BLUE, postaction={decorate}] (0.25,-0.25) -- (0,-.5) node[label={left:$j$}] {};
		\draw[draw=black, line width=2.5pt] (0.25,-0.25) -- (0.5,-0.25);
		\draw[draw=black, line width=2.5pt] (0.5,-0.25) -- (0.75,-0.25);
		\end{scope}
		\end{tikzpicture}}} \\ \nonumber
&= \sum_{i,j =1}^D \Bigg( 	
	\vcenter{\hbox{\begin{tikzpicture}[scale=1.5]\begin{scope}[decoration={markings, mark=at position 0.65 with \arrow{latex}}]
				\node[label={left:$i$}] at (0,0) {};
				\node[label={right:$i$}] at (1,0) {};
				\node[label={left:$j$}] at (0,-.5) {};
				\node[label={right:$j$}] at (1,-.5) {};
				\draw[draw=BLUE, dashed] (0,-.5) -- (0.25,-.25);
				\draw[draw=BLUE, dashed] (.25,-0.25) -- (0,0);
				\draw[draw=GREEN, dashed] (.75,-0.25) -- (1,0);	
				\draw[draw=GREEN, dashed] (.75,-0.25) -- (1,-.5);	
				\draw[draw=black, dashed] (.25,-0.25) -- (.5,-0.25) node[label=above:{\tiny $V_0$}]{};
				\draw[draw=black, dashed] (.5,-0.25) -- (.75,-0.25);
				\node[circle, fill=BLUE, inner sep=1pt, draw=BLUE] at (0.25,-.25) {};
				\node[circle, fill=GREEN, inner sep=1pt, draw=GREEN] at (0.75,-.25) {};
			\end{scope}
	\end{tikzpicture}}}
	+
	\vcenter{\hbox{\begin{tikzpicture}[scale=1.5]\begin{scope}[decoration={markings, mark=at position 0.65 with \arrow{latex}}]
		\node[label={left:$i$}] at (0,0) {};
		\node[label={right:$i$}] at (1,0) {};
		\node[label={left:$j$}] at (0,-.5) {};
		\node[label={right:$j$}] at (1,-.5) {};
		\draw[draw=BLUE ] (0,0) -- (0.25,-.25) -- (0,-.5);
		\draw[draw=GREEN ] (1,0) -- (0.75, -.25)--(1,-.5);
		\draw[draw=black, dashed] (.25,-0.25) -- (.5,-0.25) node[label=above:{\tiny $V_0$}]{};
		\draw[draw=black, dashed] (.5,-0.25) -- (.75,-0.25);
		\node[circle, fill=BLUE, inner sep=1pt, draw=BLUE] at (0.25,-.25) {};
		\node[circle, fill=GREEN, inner sep=1pt, draw=GREEN] at (0.75,-.25) {};
		\end{scope}
		\end{tikzpicture}}}
	+ \dots +
\vcenter{\hbox{\begin{tikzpicture}[scale=1.5]\begin{scope}[decoration={zigzag,segment length=3,amplitude=2,post length=0pt}]
		\node[label={left:$i$}] at (0,0) {};
		\node[label={right:$i$}] at (1,0) {};
		\node[label={left:$j$}] at (0,-.5) {};
		\node[ label={right:$j$}] at (1,-.5) {};
		\draw[draw=BLUE ] (0,-.5) -- (0.25,-.25);
		\draw[draw=BLUE ] (.25,-0.25) -- (0,0);
		\draw[draw=GREEN ] (.75,-0.25) -- (1,0);	
		\draw[draw=GREEN ] (.75,-0.25) -- (1,-.5);	
		\draw[draw=black] (.25,-0.25) -- (.5,-0.25) node[label=above:{\tiny $V_3$}]{};
		\draw[draw=black] (.5,-0.25) -- (.75,-0.25);
		\node[circle, fill=BLUE, inner sep=1pt, draw=BLUE] at (.25,-.25) {};
		\node[circle, fill=GREEN, inner sep=1pt, draw=GREEN] at (.75,-.25) {};
		\end{scope}
		\end{tikzpicture}}}
	 \Bigg) \\ \nonumber
&= (\Lambda_{V_0}^{-1})_{\ls[11]}^{\us[XY]} + (\Lambda_{V_0}^{-1})_{\ls[22]}^{\us[XY]} + (D-1) \Big( (\Lambda_{V_H}^{-1})_{\ls[11]}^{\us[XY]} + (\Lambda_{V_H}^{-1})_{\ls[22]}^{\us[XY]} + (\Lambda_{V_H}^{-1})_{\ls[33]}^{\us[XY]} \Big) \\ 
&+ \frac{(\Lambda_{V_2}^{-1})^{\us[XY]}}{2} D(D-3) + \frac{(D-1)(D-2)}{2} (\Lambda_{V_3}^{-1})^{\us[XY]}.
\end{align}
In the first line we have used \eqref{Eqn: 2pt diagram EV}, in the second we have decomposed the composite graph into the 15 atomic graphs given in Section \ref{Subsec: Feynman decomp} and in the final line we have evaluated the graphs using the Feynman rules and completed the resulting sums over products of $F$s. We would have found the same expression by setting $k=i$ and $l=j$ in \eqref{Eqn: Two point function}.

Similarly, using \eqref{Eqn: M one point function} and \eqref{Eqn: N one point function} we find the disconnected contribution is
\begin{align} \nonumber
\sum_{i,j = 1}^D \langle M_{ij} \rangle \langle N_{ij} \rangle &= \sum_{i,j = 1}^D \Bigg( \vcenter{\hbox{\begin{tikzpicture}[scale=2]
		\begin{scope}[decoration={markings, mark=at position 0.6 with \arrow{latex}}]
		\draw[draw=black, line width=2pt, dashed] (0.25,-0.25) -- (0.5,-0.25) node[circle, fill=white, inner sep=2pt, draw=black, solid ]{};;
		\draw[draw=BLUE, postaction={decorate}] (0,0) node[label={left:$i$}] {} -- (0.25,-0.25);
		\draw[draw=BLUE, postaction={decorate}] (0.25,-0.25)--(0,-0.5) node[label={left:$j$}] {};
		\end{scope}
		\end{tikzpicture}}} \Bigg)
	\Bigg(\vcenter{\hbox{\begin{tikzpicture}[scale=2]
		\begin{scope}[decoration={markings, mark=at position 0.6 with \arrow{latex}}]
		\draw[draw=black, line width=2pt, dashed] (0.25,-0.25) -- (0.5,-0.25)node[circle, fill=white, inner sep=2pt, draw=black, solid ]{};;
		\draw[draw=GREEN, postaction={decorate}] (0,0) node[label={left:$i$}] {} -- (0.25,-0.25);
		\draw[draw=GREEN, postaction={decorate}] (0.25,-0.25)--(0,-0.5) node[label={left:$j$}] {};
		\end{scope}
		\end{tikzpicture}}} \Bigg) \\ \nonumber
&= \sum_{i,j = 1}^D \Bigg(\vcenter{\hbox{\begin{tikzpicture}[scale=2]\begin{scope}[decoration={markings, mark=at position 0.65 with \arrow{latex}}]
			\node[label={left:$i$}] at (0,0) {};
			\node[label={left:$j$}] at (0,-.5) {};
			\draw[draw=BLUE, dashed] (0,0) --(0.25,-.25) -- (0,-.5);	
			\draw[draw=black, dashed] (.25,-0.25) -- (.5,-0.25) node[circle, fill=white, inner sep=2pt, draw=black, solid] {};			
			\node[circle,fill=BLUE, inner sep=1pt, draw=BLUE] at (0.25,-.25) {};
		\end{scope}
	\end{tikzpicture}}} \; + \vcenter{\hbox{\begin{tikzpicture}[scale=2]\begin{scope}[decoration={markings, mark=at position 0.65 with \arrow{latex}}]
		\node[label={left:$i$}] at (0,0) {};
		\node[label={left:$j$}] at (0,-.5) {};
		\draw[draw=BLUE ] (0,0) -- (0.25,-.25) -- (0,-.5);	
		\draw[draw=black, dashed] (.25,-0.25) -- (.5,-0.25)node[circle, fill=white, inner sep=2pt, draw=black, solid ]{};;
		\node[circle, fill=BLUE, inner sep=1pt, draw=BLUE] at (0.25,-.25) {};
		\end{scope}
		\end{tikzpicture}}} \Bigg)
		\Bigg( \vcenter{\hbox{\begin{tikzpicture}[scale=2]\begin{scope}[decoration={markings, mark=at position 0.65 with \arrow{latex}}]
		\node[label={left:$i$}] at (0,0) {};
		\node[label={left:$j$}] at (0,-.5) {};
		\draw[draw=GREEN, dashed] (0,0) --(0.25,-.25) -- (0,-.5);	
		\draw[draw=black, dashed] (.25,-0.25) -- (.5,-0.25) node[circle, fill=white, inner sep=2pt, draw=black, solid] {};			
		\node[circle,fill=GREEN, inner sep=1pt, draw=GREEN] at (0.25,-.25) {};
		\end{scope}
		\end{tikzpicture}}} \; + \vcenter{\hbox{\begin{tikzpicture}[scale=2]\begin{scope}[decoration={markings, mark=at position 0.65 with \arrow{latex}}]
		\node[label={left:$i$}] at (0,0) {};
		\node[label={left:$j$}] at (0,-.5) {};
		\draw[draw=GREEN ] (0,0) -- (0.25,-.25) -- (0,-.5);	
		\draw[draw=black, dashed] (.25,-0.25) -- (.5,-0.25)node[circle, fill=white, inner sep=2pt, draw=black, solid ]{};;
		\node[circle, fill=GREEN, inner sep=1pt, draw=GREEN] at (0.25,-.25) {};
		\end{scope}
		\end{tikzpicture}}}\Bigg) \\ \nonumber
&= \sum_{i,j = 1}^D  \Bigg( \frac{\tilde{\rho}_{\ls[1]}^{V_0}}{D} + \frac{\tilde{\rho}_{\ls[2]}^{V_0}}{\sqrt{D-1}} F(i,j) \Bigg) \Bigg( \frac{\tilde{\rho}_{\ls[3]}^{V_0}}{D} + \frac{\tilde{\rho}_{\ls[4]}^{V_0}}{\sqrt{D-1}} F(i,j) \Bigg) \\ \nonumber
&= \tilde{\rho}_{\ls[1]}^{V_0} \tilde{\rho}_{\ls[3]}^{V_0} + \sum_{i = 1}^D \frac{\tilde{\rho}_{\ls[2]}^{V_0} \tilde{\rho}_{\ls[4]}^{V_0}}{D-1} F(i,i) + \sum_{i,j = 1}^D \frac{\tilde{\rho}_{\ls[2]}^{V_0} \tilde{\rho}_{\ls[3]}^{V_0} + \tilde{\rho}_{\ls[1]}^{V_0} \tilde{\rho}_{\ls[4]}^{V_0}}{\sqrt{D-1}} F(i,j) \\ 
&= \tilde{\rho}_{\ls[1]}^{V_0} \tilde{\rho}_{\ls[3]}^{V_0} + \tilde{\rho}_{\ls[2]}^{V_0} \tilde{\rho}_{\ls[4]}^{V_0}.
\end{align}
Putting these together we have
\begin{align} \nonumber
&\sum_{i,j =1}^D \langle M_{ij} N_{ij} \rangle =  \tilde{\rho}_{\ls[1]}^{V_0} \tilde{\rho}_{\ls[3]}^{V_0} + \tilde{\rho}_{\ls[2]}^{V_0} \tilde{\rho}_{\ls[4]}^{V_0} + \sum_{i,j =1}^D \langle M_{ij} N_{ij} \rangle_{\text{conn}} \\ \nonumber
&= \tilde{\rho}_{\ls[1]}^{V_0} \tilde{\rho}_{\ls[3]}^{V_0} + \tilde{\rho}_{\ls[2]}^{V_0} \tilde{\rho}_{\ls[4]}^{V_0} + (\Lambda_{V_0}^{-1})_{\ls[11]}^{\us[XY]} + (\Lambda_{V_0}^{-1})_{\ls[22]}^{\us[XY]} + (D-1) \Big( (\Lambda_{V_H}^{-1})_{\ls[11]}^{\us[XY]} + (\Lambda_{V_H}^{-1})_{\ls[22]}^{\us[XY]} + (\Lambda_{V_H}^{-1})_{\ls[33]}^{\us[XY]} \Big) \\ 
&+ \frac{(\Lambda_{V_2}^{-1})^{\us[XY]}}{2} D(D-3) + \frac{(D-1)(D-2)}{2} (\Lambda_{V_3}^{-1})^{\us[XY]}.
\end{align}

In general we note that expectation values containing unrepeated indices must necessarily involve projecting to the trivial representation. Further to this, some computations can often be dramatically simplified by exploiting the symmetry of $(\Lambda_{V_A}^{-1})^{\us[XY]}$. So that an expectation value related to another by exchange of $M$ and $N$ can immediately be written down by substituting $(\Lambda_{V_A}^{-1})^{\us[YX]}$ for $(\Lambda_{V_A}^{-1})^{\us[XY]}$ in its representation theory expression and using the relation $\Lambda_{V_A}^{\us[YX]} = (\Lambda_{V_A}^{\us[XY]})^T$.

Similar calculations can be performed for the other quadratic expectation values. Here we list all those of mixed type
\begin{flalign} \nonumber
\sum_{i =1}^D &\langle M_{ii} N_{ii} \rangle = \frac{ \tilde{\rho}_{\ls[1]}^{V_0} \tilde{\rho}_{\ls[3]}^{V_0}}{D} + \frac{\sqrt{D-1}}{D} \Big( \tilde{\rho}_{\ls[1]}^{V_0} \tilde{\rho}_{\ls[4]}^{V_0} + \tilde{\rho}_{\ls[2]}^{V_0} \tilde{\rho}_{\ls[3]}^{V_0} \Big) + \frac{(D-1)}{D} \tilde{\rho}_{\ls[2]}^{V_0} \tilde{\rho}_{\ls[4]}^{V_0} +  \frac{(\Lambda_{V_0}^{-1})_{\ls[11]}^{\us[XY]}}{D} & \\ \nonumber
&+ \frac{(D-1)}{D} (\Lambda_{V_0}^{-1})_{\ls[22]}^{\us[XY]} + \frac{\sqrt{D-1}}{D} \Big((\Lambda_{V_0}^{-1})_{\ls[12]}^{\us[XY]} + (\Lambda_{V_0}^{-1})_{\ls[21]}^{\us[XY]} \Big) + \frac{D-1}{D} \Big( (\Lambda_{V_H}^{-1})_{\ls[11]}^{\us[XY]} + (\Lambda_{V_H}^{-1})_{\ls[22]}^{\us[XY]} \Big) \\ \nonumber
&+ \frac{(D-1)(D-2)}{D} (\Lambda_{V_H}^{-1})_{\ls[33]}^{\us[XY]} + \frac{(D-1)}{D} \Big( (\Lambda_{V_H}^{-1})_{\ls[12]}^{\us[XY]} + (\Lambda_{V_H}^{-1})_{\ls[21]}^{\us[XY]} \Big) \\ 
&+ \frac{(D-1) \sqrt{D-2}}{D} \Big( (\Lambda_{V_H}^{-1})_{\ls[13]}^{\us[XY]} + (\Lambda_{V_H}^{-1})_{\ls[31]}^{\us[XY]} + (\Lambda_{V_H}^{-1})_{\ls[23]}^{\us[XY]} + (\Lambda_{V_H}^{-1})_{\ls[32]}^{\us[XY]} \Big), 
\end{flalign}
\begin{flalign} \nonumber
\sum_{i,j =1}^D &\langle M_{ij} N_{ij} \rangle = \tilde{\rho}_{\ls[1]}^{V_0} \tilde{\rho}_{\ls[3]}^{V_0} + \tilde{\rho}_{\ls[2]}^{V_0} \tilde{\rho}_{\ls[4]}^{V_0} + (\Lambda_{V_0}^{-1})_{\ls[11]}^{\us[XY]} + (\Lambda_{V_0}^{-1})_{\ls[22]}^{\us[XY]} + (D-1) \Big( (\Lambda_{V_H}^{-1})_{\ls[11]}^{\us[XY]} & \\ 
& + (\Lambda_{V_H}^{-1})_{\ls[22]}^{\us[XY]} + (\Lambda_{V_H}^{-1})_{\ls[33]}^{\us[XY]} \Big) + \frac{(\Lambda_{V_2}^{-1})^{\us[XY]}}{2} D(D-3) + \frac{(D-1)(D-2)}{2} (\Lambda_{V_3}^{-1})^{\us[XY]},
\end{flalign}
\begin{flalign} \nonumber
\sum_{i,j=1}^D &\langle M_{ii} N_{ij} \rangle = \tilde{\rho}_{\ls[1]}^{V_0} \tilde{\rho}_{\ls[3]}^{V_0} + \sqrt{D-1} \tilde{\rho}_{\ls[2]}^{V_0} \tilde{\rho}_{\ls[3]}^{V_0} + (\Lambda_{V_0}^{-1})_{\ls[11]}^{\us[XY]} + \sqrt{D-1}(\Lambda_{V_0}^{-1})_{\ls[21]}^{\us[XY]} & \\ 
&+ (D-1) (\Lambda_{V_H}^{-1})_{\ls[22]}^{\us[XY]} + (D-1) (\Lambda_{V_H}^{-1})_{\ls[12]}^{\us[XY]} + (D-1) \sqrt{D-2} (\Lambda_{V_H}^{-1})_{\ls[32]}^{\us[XY]},
\end{flalign}
\begin{flalign} \nonumber
\sum_{i,j=1} &\langle N_{ii} M_{ij} \rangle = \tilde{\rho}_{\ls[1]}^{V_0} \tilde{\rho}_{\ls[3]}^{V_0} + \sqrt{D-1} \tilde{\rho}_{\ls[1]}^{V_0} \tilde{\rho}_{\ls[4]}^{V_0} + (\Lambda_{V_0}^{-1})_{\ls[11]}^{\us[XY]} + \sqrt{D-1}(\Lambda_{V_0}^{-1})_{\ls[12]}^{\us[XY]} & \\ 
&+ (D-1) (\Lambda_{V_H}^{-1})_{\ls[22]}^{\us[XY]} + (D-1) (\Lambda_{V_H}^{-1})_{\ls[21]}^{\us[XY]} + (D-1) \sqrt{D-2} (\Lambda_{V_H}^{-1})_{\ls[23]}^{\us[XY]},
\end{flalign}
\begin{flalign} \nonumber
\sum_{i,j = 1}^D &\langle M_{ij}  N_{ji} \rangle = \tilde{\rho}_{\ls[1]}^{V_0} \tilde{\rho}_{\ls[3]}^{V_0} + \tilde{\rho}_{\ls[2]}^{V_0} \tilde{\rho}_{\ls[4]}^{V_0} + (\Lambda_{V_0}^{-1})_{\ls[11]}^{\us[XY]} + (\Lambda_{V_0}^{-1})_{\ls[22]}^{\us[XY]} +(D-1) \big( (\Lambda_{V_H}^{-1})_{\ls[33]}^{\us[XY]} & \\ 
& + (\Lambda_{V_H}^{-1})_{\ls[12]}^{\us[XY]} + (\Lambda_{V_H}^{-1})_{\ls[21]}^{\us[XY]} \big) + (\Lambda_{V_2}^{-1})^{\us[XY]} \frac{D(D-3)}{2} + (\Lambda_{V_3}^{-1})^{\us[XY]} \frac{(D-1)(D-2)}{2},
\end{flalign}
\begin{flalign} \nonumber
\sum_{i,j =1}^D &\langle M_{ij} N_{jj} \rangle = \tilde{\rho}_{\ls[1]}^{V_0} \tilde{\rho}_{\ls[3]}^{V_0} + \sqrt{D-1} \tilde{\rho}_{\ls[1]}^{V_0} \tilde{\rho}_{\ls[4]}^{V_0} + (\Lambda_{V_0}^{-1})_{\ls[11]}^{\us[XY]} + \sqrt{D-1} (\Lambda_{V_0}^{-1})_{\ls[12]}^{\us[XY]} & \\
&+  (D-1) (\Lambda_{V_H}^{-1})_{\ls[11]}^{\us[XY]} + (D-1) (\Lambda_{V_H}^{-1})_{\ls[12]}^{\us[XY]} + (D-1) \sqrt{D-2} (\Lambda_{V_H}^{-1})_{\ls[13]}^{\us[XY]},
\end{flalign}
\begin{flalign} \nonumber
\sum_{i,j =1}^D &\langle N_{ij} M_{jj} \rangle = \tilde{\rho}_{\ls[1]}^{V_0} \tilde{\rho}_{\ls[3]}^{V_0} + \sqrt{D-1} \tilde{\rho}_{\ls[1]}^{V_0} \tilde{\rho}_{\ls[4]}^{V_0} + (\Lambda_{V_0}^{-1})_{\ls[11]}^{\us[XY]} + \sqrt{D-1} (\Lambda_{V_0}^{-1})_{\ls[21]}^{\us[XY]} & \\
&+  (D-1) (\Lambda_{V_H}^{-1})_{\ls[11]}^{\us[XY]} + (D-1) (\Lambda_{V_H}^{-1})_{\ls[21]}^{\us[XY]} + (D-1) \sqrt{D-2} (\Lambda_{V_H}^{-1})_{\ls[31]}^{\us[XY]},
\end{flalign}
\begin{flalign} \nonumber
\sum_{i,j =1}^D &\langle M_{ii} N_{jj} \rangle = \tilde{\rho}_{\ls[1]}^{V_0} \tilde{\rho}_{\ls[3]}^{V_0} + \sqrt{D-1} \Big( \tilde{\rho}_{\ls[1]}^{V_0} \tilde{\rho}_{\ls[4]}^{V_0} + \tilde{\rho}_{\ls[2]}^{V_0} \tilde{\rho}_{\ls[3]}^{V_0} \Big) + (D-1) \tilde{\rho}_{\ls[2]}^{V_0} \tilde{\rho}_{\ls[4]}^{V_0} + (\Lambda_{V_0}^{-1})_{\ls[11]}^{\us[XY]} & \\
&+ (D-1) (\Lambda_{V_0}^{-1})_{\ls[22]}^{\us[XY]} + \sqrt{D-1} \Big( (\Lambda_{V_0}^{-1})_{\ls[12]}^{\us[XY]} + (\Lambda_{V_0}^{-1})_{\ls[21]}^{\us[XY]} \Big),
\end{flalign}
\begin{flalign} 
\sum_{i,j,k =1}^D &\langle M_{ij} N_{ik} \rangle = D \tilde{\rho}_{\ls[1]}^{V_0} \tilde{\rho}_{\ls[3]}^{V_0} + D (\Lambda_{V_0}^{-1})_{\ls[11]}^{\us[XY]} + D (D-1) (\Lambda_{V_H}^{-1})_{\ls[22]}^{\us[XY]}, &
\end{flalign}
\begin{flalign} 
\sum_{i,j,k =1}^D &\langle M_{ij} N_{kj} \rangle = D \tilde{\rho}_{\ls[1]}^{V_0} \tilde{\rho}_{\ls[3]}^{V_0} + D (\Lambda_{V_0}^{-1})_{\ls[11]}^{\us[XY]} + D (D-1) (\Lambda_{V_H}^{-1})_{\ls[11]}^{\us[XY]}, &
\end{flalign}
\begin{flalign} 
\sum_{i,j,k =1}^D &\langle M_{ij} N_{jk} \rangle = D \tilde{\rho}_{\ls[1]}^{V_0} \tilde{\rho}_{\ls[3]}^{V_0} +  D (\Lambda_{V_0}^{-1})_{\ls[11]}^{\us[XY]} + D (D-1)(\Lambda_{V_H}^{-1})_{\ls[12]}^{\us[XY]}, &
\end{flalign}
\begin{flalign} 
\sum_{i,j,k =1}^D &\langle N_{ij} M_{jk} \rangle = D \tilde{\rho}_{\ls[1]}^{V_0} \tilde{\rho}_{\ls[3]}^{V_0} +  D (\Lambda_{V_0}^{-1})_{\ls[11]}^{\us[XY]} + D (D-1)(\Lambda_{V_H}^{-1})_{\ls[21]}^{\us[XY]}, &
\end{flalign}
\begin{flalign}
\sum_{i,j,k =1}^D &\langle M_{ii} N_{jk} \rangle = D \tilde{\rho}_{\ls[1]}^{V_0} \tilde{\rho}_{\ls[3]}^{V_0} + D \sqrt{D-1} \tilde{\rho}_{\ls[2]}^{V_0} \tilde{\rho}_{\ls[3]}^{V_0} + D (\Lambda_{V_0}^{-1})_{\ls[11]}^{\us[XY]} + D \sqrt{D-1} (\Lambda_{V_0}^{-1})_{\ls[21]}^{\us[XY]}, &
\end{flalign}
\begin{flalign}
\sum_{i,j,k =1}^D &\langle N_{ii} M_{jk} \rangle =  D \tilde{\rho}_{\ls[1]}^{V_0} \tilde{\rho}_{\ls[3]}^{V_0} + D \sqrt{D-1} \tilde{\rho}_{\ls[2]}^{V_0} \tilde{\rho}_{\ls[3]}^{V_0} + D (\Lambda_{V_0}^{-1})_{\ls[11]}^{\us[XY]} + D \sqrt{D-1} (\Lambda_{V_0}^{-1})_{\ls[12]}^{\us[XY]}, &
\end{flalign}
\begin{flalign}
\sum_{i,j,k,l =1}^D &\langle M_{ij} N_{kl} \rangle = D^2 \tilde{\rho}_{\ls[1]}^{V_0} \tilde{\rho}_{\ls[3]}^{V_0} + D^2 (\Lambda_{V_0}^{-1})_{\ls[11]}^{\us[XY]}. &
\end{flalign}

\subsection{Cubic expectation values}
Again, we perform a detailed calculation of one cubic expectation value, $\sum_{i,j} \langle M_{ij}^2 N_{ij} \rangle$, as an example before listing other cubic results. Using Wick's theorem we write the cubic expectation value as a sum of products of quadratic and linear expectation values
\begin{align} \label{Eqn: Wick expansion cubic example}
\sum_{i,j=1}^D \langle M_{ij}^2 N_{ij} \rangle = \sum_{i,j=1}^D \Big( \langle M_{ij} \rangle^2 \langle N_{ij} \rangle + 2 \langle M_{ij} N_{ij} \rangle \langle M_{ij} \rangle  +  \langle M_{ij}^2 \rangle \langle N_{ij} \rangle \Big).
\end{align}
The disconnected piece of this expression is given by
\begin{align} \nonumber
\sum_{i,j=1}^D  \langle M_{ij} \rangle^2 \langle N_{ij} \rangle &=  \sum_{i,j = 1}^D  \Bigg( \frac{\tilde{\rho}_{\ls[1]}^{V_0}}{D} + \frac{\tilde{\rho}_{\ls[2]}^{V_0}}{\sqrt{D-1}} F(i,j) \Bigg)^2 \Bigg( \frac{\tilde{\rho}_{\ls[3]}^{V_0}}{D} + \frac{\tilde{\rho}_{\ls[4]}^{V_0}}{\sqrt{D-1}} F(i,j) \Bigg) \\
&=  \frac{1}{D} \Bigg( (\tilde{\rho}_{\ls[1]}^{V_0})^2 \tilde{\rho}_{\ls[3]}^{V_0} + (\tilde{\rho}_{\ls[2]}^{V_0})^2 \tilde{\rho}_{\ls[3]}^{V_0} + \frac{(D-2)}{\sqrt{D-1}}(\tilde{\rho}_{\ls[2]}^{V_0})^2 \tilde{\rho}_{\ls[4]}^{V_0} + \frac{2}{D} \tilde{\rho}_{\ls[1]}^{V_0} \tilde{\rho}_{\ls[2]}^{V_0}  \tilde{\rho}_{\ls[4]}^{V_0}   \Bigg).
\end{align}

\subsubsection{$\sum_{i,j=1}^D \langle M_{ij} N_{ij} \rangle_{\text{conn}} \langle M_{ij} \rangle$ Contribution from the $V_0$ channel}

\begin{align} \nonumber
\sum_{i,j=1}^D \langle M_{ij} N_{ij} \rangle_{\text{conn}}^{V_0} \langle M_{ij} \rangle &= \sum_{i,j=1}^D \Bigg(  \frac{\tilde{\rho}_{\ls[1]}^{V_0}}{D} + \frac{\tilde{\rho}_{\ls[2]}^{V_0}}{\sqrt{D-1}} F(i,j)  \Bigg) \Bigg( \frac{1}{D^2} (\Lambda_{V_0}^{-1})_{\ls[11]}^{\us[XY]} + \frac{(\Lambda_{V_0}^{-1})_{\ls[22]}^{\us[XY]}}{D-1} F(i,j) F(i,j) \\ \nonumber
&+ \frac{(\Lambda_{V_0}^{-1})_{\ls[12]}^{\us[XY]}}{D \sqrt{D-1}} F(i,j) +  \frac{(\Lambda_{V_0}^{-1})_{\ls[21]}^{\us[XY]}}{D \sqrt{D-1}} F(i,j) \Bigg) \\ \nonumber
&= \frac{1}{D} \Bigg( \tilde{\rho}_{\ls[1]}^{V_0} (\Lambda_{V_0}^{-1})_{\ls[11]}^{\us[XY]} +  \tilde{\rho}_{\ls[1]}^{V_0} (\Lambda_{V_0}^{-1})_{\ls[22]}^{\us[XY]} + \frac{(D-2)}{\sqrt{D-1}}  \tilde{\rho}_{\ls[2]}^{V_0} (\Lambda_{V_0}^{-1})_{\ls[22]}^{\us[XY]} \\
&+ \tilde{\rho}_{\ls[2]}^{V_0} \Big( (\Lambda_{V_0}^{-1})_{\ls[12]}^{\us[XY]} + (\Lambda_{V_0}^{-1})_{\ls[21]}^{\us[XY]} \Big) \Bigg) 
\end{align}

\subsubsection{$\sum_{i,j=1}^D \langle M_{ij} N_{ij} \rangle_{\text{conn}} \langle M_{ij} \rangle$ Contribution from the $V_H$ channel}

\begin{align} \nonumber
\sum_{i,j=1}^D \langle M_{ij} N_{ij} \rangle_{\text{conn}}^{V_H}& \langle M_{ij} \rangle = \sum_{i,j = 1}^D  \Bigg( \frac{\tilde{\rho}_{\ls[1]}^{V_0}}{D} + \frac{\tilde{\rho}_{\ls[2]}^{V_0}}{\sqrt{D-1}} F(i,j)  \Bigg) \Bigg( \frac{(\Lambda_{V_H}^{-1})_{\ls[11]}^{\us[XY]}}{D} F(j,j) + \frac{(\Lambda_{V_H}^{-1})_{\ls[22]}^{\us[XY]}}{D} F(i,i) \\ \nonumber
&+ \frac{D(\Lambda_{V_H}^{-1})_{\ls[33]}^{\us[XY]}}{(D-2)} \sum_{p,q = 1}^{D} F(i,p) F(j,p) F(i,q) F(j,q) F(p,q) \\ \nonumber
&+ \frac{(\Lambda_{V_H}^{-1})_{\ls[12]}^{\us[XY]}}{D} F(j,i) + \frac{(\Lambda_{V_H}^{-1})_{\ls[21]}^{\us[XY]}}{D} F(i,j) + \frac{(\Lambda_{V_H}^{-1})_{\ls[13]}^{\us[XY]}}{\sqrt{D-2}} \sum_{p = 1}^{D} F(j,p) F(i,p) F(j,p) \\ \nonumber
&+ \frac{(\Lambda_{V_H}^{-1})_{\ls[31]}^{\us[XY]}}{\sqrt{D-2}} \sum_{p = 1}^{D} F(i,p) F(j,p) F(j,p) + \frac{(\Lambda_{V_H}^{-1})_{\ls[23]}^{\us[XY]}}{\sqrt{D-2}} \sum_{p = 1}^{D} F(i,p) F(i,p) F(j,p) \\ \nonumber
&+ \frac{(\Lambda_{V_H}^{-1})_{\ls[32]}^{\us[XY]}}{\sqrt{D-2}} \sum_{p = 1}^{D} F(i,p) F(j,p) F(i,p) \Bigg) \\ \nonumber
&= \frac{1}{D} \Bigg( (D-1)  \tilde{\rho}_{\ls[1]}^{V_0}  \Big( (\Lambda_{V_H}^{-1})_{\ls[11]}^{\us[XY]} + (\Lambda_{V_H}^{-1})_{\ls[22]}^{\us[XY]} + (\Lambda_{V_H}^{-1})_{\ls[33]}^{\us[XY]} \Big) \\ \nonumber
&+ \tilde{\rho}_{\ls[2]}^{V_0} \Big(  \sqrt{D-1} (D-3) (\Lambda_{V_H}^{-1})_{\ls[33]}^{\us[XY]} + \sqrt{D-1} \big( (\Lambda_{V_H}^{-1})_{\ls[12]}^{\us[XY]} + (\Lambda_{V_H}^{-1})_{\ls[21]}^{\us[XY]} \big) \\
&+ \sqrt{D-1} \sqrt{D-2} \big( (\Lambda_{V_H}^{-1})_{\ls[13]}^{\us[XY]} + (\Lambda_{V_H}^{-1})_{\ls[31]}^{\us[XY]} + (\Lambda_{V_H}^{-1})_{\ls[23]}^{\us[XY]} + (\Lambda_{V_H}^{-1})_{\ls[32]}^{\us[XY]} \big)  \Big) \Bigg).
\end{align}
Where we have used the fact that
\begin{equation} \label{Eqn: Cubic F ident 1}
\sum_{i,j,p,q = 1}^{D} F(i,p) F(j,p) F(i,q) F(j,q) F(p,q) F(i,j) = \frac{(D-1)(D-2)(D-3)}{D^2}.
\end{equation}

\subsubsection{$\sum_{i,j=1}^D \langle M_{ij} N_{ij} \rangle_{\text{conn}} \langle M_{ij} \rangle$ Contribution from the $V_2$ channel}

The $V_2$ contribution includes the same sum over $F$s given by \eqref{Eqn: Cubic F ident 1}.
\begin{align} \nonumber
\sum_{i,j=1}^D \langle M_{ij} N_{ij} \rangle_{\text{conn}}^{V_2} &\langle M_{ij} \rangle = \sum_{i,j = 1}^D  \Bigg(  \frac{\tilde{\rho}_{\ls[1]}^{V_0}}{D} + \frac{\tilde{\rho}_{\ls[2]}^{V_0}}{\sqrt{D-1}} F(i,j)  \Bigg) \Bigg( (\Lambda_{V_2}^{-1})^{\us[XY]} \Big( \frac{1}{2} F(i,i) F(j,j) + \frac{1}{2} F(i,j) F(j,i) \\ \nonumber
&- \frac{D}{D-2} \sum_{p,q = 1}^D F(i,p) F(j,p) F(i,q) F(j,q) F(p,q) - \frac{1}{(D-1)} F(i,j) F(i,j) \Big) \Bigg) \\ \nonumber
&= \frac{(D-3)}{2} \tilde{\rho}_{\ls[1]}^{V_0} (\Lambda_{V_2}^{-1})^{\us[XY]} + \frac{1}{\sqrt{D-1}} \tilde{\rho}_{\ls[2]}^{V_0} (\Lambda_{V_2}^{-1})^{\us[XY]} \Bigg(  \frac{(D-1)(D-2)}{2D} \\ \nonumber
&- \frac{(D-1)(D-3)}{D} - \frac{(D-2)}{D}  \Bigg) \\ 
&= \frac{(D-3)}{2} \tilde{\rho}_{\ls[1]}^{V_0} (\Lambda_{V_2}^{-1})^{\us[XY]} - \frac{(D-3)}{2 \sqrt{D-1}} \tilde{\rho}_{\ls[2]}^{V_0} (\Lambda_{V_2}^{-1})^{\us[XY]}.
\end{align}

\subsubsection{$\sum_{i,j=1}^D \langle M_{ij} N_{ij} \rangle_{\text{conn}} \langle M_{ij} \rangle$ Contribution from the $V_3$ channel}

\begin{align} \nonumber
\sum_{i,j=1}^D \langle M_{ij} N_{kl} \rangle_{\text{conn}}^{V_3} &\langle M_{ij} \rangle = \sum_{i,j = 1}^D  \Bigg(  \frac{\tilde{\rho}_{\ls[1]}^{V_0}}{D} + \frac{\tilde{\rho}_{\ls[2]}^{V_0}}{\sqrt{D-1}} F(i,j)  \Bigg) \Bigg( \frac{(\Lambda_{V_3}^{-1})^{\us[XY]}}{2} \big( F(i,i) F(j,j) - F(i,j) F(j,i) \big) \Bigg) \\ 
&= \frac{(D-1)(D-2)}{2D} \tilde{\rho}_{\ls[1]}^{V_0} (\Lambda_{V_3}^{-1})^{\us[XY]} - \frac{\sqrt{D-1}(D-2)}{2D} \tilde{\rho}_{\ls[2]}^{V_0} (\Lambda_{V_3}^{-1})^{\us[XY]}. 
\end{align}

\subsubsection{$\sum_{i,j=1}^D \langle M_{ij} N_{ij} \rangle_{\text{conn}} \langle M_{ij} \rangle$ Summing all channels}
\begin{align} \label{Eqn: Intermediate 2 point cubic result} \nonumber
\sum_{i,j=1}^D \langle M_{ij} N_{ij} \rangle_{\text{conn}} &\langle M_{ij} \rangle = \frac{1}{D} \Bigg( \tilde{\rho}_{\ls[1]}^{V_0} (\Lambda_{V_0}^{-1})_{\ls[11]}^{\us[XY]} +  \tilde{\rho}_{\ls[1]}^{V_0} (\Lambda_{V_0}^{-1})_{\ls[22]}^{\us[XY]} + \frac{(D-2)}{\sqrt{D-1}}  \tilde{\rho}_{\ls[2]}^{V_0} (\Lambda_{V_0}^{-1})_{\ls[22]}^{\us[XY]} \\ \nonumber
&+ \tilde{\rho}_{\ls[2]}^{V_0} \Big( (\Lambda_{V_0}^{-1})_{\ls[12]}^{\us[XY]} + (\Lambda_{V_0}^{-1})_{\ls[21]}^{\us[XY]} \Big) \Bigg) \\ \nonumber
&+ \frac{1}{D} \Bigg( (D-1)  \tilde{\rho}_{\ls[1]}^{V_0}  \Big( (\Lambda_{V_H}^{-1})_{\ls[11]}^{\us[XY]} + (\Lambda_{V_H}^{-1})_{\ls[22]}^{\us[XY]} + (\Lambda_{V_H}^{-1})_{\ls[33]}^{\us[XY]} \Big) \\ \nonumber
&+ \tilde{\rho}_{\ls[2]}^{V_0} \Big(  \sqrt{D-1} (D-3) (\Lambda_{V_H}^{-1})_{\ls[33]}^{\us[XY]} \sqrt{D-1} \big( (\Lambda_{V_H}^{-1})_{\ls[12]}^{\us[XY]} + (\Lambda_{V_H}^{-1})_{\ls[21]}^{\us[XY]} \big) \\ \nonumber
&+ \sqrt{D-1} \sqrt{D-2} \big( (\Lambda_{V_H}^{-1})_{\ls[13]}^{\us[XY]} + (\Lambda_{V_H}^{-1})_{\ls[31]}^{\us[XY]} + (\Lambda_{V_H}^{-1})_{\ls[23]}^{\us[XY]} + (\Lambda_{V_H}^{-1})_{\ls[32]}^{\us[XY]} \big)  \Big) \Bigg) \\ \nonumber
&+ \frac{(D-3)}{2} \tilde{\rho}_{\ls[1]}^{V_0} (\Lambda_{V_2}^{-1})^{\us[XY]} - \frac{(D-3)}{2 \sqrt{D-1}} \tilde{\rho}_{\ls[2]}^{V_0} (\Lambda_{V_2}^{-1})^{\us[XY]} \\
&+ \frac{(D-1)(D-2)}{2D} \tilde{\rho}_{\ls[1]}^{V_0} (\Lambda_{V_3}^{-1})^{\us[XY]} - \frac{\sqrt{D-1}(D-2)}{2D} \tilde{\rho}_{\ls[2]}^{V_0} (\Lambda_{V_3}^{-1})^{\us[XY]}.
\end{align}

\subsubsection{$\sum_{i,j=1}^D \langle M_{ij}^2\rangle_{\text{conn}} \langle N_{ij} \rangle$}
This is a similar form of expression to \eqref{Eqn: Intermediate 2 point cubic result}. Making the appropriate changes under a swap of $M$ and $N$ leaves us with
\begin{align} \nonumber
\sum_{i,j=1}^D \langle M_{ij}^2 \rangle_{\text{conn}} &\langle N_{ij} \rangle = \frac{1}{D} \Bigg( \tilde{\rho}_{\ls[3]}^{V_0} (\Lambda_{V_0}^{-1})_{\ls[11]}^{\us[X]} +  \tilde{\rho}_{\ls[3]}^{V_0} (\Lambda_{V_0}^{-1})_{\ls[22]}^{\us[X]} + \frac{(D-2)}{\sqrt{D-1}}  \tilde{\rho}_{\ls[4]}^{V_0} (\Lambda_{V_0}^{-1})_{\ls[22]}^{\us[X]} \\ \nonumber
&+ 2 \tilde{\rho}_{\ls[4]}^{V_0}  (\Lambda_{V_0}^{-1})_{\ls[12]}^{\us[X]} \Bigg) + \frac{1}{D} \Bigg( (D-1)  \tilde{\rho}_{\ls[3]}^{V_0}  \Big( (\Lambda_{V_H}^{-1})_{\ls[11]}^{\us[X]} + (\Lambda_{V_H}^{-1})_{\ls[22]}^{\us[X]} + (\Lambda_{V_H}^{-1})_{\ls[33]}^{\us[X]} \Big) \\ \nonumber
&+ \tilde{\rho}_{\ls[4]}^{V_0} \Big(  \sqrt{D-1} (D-3) (\Lambda_{V_H}^{-1})_{\ls[33]}^{\us[X]} + 2 \sqrt{D-1}  (\Lambda_{V_H}^{-1})_{\ls[12]}^{\us[X]} \\ \nonumber
&+ 2 \sqrt{D-1} \sqrt{D-2} \big( (\Lambda_{V_H}^{-1})_{\ls[13]}^{\us[X]} + (\Lambda_{V_H}^{-1})_{\ls[23]}^{\us[X]} \big)  \Big) \Bigg) \\ \nonumber
&+ \frac{(D-3)}{2} \tilde{\rho}_{\ls[3]}^{V_0} (\Lambda_{V_2}^{-1})^{\us[X]} - \frac{(D-3)}{2 \sqrt{D-1}} \tilde{\rho}_{\ls[4]}^{V_0} (\Lambda_{V_2}^{-1})^{\us[X]} \\
&+ \frac{(D-1)(D-2)}{2D} \tilde{\rho}_{\ls[3]}^{V_0} (\Lambda_{V_3}^{-1})^{\us[X]} - \frac{\sqrt{D-1}(D-2)}{2D} \tilde{\rho}_{\ls[4]}^{V_0} (\Lambda_{V_3}^{-1})^{\us[X]}.
\end{align}
Our final result is then composed of the relevant sum of the above results, given by \eqref{Eqn: Wick expansion cubic example}. This is listed below along with a small collection of other cubic results
\begin{flalign} \nonumber
\sum_{i=1}^D \langle M_{ii}^2 N_{ii} \rangle &=  \frac{1}{D^2} \Bigg( (\tilde{\rho}_{\ls[1]}^{V_0})^2 \tilde{\rho}_{\ls[3]}^{V_0} + \sqrt{D-1} (\tilde{\rho}_{\ls[1]}^{V_0})^2 \tilde{\rho}_{\ls[4]}^{V_0} + (D-1) (\tilde{\rho}_{\ls[2]}^{V_0})^2 \tilde{\rho}_{\ls[3]}^{V_0} & \\ \nonumber
&+ (D-1)^{\frac{3}{2}} (\tilde{\rho}_{\ls[2]}^{V_0})^2 \tilde{\rho}_{\ls[4]}^{V_0} + 2\sqrt{D-1} \tilde{\rho}_{\ls[1]}^{V_0} \tilde{\rho}_{\ls[2]}^{V_0}  \tilde{\rho}_{\ls[3]}^{V_0} + 2 (D-1) \tilde{\rho}_{\ls[1]}^{V_0} \tilde{\rho}_{\ls[2]}^{V_0}  \tilde{\rho}_{\ls[4]}^{V_0}   \Bigg) \\ \nonumber
&+ 2 \Bigg( \frac{\tilde{\rho}_{\ls[1]}^{V_0}}{D}+ \frac{\sqrt{D-1}}{D} \tilde{\rho}_{\ls[2]}^{V_0} \Bigg) \Bigg( \frac{(\Lambda_{V_0}^{-1})_{\ls[11]}^{\us[XY]}}{D} + \frac{(D-1)}{D} (\Lambda_{V_0}^{-1})_{\ls[22]}^{\us[XY]} \\ \nonumber
&+ \frac{\sqrt{D-1}}{D} \Big((\Lambda_{V_0}^{-1})_{\ls[12]}^{\us[XY]} + (\Lambda_{V_0}^{-1})_{\ls[21]}^{\us[XY]} \Big) + \frac{D-1}{D} \Big( (\Lambda_{V_H}^{-1})_{\ls[11]}^{\us[XY]} + (\Lambda_{V_H}^{-1})_{\ls[22]}^{\us[XY]} \Big) \\ \nonumber
&+ \frac{(D-1)(D-2)}{D} (\Lambda_{V_H}^{-1})_{\ls[33]}^{\us[XY]} + \frac{(D-1)}{D} \Big( (\Lambda_{V_H}^{-1})_{\ls[12]}^{\us[XY]} + (\Lambda_{V_H}^{-1})_{\ls[21]}^{\us[XY]} \Big) \\ \nonumber
&+ \frac{(D-1) \sqrt{D-2}}{D} \Big( (\Lambda_{V_H}^{-1})_{\ls[13]}^{\us[XY]} + (\Lambda_{V_H}^{-1})_{\ls[31]}^{\us[XY]} + (\Lambda_{V_H}^{-1})_{\ls[23]}^{\us[XY]} + (\Lambda_{V_H}^{-1})_{\ls[32]}^{\us[XY]} \Big) \Bigg) \\ \nonumber
&+ \Bigg( \frac{\tilde{\rho}_{\ls[3]}^{V_0}}{D} + \frac{\sqrt{D-1}}{D} \tilde{\rho}_{\ls[4]}^{V_0} \Bigg) \Bigg( \frac{(\Lambda_{V_0}^{-1})_{\ls[11]}^{\us[X]}}{D} + \frac{(D-1)}{D} (\Lambda_{V_0}^{-1})_{\ls[22]}^{\us[X]}+ \frac{2 \sqrt{D-1}}{D} (\Lambda_{V_0}^{-1})_{\ls[12]}^{\us[X]} \\ \nonumber
& + \frac{D-1}{D} \Big( (\Lambda_{V_H}^{-1})_{\ls[11]}^{\us[X]} + (\Lambda_{V_H}^{-1})_{\ls[22]}^{\us[X]} \Big) + \frac{(D-1)(D-2)}{D} (\Lambda_{V_H}^{-1})_{\ls[33]}^{\us[X]} + \frac{2 (D-1)}{D} (\Lambda_{V_H}^{-1})_{\ls[12]}^{\us[X]} \\ 
&+ \frac{2 (D-1) \sqrt{D-2}}{D} \Big( (\Lambda_{V_H}^{-1})_{\ls[13]}^{\us[X]} + (\Lambda_{V_H}^{-1})_{\ls[23]}^{\us[X]} \Big) \Bigg),
\end{flalign}

\begin{flalign} \nonumber
\sum_{i,j=1}^D \langle M_{ij}^2 N_{ij} \rangle &=  \frac{1}{D} \Bigg( (\tilde{\rho}_{\ls[1]}^{V_0})^2 \tilde{\rho}_{\ls[3]}^{V_0} + (\tilde{\rho}_{\ls[2]}^{V_0})^2 \tilde{\rho}_{\ls[3]}^{V_0} + \frac{(D-2)}{\sqrt{D-1}}(\tilde{\rho}_{\ls[2]}^{V_0})^2 \tilde{\rho}_{\ls[4]}^{V_0} + \frac{2}{D} \tilde{\rho}_{\ls[1]}^{V_0} \tilde{\rho}_{\ls[2]}^{V_0}  \tilde{\rho}_{\ls[4]}^{V_0}   \Bigg) & \\ \nonumber
&+ \frac{2}{D} \Bigg( \tilde{\rho}_{\ls[1]}^{V_0} (\Lambda_{V_0}^{-1})_{\ls[11]}^{\us[XY]} +  \tilde{\rho}_{\ls[1]}^{V_0} (\Lambda_{V_0}^{-1})_{\ls[22]}^{\us[XY]} + \frac{(D-2)}{\sqrt{D-1}}  \tilde{\rho}_{\ls[2]}^{V_0} (\Lambda_{V_0}^{-1})_{\ls[22]}^{\us[XY]} \\ \nonumber
&+ \tilde{\rho}_{\ls[2]}^{V_0} \Big( (\Lambda_{V_0}^{-1})_{\ls[12]}^{\us[XY]} + (\Lambda_{V_0}^{-1})_{\ls[21]}^{\us[XY]} \Big) \Bigg) \\ \nonumber
&+ \frac{2}{D} \Bigg( (D-1)  \tilde{\rho}_{\ls[1]}^{V_0}  \Big( (\Lambda_{V_H}^{-1})_{\ls[11]}^{\us[XY]} + (\Lambda_{V_H}^{-1})_{\ls[22]}^{\us[XY]} + (\Lambda_{V_H}^{-1})_{\ls[33]}^{\us[XY]} \Big) \\ \nonumber
&+ \tilde{\rho}_{\ls[2]}^{V_0} \Big(  \sqrt{D-1} (D-3) (\Lambda_{V_H}^{-1})_{\ls[33]}^{\us[XY]} \sqrt{D-1} \big( (\Lambda_{V_H}^{-1})_{\ls[12]}^{\us[XY]} + (\Lambda_{V_H}^{-1})_{\ls[21]}^{\us[XY]} \big) \\ \nonumber
&+ \sqrt{D-1} \sqrt{D-2} \big( (\Lambda_{V_H}^{-1})_{\ls[13]}^{\us[XY]} + (\Lambda_{V_H}^{-1})_{\ls[31]}^{\us[XY]} + (\Lambda_{V_H}^{-1})_{\ls[23]}^{\us[XY]} + (\Lambda_{V_H}^{-1})_{\ls[32]}^{\us[XY]} \big)  \Big) \Bigg) \\ \nonumber
&+ (D-3) \tilde{\rho}_{\ls[1]}^{V_0} (\Lambda_{V_2}^{-1})^{\us[XY]} - \frac{(D-3)}{\sqrt{D-1}} \tilde{\rho}_{\ls[2]}^{V_0} (\Lambda_{V_2}^{-1})^{\us[XY]} \\ \nonumber
&+ \frac{(D-1)(D-2)}{D} \tilde{\rho}_{\ls[1]}^{V_0} (\Lambda_{V_3}^{-1})^{\us[XY]} - \frac{\sqrt{D-1}(D-2)}{D} \tilde{\rho}_{\ls[2]}^{V_0} (\Lambda_{V_3}^{-1})^{\us[XY]} \\ \nonumber
&+ \frac{1}{D} \Bigg( \tilde{\rho}_{\ls[3]}^{V_0} (\Lambda_{V_0}^{-1})_{\ls[11]}^{\us[X]} +  \tilde{\rho}_{\ls[3]}^{V_0} (\Lambda_{V_0}^{-1})_{\ls[22]}^{\us[X]} + \frac{(D-2)}{\sqrt{D-1}}  \tilde{\rho}_{\ls[4]}^{V_0} (\Lambda_{V_0}^{-1})_{\ls[22]}^{\us[X]} \\ \nonumber
&+ 2 \tilde{\rho}_{\ls[4]}^{V_0}  (\Lambda_{V_0}^{-1})_{\ls[12]}^{\us[X]} \Bigg) + \frac{1}{D} \Bigg( (D-1)  \tilde{\rho}_{\ls[3]}^{V_0}  \Big( (\Lambda_{V_H}^{-1})_{\ls[11]}^{\us[X]} + (\Lambda_{V_H}^{-1})_{\ls[22]}^{\us[X]} + (\Lambda_{V_H}^{-1})_{\ls[33]}^{\us[X]} \Big) \\ \nonumber
&+ \tilde{\rho}_{\ls[4]}^{V_0} \Big(  \sqrt{D-1} (D-3) (\Lambda_{V_H}^{-1})_{\ls[33]}^{\us[X]} + 2 \sqrt{D-1}  (\Lambda_{V_H}^{-1})_{\ls[12]}^{\us[X]} \\ \nonumber
&+ 2 \sqrt{D-1} \sqrt{D-2} \big( (\Lambda_{V_H}^{-1})_{\ls[13]}^{\us[X]} + (\Lambda_{V_H}^{-1})_{\ls[23]}^{\us[XY]} \big)  \Big) \Bigg) \\ \nonumber
&+ \frac{(D-3)}{2} \tilde{\rho}_{\ls[3]}^{V_0} (\Lambda_{V_2}^{-1})^{\us[X]} - \frac{(D-3)}{2 \sqrt{D-1}} \tilde{\rho}_{\ls[4]}^{V_0} (\Lambda_{V_2}^{-1})^{\us[X]} \\
&+ \frac{(D-1)(D-2)}{2D} \tilde{\rho}_{\ls[3]}^{V_0} (\Lambda_{V_3}^{-1})^{\us[X]} - \frac{\sqrt{D-1}(D-2)}{2D} \tilde{\rho}_{\ls[4]}^{V_0} (\Lambda_{V_3}^{-1})^{\us[X]},
\end{flalign}

\begin{flalign} \nonumber
\sum_{i,j,k=1}^D \langle M_{ij} M_{jk} N_{ki} \rangle &= ( \tilde{\rho}_{\ls[1]}^{V_0} )^2 \tilde{\rho}_{\ls[3]}^{V_0} + \frac{( \tilde{\rho}_{\ls[2]}^{V_0} )^2 \tilde{\rho}_{\ls[4]}^{V_0}}{\sqrt{D-1}} + \tilde{\rho}_{\ls[3]}^{V_0} \Big( (\Lambda_{V_0}^{-1})_{\ls[11]}^{\us[X]} + (D-1) (\Lambda_{V_H}^{-1})_{\ls[12]}^{\us[X]}  \Big) & \\ \nonumber
&+  \frac{\tilde{\rho}_{\ls[4]}^{V_0}}{\sqrt{D-1}} \Big( (\Lambda_{V_0}^{-1})_{\ls[22]}^{\us[X]} + (D-1) (\Lambda_{V_H}^{-1})_{\ls[33]}^{\us[X]} + (D-1) (\Lambda_{V_H}^{-1})_{\ls[12]}^{\us[X]}  \\ \nonumber
&+ \frac{D(D-3)}{2} (\Lambda_{V_2}^{-1})^{\us[X]} - \frac{(D-1)(D-2)}{2}  (\Lambda_{V_3}^{-1})^{\us[X]} \Big)  \\ \nonumber
&+ \tilde{\rho}_{\ls[1]}^{V_0} \Big( 2 (\Lambda_{V_0}^{-1})_{\ls[11]}^{\us[XY]} + (D-1) (\Lambda_{V_H}^{-1})_{\ls[21]}^{\us[XY]} + (D-1) (\Lambda_{V_H}^{-1})_{\ls[12]}^{\us[XY]}  \Big) \\ \nonumber
&+ \frac{\tilde{\rho}_{\ls[2]}^{V_0}}{\sqrt{D-1}} \Big( 2 (\Lambda_{V_0}^{-1})_{\ls[22]}^{\us[XY]} + 2 (D-1) (\Lambda_{V_H}^{-1})_{\ls[33]}^{\us[XY]} + (D-1) (\Lambda_{V_H}^{-1})_{\ls[12]}^{\us[XY]} \\ 
&+ (D-1) (\Lambda_{V_H}^{-1})_{\ls[21]}^{\us[XY]} + D (D-3) (\Lambda_{V_2}^{-1})^{\us[XY]} - (D-1)(D-2) (\Lambda_{V_3}^{-1})^{\us[XY]} \Big), 
\end{flalign}

\begin{flalign}
\sum_{i_1, \dots, i_6}^D \langle M_{i_1 i_2} M_{i_3 i_4} N_{i_5 i_6} \rangle &= D^4 \Big( (\Lambda_{V_0}^{-1})_{\ls[11]}^{\us[X]} \tilde{\rho}_{\ls[3]}^{V_0} + 2 (\Lambda_{V_0}^{-1})_{\ls[11]}^{\us[XY]} \tilde{\rho}_{\ls[1]}^{V_0} \Big) + D^3 (\tilde{\rho}_{\ls[1]}^{V_0})^2 \tilde{\rho}_{\ls[3]}^{V_0}. 
\end{flalign}

\subsection{Quartic expectation values}

For the sake of brevity we consider one of the simplest four-point expectation values as our example: One in which all matrix indices are distinct. As there are no repeated indices in the summand, all terms in \eqref{Eqn: M one point function}, \eqref{Eqn: N one point function}, \eqref{Eqn: Two point function} and \eqref{Eqn: Two point function MM} involving a projection to $V_H$ do not contribute. This is due to the fact that the projection to $V_0$ performed by the sums does not overlap with the initial $V_H$ projection. In practice this means we can ignore all but the first terms in each of these expansions - that is only the terms containing $\tilde{\rho}_{\ls[1]}^{V_0}, \tilde{\rho}_{\ls[3]}^{V_0}, (\Lambda_{V_0}^{-1})^{\us[X]}_{\ls[11]}, (\Lambda_{V_0}^{-1})^{\us[Y]}_{\ls[11]}, (\Lambda_{V_0}^{-1})^{\us[XY]}_{\ls[11]}$ contribute. Performing the necessary Wick contractions we find
\begin{align} \nonumber
\sum_{i_1, \dots , i_8} \langle M_{i_1 i_2} M_{i_3 i_4} N_{i_5 i_6} N_{i_7 i_8} \rangle &= \sum_{i_1, \dots , i_8} \Big( \langle M_{i_1 i_2} \rangle \langle M_{i_3 i_4} \rangle \langle N_{i_5 i_6} \rangle \langle N_{i_7 i_8} \rangle \\ \nonumber
&+ \langle M_{i_1 i_2} M_{i_3 i_4} \rangle \langle N_{i_5 i_6}  N_{i_7 i_8} \rangle + 2 \langle M_{i_1 i_2} N_{i_5 i_6} \rangle \langle M_{i_3 i_4} N_{i_7 i_8} \rangle  \\ \nonumber
&+  \langle M_{i_1 i_2} M_{i_3 i_4} \rangle \langle N_{i_5 i_6} \rangle \langle N_{i_7 i_8} \rangle + \langle N_{i_5 i_6}  N_{i_7 i_8} \rangle  \langle M_{i_1 i_2} \rangle \langle M_{i_3 i_4} \rangle \\ \nonumber
&+ 4  \langle M_{i_1 i_2} N_{i_5 i_6} \rangle  \langle M_{i_3 i_4} \rangle \langle N_{i_7 i_8} \rangle \Big) \\ \nonumber
&= D^4 \Big( (\tilde{\rho}_{\ls[1]}^{V_0})^2 (\tilde{\rho}_{\ls[3]}^{V_0})^2 + (\Lambda_{V_0}^{-1})^{\us[X]}_{\ls[11]} (\tilde{\rho}_{\ls[3]}^{V_0})^2 + (\Lambda_{V_0}^{-1})^{\us[Y]}_{\ls[11]} (\tilde{\rho}_{\ls[1]}^{V_0})^2 \\ 
&+ 4 (\Lambda_{V_0}^{-1})^{\us[XY]}_{\ls[11]} \tilde{\rho}_{\ls[1]}^{V_0} \tilde{\rho}_{\ls[3]}^{V_0} + (\Lambda_{V_0}^{-1})^{\us[X]}_{\ls[11]} (\Lambda_{V_0}^{-1})^{\us[Y]}_{\ls[11]} + 2 \big( (\Lambda_{V_0}^{-1})^{\us[XY]}_{\ls[11]} \big)^2 \Big).
\end{align}

Finally, we list a small collection of quartic results
\begin{flalign} \nonumber
\sum_{i_1, \dots, i_8}^D \langle M_{i_1 i_2} M_{i_3 i_4} N_{i_5 i_6} N_{i_7 i_8} \rangle &= D^4 \Big( (\tilde{\rho}_{\ls[1]}^{V_0})^2 (\tilde{\rho}_{\ls[3]}^{V_0})^2 + (\Lambda_{V_0}^{-1})^{\us[X]}_{\ls[11]} (\tilde{\rho}_{\ls[3]}^{V_0})^2 + (\Lambda_{V_0}^{-1})^{\us[Y]}_{\ls[11]} (\tilde{\rho}_{\ls[1]}^{V_0})^2 \\ 
&+ 4 (\Lambda_{V_0}^{-1})^{\us[XY]}_{\ls[11]} \tilde{\rho}_{\ls[1]}^{V_0} \tilde{\rho}_{\ls[3]}^{V_0} + (\Lambda_{V_0}^{-1})^{\us[X]}_{\ls[11]} (\Lambda_{V_0}^{-1})^{\us[Y]}_{\ls[11]} + 2 \big( (\Lambda_{V_0}^{-1})^{\us[XY]}_{\ls[11]} \big)^2 \Big)
\end{flalign}

\begin{flalign} \nonumber
\sum_{i_1, \dots, i_7}^D \langle M_{i_1 i_2} M_{i_3 i_4} N_{i_1 i_5} N_{i_6 i_7} \rangle &= D^3 \Big( (\tilde{\rho}_{\ls[1]}^{V_0})^2 (\tilde{\rho}_{\ls[3]}^{V_0})^2 + (\tilde{\rho}_{\ls[3]}^{V_0})^2 (\Lambda_{V_0}^{-1})^{\us[X]}_{\ls[11]} + (\tilde{\rho}_{\ls[1]}^{V_0})^2 (\Lambda_{V_0}^{-1})^{\us[Y]}_{\ls[11]} & \\ \nonumber 
&+ \tilde{\rho}_{\ls[1]}^{V_0} \tilde{\rho}_{\ls[3]}^{V_0} \big( 4 (\Lambda_{V_0}^{-1})^{\us[XY]}_{\ls[11]} + (D-1) (\Lambda_{V_H}^{-1})^{\us[XY]}_{\ls[22]} \big) + 2 \big( (\Lambda_{V_0}^{-1})^{\us[XY]}_{\ls[11]} \big)^2 \\
&+  (\Lambda_{V_0}^{-1})^{\us[X]}_{\ls[11]} (\Lambda_{V_0}^{-1})^{\us[Y]}_{\ls[11]} + (D-1) (\Lambda_{V_0}^{-1})^{\us[XY]}_{\ls[11]} (\Lambda_{V_H}^{-1})^{\us[XY]}_{\ls[22]} \Big)
\end{flalign}

\begin{flalign} \nonumber
\sum_{i_1, \dots, i_6}^D \langle M_{i_1 i_2} M_{i_2 i_3} N_{i_4 i_5} N_{i_5 i_6} \rangle &= D^2 \Big( (\tilde{\rho}_{\ls[1]}^{V_0})^2 (\tilde{\rho}_{\ls[3]}^{V_0})^2 + \big( (\Lambda_{V_0}^{-1})^{\us[X]}_{\ls[11]} + (D-1) (\Lambda_{V_H}^{-1})^{\us[X]}_{\ls[12]}  \big) (\tilde{\rho}_{\ls[3]}^{V_0})^2 & \\ \nonumber
&+ \big( (\Lambda_{V_0}^{-1})^{\us[Y]}_{\ls[11]} + (D-1) (\Lambda_{V_H}^{-1})^{\us[Y]}_{\ls[12]}  \big) (\tilde{\rho}_{\ls[1]}^{V_0})^2 + 4 (\Lambda_{V_0}^{-1})^{\us[XY]}_{\ls[11]} \tilde{\rho}_{\ls[1]}^{V_0} \tilde{\rho}_{\ls[3]}^{V_0} \\ \nonumber
&+ 2 \big( (\Lambda_{V_0}^{-1})^{\us[XY]}_{\ls[11]} \big)^2 + (D-1) \big( (\Lambda_{V_H}^{-1})^{\us[XY]}_{\ls[12]} (\Lambda_{V_H}^{-1})^{\us[XY]}_{\ls[21]} \\ \nonumber 
&+ (\Lambda_{V_H}^{-1})^{\us[XY]}_{\ls[11]} (\Lambda_{V_H}^{-1})^{\us[XY]}_{\ls[22]}  \big) + \big( (\Lambda_{V_0}^{-1})^{\us[X]}_{\ls[11]} + (D-1) (\Lambda_{V_H}^{-1})^{\us[X]}_{\ls[12]} \big) (\Lambda_{V_0}^{-1})^{\us[Y]}_{\ls[11]} \\
&+ (D-1) \big( ( \Lambda_{V_0}^{-1} )^{\us[X]}_{\ls[11]} + (D-1)(\Lambda_{V_H}^{-1})^{\us[X]}_{\ls[12]} \big) (\Lambda_{V_H}^{-1})^{\us[Y]}_{\ls[12]} \Big)
\end{flalign}

\begin{flalign} \nonumber
\sum_{i_1, \dots, i_5}^D \langle M_{i_1 i_2} M_{i_2 i_3} N_{i_3 i_4} N_{i_4 i_5} \rangle &= D \Big( (\tilde{\rho}_{\ls[1]}^{V_0})^2 (\tilde{\rho}_{\ls[3]}^{V_0})^2 + (\tilde{\rho}_{\ls[3]}^{V_0})^2 \big( (\Lambda_{V_0}^{-1})^{\us[X]}_{\ls[11]} +(D-1)(\Lambda_{V_H}^{-1})^{\us[X]}_{\ls[12]} \big) & \\ \nonumber
&+ (\tilde{\rho}_{\ls[1]}^{V_0})^2 \big( (\Lambda_{V_0}^{-1})^{\us[Y]}_{\ls[11]} +(D-1)(\Lambda_{V_H}^{-1})^{\us[Y]}_{\ls[12]} \big) + \tilde{\rho}_{\ls[1]}^{V_0} \tilde{\rho}_{\ls[3]}^{V_0} \big( 4 (\Lambda_{V_0}^{-1})^{\us[XY]}_{\ls[11]} \\ \nonumber 
&+ (D-1)(\Lambda_{V_H}^{-1})^{\us[XY]}_{\ls[12]}  \big) + \tilde{\rho}_{\ls[2]}^{V_0} \tilde{\rho}_{\ls[4]}^{V_0} (\Lambda_{V_H}^{-1})^{\us[XY]}_{\ls[12]} + 2 \big((\Lambda_{V_0}^{-1})^{\us[XY]}_{\ls[11]}\big)^2 \\ \nonumber
&+ (D-1) \big( (\Lambda_{V_H}^{-1})^{\us[XY]}_{\ls[12]} (\Lambda_{V_H}^{-1})^{\us[XY]}_{\ls[21]} + (\Lambda_{V_H}^{-1})^{\us[XY]}_{\ls[11]} (\Lambda_{V_H}^{-1})^{\us[XY]}_{\ls[22]} + (\Lambda_{V_H}^{-1})^{\us[XY]}_{\ls[13]} (\Lambda_{V_H}^{-1})^{\us[XY]}_{\ls[32]} \\ \nonumber
&+ (\Lambda_{V_H}^{-1})^{\us[XY]}_{\ls[12]} (\Lambda_{V_H}^{-1})^{\us[XY]}_{\ls[33]} \big) + \frac{D(D-3)}{2} (\Lambda_{V_H}^{-1})^{\us[XY]}_{\ls[12]} (\Lambda_{V_2}^{-1})^{\us[XY]} \\ \nonumber
&- \frac{(D-1)(D-2)}{2} (\Lambda_{V_H}^{-1})^{\us[XY]}_{\ls[12]} (\Lambda_{V_3}^{-1})^{\us[XY]} + (\Lambda_{V_H}^{-1})^{\us[XY]}_{\ls[12]} \big( (D-1)(\Lambda_{V_0}^{-1})^{\us[XY]}_{\ls[11]} \\ \nonumber
&+ (\Lambda_{V_0}^{-1})^{\us[XY]}_{\ls[22]} \big) + (\Lambda_{V_0}^{-1})^{\us[Y]}_{\ls[11]} \big( (\Lambda_{V_0}^{-1})^{\us[X]}_{\ls[11]} + (D-1) (\Lambda_{V_H}^{-1})^{\us[X]}_{\ls[12]} \big) \\ \nonumber
&+ (D-1) (\Lambda_{V_H}^{-1})^{\us[Y]}_{\ls[12]} \big( (\Lambda_{V_0}^{-1})^{\us[X]}_{\ls[11]} + (D-1) (\Lambda_{V_H}^{-1})^{\us[X]}_{\ls[12]} \big) \\ 
&+ \sqrt{D-1} (\Lambda_{V_H}^{-1})^{\us[XY]}_{\ls[12]} \big( \tilde{\rho}_{\ls[2]}^{V_0} \tilde{\rho}_{\ls[3]}^{V_0} + \tilde{\rho}_{\ls[1]}^{V_0} \tilde{\rho}_{\ls[4]}^{V_0} + (\Lambda_{V_0}^{-1})^{\us[XY]}_{\ls[12]} + (\Lambda_{V_H}^{-1})^{\us[XY]}_{\ls[21]} \big) \Big)
\end{flalign}

Performing the calculation of expectation values of cubic order and higher by hand can be a technical process prone to error. To aid any future comparison of these results with experiment we provide a Sage program capable of calculating expectation values of observables. The details of this algorithm can be found in Appendix \ref{apx: The Algorithm} and instructions for use are contained within the code itself.

\section{Summary and Outlook } 

This paper furthers the Linguistic Matrix Theory programme begun in \cite{LMT} and continued in \cite{PIGMM} and \cite{GTMDS}. The 13-parameter Gaussian 1-matrix model solved in \cite{PIGMM} has been applied in computational linguistics to study the statistics of adjectives and verbs. Here we employed similar techniques, rooted in the representation theory of $S_D$, to solve the most general 13+13+15 parameter Gaussian 2-matrix model. This model will be useful to study the invariant moments of adjectives along with verbs in the context of type-driven compositional distributional semantics, using for example the datasets built in \cite{LMT}. To aid comparison with experiment we listed explicit results for all quadratic moments and a selection of cubic and quartic moments. In addition to this we have written a Sage script (available at \href{https://github.com/adrianpadellaro/PIG2MM}{\it Link to GitHub repository for this paper}) that can be used to generate results for any expectation value, without recourse to explicit manual calculations. The central results of the paper can be divided into three pieces: observables, action and expectation values,  computer algorithms.

We further developed the correspondence observed in \cite{LMT}, between $S_D$-invariant 1-matrix polynomials (observables) and directed graphs at large $D$. We began with an extension of the representation theoretic counting of observables to the 2-matrix case and found that the counting exhibits the same stability property as the 1-matrix case for large $D$. The stability is an indication that the 2-matrix observables are in one-to-one correspondence with directed two-colored graphs. We counted directed colored graphs and compared to the representation theoretic counting. The comparison revealed that the correspondence holds for any $D$, if we restrict to graphs with $D$ vertices. Having found a bijective correspondence between observables and directed colored graphs, we developed a method for counting and constructing graphs. This method exploits a correspondence between equivalence classes of permutations and graphs using a diagrammatic realisation of permutations, which links to D-brane physics (e.g. strings between giant gravitons as in \cite{deMelloKoch:2012ck}) and also to the use of diagrams in knot theory (e.g. as in \cite{RT}). By using permutation groups related to the local structure of the graphs at the vertices, graphs and therefore observables, were put in one-to-one correspondence with double cosets of permutation groups. Double cosets can be counted and constructed efficiently using group theoretical techniques. This framework admits  extension to permutation invariant observables in $k$-matrix models. As a whole, Section \ref{Section: Observables and multi-graphs} shows that the discussion of observables lies at a rich intersection of representation theory, combinatorics, graph theory and group theory. The  explicit formulae and theoretical perspective developed here for the enumeration of graphs can potentially  be useful in other applications of graphs within theoretical physics: for example an interesting recent application of graphs is in jet algorithms \cite{Thaler}.    

The partition function is most efficiently solved using a representation theory basis, distinct from the graph basis developed for observables.
Linear and quadratic moments are simple (block diagonal) in the representation basis.
However, in the graph basis there is non-trivial mixing and the quadratic moments can contain up to 15 independent terms.
We gave a diagrammatic description of the representation theoretic structure of the linear and quadratic moments.
In combination with Wick's theorem, the diagrams can be used as Feynman rules to diagrammatically compute higher-order expectation values.
The contributions to expectation values of observables in the graph basis can be expressed as products of matrix elements $F(i,j)$ (associated with a projector) with all indices summed.
In general the contributions, which are Laurent polynomials in $D$, are non-trivial to compute.
At first sight, the computational complexity of evaluating the index sums scales with $D$. 
We gave a rule for translating a general product of $F$s (with all indices summed) to an undirected graph ("closed $F$-graph").
Using the graph interpretation, we mapped the problem of evaluating the Laurent polynomial into a graph problem.
The computational complexity of the graph algorithm which solves this problem is manifestly independent of $D$, it only depends on the number of vertices in the closed $F$-graph. 

The Sage script for computing expectation values of observables was largely inspired by these two theoretical results. In particular, it relies heavily on the graph interpretation of observables and the correspondence between their expectation values and $F$-graphs. In addition, the GitHub repository contains a short Sage script for generating all the distinct observables at fixed degree. The output is combinatoric data to be used as input to the expectation value algorithm.

A simple generalisation of the results presented in this paper could be used to solve permutation invariant Gaussian $k$-matrix models, containing $k$ distinct matrices $M_1, M_2, \dots, M_k$. These models are defined by an $S_D$-invariant action containing two linear terms for each of the $k$ matrices, 11 terms for each of the $k$ quadratic matrix combinations of the form $M_i^2$, along with 15 terms from each of the $\frac{1}{2} k(k-1)$ combinations of the form $M_i M_j$ with $i \neq j$. The change of variables required to factorise the partition function is equivalent to that performed in this paper and leads to the definition of analogous symmetric matrices $\Lambda_{V_A}$ containing the quadratic couplings of representation theoretic parameters associated with all $k$ matrices. Again the model is solved by the inversion of these coupling matrices. Expectation values of the original $M_k$ matrices can be evaluated with Wick's theorem and expressions analogous to \eqref{Eqn: Two point function} and \eqref{Eqn: Two point function MM}.

A natural future direction following this work would be to extend the results of this paper  to permutation invariant models for tensor variables $T_{ijk}$ transforming as $ V_D \otimes V_D \otimes V_D$. This is further motivated by the application of these models in type-driven compositional distributional semantics in which three-index tensors are used to represent transitive verbs \cite{CSC}. Explicit machine-learning algorithms for constructing ensembles of  these three-index tensors, as well as 2-matrix ensembles, from natural language data 
 have been  designed \cite{GDZS, PRC, WSC}.

\vskip.5cm 

\begin{center} 
{ \bf Acknowledgements} 
\end{center} 
SR is supported by the STFC consolidated grant ST/P000754/1 `` String Theory, Gauge Theory \& Duality” and  a Visiting Professorship at the University of the Witwatersrand, funded by a Simons Foundation grant (509116)  awarded to the Mandelstam Institute for Theoretical Physics. We are grateful for conversations on the subject of this paper to Joseph Bengeloun, Robert de Mello Koch, Mehrnoosh Sadrzadeh.

\vskip.5cm 

\appendix

\appendix
\section{Multi-dimensional Gaussian integration} \label{Section: Gaussian integration of matrices}
We are interested in solving the following type of Gaussian integral
\begin{equation}
\int_{-\infty}^{+\infty} \cdots \int_{-\infty}^{+\infty} e^{- \frac{1}{2} \textbf{x}^T \Lambda\textbf{x} + \textbf{J}^T \textbf{x}} dx_1 dx_2 \dots dx_n 
\end{equation}
where $\textbf{x}$ and $\textbf{J}$ are n-dimensional vectors, and $\Lambda$ is a symmetric, non-singular, $n \times n$ matrix. We have $n$ gaussian integrals, all coupled to each other through the action of $\Lambda$. In order to perform this integration we must decouple each $x_i$ integral so that they can be dealt with independently using the basic one dimensional result
\begin{equation}
\int_{-\infty}^{+\infty} e^{-\frac{1}{2}a x^2}dx = \sqrt{\frac{2 \pi}{a}}.
\end{equation}
Performing this procedure leads to the standard result \cite{Straub, Zee}
\begin{equation} \label{Eqn: multi-dim gaussian result}
\int_{-\infty}^{+\infty} \cdots \int_{-\infty}^{+\infty} e^{- \frac{1}{2} \textbf{x}^T \Lambda\textbf{x} + \textbf{J}^T \textbf{x}} dx_1 dx_2 \dots dx_n = \frac{(2\pi)^{n/2}}{(\text{det} \Lambda)^{1/2}}e^{\frac{1}{2} \textbf{J}^T \Lambda^{-1} \textbf{J}}.
\end{equation}

\section{ Background  $S_D$ representation theory for $V_D$ and its tensor products. } \label{Appendix: Sup formulae}

We consider the natural representation of the symmetric group, $V_D$, as a span of $D$ basis vectors $\{e_1, e_2, \dots, e_D \}$ and a set of linear operators $\rho_{V_D} (\sigma)$, with $\sigma \in S_D$, acting on this basis as
\begin{equation}
\rho_{V_D}(\sigma) e_i = e_{\sigma^{-1}(i)}.
\end{equation}
We can form linear combinations of these elements as follows
\begin{align} \nonumber
E_0 &= \frac{1}{\sqrt{D}} (e_1 + e_2 + \dots + e_D), \\
E_a &= \frac{1}{\sqrt{a(a+1)}} (e_1 + e_2 + \dots + e_{a} - a e_{a+1}), \quad 1 \leq a \leq D-1.
\end{align}
$E_0$ is an invariant under the action of $S_D$, and the $E_a$ form an invariant subspace of $V_D$, these invariant subspaces correspond to the two terms in the decomposition
\begin{equation}
V_D = V_0 \oplus V_H.
\end{equation}
What is more, it is easily checked that the $E_a$ form an orthonormal basis of $V_H$ by the orthonormality of the $e_i$. We define the overlap of these bases by $C_{a,i}$, 
\begin{equation}
C_{a,i} = \langle E_a | e_i \rangle = \frac{1}{\sqrt{a(a+1)}} \Big( -a \delta_{i,a+1} + \sum_{j=1}^a \delta_{ij} \Big).
\end{equation}
The  $V_0$ overlap with the original basis is given by
 \begin{equation}
C_{0,i} = \langle E_0 | e_i \rangle = \frac{1}{\sqrt{D}}.
\end{equation}
From
\begin{equation}
\sum_{A=0}^{D-1} C_{A,i} C_{A,j} = C_{0,i} C_{0,j} + \sum_{a=1}^{D-1} C_{a,i} C_{a,j} = \delta_{ij}
\end{equation}
we find
\begin{equation} \label{Eqn: f(i,j)}
\sum_{a=1}^{D-1} C_{a,i} C_{a,j} = \Big( \delta_{ij} - \frac{1}{D} \Big) = F(i,j).
\end{equation}
This $F(i,j)$ plays a central role in this paper. It is the projector in $V_D$ for $V_H$. 

Further to this we frequently use the projector in $V_H \otimes V_H$ for $V_H$ and the related Clebsch-Gordan coefficients, $C^{V_H V_H \rightarrow V_H}_{a,b; \quad c}$. It is a convenient fact that these can again be written in terms of of the projectors in $V_D$ for $V_H$
\begin{equation}
C^{V_H V_H \rightarrow V_H}_{a,b; \quad c} = \sqrt{\frac{D}{(D-2)}} \sum_i C_{a,i} C_{b,i} C_{c,i}.
\end{equation}
Details of this procedure are contained within the appendices of \cite{PIGMM}.

The above can be used to write down an expansion of the $M_{ij}$ variables in terms of the $S_D$ diagonal $X^{V_A}$, this is given by
\begin{align} \label{Eqn: M diagonal variable expansion} \nonumber
M_{ij} &= \Bigg( \frac{1}{D} X^{V_0}_{1} + \frac{1}{\sqrt{D-1}} \sum_{a=1}^{D-1} C_{a,i} C_{a,j} X^{V_0}_{2} \Bigg) \\ \nonumber
&+ \Bigg( \frac{1}{\sqrt{D}} \sum_{a=1}^{D-1} C_{a,j} X^{V_H}_{1,a} + \frac{1}{\sqrt{D}} \sum_{a=1}^{D-1} C_{a,i} X^{V_H }_{2,a} + \sum_{a,b,c = 1}^{D-1} C_{a,i} C_{b,j} C_{a,b; \quad c}^{V_H V_H \rightarrow V_H} X^{V_H}_{3,c} \Bigg) \\
&+ \sum_{a,b = 1}^{D-1} \sum_{c=1}^{\text{Dim}V_2} C_{a,i} C_{b,j} C_{a,b; \quad c}^{V_H V_H \rightarrow V_2} X_c^{V_2} + \sum_{a,b = 1}^{D-1} \sum_{c=1}^{\text{Dim}V_3} C_{a,i} C_{b,j} C_{a,b; \quad c}^{V_H V_H \rightarrow V_3} X_c^{V_3}.
\end{align}
Again, the details of this procedure are given in \cite{PIGMM}. The analogous expression for $N_{ij}$ is given by making the substitution $X \rightarrow Y$.

\section{Algorithm for computation of expectation values of observables}
\label{apx: The Algorithm}
\begin{figure}[h]
	\centering
	\scalebox{0.5}[0.5]{
		\begin{tikzpicture}
		\draw (0,0) node[draw, rounded rectangle, fill=white, anchor=south] {Start} -- 
		(0,-4)
		node[draw, rectangle, align= center, fill=white, text width=8cm, minimum height=5cm] {
			Step 1: Inputs
			\begin{itemize}
			\item Set of pair of integers $\mathit{Mtuples}$.
			\item Set of pair of integers $\mathit{Ntuples}$.
			\item Set partition $\mathit{invariants}$.
			\end{itemize}} -- 
		(10,-4) 
		node[draw, rectangle, align= center, fill=white, text width=8cm, minimum height=5cm] {
			Step 2: Wick's theorem
			\begin{itemize}
			\item $\mathit{Tuples} = \mathit{Mtuples+Ntuples}$.
			\item $\text{Construct set partitions of \textit{Tuples}}$ with blocks of size one or two.
			\end{itemize}} -- 
		(10,-10)
		node[draw, rectangle, align= center, fill=white,text width=8cm, minimum height=5cm] {
			Step 3: Wick's theorem continued
			\begin{itemize}
			\item Using the Feynman rules for degree one and two expectation values, turn each block into a sum of open $F$-graphs.
			\item For each set partition, multiply the blocks by combining $F$-graphs.
			\item Sum up the products of blocks for different set partitions.
			\end{itemize}} -- 
		(0,-10) 
		node[draw, rectangle, align= center, fill=white, text width=8cm, minimum height=5cm] {
			Step 4: Convert into F-graphs
			\begin{itemize}
			\item Using the input $\mathit{invariants}$ and concatenation of lists, turn the open $F$-graphs into closed $F$-graphs.
			\end{itemize}}  -- 
		(0,-16) 
		node[draw, rectangle, align= center, fill=white, text width=8cm, minimum height=5cm] {
			Step 5: Evaluate graphs
			\begin{itemize}
			\item Use the graph interpretation of equation \eqref{eq:kappagraphpolynomial} to evaluate the Laurent polynomial for the sum of graphs in the previous step.
			\item Print the polynomial.
			\end{itemize}} --
		(0,-20)
		node[draw, rounded rectangle, fill=white, anchor=south] {End}
		;
		\end{tikzpicture}}
	\caption{Structure of algorithm.}
	\label{fig: Algorithm Flow Chart}
\end{figure}
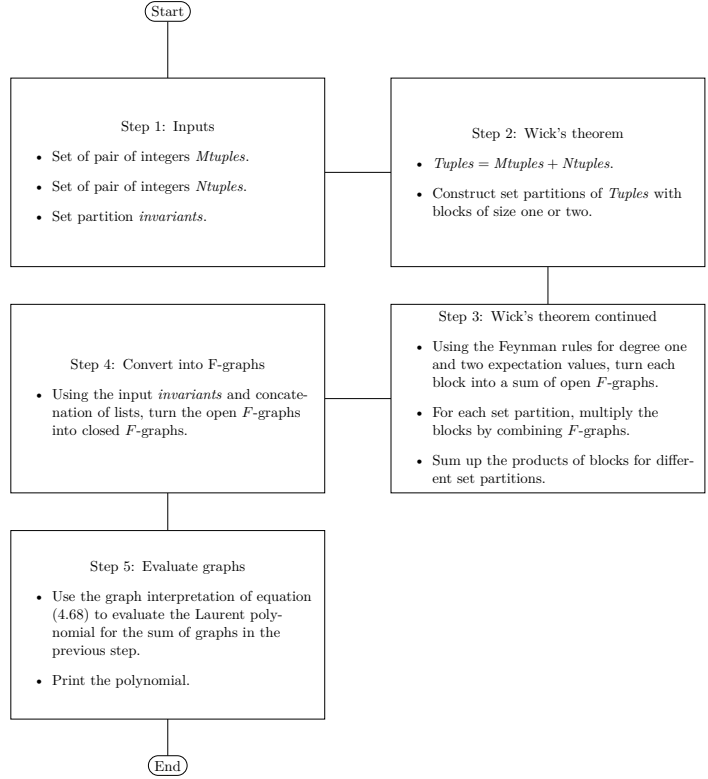

In this appendix we will describe an algorithm for computing expectation values of observables with any number of $M$ and $N$. The algorithm is currently implemented in Sage as a Jupyter Notebook (\href{https://github.com/adrianpadellaro/PIG2MM}{\it Link to GitHub repository for this paper}).
The rough outline of the algorithm is presented as a flowchart in Figure \ref{fig: Algorithm Flow Chart}. The details of each step will be presented in the following subsections.
The steps in the flowchart can be summarized as
\begin{enumerate}
	\item The three inputs are, a set of ordered pairs of integers from the set $\{1,\dots, 2m\}$ called $\textit{Mtuples}$, a set of ordered pairs of integers from the set $\{2m+1,\dots,2m+2n\}$ called $\textit{Ntuples}$ and a set partition (set of subsets) of $\{1,\dots,2m+2n\}$ called $\textit{invariants}$. Together they specify a particular observable. The connection between the input data and graphs is described in Section \ref{apx: alg_input}.
	\item The second step is about performing the combinatoric part of Wick's theorem. The pairs in $\textit{Mtuples}, \textit{Ntuples}$ are partitioned into sets of size one or two. All possible ways to partition the pairs is stored in a list. This sets us up for the next step in Wick's theorem, which is about evaluating and multiplying together the linear and quadratic (connected) expectation values.
	\item Every part of size one(two) in a set partition appearing in Wick's theorem corresponds to a linear(quadratic) expectation value. Each linear(quadratic) expectation value has a Feynman rule expressible in terms of $V_H$ projectors $F(i,j)$. Sums of products of $F$'s are associated with open or closed ``F-graphs''  as explained in Section \ref{Subsec: Feynman decomp}. The detailed role of these graphs in the algorithm is described in 
Section \ref{apx: multiplication_graphs} and Section \ref{apx: specifying observable}. 
	\item By the end of step 3 we have not performed the sums over matrix indices corresponding to a particular observable. Step 4 uses the input $\textit{invariants}$ to turn the open $F$-graphs into closed $F$-graphs. Closed $F$-graphs can be evaluated as Laurent polynomials in $D$ using the methods described in Section \ref{subsection: F-graph Evaluation}. Open $F$-graphs are turned into closed $F$-graphs by adding the set of subsets $\textit{invariants}$ to every open $F$-graph (which are implemented using lists of lists). The method $\textit{keepElements}$ selects a subset of elements of $\textit{invariants}$. This step is motivated and described in Section \ref{apx: specifying observable}.
%
	\item The fifth and last step is to evaluate the Laurent polynomials of all the $F$-graphs coming from step 4. The list of graphs produced in step 4 is fed into the function $GP$, which calculates the Laurent polynomial for every graph and sums them up, using the method described in Section \ref{subsection: F-graph Evaluation}. The resulting Laurent polynomial is the output of the algorithm and corresponds to the expectation value of the observable.
\end{enumerate}

\subsection{Input data}
\label{apx: alg_input}
The expectation value
\begin{equation}
\langle M_{i_1 i_2} \dots M_{i_{2m{-}1} i_{2m}} N_{i_{2m{+}1} i_{2m{+}2}} \dots N_{i_{2m{+}2n{-}1} i_{2m{+}2n}} \rangle,
\end{equation}
is specified by two lists of ordered pairs of integers, for $M$ and $N$ matrices respectively. Specifically, since every matrix $M_{i_p i_{p+1}}$ has a pair of indices we label these using pairs of integers $(p,q)$. For example, the expectation value
\begin{equation}
\langle M_{i_1 i_2} M_{i_3 i_4} N_{i_5 i_6} N_{i_7 i_8} \rangle,
\end{equation}
can be associated with the two lists
\begin{align}
&\textit{Mtuples}=[(1,2),(3,4)],\\
&\textit{Ntuples}=[(5,6),(7,8)].
\end{align}
Secondly, each invariant is associated with a set partition of the set of indices. That is, a set partition of $\{1,\dots,2m{+}2n\}$. For example, the observable
\begin{equation}
\sum_{i,j,k} \langle M_{i i} M_{i j} N_{j k} N_{k k} \rangle,
\end{equation}
can be specified by the following three inputs
\begin{align}
&\textit{Mtuples}=[(1,2),(3,4)],\\
&\textit{Ntuples}=[(5,6),(7,8)], \\
&\textit{invariant} = [\underbrace{[1,2,3]}_{i},\underbrace{[4,5]}_{j},\underbrace{[6,7,8]}_{k}].
\end{align}

\begin{figure}
	\centering
	\subcaptionbox{
		\label{fig: Half Edge Cut of an Observables a}}[0.4\textwidth]
	{
		\begin{tikzpicture}[scale=2]
		\definecolor{GREEN}{rgb}{0.0,0.70,0.24}
		\definecolor{BLUE}{rgb}{0.0,0.24,0.70}
		\begin{scope}[decoration={markings, mark=at position 0.75 with {\arrow[scale=0.6]{latex}}}]
		\draw[draw=BLUE, postaction={decorate}] (0,0) node[circle, fill, inner sep=1pt] {} to[out=135, in=180] ++(0,0.5) to[out=0,in=45] (0,0);
		\draw[draw=BLUE, postaction={decorate}] (0,0) node[circle, fill, inner sep=1pt] {} to[out=45, in=135] ++(1,0) node[circle, fill, inner sep=1pt] {};
		\draw[draw=GREEN, postaction={decorate}] (1,0) node[circle, fill, inner sep=1pt] {} to[out=45, in=135] ++(1,0) node[circle, fill, inner sep=1pt] {};
		\draw[draw=GREEN, postaction={decorate}] (2,0) node[circle, fill, inner sep=1pt] {} to[out=135, in=180] ++(0,0.5) to[out=0,in=45] (2,0);
		\end{scope}
		\end{tikzpicture}}
	\subcaptionbox{\label{fig: Half Edge Cut of an Observables b}}[0.4\textwidth]
	{
		\begin{tikzpicture}[scale=5]
		\definecolor{GREEN}{rgb}{0.0,0.70,0.24}
		\definecolor{BLUE}{rgb}{0.0,0.24,0.70}
		\begin{scope}[decoration={markings, mark=at position 0.55 with {\arrow[scale=0.6]{latex}}}]
		\draw[draw=BLUE, postaction={decorate}] (0,0) node[circle, fill, inner sep=1pt] {} to[out=135, in=180] node[left, pos=.7] {\tiny $1$} (0,0.25) node[cross=2pt] {};
		\draw[draw=BLUE, postaction={decorate}] (0,0.25) to[out=0,in=45] node[right, pos=.3] {\tiny $2$} (0,0);
		\draw[draw=BLUE, postaction={decorate}] (0,0) node[circle, fill, inner sep=1pt] {} to[out=45, in=180] node[above, pos=.7] {\tiny $3$} (0.25,0.1) node[cross=2pt] {};
		\draw[draw=BLUE, postaction={decorate}] (0.25,0.1) to[out=0,in=135] node[above, pos=.3] {\tiny $4$} (.5,0);
		\draw[draw=GREEN, postaction={decorate}] (.5,0) node[circle, fill, inner sep=1pt] {} to[out=45, in=180] node[above, pos=.7] {\tiny $5$} (.75,0.1) node[cross=2pt] {};
		\draw[draw=GREEN, postaction={decorate}] (.75, 0.1) to[out=0,in=135] node[above, pos=.3] {\tiny $6$} (1,0);
		\draw[draw=GREEN, postaction={decorate}] (1,0) node[circle, fill, inner sep=1pt] {} to[out=135, in=180] node[left, pos=.7] {\tiny $7$} (1,0.25) node[cross=2pt] {};
		\draw[draw=GREEN, postaction={decorate}] (1,0.25) to[out=0,in=45] node[right, pos=.3] {\tiny $8$} (1,0)  node[circle, fill, inner sep=1pt] {};		
		\end{scope}
		\end{tikzpicture}}
	\caption{(a) Example of quartic observable graph (b) Description of observable graph in terms of $\Sigma_{\cross}$ and $\Sigma_{v}$.}
	\label{fig: Half Edge Cut of an Observables}
\end{figure}
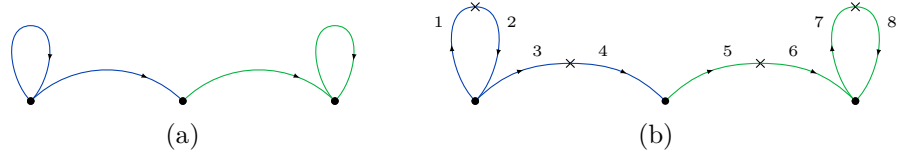

The set of input data is directly related to the double cosets in Section \ref{Double cosets and multi-graphs} in the same way the $(\Sigma_0, \Sigma_1)$ data for undirected graphs in Section \ref{subsection: maps between graphs} were related to double cosets. We start by labeling outgoing edges using odd numbers and incoming edges using even numbers. This defines the symmetric groups appearing in the double coset as the following permutation groups
\begin{equation}
\begin{aligned}
S_m^+ \times S_m^- \times S_n^+ \times S_n^- &\cong S_{\{1,\dots,2m-1\}} \times S_{\{2,\dots,2m\}} \times S_{\{2m+1,\dots, 2m+2n-1\}} \times S_{\{2m+2,\dots,2m+2n\}} \\
&\equiv S_m^{\text{odd}} \times S_m^{\text{even}} \times S_n^{\text{odd}}\times S_n^{\text{even}}.
\end{aligned}
\end{equation}
To keep track of identifications, we introduce a set of ordered pairs of integers
\begin{equation}
\Sigma_{\cross} = (1,2), \dots, (2m-1,2m); (2m+1,2m+2),\dots(2m+2n-1,2m+2n),
\end{equation}
where $(i,j)$ means that the outgoing edge labeled by $i$ is connected to the incoming edge $j$.
The vertices are then described by a set partition $\Sigma_v$, which partitions $\{1,\dots,2m+2n\}$ into sets. For example, the data
\begin{align} \nonumber
\Sigma_{\cross} &= (1,2), (3,4);\, (5,6), (7,8),\\
\Sigma_{v}&=\langle 123 \rangle, \langle 45 \rangle, \langle 678 \rangle,
\end{align}
describes the graph in Figure \ref{fig: Half Edge Cut of an Observables a}. The construction corresponds to introducing auxiliary vertices at the midpoint of every edge as in Figure \ref{fig: Half Edge Cut of an Observables b}. An edge between a vertex and auxiliary vertex is called a half-edge.

The group $ S_m^{\text{odd}} \times S_m^{\text{even}} \times S_n^{\text{odd}}\times S_n^{\text{even}}$ acts on $\Sigma_v$ to generate all graphs.\footnote{It generates all graphs of the same type, corresponding to the vector partitions discussed in \ref{Double cosets and multi-graphs}.} \footnote{An analogous description exists where we let the group act on $\Sigma_{\cross}$ instead. In this setting $G(\vec{m}^+,\vec{m}^-;\vec{n}^+,\vec{n}^-)$ and $\diag(S_m) \times \diag(S_n)$ exchange roles. That is, we consider the set of all $\Sigma_{\cross}$ arranged into orbits of $G(\vec{m}^+,\vec{m}^-;\vec{n}^+,\vec{n}^-)$ while $\Sigma_v$ remains fixed.} Permutations in the same double coset in
\begin{equation}
G(\vec{m}^+,\vec{m}^-;\vec{n}^+,\vec{n}^-) \left\backslash (S_m^+ \times S_m^- \times S_n^+ \times S_n^-) \right/ ( \diag(S_m) \times \diag(S_n) ) 
\end{equation}
generate equivalent graphs. Note that $G(\vec{m}^+,\vec{m}^-;\vec{n}^+,\vec{n}^-)$ is precisely the subgroup that stabilizes $\Sigma_v$. By the orbit-stabilizer theorem, elements in the orbit of $\Sigma_v$ are counted by cosets in
\begin{equation}
G(\vec{m}^+,\vec{m}^-;\vec{n}^+,\vec{n}^-) \left\backslash (  S_m^+ \times S_m^- \times S_n^+ \times S_n^-) \right.
\end{equation}
By further arranging the set of elements in the orbit of $\Sigma_v$ into orbits of $\diag(S_m) \times \diag(S_n)$ we complete the double coset. At the level of groups, we are arranging a set of representatives of the coset into orbits under $\diag(S_m) \times \diag(S_n)$. This gives a simple algorithm for generating a full set of in-equivalent input data.
\begin{itemize}
	\item Generate the set of all set partitions of $\{1,\dots,2m+2n\}$.
	\item Organize the set of set partitions into orbits of $\diag(S_m) \times \diag(S_n)$.
	\item Any collection of representatives (one for each orbit) corresponds to a collection of distinct input observables together with the list $$\Sigma_{\cross} = (1,2), \dots, (2m-1,2m); (2m+1,2m+2),\dots(2m+2n-1,2m+2n).$$
\end{itemize}
These orbits have been implemented in Sage (using GAP).

\subsection{Wick's theorem: set partitions}
\label{apx: wicksthrm}
Wick's theorem is described in Section \ref{subsection: wicks theorem}. Here we describe the implementation used in the algorithm. Given the input data described above, the steps are
\begin{enumerate}
	\item Combine the two lists of ordered pairs into one list.
	\begin{equation}
	\textit{Tuples} = \textit{Mtuples} + \textit{Ntuples}.
	\end{equation}
	\item Each term in Wick's theorem for the expectation value of a $m+n$ degree observable is a partition of the set of matrices into parts of size $k=1,2$. Equivalently, each term is a set partition of $\{1, \dots, m{+}n\}$ into parts of size $k=1,2$. Each integer is associated with a particular matrix in the expectation value through indexing of $\textit{Tuples}$. For the full expectation value we need all set partitions of $\{1, \dots, m{+}n\}$. To this end, construct the list
	\begin{equation}
	\textit{wicks} = [\text{Set partitions of $[1,\dots,m{+}n]$ with parts smaller than $3$}].
	\end{equation}
\end{enumerate}
This is used in the next step, which involves multiplication of one-point and two-point functions.

\subsection{Wick's theorem: combining  open $F$-graphs }
\label{apx: multiplication_graphs}
In the algorithm an $F$-graph is implemented using an ordered pair $(c, G)$, where $c$ is a coefficient and $G$ a graph.
The graph $G$ is a list of vertices $[v_1, \dots, v_k]$, where the vertices $v_i$ are themselves lists of integers, labeling the end points.
For example,
\begin{equation}
G = [[1,2], [3,4,5]] = \vcenter{\hbox{\begin{tikzpicture}
		\begin{scope}[decoration={markings}]
		\draw[postaction={decorate}] (0,0) node[circle, fill, inner sep=1pt] {} -- ++(-0.25,1) node[above] {\tiny $1$};
		\draw[postaction={decorate}] (0,0) -- ++(0.25,1) node[above] {\tiny $2$};
		\draw[postaction={decorate}] (1,0) node[circle, fill, inner sep=1pt] {}-- ++(-0.25,1) node[above] {\tiny $3$};
		\draw[postaction={decorate}] (1,0) -- ++(0,1) node[above] {\tiny $4$};
		\draw[postaction={decorate}] (1,0) -- ++(0.25,1) node[above] {\tiny $5$};
		\end{scope}
		\end{tikzpicture}}}
\end{equation}
is an open $F$-graph with two vertices, one with two labeled edges, one with three labeled edges.

We are interested in linear combinations of graphs. A linear combination is simply a list of ordered pairs $(c,G)$.
For example,
\begin{equation}
[(5,[[1,2],[3,4,5]]),(3,[[6,7]])] = 5\vcenter{\hbox{\begin{tikzpicture}
		\begin{scope}[decoration={markings}]
		\draw[postaction={decorate}] (0,0) node[circle, fill, inner sep=1pt] {} -- ++(-0.25,1) node[above] {\tiny $1$};
		\draw[postaction={decorate}] (0,0) -- ++(0.25,1) node[above] {\tiny $2$};
		\draw[postaction={decorate}] (1,0) node[circle, fill, inner sep=1pt] {} -- ++(-0.25,1) node[above] {\tiny $3$};
		\draw[postaction={decorate}] (1,0) -- ++(0,1) node[above] {\tiny $4$};
		\draw[postaction={decorate}] (1,0) -- ++(0.25,1) node[above] {\tiny $5$};
		\end{scope}
		\end{tikzpicture}}}~+~3\vcenter{\hbox{\begin{tikzpicture}
		\begin{scope}[decoration={markings}]
		\draw[postaction={decorate}] (0,0) node[circle, fill, inner sep=1pt] {} -- ++(-0.25,1) node[above] {\tiny $6$};
		\draw[postaction={decorate}] (0,0) -- ++(0.25,1) node[above] {\tiny $7$};
		\end{scope}
		\end{tikzpicture}}}.
\end{equation}
For expectation values of monomials of degree three and higher we need a notion of multiplication of graphs. Multiplication of graphs corresponds to multiplying products of $F$.
The product of two graphs $g_1=(c_1,G_1), g_2=(c_2, G_2)$ is a graph $(c_1 c_2, G_1 \cup G_2)$, which is implemented using the method \textit{diagramprod}($g_1$, $g_2$).
It takes two ordered pairs and returns a pair $(c_1c_2, G_1 + G_2)$, where addition of two lists is given by concatenation (combining two lists).
For example,
\begin{equation}
\textit{diagramprod}((c_1, [[1,2,3]]), (c_2,[[4,5]])) = (c_1c_2, [[1,2,3],[4,5]]).
\end{equation}
The method $\textit{algprod}$ implements the product for linear combinations of graphs by distributing the product over the sums.
For more than two linear combinations, the product is implemented recursively through the method $\textit{recurseprod}$. Two labeled edges are connected if the same integer labeling the end points appears in a graph twice. This allows for the description of closed $F$-graphs using lists where all integers appear twice.

The multiplication of one-point and two-point functions is implemented in the algorithm using the following steps
\begin{enumerate}
	\item For every set partition $\textit{wick}$ in $\textit{wicks}$, iterate over the parts:\footnote{In Python we can loop through $\textit{wick}$ as $\textit{matrix1}, *\textit{matrix2}$. Then $\textit{matrix2}$ will be an empty list if $\textit{wick}$ is a part of size one. Otherwise it will be the second integer in the part.}
	\begin{enumerate}
		\item If the length of a part is one, determine if it is a $\langle M\rangle$ or $\langle N \rangle$ correlator.\footnote{For example, by comparing $\textit{matrix1}$ to the length of $\textit{Mtuples}$.}
		Calculate the corresponding linear combination of graphs, using either the method $\textit{OnePointM}$ or $\textit{OnePointN}$. These methods return a linear combination of $F$-graphs, determined by the linear Feynman rules discussed in Section \ref{Section: Feynman Graphs}, in terms of lists of pairs $(c,G)$. The inputs to these methods define which integers are to be used to label the end points the edges. In terms of equations, this corresponds to choosing the correct index labels for the $F(i_a,j_b)$. Add the result to a list $\textit{onepts}$.
		\item If the length of a part is two, determine if it is a $\langle MM \rangle, \langle MN \rangle$ or $\langle NN \rangle$ correlator. 
		Calculate the corresponding linear combination of $F$-graphs, using either the method $\textit{TwoPointMM}, \textit{TwoPointNN}$ or $\textit{TwoPointMN}$. These methods do the analogue of the above methods for the quadratic Feynman rules. Add the result to a list $\textit{twopts}$. Increase the internal index counter.\footnote{Two point expectation values have an internal index/edge. It is important that we give the two point expectation values a distinct integer, different from those in the list of ordered pairs associated with matrices, to use for this purpose.}
	\end{enumerate}
	\item The product of expectation values is given by the product of graphs in the lists $\textit{onepts}$ and $\textit{twopts}$. Add the result of $\textit{recurseprod}$($\textit{onepts}+\textit{twopts}$) into a list $\textit{contractions}$.
	\item Clear the lists $\textit{onepts}$, $\textit{twopts}$.
\end{enumerate}

\subsection{Specifying observable: from open to closed $F$-graphs}
\label{apx: specifying observable}
At the end of step 2 we will have a list of ordered pairs, $[(c_1, G_1), \dots]$, corresponding to a linear combination of open $F$-graphs. This comes from explicitly performing the Wick contractions and mapping the results to expressions with (sums of) products of the $V_H$ projector $F(i_a,i_b)$ and interpreting the terms as graphs.

From the linear combination of open $F$-graphs ($\textit{contractions}$), we construct a linear combination of closed $F$-graphs with connected edges using the input $\textit{invariants}$. This is done as follows. For every graph term $(c,g)$ in $\textit{contractions}$, produce the graph term $(c,g')$ where $g'$ is $\textit{invariant}$ added to $g$ with the integers in $\textit{invariant}$ which are not in the $g$ removed. This is done through the method $\textit{keepElements}$, which removes all integers in the first input which are not in the second. For example
\begin{equation}
	\textit{keepElements}([[1,2,3],[4,5],[6,7]], [[2,3,4],[7]]) = [[2,3],[4],[7]].
\end{equation}
The reason we need to do this is because some indices which appear on the LHS of a linear or quadratic expectation value do not appear on the RHS. For example, this happens when the Feynman graph has a dashed external line. Note that this will give us lists where all integers appear twice.

\subsection{Evaluating closed  $F$-graphs}
We now have a list of closed $F$-graphs. The last step is to calculate the contribution to the Laurent polynomial from each one and sum them up. The method $\textit{GraphPolynomialData}$ implements the graph interpretation of equation \eqref{eq:kappagraphpolynomial}. It takes a closed $F$-graph and produces the data necessary to reproduce every term in the corresponding Laurent polynomial. Specifically, it returns a list $[l, l_3]$, where $l_3$ is the total number of connected edges in the graph. $l$ is a list of pairs $[k, N_e]$, which determine a term in the Laurent polynomial \eqref{eq:kappagraphpolynomial} through $(D)_k(1-D)^{N_e}$, where $(D)_k$ is the falling factorial (implemented using the method $\textit{DFactorialk}$). To produce the explicit Laurent polynomial we feed the data into the method $\textit{GraphPolynomialFromData}$. For linear combinations of graphs, the map is extended linearly. The linear extension is implemented through the method $\textit{GP}$.

\section{Table of closed $F$-graph polynomials}
In this appendix we have collected the results necessary to evaluate any graph with up to six edges, and some with more than six edges. The graphs with up to six edges that are not listed can be obtained by adding loops on vertices or taking the disjoint union of graphs. There is another operation, which involves adding vertices in the middle of an edge, called splitting. Splitting an edge does not change the value of the polynomial, because of the property
\begin{equation}
	\sum_k F(i,k)F(k,j) = F(i,j).
\end{equation}
These three operations correspond to simple operations at the level of polynomials: multiply by $\tfrac{1}{D}(D-1)$ for every loop, multiply by 1 for every splitting, multiply the polynomials of disjoint graphs. The reason loops give this contribution is that they correspond to tensors $F(i,i) = (1-\frac{1}{D})$, which can be brought out of the sum to give a multiplicative factor.

The simplest example involves graphs with two vertices and any number of edges connecting the two. There are only two partitions: $P=12$ and $P=1|2$, which give back the original graph and the graph with no edges, respectively. The polynomial for $A$ edges on two vertices takes the form
\begin{equation}
	\frac{1}{(-D)^A}\qty(D(1-D)^A + D(D-1)(1-D)^0) = \frac{D(D-1)}{(-D)^A}\qty(1-(1-D)^{A-1})
\end{equation}
The first non-trivial example are graphs with three vertices. For example, the polynomial for the graph
\begin{equation}
	\vcenter{\hbox{\begin{tikzpicture}[scale=2, baseline]
			\coordinate (v1) at (-0.5,0);
			\coordinate (v2) at (0,0);
			\coordinate (v3) at (-0.25,-0.25);
			\node[circle, fill,inner sep=1pt, label={\tiny $1$}] at (v1) {};
			\node[circle, fill,inner sep=1pt, label={\tiny $2$}] at (v2) {};
			\node[circle, fill,inner sep=1pt, label={below: \tiny $3$}] at (v3) {};
			\draw (v1) -- (v2);
			\draw (v1) to[bend left] (v3) to[bend left] (v1);
			\draw (v2) to[bend left] (v3) to[bend left] (v2);
			\end{tikzpicture}}}
\end{equation}
with 5 edges on 3 vertices is computed as follows

\begin{align}
	G_{123} &= \vcenter{\hbox{\begin{tikzpicture}[scale=2,baseline]
			\coordinate (v1) at (-0.5,0);
			\coordinate (v2) at (0,0);
			\coordinate (v3) at (-0.25,-0.25);
			\node[circle, fill,inner sep=1pt, label={\tiny $1$}] at (v1) {};
			\node[circle, fill,inner sep=1pt, label={\tiny $2$}] at (v2) {};
			\node[circle, fill,inner sep=1pt, label={below: \tiny $3$}] at (v3) {};
			\draw (v1) -- (v2);
			\draw (v1) to[bend left] (v3) to[bend left] (v1);
			\draw (v2) to[bend left] (v3) to[bend left] (v2);
			\end{tikzpicture}}} \longrightarrow \frac{-D}{D^5}(1-D)^5 \\
	G_{12|3}&=\vcenter{\hbox{\begin{tikzpicture}[scale=2,baseline]
			\coordinate (v1) at (-0.5,0);
			\coordinate (v2) at (0,0);
			\coordinate (v3) at (-0.25,-0.25);
			\node[circle, fill,inner sep=1pt, label={\tiny $1$}] at (v1) {};
			\node[circle, fill,inner sep=1pt, label={\tiny $2$}] at (v2) {};
			\node[circle, fill,inner sep=1pt, label={below: \tiny $3$}] at (v3) {};
			\draw (v1) -- (v2);
			\end{tikzpicture}}} \longrightarrow \frac{-D(D-1)}{D^5}(1-D)^1 \\
	G_{13|2}&=\vcenter{\hbox{\begin{tikzpicture}[scale=2,baseline]
			\coordinate (v1) at (-0.5,0);
			\coordinate (v2) at (0,0);
			\coordinate (v3) at (-0.25,-0.25);
			\node[circle, fill,inner sep=1pt, label={\tiny $1$}] at (v1) {};
			\node[circle, fill,inner sep=1pt, label={\tiny $2$}] at (v2) {};
			\node[circle, fill,inner sep=1pt, label={below: \tiny $3$}] at (v3) {};
			\draw (v1) to[bend left] (v3) to[bend left] (v1);
			\end{tikzpicture}}} \longrightarrow \frac{-D(D-1)}{D^5}(1-D)^2 \\
	G_{23|1}&=\vcenter{\hbox{\begin{tikzpicture}[scale=2,baseline]
			\coordinate (v1) at (-0.5,0);
			\coordinate (v2) at (0,0);
			\coordinate (v3) at (-0.25,-0.25);
			\node[circle, fill,inner sep=1pt, label={\tiny $1$}] at (v1) {};
			\node[circle, fill,inner sep=1pt, label={\tiny $2$}] at (v2) {};
			\node[circle, fill,inner sep=1pt, label={below: \tiny $3$}] at (v3) {};
							\draw (v2) to[bend left] (v3) to[bend left] (v2);
			\end{tikzpicture}}} \longrightarrow \frac{-D(D-1)}{D^5}(1-D)^2 \\
	G_{1|2|3}&=\vcenter{\hbox{\begin{tikzpicture}[scale=2,baseline]
			\coordinate (v1) at (-0.5,0);
			\coordinate (v2) at (0,0);
			\coordinate (v3) at (-0.25,-0.25);
			\node[circle, fill,inner sep=1pt, label={\tiny $1$}] at (v1) {};
			\node[circle, fill,inner sep=1pt, label={\tiny $2$}] at (v2) {};
			\node[circle, fill,inner sep=1pt, label={below: \tiny $3$}] at (v3) {};
			\end{tikzpicture}}} \longrightarrow \frac{-D(D-1)(D-2)}{D^5}(1-D)^0
\end{align}
When summed up, the contributions give
\begin{equation}
	\vcenter{\hbox{\begin{tikzpicture}[scale=2, baseline]
			\coordinate (v1) at (-0.5,0);
			\coordinate (v2) at (0,0);
			\coordinate (v3) at (-0.25,-0.25);
			\node[circle, fill,inner sep=1pt] at (v1) {};
			\node[circle, fill,inner sep=1pt] at (v2) {};
			\node[circle, fill,inner sep=1pt] at (v3) {};
			\draw (v1) -- (v2);
			\draw (v1) to[bend left] (v3) to[bend left] (v1);
			\draw (v2) to[bend left] (v3) to[bend left] (v2);
			\end{tikzpicture}}} = \frac{(D-1)(D-2)^2}{D^2}
\end{equation}
We list the result of using this method for several graphs, with up to six edges, below.

\noindent Two Edges
\begin{equation}
	\vcenter{\hbox{\begin{tikzpicture}[scale=2]
				\draw (-0.5,0) node[circle, fill,inner sep=1pt] {}  to[bend left] (0,0) node[circle, fill,inner sep=1pt] {};
				\draw (-0.5,0)  to[bend right] (0,0);
	\end{tikzpicture}}} = \sum_{i_1,i_2} F(i_1,i_2)^2 = (D-1)
\end{equation}
Three edges
\begin{align}
	\vcenter{\hbox{\begin{tikzpicture}[scale=2]
				\draw (-0.5,0) node[circle, fill,inner sep=1pt] {}  to[bend left] (0,0) node[circle, fill,inner sep=1pt] {};
				\draw (-0.5,0)  to[bend right] (0,0);
				\draw (-0.5,0)  -- (0,0);
	\end{tikzpicture}}} &= \sum_{i_1,i_2} F(i_1,i_2)^3 = \frac{(D-1)(D-2)}{D}
\end{align}
Four Edges
\begin{align}
	\vcenter{\hbox{\begin{tikzpicture}[scale=2]
				\draw (-0.5,0) node[circle, fill,inner sep=1pt] {}  to[bend left] (0,0) node[circle, fill,inner sep=1pt] {};
				\draw (-0.5,0)  to[bend right] (0,0);
				\draw (-0.5,0)  to[out=90, in=90] (0,0);
				\draw (-0.5,0)  to[out=-90, in=-90] (0,0);
	\end{tikzpicture}}}  &= \sum_{i_1,i_2} F(i_1,i_2)^4 = \frac{(D^2-3D+3)(D-1)}{D^2}
\end{align}
Five Edges
\begin{align}
	\vcenter{\hbox{\begin{tikzpicture}[scale=2]
				\draw (-0.5,0) node[circle, fill,inner sep=1pt] {}  to[bend left] (0,0) node[circle, fill,inner sep=1pt] {};
				\draw (-0.5,0)  to[bend right] (0,0);
				\draw (-0.5,0)  to[out=90, in=90] (0,0);
				\draw (-0.5,0)  to[out=-90, in=-90] (0,0);
				\draw (-0.5,0) -- (0,0);
	\end{tikzpicture}}}  &=\sum_{i_1, i_2}  F(i_1,i_2)^5 = \frac{(D^2 - 2D + 2)(D - 1)(D - 2)}{D^3} \\
	\vcenter{\hbox{\begin{tikzpicture}[scale=2]
				\draw (-0.5,0) node[circle, fill,inner sep=1pt] {}  -- (0,0) node[circle, fill,inner sep=1pt] {};
				\draw (-0.25,-0.25) node[circle, fill,inner sep=1pt] {} to[bend left] (0,0);
				\draw (-0.25,-0.25)  to[bend right] (0,0);		
				\draw (-0.25,-0.25) to[bend left] (-0.5,0);
				\draw (-0.25,-0.25) to[bend right] (-0.5,0);
	\end{tikzpicture}}}&=\sum_{i_1, i_2,i_3}  F(i_1,i_3)^2F(i_2,i_3)^2F(i_1,i_2) = \frac{(D-1)(D-2)^2}{D^2} 
\end{align}
Six edges
\begin{align}
	\vcenter{\hbox{\begin{tikzpicture}[scale=2]
				\draw (-0.5,0) node[circle, fill,inner sep=1pt] {}  to[bend left] (0,0) node[circle, fill,inner sep=1pt] {};
				\draw (-0.5,0)  to[bend right] (0,0);
				\draw (-0.5,0)  to[out=90, in=90] (0,0);
				\draw (-0.5,0)  to[out=-90, in=-90] (0,0);
				\draw (-0.5,0) to[out=10,in=170] (0,0);
				\draw (-0.5,0) to[out=-10,in=190] (0,0);
	\end{tikzpicture}}}  &=\sum_{i_1, i_2}  F(i_1,i_2)^6 = \frac{(D^4 - 5D^3 + 10D^2 - 10D + 5)(D - 1)}{D^4} \\
	\vcenter{\hbox{\begin{tikzpicture}[scale=1]
				\draw (0,0) node[circle, fill,inner sep=1pt] {} to[bend left] (1,0);
				\draw (0,0)  to[bend right] (1,0) node[circle, fill,inner sep=1pt] {};
				\draw (1,0) node[circle, fill,inner sep=1pt] {} to[bend left] (.5,.7);
				\draw (1,0)  to[bend right] (.5,.7) node[circle, fill,inner sep=1pt] {};
				\draw (0,0) to[bend left] (.5,.7);
				\draw (0,0)  to[bend right] (.5,.7);
	\end{tikzpicture}}}  &=\sum_{i_1, i_2,i_3}  F(i_1,i_2)^2F(i_2,i_3)^2F(i_3,i_1)^2 = \frac{(D^2-2D+2)(D-1)(D-2)}{D^3} \\
	\vcenter{\hbox{\begin{tikzpicture}[scale=1]
				\draw (0,0) node[circle, fill,inner sep=1pt] {} --(1,0) node[circle, fill,inner sep=1pt] {};
				\draw (1,0) node[circle, fill,inner sep=1pt] {} to[bend left] (.5,.7);
				\draw (1,0)  to[bend right] (.5,.7) node[circle, fill,inner sep=1pt] {};
				\draw (0,0) to[bend left] (.5,.7);
				\draw (0,0)  to[bend right] (.5,.7);
				\draw (0,0)  -- (.5,.7);
	\end{tikzpicture}}}  &=\sum_{i_1, i_2,i_3}  F(i_1,i_2)^3F(i_2,i_3)^2F(i_3,i_1) = \frac{(D^2 - 3D + 3)(D - 1)(D - 2)}{D^3} \\
	\vcenter{\hbox{\begin{tikzpicture}[scale=2]
				\draw (-0.5,0) node[circle, fill,inner sep=1pt] {}  to[bend left] (0,0) node[circle, fill,inner sep=1pt] {};
				\draw (-0.5,0)  to[bend right] (0,0);
				\draw (-0.5,0)  -- (0,0);
				\draw (0,0) node[circle, fill,inner sep=1pt] {}  to[bend left] (0.5,0) node[circle, fill,inner sep=1pt] {};
				\draw (0,0)  to[bend right] (0.5,0);
				\draw (0,0)  -- (0.5,0);
	\end{tikzpicture}}} &= \sum_{i_1,i_2,i_3} F(i_1,i_2)^3F(i_2,i_3)^3 =\frac{(D - 1)^2(D - 2)^2}{D^3}\\
	\vcenter{\hbox{\begin{tikzpicture}[scale=2]
				\draw (-0.5,0) node[circle, fill,inner sep=1pt] {} -- (0,0) node[circle, fill,inner sep=1pt] {} -- (0,-0.5) node[circle, fill,inner sep=1pt] {} -- (-0.5,-0.5) node[circle, fill,inner sep=1pt] {} -- (-0.5,0);
				\draw (-0.5,0) to[bend left] (0,0);
				\draw (-0.5,-0.5) to[bend right] (0,-0.5);
	\end{tikzpicture}}}&=\sum_{i_1,i_2,i_3,i_4}F(i_1,i_2)^2F(i_2,i_3)F(i_3,i_4)^2F(i_4,i_1) = \frac{(D - 1)(D - 2)^2}{D^2}
\end{align}

\end{document}